# Self-Assembly, Interfacial Properties, Interactions with Macromolecules and Molecular Modelling and Simulation of Microbial Bio-based Amphiphiles (Biosurfactants). A Tutorial Review.

Niki Baccile,[a,*] Chloé Seyrig,[a] Alexandre Poirier,[a] Silvia Alonso-de Castro,[a] Sophie L. K. W. Roelants,[b,c] Stéphane Abel[d]

[a] Sorbonne Université, Centre National de la Recherche Scientifique, Laboratoire de Chimie de la Matière Condensée de Paris, LCMCP, F-75005 Paris, France
E-mail: niki.baccile@sorbonne-universite.fr
[b] Ghent University, Centre for Industrial Biotechnology and Biocatalysis (InBio.be), Coupure Links 653, B-9000 Gent, Belgium
[c] Bio Base Europe Pilot Plant, Rodenhuizekaai 1, B-9000 Gent, Belgium
[d] Université Paris-Saclay, CEA, CNRS, Institute for Integrative Biology of the Cell (I2BC), 91198, Gif-sur-Yvette, France

**Abstract**

Chemical surfactants are omnipresent in consumers' products but they suffer from environmental concerns. For this reason, complete replacement of petrochemical surfactants by biosurfactants constitute a holy grail but this is far from occurring any soon. If the "biosurfactants revolution" has not occurred, yet, mainly due to the higher cost and lower availability of biosurfactants, another reason explains this fact: the poor knowledge of their properties in solution.

This tutorial review aims at reviewing the self-assembly properties and phase behavior, experimental (sections 2.3 and 2.4) and from molecular modelling (section 5), in water of the most important microbial biosurfactants (sophorolipids, rhamnolipids, surfactin, cellobioselipids, glucolipids) as well as their major derivatives. A critical discussion of such properties in light of the well-known packing parameter of surfactants is also provided (section 2.5). The relationship between the nanoscale self-assembly and macroscopic materials properties, including hydrogelling, solid foaming, templating or encapsulation is specifically discussed (section 2.7). We also present their self-assembly and adsorption at flat and complex air/liquid (e.g., foams), air/solid (adhesion), liquid/solid (nanoparticles) and liquid/liquid (e.g., emulsions) interfaces (section 3). A critical discussion on the use of biosurfactants as capping agents for the development of stable nanoparticles is specifically provided (section 3.2.4). Finally, we discuss the major findings involving biosurfactants and macromolecules, including proteins, enzymes, polymers and polyelectrolytes.

# Table of contents

## 0. Introduction

Surfactants are a diverse class of chemicals applied in a vast array of applications and markets reaching production volumes of about 20 million tons per year.[1,2] The word "surfactant" is the contraction of "surface active agent", indicating that surfactants act at and between interfaces and more specifically have the specific property of lowering the surface tension in liquids. This behavior is attributed to their 'amphiphilic' nature defined as molecules with a hydrophilic ("water-loving") and a hydrophobic ("water-hating") part. Due to the widespread use and application of surfactants, research on/about surfactants constitutes a science of its own.

Surfactants have played a decisive role in shaping the concepts of sustainability and environmentally friendly green chemistry. Fatty acid soaps guarantee cleanliness and hygiene since time immemorial. Surfactants are involved in the environmentally friendly production of rubber, plastics, paints and adhesives in the aqueous phase. In the field of polymer synthesis, surfactants make this possible in water, thus lowering, or even eliminating, the risks of these processes, such as fire hazards. Toxic emissions are reduced towards zero and occupational safety is increased.

However, their ubiquitous use in our everyday lives also has some drawbacks. Surfactants have been associated with pollution problems, but also with dermatological issues such as skin irritation and even allergic reactions. Moreover, many of the produced surfactants are derived from petrochemical resources and associated with harsh and/or polluting production processes. Many products have already been banned for reasons of toxicity and/or pollution in the past 30 years and more are expected to follow. For these reasons, investigations aiming at finding non-toxic, benign, products and more specifically natural biobased alternatives to petrochemical surfactants started as a sub-field in surfactant's science since the 1960's, and developed as a field *per se* since the 1970's, motivated by the oil crisis and raising of oil costs.[3–5] Employment of linear alkylbenzene sulphonates and methyl ester sulphonates instead of their branched counterparts, use natural fatty alcohol alternatives to synthetic alcohol polyglycolethers or sulphates or use of green fatty alcohol (or guerbet) alcohol polyglycol ethers, -ethersulphate, -phosphates, and -sulphosuccinate surfactants instead of alkylphenol polyglycolethers are some of the common strategies employed by industry to develop more benign molecules.[6,7] The quest of more ecofriendly surfactants is then just a natural consequence of this long-date trend.

Bio-based surfactants, or biosurfactants, are defined as molecules that are fully based on biomass such as sugars, plant oils, amino acids, etc. This field is characterized by two different production approaches. In the first, chemical, approach, biobased hydrophilic and hydrophobic molecules are covalently linked through organic[8,9] chemistry. In the second, biological, approach, biosurfactants are either extracted from plants or produced through biocatalytical (use of enzymes) or microbial processes. Although the frontiers between and within these approaches are sometimes blurry, and for this reason clarified later on in section 2, a broad community agrees on employing the word "biosurfactants" in relationship to amphiphilic surface active agents produced by a microbial fermentation process.[10–12] For this reason, the term "*biosurfactants*" is employed throughout this review as a synonym of "*microbial biosurfactants*" and we will employ them as synonyms. If we will also discuss derivatives of microbial biosurfactants, we will not address biosurfactants extracted from plants (e.g., saponins).

Research on microbial biosurfactants is known since the 1960's[13,14] but it is becoming a trendy topic since two decades. A number of review papers and books have been published on this topic. They commonly address the topic of microbial biosurfactants' classification, the synthesis' strategy, derivatization and genetic modification towards development of new chemistry,[15–17] their aqueous and antimicrobial properties and their application potential in various fields. A non-exhaustive list is given in Ref. [11,12,17–24]. For this reason, these topics will not be covered in the present work and we direct the reader towards them for a comprehensive understanding of the world of biosurfactants.[10]

This review focuses on the physicochemical properties of microbial biosurfactacts in aqueous solutions, at interfaces and in the presence of macromolecules. Self-assembly will be an important aspect and this work will update and extend previous reviews on a similar topic.[19,25] However, the aim of this review is to step out of the domain of microbial biosurfactants as a topic of its own but to connect it to the more extended field of surfactant and colloids science. In fact, considering microbial biosurfactants as simple surfactants is highly reductive, because in many cases their behavior cannot be classified as that of a surface active agent. Microbial biosurfactants behave both as surfactants and lipids, according to the physicochemical conditions of the medium and in this regard, they should be rather addressed as bioamphiphiles. Such distinction is crucial for applications, as discussed below.

Controlling the physicochemical properties of surfactants in general, and microbial biosurfactants in particular, is of extreme importance to design a given application. Knowledge of critical micelle concentration (cmc) and surface tension is crucial, but not enough. The structure, morphology and size of the assembled molecules is also very important because the properties of the supramolecular aggregates depend on them and may even determine whether a given molecule has a surfactant or lipid behavior. In the former case, the molecule can be used for applications needing reduction in surface tension, such as a detergent, emulsifier or foaming agent. In the latter case, the molecule could be rather employed in biomedical applications. Such a distinction is quite obvious in the field of surfactant science,[26] but much less in the field of microbial biosurfactants. The field of surfactant science has rationalized the behavior of surfactants in solution for classical head-tail molecules.[2] However, most microbial biosurfactants do not have such a structure but rather a bolaform (head-tail-head) shape and they rather follow the behavior of bolaamphiphiles.[27] In fact, the molecular complexity of microbial biosurfactants including chemical asymmetry between the headgroups, branching, plethora of labile hydrogen atoms, sensitivity to pH, make them fascinating molecules with unexpected physicochemical properties.

This review tries to merge the field of microbial biosurfactants with the field of surfactant and colloid science and it has the double goal of interpreting the data produced in the field of biosurfactants with the current knowledge of colloid science. At the same time, it intends to show that colloid science, built on model molecular compounds, is still not able to predict the behavior of many microbial biosurfactants in solution and at interfaces. For this reason, this review is structured as follows. Section 1 will summarize the classical knowledge of surfactants in solution. We will review basic concepts such as hydrophilic lipophilic balance (HLB), cmc and surface tension, but also more advanced concepts like self-assembly of amphiphiles, including classical and more recent interpretation of the theory. Section 2 will focus on the solution properties of microbial biosurfactants including basic (cmc, surface tension) and more advanced (self-assembly) aspects of surfactants' physical chemistry. This section will be put in perspective to the general knowledge of surfactant science discussed in Ch. 1, namely on the application of the packing parameter appraoch to biosurfactants. This section will also briefly present the recent advances in the processing of microbial biosurfactants into materials. Section 3 will focus on the properties of microbial biosurfactants at various interfaces, and in particular at the solid-air, solid-liquid and liquid-liquid interfaces, important for wetting, adhesion or emulsification. Section 4 will concern the interaction of microbial biosurfactants with macromolecules, synthetic (polymers, polyelectrolytes) and biological (e.g., proteins, enzymes), such a topic being of paramount important for the development of commercial formulated products. Section 5 will concern the numerical modelling of microbial biosurfactants, being this topic very important towards a better understanding of the structure-function relationships, mostly unpredictable at the current state of the art. The review will end with some perspective remarks on the current state of the art and future perspectives.

Figure 1 summarizes the content of the review, from nature to bioamphiphiles-based materials and interfaces.

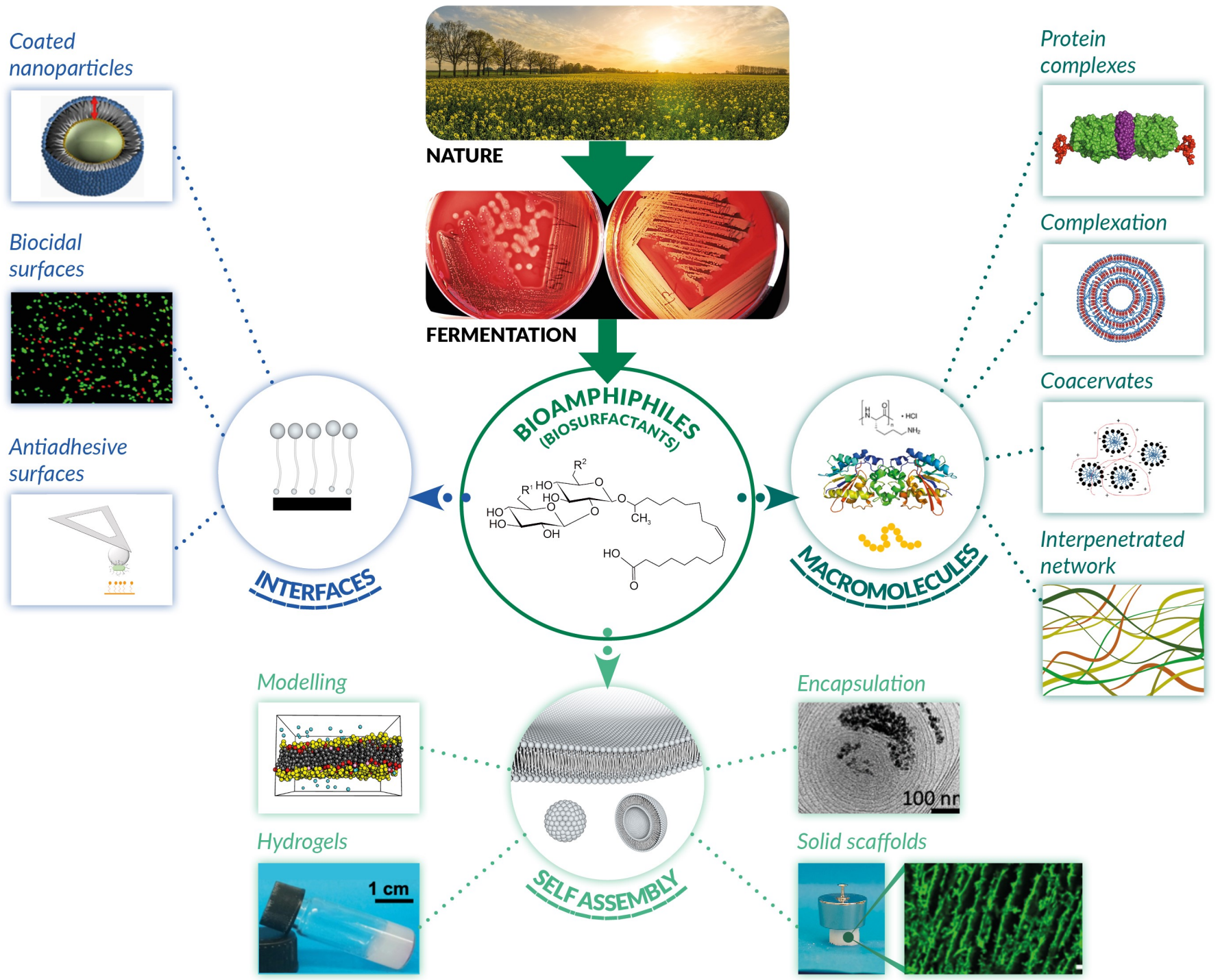

Figure 1 – Schematic view of the topic of this review, from nature to bioamphiphiles through fermentation: self-assembly, interfaces and interactions with macromolecules (selected images, courtesy of D. Otzen, J. S. Pedersen, J. D. Kasperen, Aarhus University, Denmark; A. Lassenberger, Fribourg University, Switzerland; C. Valotteau, Aix-Marseille University, France)

# 1. Surfactants in solution

## 1.1. General Introduction

Amphiphiles in general, and surfactants in particular, are omnipresent molecules in nature, including in living organisms. Bile salts are important to solubilize hydrophobic molecules in blood, pulmonary surfactant reduce the surface tension at the air/liquid interface in the lung, or phospholipids, constituting the outer membrane or cells, are just some crucial examples. This class of molecules also constitutes an important share of manufactured compounds and which have certainly contributed to the development of modern societies. Surfactants are employed in cleaning formulations, home-care products, cosmetics, oil recovery applications, lubricants and much more. It is then obvious that the "green revolution" invests this field, as well. The science of surfactants in solutions is well-developed since more than half a century and we address the readers to a more consequent set of data for an extensive overview of surfactants science.[1,2] The ambition of this section is simply to recall the major concepts in surfactants science and which help understanding the role and properties of surfactants in relationship to their molecular structure. These concepts will be necessary to understand the properties of biosurfactants in solution or at interfaces, but also to engage some considerations on the structure-property relationship in biosurfactants. One should note that we voluntarily focus only on low-molecular weight amphiphiles, thus excluding block copolymer systems.

## 1.2. Hydrophilic-Lipophilic Balance (HLB) and Hydrophilic-Lipophilic Difference (HLD)

One of the crudest approach to forecast the properties of surfactants has been developed more than half a century ago by Griffin[28] and Davies[29] and improved over the years.[30] The HLB was conceived to create an empirical relationship between the surfactants' properties (e.g., oil-in-water or water-in-oil emulsifier, wetting agent or detergent) and their structure, whereas the latter is generally expressed in terms of the balance of the hydrophilic and hydrophobic portions of the molecule. Initially developed for polyoxyethylene-type surfactants, HLB has been widened to a much broader class of molecules by including the contribution of specific chemical groups, which have a strong influence on the properties.

The first version of HLB was developed for a small family of compounds and considered a numeric index between 0 and 20, the higher number identifying a more water-soluble molecule. For instance, the HLB increases between 0 and 10 for antifoaming agents, emulsifiers (for w/o emulsions) and wetting agents; HLB between 10 and 20 identify emulsifiers (o/w), detergents and solubilizers. For polyoxyethylene-type surfactants, the HLB is simply $\mathrm{HLB} = \frac{\mathrm{POE\ (wt\%)}}{5}$, with POE being the weight fraction of the polyoxyethylene segment. In the later versions of HLB, both the hydrophilic and hydrophobic moieties were included in the calculation. Davies' method states that HLB= 7 + Σ(unit value of hydrophilic groups) + Σ(unit value of hydrophobic groups), with unit values varying between specific chemical groups: 19.1 for carboxylate or -0.475 for methyl groups. More recent developments like the organic conceptual diagrammes, intermolecular forces started to be integrated in the calculation of HLB values, although the overall objective is the same.[30]

Despite its astonishing simplicity, the HLB method has been employed in the surfactant industry for years and it works nicely on well-established molecules, like non-ionic surfactants. Nonetheless, this method fails for a number of systems because it does not take into consideration the effect of temperature, electrolytes and ionic strength, impurities, and additives in general, very common in commercial-grade surfactants, such as dodecanol in sodium dodecyl sulfate. The intrinsic problem of the HLB method is the diversity of approaches to calculate it and one can easily show that for a given surfactant, the HLB values can vary by a factor close to 10.[30] It is also not clear how HLB can be reliably estimated and connected to the properties of more exotic molecular structures like divalent, gemini, branched or bolaform amphiphiles. Finally, HLB is only meant to predict a macroscopic property but not the actual morphology, structure and size of a given assembled form of an amphiphile above its critical micellar concentration. To overcome these drawbacks, more advanced experimental and theoretical approaches have been developed from the late 70's, such as the idea of Hydrophilic Lipophilic Difference (HLD).

HLD, developed by Salager,[31–33] is a much less known concept than HLB but it constitutes a more interesting expanded approach. The HLD equation is defined as HLD= $S\text{-}k\cdot EACN\text{-}\alpha(T\text{-}25)+Cc$, with $S$ being related to the salinity of the medium, $EACN$ the effective alkane carbon number, that is the oil contribution, where the prefactor $k$ is empirical, $T$ the temperature, the prefactor $\alpha$ also being surfactant-dependent. Finally, $Cc$ is a characteristic value associated to the hydrophilic/hydrophobic balance of the surfactant. In respect to HLB, HLD has the advantage to include salinity, temperature and nature of the oil, except for the contribution of a cosurfactant (e.g., long-chain alcohol), which is too complex to estimate. In the HLD approach, one must identify neutrality, HLD= 0, indicating the balance between the hydrophilic and hydrophobic contributions to the formulation, and start to work from this point by playing with each component in order to prepare stable o/w or w/o emulsions, for instance. The HLD approach is highly practical and useful in surfactants formulation but it also suffers from the same drawback as any highly oversimplified models, it relies on empirical parameters which are easily determined for classical ionic and non-ionic surfactants but they cannot be generalized to complex amphiphilic molecules.

### 1.3. Surface Tension and critical micelle concentration (CMC)

The surface tension is a parameter of paramount importance in a number of physical phenomena like adsorption, wetting, catalysis, distillation and much more, with direct involvement in the conception of industrial products in coating, food, detergents, cosmetics and so on. Surface tension is defined as the energy required to create a unit area of interphase[34] and surfactants play a crucial role in lowering the surface tension of water at the water-air interface from about 70 mN/m to about 25 to 40 mN/m. Upon mixing micromolar amounts of a surfactant in water, the water-air interface is occupied by surfactant monomers, pointing the hydrophilic headgroup towards water and the hydrophobic chain towards air. This phenomenon is at the origin of the reduction in surface tension and to the increase in surfactant packing at the interface.[26]

When the surfactant reaches the conditions of maximum packing, it will start aggregating in the bulk solution into spheroidal aggregates, called micelles. The concentration at which aggregation occurs is called critical micelle concentration, widely known as CMC,[2] and also referred to as CMC1, in opposition to CMC2, the concentration value above which micellar growth is rapidly implemented.[35] CMC is classically determined by the inflection point in surface tension vs. concentration experiments, although many other techniques, such as turbidity, self-diffusion NMR, solubilization, pyrene fluorescence and many others can be equally used. The typical CMC1 values for a broad set of surfactants settles in the order of the mM range, although the dispersion is broad (between $10^{-5}$ and $10^{-1}$ M) and it strongly depends on the chemical structure of the surfactant, where type of headgroup and chain length are critical parameters.[36]

There are four main families of classical head-tail surfactants and they are classified on the basis of their headgroup: cationic, anionic, non-ionic and zwitterionic. Anionic surfactants, generally characterized by carboxylate, sulfate, sulfonate or phosphate polar groups, are sensitive to hard water. They are by far the ones produced in largest amount. Non-ionic surfactants are also very popular for their lack of sensitivity to hard water and compatibility with ionic surfactants. They are the second largest class and their hydrophilic part is often constituted of oxyethylene units. Cationic surfactants, the third largest class, are amine and quaternary ammonium-derivatives and for this reason they are also sensitive to hard water and not compatible with anionic surfactants. Finally, zwitterionic surfactants contain coexisting negative (variable chemical group) and positive (generally ammonium group) charges and they are often characterized by an isoelectric point (IEP, the point at which the number of negative charges equals the number of positive charges), where they behave as non-ionic surfactants. The properties of these surfactants then strongly vary with pH. Betaine and amine oxide derivatives are classical zwitterionic surfactants.[26]

Whichever the chemical nature of the head group, the CMC decreases with increasing the length of the alkyl chain, where the decrease is more pronounced for non-ionics than for ionics, respectively a factor 3 and 2 upon addition a methylene group in the aliphatic chain. The CMC values of non-ionic surfactants are about two orders of magnitude lower than the values of ionic surfactants. Interestingly, among ionic surfactants, the difference in CMC is milder, with cationics having higher CMC values than anionics, while among non-ionics, CMC slightly increases with bulkiness of headgroup. Other parameters have an important influence on CMC such as the valency of counterions for ionic surfactants (the higher the valency, the lower the CMC), branching, unsaturation, cosolutes. Temperature is also an important parameter, which however has a much stronger impact on the surfactant's solubility itself through the Kraft phenomenon. The Kraft point is defined as the temperature below which the surfactant is insoluble and above which solubility experiences an exponential increase.[36]

One last remark concerns the estimation of CMC for bolaamphiphiles (bolas). Bolas have attracted a lot of attention in the past years,[27] but they have been studied in a less rational manner than single chain surfactants. For these reason, to the best of our knowledge, no general experimental trend in their CMC has been reported so far. Nonetheless, Nagarajan has calculated, and compared to experiments, the CMC values for bolaform surfactants and found that higher values than single-head amphiphiles are expected, the second headgroup in bolas improving the monomer solubility in water. Depending on the nature of the headgroup (ionic or non-ionic), he gives values in the order of $10^{-2}$ M.[37]

### 1.4. Self-assembly of amphiphiles in solution: the packing parameter (*PP*)

HLB, surface tension and CMC are empirical macroscopic approaches to study the structure-properties relationship in surfactants. However, if they are useful in understanding some specific systems, they lack from a mesoscale, molecular, understanding of the self-assembly phenomenon. An important advance in this field was proposed by the combined efforts of Tanford[38] and Israelachvili, Mitchell and Ninham (IMN),[39] who proposed a thermodynamic description of amphiphile self-assembly. In this section, we illustrate this approach, which drove the establishment of what's widely known as the packing parameter (*PP*) concept.

The *PP* idea is extremely simple and powerful and it has been successfully used in predicting the morphology of the aggregates for a large number of amphiphiles; it has also successfully been able to explain a broad number of experimental evidence in phase transitions in surfactant science and beyond. However, even the classical (IMN) description of *PP* has some drawbacks. For this reason, since the late 80's, several authors have developed complementary, and even alternative, models in the effort to build a generalized theory of self-assembly: Svenson has reviewed the major ones some time ago[40] and we address the more experienced reader to the works of Eriksson (1985 and onward),[41,42] Blankschtein (1990 and onward),[43] Nagarajan (1991 and onward)[44–46] and, more recently, Bergström (2000 and onward),[47,48] who tried to build a general micelle model.[49,50] A general, less technical, discussion on the self-assembly of surfactants can be found in several books[1,2] and a more general consideration on the limits of modelling the self-assembly of amphiphiles is presented in ref. [43]. A selection of the IMN and later approaches will be illustrated in their essential aspects below and we will critically discuss it within the frame of self-assembly of biosurfactants in section 2.5.

*From intermolecular forces to the classical theory of the packing parameter*. The classical morphologies observed when amphiphiles are dispersed in water above their CMC are spherical, or cylindrical, micelles and bilayers, the latter in the shape of vesicles or flat lamellae. The energy required to associate a number of amphiphiles from a dispersed to an aggregated, micellar, state is defined as the standard free energy of micellization. This approach was first described by Tanford[38] and it revealed to be the most appropriate attempt to describe micellization, where micelles are constituted of a finite number of amphiphiles, they are distinct species and they are characterized by their own self free energy.

In its simplest form, the free energy of micellization could be described with only three main contributions: 1) a negative, attractive, force resulting from the unfavourable interaction between the hydrocarbon moiety and water and for this reason referred to as the transfer, or hydrophobic, contribution. 2) A positive, repulsive, contribution from the interfacial free energy between the micellar surface and water and due to the residual contact between the aliphatic chain and water. 3) A positive contribution coming from the repulsive interaction between the polar headgroups. This contribution has a steric component, to which an electrostatic term should be included when the headgroups are charged.

Given the above, the so-called tail-transfer contribution (n. 1) drives the micellar formation and it mainly affects the value of CMC. Both repulsive contributions, on the contrary, are sensitive to the headgroup surface area and they are responsible for the micellar growth. In particular, contribution n.2 is directly proportional to the headgroup area, meaning that micellization is favoured for small headgroups. Contribution n. 3, being on the contrary dependent on steric and electrostatic repulsion, is inversely proportional to the heagroup size, meaning that micellization becomes unfavourable when headgroups come too close together. The headgroup size is then crucial in contributions n.2 and n.3, which consequently play an opposite role: for decreasing headgroup size, contribution n.2 favours micellization and micellar growth while contribution n.3 has the opposite effect, it limits micellization and growth. In summary, according to the thermodynamic description of self-assembly, the alkyl chain drives self-assembly, while the headgroup controls the size and shape of the aggregate, as summarized in Figure 2. This description constitutes the background of the IMN approach to develop the packing parameter, a simple tool to predict the shape of self-assembled aggregates of amphiphiles.[39]

Israelachvili and coworkers made the hypothesis that the hydrophobic region of the micelles is liquid, that is it has a similar density as hydrocarbon liquids. This is crucial because it means that the aliphatic chain of the amphiphile homogeneously fills the micellar core. The consequence of these hypotheses is stronger than it seems as one can now state that the volume of the micellar core is equal to the volume of a single aliphatic chain, well-known or easy to estimate, times the aggregation number. In other words, a mathematical entity, the volume, can be linked to the molecular scale, with the constraint that at least one of the dimensions of the aggregate cannot exceed the extended length of the chain. Consider the simple case of a micelle of spherical morphology with core radius (radius of the hydrophobic region) $R$, volume $V_{core} = \frac{4}{3}\pi R^3$ and surface area $A_s = 4\pi R^2$. On the basis of the IMN model, the micelle is entirely filled with $N$ molecules each having an aliphatic chain of length $L$ and individual volume $V$ and equilibrium headgroup of surface area $A_e$ (the sense of the equilibrium area will be explained below and in the *Note 1* below), shown in Figure 2a. The core volume can then also be written as $V_{core} = \frac{4}{3}\pi R^3 = NV$ while the surface area will be $A_s = 4\pi R^2 = NA_e$. If, according to the model, one imposes that $R$ cannot exceed $L$, that is $R \equiv L$, one finds the condition that $\frac{V}{LA_e} = \frac{1}{3}$, where $\frac{1}{3}$ is the maximum value that the quantity $\frac{V}{LA_e}$ can assume. Such quantity is exactly the definition of the packing parameter, $PP = \frac{V}{LA_e}$.[39] Figure 2b summarizes the classical limits of the *PP* for the major morphologies found in amphiphilic systems: spheres, cylinders and bilayers.

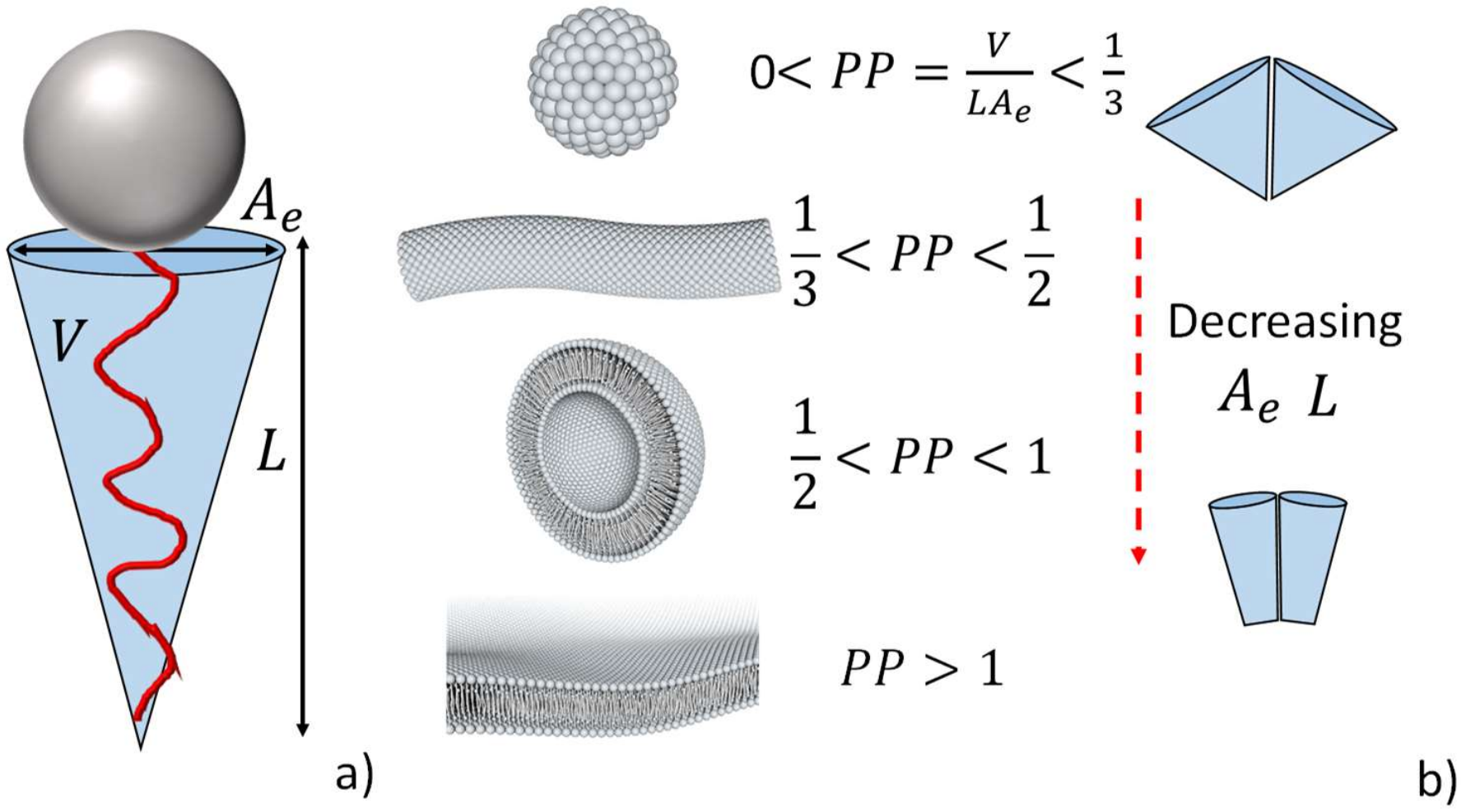

Figure 2 – a) Scheme of an amphiphile molecule and its main geometrical parameters according to the IMN model, $A_e$= equilibrium surface area of the headgroup, $L$= length of the tail, $V$= volume occupied by the tail. b) Typical values of the packing parameter, *PP*, and corresponding amphiphile morphology

*Considerations over the PP: from the classical theory to improved models.* The packing parameter is extremely simple in its final form and it has astonishingly been verified in so many systems that is it pointless to summarize them here. One can refer to many reviews and books to make oneself a more precise idea.[1,51] However, there are many bias in the classical version of the *PP*.[43] One the main problems is that the exact interpretation of the molecular parameters has been often limited to their simple geometry, such as considering $A_e$ as equivalent to the steric hindrance of the surfactant's headgroup,[52] or the homogeneity of the tail's density.[52] Entropy is not considered, although it has an important role,[53] and so on. The *PP* model, for instance, can neither describe microemulsions and explain the existence of a second CMC (CMC2),[50] nor explain the formation of ribbons with specific counterions.[54] If reviewing all inconvenients and drawbacks in the general *PP* approach has been done before[40] and it would be out of the scope of this review, we address hereafter a simple description of the major ones and which could be used to better understand the self-assembly of biosurfactants. Finally, we should say that if a general micelle model was proposed in recent years,[35,49,50] none of the existing models has gained the same success and broad employment as the classical *PP* theory, which is still by far used to interpret, both qualitatively and quantitatively, the self-assembly properties of amphiphiles in solution by less advised researchers. This is also the case for biosurfactants, which constitute a typical case where the limits of employing the *PP* approach to explain, and predict, their self-assembly properties is self-evident. To this regard, the extension of the *PP* approach to the self-assembly of bolaamphiphiles will be mentioned later in this section, while more specific considerations of the *PP* concept applied to biosurfactants will be given in section 2.5.

*Surface area*. The surface area $A_e$ has a crucial impact on the morphology of the aggregate and it is often associated to the steric hindrance of the headgroups. If this is not the original meaning of the surface area in the Tanford and IMN models, such a simplification is still considered in many papers by less experienced authors.[36,55] $A_e$ is not the simple bulkiness of the headgroup but it rather characterizes, from the free energy description, an equilibrium area per molecule, obtained from the minimization of the free energy of micellization, and proportional to the ratio between an "interaction parameter" and the interfacial free energy (Refer to *Note 1* below for more details). This means that $A_e$ includes the bulkiness but also the effects coming from the electrostatic interaction, hidden in the expression "interaction parameter", and more difficult to evaluate. From a qualitative point of view, $A_e$ is indirectly proportional to the ionic strength of the solution, having as a consequence that *PP* increases with ionic strength (*Note 1* below). Conceptually, this is logical because electrostatic repulsions between headgroups are screened at high ionic strength, an event which allows a closer contact between the headgroups (smaller $A_e$) and a transition from high-curvature (micelles) to low-curvature (cylinders, vesicles) morphologies (Figure 2). This is commonly verified experimentally and it is at the basis of detergent formulations, where salt induces the formation of elongated micelles, responsible for an increase in viscosity.

Unfortunately, the number of parameters influencing $A_e$ and *PP* are not limited to the ionic strength, they are not explicitly included in the classical theory and required, in some cases, the development of new models. The chemical nature of the headgroup, the binding affinity of counterions,[56,57] the hydration of counterions (chao/kosmotropic effects),[56,57] the hydration of the headgroup, the hydrophobicity of some headgroups are just the most relevant ones. These parameters have a strong impact on $A_e$, but their effect can hardly be quantified and this is the reason why many systems are treated in an empirical, and not analytical, way. This means that actual value of *PP*, hence of the aggregate shape, can be much different than predicted on the basis of the sole bulkiness of the headgroup. This was illustrated, among others, on cationic alkylammonium surfactants, where the phase behavior of compounds with bulky headgroups (tripropyl, tributyl) was unexpected. Tributyl headgroups are more chaotropic: they are bulky, hydrophobic, their charge is screened, their binding to the counterions is less pronounced and they were even suggested to be partly solubilized in the micellar core, thus even affecting the solubility of the entire molecule.[52] The role of counterions has also been investigated for years and sometimes successfully associated to the Hofmeister series, with the idea that some counterions strongly bind to the headgroup than others (e.g., iodide binds stronger than chloride), while other counterions are more hydrated than others.[52] In all cases, the effective headgroup surface area should include these effects, which modify its actual value and, eventually, *PP*.[56,57] Interestingly enough, the thermodynamic expression for $A_e$ is independent of the chain length and volume, which constitutes a serious limitation. To overcome this inconvenient, a more recent model gave more weight to the influence of the tail on *PP*.[45,46]

*Note 1: Free energy and equilibrium surface area using the Tanford model.*[38,39,46]

The free energy of micellization is given in Eq. 1.1, with $\mu_N^0$ being the standard state chemical potential of an isolated micelle of aggregation number $N$, $k$ the Boltzmann constant and $T$ the temperature. The *Tr*, *Int* and *Head* subscripts refer to the tail transfer, interfacial and headgroup components of the free energy, described in the text. The *Int* and *Head* components can be related to the surface area, $A$, through the contact free energy per unit area (or interfacial free energy), $\sigma$, and the headgroup repulsion parameter, $\alpha$, as shown in Eq. 1.2

$$\left(\frac{\Delta\mu_N^0}{kT}\right) = \left(\frac{\Delta\mu_N^0}{kT}\right)_{Tr} + \left(\frac{\Delta\mu_N^0}{kT}\right)_{Int} + \left(\frac{\Delta\mu_N^0}{kT}\right)_{Head} \quad \text{Eq. 1.1}$$

$$\left(\frac{\Delta\mu_N^0}{kT}\right) = \left(\frac{\Delta\mu_N^0}{kT}\right)_{Tr} + \left(\frac{\sigma}{kT}\right)A + \left(\frac{\alpha}{kT}\right)\frac{1}{A} \quad \text{Eq. 1.2}$$

$\alpha$ is explicitly given in Eq. 1.3 and it is shown to depend on the electron charge, $e$, the dielectric constant of the solvent (generally water), $\epsilon$, the capacitor distance in the double layer model, $d$, the alkyl chain length, $L$, and the inverse Debye length, $\kappa$. $\kappa$ is given in Eq. 1.4 and is directly proportional to the ionic strength, $n_0$, the counterion concentration in solution.

$$\alpha = \frac{2\pi e^2 d}{\epsilon}\left(\frac{1}{1+\kappa L}\right) \quad \text{Eq. 1.3}$$

$$\kappa = \left(\frac{8\pi e^2 n_0}{\epsilon kT}\right)^{1/2} \quad \text{Eq. 1.4}$$

Minimization of the free energy with respect to the surface area yields the equilibrium surface area, $A_e$, given in Eq. 1.5

$$A_e = \left(\frac{\alpha}{\sigma}\right)^{\frac{1}{2}} \quad \text{Eq. 1.5}$$

$$PP = \frac{V}{LA_e} \quad \text{Eq. 1.6}$$

Using Eq. 1.3 and Eq. 1.4 one finds that $\alpha$ has an inverse dependence on $n_0$, which, introduced in Eq. 1.5, one finds an inverse relationship between $A_e$ and the ionic strength, $A_e \sim \frac{1}{n_0}$. Considering the packing parameter, $PP$ (Eq. 1.6), one easily finds that, $PP \sim n_0$, that is the packing parameter increases with ionic strength. In other words, increasing the ionic strength in charged surfactant systems provides a transition spheres → cylinders → bilayers.

*The role of the tail*. The tail volume, $V$, of uniform density is the main hypothesis of the *PP* model. However, this is a strong approximation, as shown by Nagarajan some years ago.[45,46] In brief, similarly as described in block copolymers,[46] he considered that the tail must deform in a non-uniform manner along its length in order to fill the aggregate core with uniform density. The elasticity of the chain introduces an additional, packing (or elastic) energy, term in the free energy of micellization. The expression of this term was described to be dependent both on surface area and on the chain length in such a way that minimization of the free energy yields a more complex expression of the equilibrium area, which now depends on the surface area and "interaction parameter", as described above for $A_e$, but also on the tail length: the longer the tail, the larger $A_e$ (see *Note 2* below). The effects of this refined model on the calculation of the *PP* is shown to be significant: *PP* is calculated to decrease with increasing chain length. Whether or not the refined *PP* has an actual effect on the aggregate structure depends on the overall magnitude of *PP*, but Nagarajan has shown that both spherical and cylindrical aggregates could be expected for a given headgroup, depending on whether tail stretching is considered or not in the model. Refinement of the model by Nagarajan was shown to better describe experimental evidence in various systems including solubilization, solubilizate-induced phase transitions, aggregation of gemini surfactants and mixed surfactant systems and so on.[46] Similar conclusions, although adopting alternate approaches, were shown by others.[43]

*Note 2: Free energy and equilibrium surface area using the Nagarajan model.*[45,46]

The free energy micellization in the Nagarajan model is given in Eq. 2.1, where all terms have been defined in *Note 1* above, except for the *Pack* term (in red for more clarity), and which refers to the contribution coming from the surfactant's tail packing (or inhomogeneous elastic deformation). The *Pack* term is inversely proportional to the surface area and to a quantity called $Q$, with $Q = cVL$, and explicitly shown in Eq. 2.2, all other terms in Eq. 2.2 have been defined in *Note 1* above. $Q$ depends on the tail volume ($V$) and length ($L$) through a constant, $c$, which assumes different finite values[45,46] depending the micellar morphology (sphere, cylinder, bilayer).

$$\left(\frac{\Delta\mu_N^0}{kT}\right) = \left(\frac{\Delta\mu_N^0}{kT}\right)_{Tr} + \left(\frac{\Delta\mu_N^0}{kT}\right)_{Int} + \left(\frac{\Delta\mu_N^0}{kT}\right)_{Head} + \left(\frac{\Delta\mu_N^0}{kT}\right)_{Pack} \qquad \text{Eq. 2.1}$$

$$\left(\frac{\Delta\mu_N^0}{kT}\right) = \left(\frac{\Delta\mu_N^0}{kT}\right)_{Tr} + \left(\frac{\sigma}{kT}\right)A + \left(\frac{\alpha}{kT}\right)\frac{1}{A} + \frac{Q}{A^2} \qquad \text{Eq. 2.2}$$

Eq. 2.2 then shows the direct relationship between the tail length and the free energy, of which the minimization provides a direct relationship between the equilibrium surface area, $A_e$, and $Q$, that is the tail length (Eq. 2.3, of which all terms are defined in *Note 1* above).

$$A_e = \left(\frac{\alpha}{\sigma} + \frac{2Q}{A_0}\frac{kT}{\sigma}\right)^{\frac{1}{2}} \qquad \text{Eq. 2.3}$$

Eq. 2.3 shows a direct proportionality between $A_e$ and the tail length, $A_e \sim L$, which, introduced in the $PP$ expression (Eq. 1.6), shows a reinforcement of the indirect proportionality between $PP$ and $L$, $PP \sim \frac{1}{L}$. In other words, increasing the chain length has the effect of decreasing the packing parameter and promoting a transition bilayers → cylinders → spheres.

*The general micelle model*. A collective, rather than individual, description of supramolecular association into plate-like micelles started to be considered by Bergström in the early 2000.[47,48] The major novelty concerns the introduction of the bending elasticity in the free energy of micellization.[35,49,50] Differently than the thermodynamic approach described above, centered on the aggregation of individual molecules, Bergström considers the collective aspect of surfactant aggregation and it employs the Helfrich description of surfactant membranes.[58] A surfactant film is considered as a continuous surface of arbitrary curvature, where each point can be described with two radii of curvature ($R_1$, $R_2$) and two corresponding curvatures ($c_1$, $c_2$), shown and defined in Figure 3a. For such a system one can define a surface mean, $H$ (Eq. 3.2, *Note 3* below), and Gaussian, $K$ (Eq. 3.3, *Note 3* below), curvatures, of which the sign has the property of shaping the surface, as shown in Figure 3b. The free energy per unit area, $\gamma$, at the surface of the surfactant monolayer can then be expressed as a function of $H$ and $K$ through the equilibrium curvature ($H_0$), bending rigidity ($k_c$) and saddle splay ($\bar{k}_c$) constants (Eq. 3.1, *Note 3* below), which altogether identify the intrinsic elastic properties of the surfactant film at a given temperature (Refer to *Note 3* below for more details).[58]

On this basis, Bergström derives an expression for both the micellar volume fraction and $A_e$ linking them to the elastic constants of the surfactant's film (Eq. 3.4, *Note 3* below).[47–50,59] More interestingly, he conciliates the thermodynamic and the elastic approaches by specifically connecting the elastic constants to the typical intermolecular interactions found in micellar and bilayer surfactant systems.[59] The advantage of this generalized model, which makes the classical model a special case, consists in its ability to describe more complex systems than spheres and bilayers. The generalized model could satisfactorily describe micellar growth, and in particular the existence of a CMC2, microemulsions and reverse micellization in a hydrophobic solvent.[35] The validity of this model was experimentally proven by measuring the elastic constants of micellar aggregates composed of cationic surfactants, but also by the prediction of CMC2 in gemini surfactants.[35] This theory could certainly be more appropriate to describe the self-assembly behavior of complex surfactants, like bolaamphiphiles and, in particular, biosurfactants. Nonetheless, the major drawback is certainly its complexity, if compared to the *PP* approach.

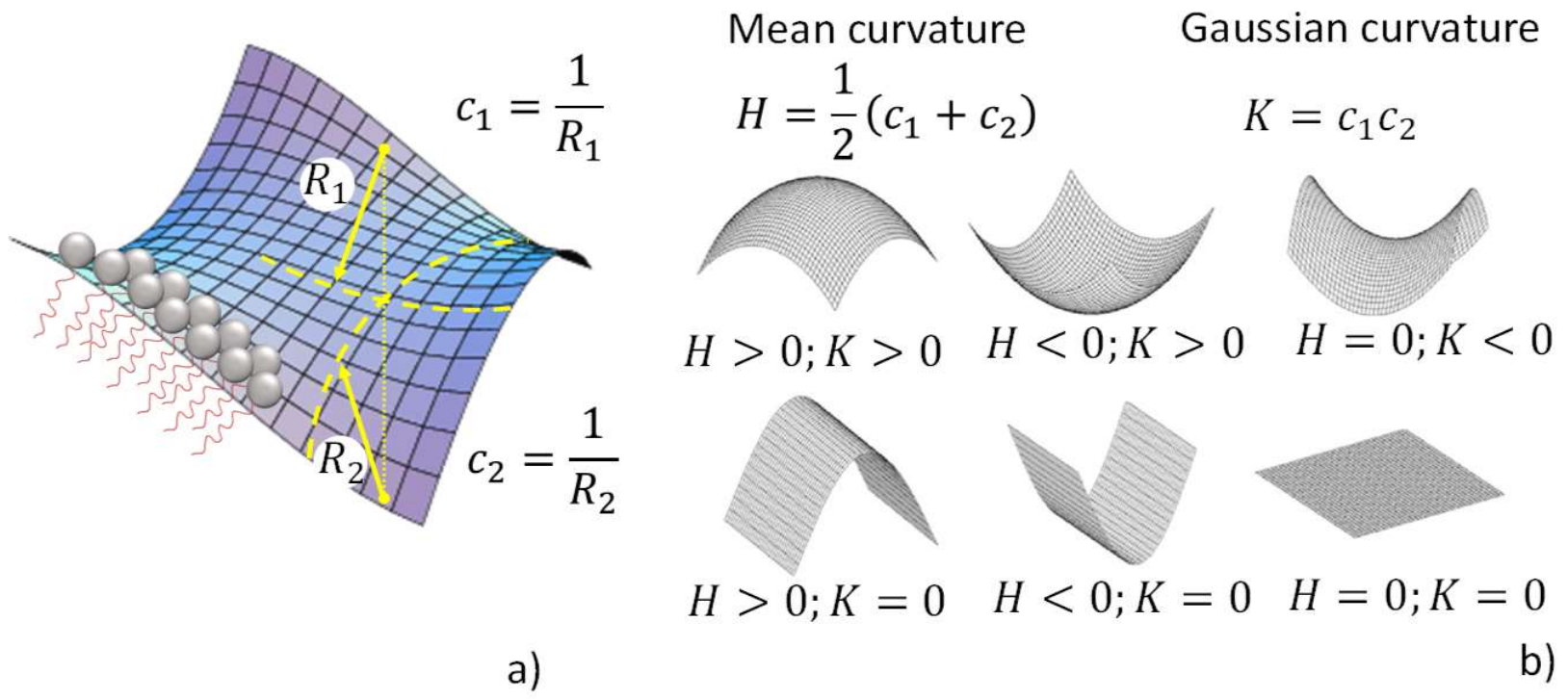

Figure 3 – a) Relationship between the radii of curvature, $R_1$ et $R_2$ and curvatures $c_1$ and $c_2$ at a surface composed of a surfactant film. b) Definition of the mean and Gaussian curvatures and the effect of their sign on a surface

*Note 3: Modelling micellization using the elastic free energy: the Bergström model.*[47–50,59]

Bergström considers that a surfactant aggregate, may it be a spherical micelle or a more complex morphology, should be described as a collective entity, which can be described with an elastic free energy per unit area, $\gamma$, as defined by Helfrich[58] and summarized in Eq. 3.1. $k_c$ (always positive) and $\bar{k}_c$ (positive or negative) are respectively the bending rigidity and the saddle splay constants and they characterize the rigidity of the surfactant's layer. For large and positive values of these constants, the elastic free energy increases, indicating a stiffer surface. $H$ and $K$ respectively identify the mean and Gaussian curvatures, defined in Eq. 3.2 and Eq. 3.3, where $c_1$ and $c_2$ are the inverse of the curvature radii at a single point on the aggregate's surface, conveniently defined at the core/solvent, e.g. hydrocarbon/water, interface. Finally, $\gamma_0$ is the elastic free energy per unit area at $H = H_0$ and $K = 0$, where $H_0$ (positive or negative) is the spontaneous curvature of the film, connected to the planar interfacial tension, $\sigma$, mentioned in *Note 1* above and showing that the Tanford model is a special case of the present theory.[47–50,59]

$$\gamma = \gamma_0 + 2k_c(H - H_0)^2 + \bar{k}_c K \qquad \text{Eq. 3.1}$$

$$H = \frac{1}{2}(c_1 + c_2) \qquad \text{Eq. 3.2}$$

$$K = c_1 c_2 \qquad \text{Eq. 3.3}$$

By employing the same approach as Tanford,[38] Eriksson,[42] Blankschtein[43] or Nagarajan,[44] Bergström defines a free energy expression as a sum of various contributions, all associated to the classical intermolecular and intramolecular parameters (transfer, interfacial, electrostatic, tail, headgroup); however, differently than all others, each contribution is now dependent on $H$ and $K$, that is on the curvature of the surfactant's film. Minimization of the free energy yields an expression of the equilibrium area, $A_e$, which is also dependent on $H$ and $K$ $(\text{in red in Eq. 3.4})$ .[59] In Eq. 3.4, $\xi$ and $V$ are respectively the thickness of the hydrocarbon part of the surfactant film (equivalent to the previously defined tail length, $L$) and volume of the surfactant tail.

$$\frac{1}{A_e} = \frac{\xi}{V}\left(1 - \xi H + \frac{\xi^2}{3}K\right);\ \xi(H,K), V(H,K) \qquad \text{Eq. 3.4}$$

The dependency of $\xi$ and $V$ in particular, and $A_e$ in general, on the curvatures, $H$ and $K$, is quite complex, it will not be detailed here but it can be found elsewhere.[59] Bergström certainly derives a more complex expression of $A_e$ than before, but it now can be related to the elastic properties of the surfactant film. Additionally, the elastic constants of the film can be specifically connected to all contributions (electrostatic, headgroup, ….) to the free energy.

## 1.5. Self-assembly theories applied to bolaamphiphiles.

The models to understand and predict the self-assembly properties of amphiphiles have mainly been developed on the basis of head-tail amphiphiles. However, they should be valid, in principle, for more complex systems such as gemini surfactants or bolaamphiphiles,[35] although this is not the case.[40] In fact, considering the specific structure of the latter, which often requires a multi-step organic synthesis scheme, thus limiting their commercial potential, the amount of theoretical work associated to complex structures is clearly less abundant. Nevertheless, since the mid-80's there has been an effort to verify the validity of self-assembly models for more exotic amphiphiles.

Nagarajan has shown, already in the late 80's, that the Tanford theory of self-assembly and the packing parameter approach are applicable to bolaamphiphiles.[37] His theoretical approach considers a double value for the headgroup surface area, because bolas have two headgroups, the same volume of the tail and the assumption that the smallest dimension cannot exceed half of the chain length. In the latter case, Nagarajan considers that bolaamphiphile either cross or bend in the aggregated object, of which the smallest cross section necessary corresponds to the entire length of the bola, $L$, (that is $R = \frac{L}{2}$), differently than in single-chain amphiphiles ($R = L$). Interestingly, he shows that the expression of the packing parameter is unchanged and so are the lower and upper limits with respect to the morphologies, as shown in Figure 2 for head-tail surfactants; he also finds that the predicted aggregation numbers are smaller in the case of bolas. Upon using numerical values and comparing with the relatively few data available in the literature at that time, he finds that bolas have a smaller tendency to aggregate, they are more soluble, if they form micelles, these are smaller in size, and in general they are expected to form flatter geometries, like vesicles or bilayers, if compared to the a head-tail amphiphile of similar structure.

If part of these predictions are interesting and verified for some specific examples available in the literature,[60] the broader amount of work published on the self-assembly of bolaamphiphiles along the years has demonstrated that these molecules have a much richer, and more complex, phase behavior,[25,27] depending on a broader number of parameters, which have never been rationalized so far. In particular, many bolas have a spontaneous tendency to form semicrystalline fibers or lipid nanotubes,[25,27] which are morphologies that are not predicted by any of the theoretical models. This illustrates the limits of the Tanford and IMN approach to satisfactorily describe the self-assembly of bolas in particular, and of complex amphiphiles in general.[40] In a more recent work,[46] Nagarajan evokes the use of refined models (e.g., the tail model described above and in *Note 2* above) for complex DNA, peptide or polyoxometallate amphiphiles (but not bolaamphiles), and Bergström has shown the validity of the general micelle model to gemini surfactants.[49] However, the amount of available data where refined models were successfully applied to a broad range of new amphiphiles is still too small to generate a trustable and generalize picture of the structure-property relationship in more advanced systems.

## 1.6. General remarks on amphiphile self-assembly

The question of predicting the self-assembly of non-polymeric amphiphilic molecules in solution has generated a worldwide interest, both for the understanding of lipid membranes in living organisms and to prepare soft and inorganic materials. A number of thermodynamic models exist since the 70's, whereas the most successful and widespread one is the Tanford approach, which lead to the definition of the packing parameter, simply-expressed in terms of geometrical attributes of the single amphiphile: headgroup and tail volume and length. Refinement of the packing parameter model has occurred over the years, including specific effects of entropy, tail stretching, physicochemical environment, membrane elasticity, etc… The common denominator among all models is the addition of free energy terms, each expressed for a specific interaction and refined models were particularly successful to fill the gaps of the IMN approach. However, many unsolved issues were already pointed out by Svenson in 2004[40] and they still exist nowadays:

° Refined models, and in particular the Bergtröm approach, could be of help in some cases. However, they are often too complex to be handle by non-experienced users. For this reason, the classical IMN packing parameter approach is still used to explain the self-assembly properties of amphiphiles in solution, even if it is still misused, especially in considering the effect of the headgroup surface area resulting from a mere steric hindrance effect.

° An impressive number of advanced molecular systems with interesting self-assembly properties exist nowadays in the literature. Bolaamphiphiles are just an example but amphiphiles based on peptides, DNA, ionic liquids or nanoparticles, just to cite some, are becoming common, as recently discussed in Ch. 4 of Ref. [61]. Very often, the self-assembly properties of these systems are interpreted on the basis on their calculated packing parameter, but in many cases such approach fails.

° Failure of the packing parameter approach depends on many variables,[40] of which a tentative non-exhaustive list can be:

- A complex chemical structure introduces a large number of simultaneous inter- and intra-molecular interactions. For instance, polyols like sugars bring an additional intra- and intermolecular H-bonding network, stronger hydration, orientational interactions, just to cite some.[62]
- Strong binding effects of the counterions as well as counterion-induced assembly and orientation.[54]
- Presence of stimuli-responsive chemical groups which react with physicochemical parameters like pH, light, temperature, etc…, often creating mixtures of two amphiphiles of the same structure but of different properties.[60]

- Kinetic effects can be important. In fact, all models mentioned above are based on thermodynamic equilibrium, while many systems can self-assemble under non-equilibrium conditions. This aspect was discussed in part in Ch. 11 in Ref. [61]

In this review, we will try to outline which models have been used to explain the self-assembly of biosurfactants and which models could be actually used instead in order to account for unexpected self-assembled properties.

### 1.7. Analytical techniques

The self-assembly of amphiphiles can be studied with a number of experimental techniques. This short section, which we recently proposed in ref. [63] and reported here for a matter of self-consistence, provides a list of the major ones employed to study surfactants and colloids in solution; we also indicate which are the most common and the most suitable ones. We also indicate the level of accessibility and experience required to perform a fruitful analysis. Since this section could be a review on its own, we intend to give the essential information but keeping it as short as possible.

As far as the evaluation of CMC is concerned, one can measure several physical properties (e.g., turbidity, surface tension or diffusion coefficients), qualitatively presented in Figure 4 (a detailed description on their advantages and disadvantages can be found in Ch. 2 of Ref. [36]), using a number of equivalent techniques (e.g., tensiometry, nuclear magnetic resonance, light scattering, etc…). Choosing to measure a given property is often adapted to the type of surface-active compound; for instance, surface tension gives reliable results when studying the CMC of long chain amphiphiles, which generally associate faster than small-chain ones. Surface tension and conductivity are the most common and appropriate properties to measure the CMC of ionic compounds.[64,65] However, for long equilibration processes, surface tension can be limiting due to evaporation. To overcome this problem, and for those who do not possess any of the different types of existing tensiometers, many other properties and experimental techniques offer equally good results. These include: self-diffusion NMR, measuring the difference in diffusion coefficient of single and aggregated surfactants; fluorescence spectroscopy, measuring the intensity ratio between the first and third vibronic peaks ($I_1/I_3$) of an internalized hydrophobic probe like pyrene, of which the emission properties are strongly affected by the medium polarity; static light scattering, measuring turbidity, and so on (Figure 4). In terms of convenience, turbidity is highly practical because it can be measured at a fixed angle and wavelength using a common UV-Vis spectrometer or dynamic light scattering (DLS) apparatus, very common instruments in many chemistry and biology laboratories. However, turbidity is less precise than surface tension measurements. Self-diffusion NMR is precise and interesting for the spectroscopic resolution in case of mixtures, but experiments can be quite long at low concentrations due to the intrinsically low sensitivity of NMR, it requires a NMR spectrometer and enough technical experience of the user. Measuring the spectral emission of an external probe may then be preferential, as it only requires access to a spectrofluorometer or even to UV-Vis spectrometer, two equipments widely found in many labs, and use of a low-cost molecular probe like pyrene. This approach is also interesting because it provides information about the local polarity around the probe and pieces of information such as permeability to water of micellar aggregates.[66] However, as many conceptually-similar methods, one should be aware of the fact that the probe could potentially perturb the self-assembly conditions. Overall, measuring CMC is quite a straightforward experiment, which can be performed in most chemistry and biology laboratories, often employing already existing instrumentation. However, surface tension and CMC do not constitute satisfactory pieces of data to describe the aggregation behaviour of amphiphiles.

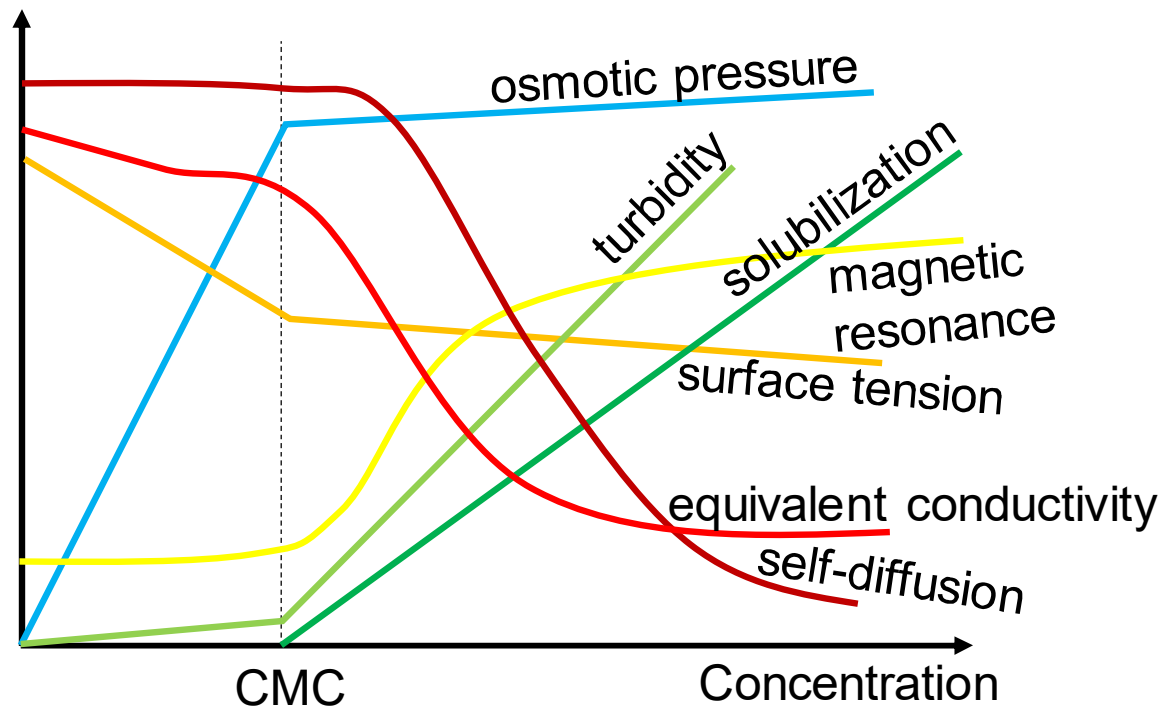

Figure 4 - Major properties employed in the study of critical micelle concentration. Image redrawn from B. Lindman and H. Wennerström, Topics in Current Chemistry, Vol. 87, Springer-Verlag, Berlin, 1980, p. 6.

Evaluation of size and morphology of diluted self-assembled amphiphiles, but also the structure and interactions of soft condensed aggregates, requires a more complex analytical approach. Access to the necessary experimental tools is often limited, data analysis and interpretation is rarely accessible to beginners and crossing the results from at least two complementary techniques is often a requirement. The most common techniques employed in the advanced study of the solution self-assembly properties of surface-active compounds can be divided into four main categories: scattering/diffraction, spectroscopy, microscopy and thermodynamics (Figure 5). Some of the techniques depicted in Figure 5 are strongly advised (in green); however, other ones (in red), like the popular scanning electron microscopy (SEM), have some intrinsic conditions of use, which are not generally compatible with the study of self-assembly in solution. If, at a first glance, they could bring some piece of information, they are actually not advised or, if employed, the result should be interpreted with caution and combination with at least another more appropriate technique is necessary. Table 1 also classifies these same techniques by their functional use; provides the type of information; the typical size range accessible; the physical state of the sample as well as the limitations in terms of accessibility to the equipment and complexity in terms of data treatment and interpretation. Some of these techniques have been reviewed in depth by Yu et al. (2013) within the context of soft materials.[67]

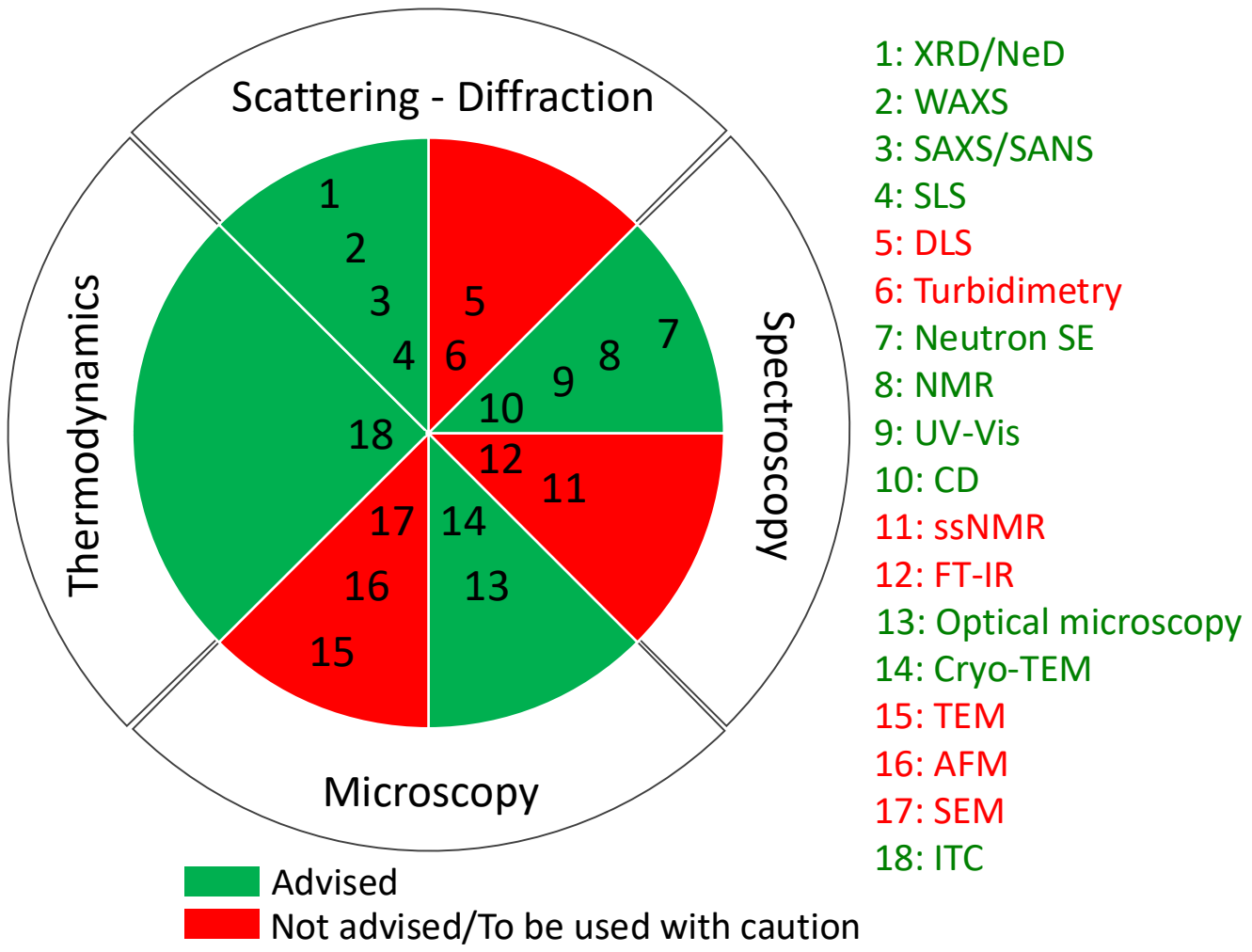

Figure 5 – List of the most common techniques, classified in four domains, employed in the study of amphiphile self-assembly in solution. Acronyms: SAXS (small-angle X-ray scattering), SANS (small-angle neutron scattering), SLS (static light scattering), Cryo-TEM (cryogenic transmission electron microscopy), TEM (transmission electron microscopy), SEM (scanning electron microscopy), AFM (atomic force microscopy), DLS (dynamic light scattering), XRD (X-ray diffraction), NeD (neutron diffraction), WAXS (wide-angle X-ray scattering), PL (polarized light microscopy), CD (circular dichroism), FT-IR (Fourier transform infrared spectroscopy), Neutron SE (neutron spin-echo), NMR (nuclear magnetic resonance), ITC (isothermal titration calorimetry), UV-Vis (ultra violet visible spectroscopy), ssNMR (solid state nuclear magnetic resonance)

For a representative insight of self-assembled structures in solution, several aspects should be respected:
- To perform the study in the parent solvent
- To keep the concentration unchanged
- To avoid drying
- To avoid sample degradation
- To adapt the technique to the dynamics of self-assembly
- To be aware of common artifacts
- To cross several techniques

*Size and morphology.* The most reliable way to measure size and morphology of self-assembled aggregates, and colloidal structures in general, is the combination of small angle X-ray/neutron scattering (SAXS/SANS respectively) or static light scattering (SLS) with cryogenic transmission electron microscopy (cryo-TEM).[68] SAXS generally provides sufficient information and access to SAXS instruments, although limited, is still more convenient than access to neutron facilities. However, SANS is sometimes preferred if the contrast between the electron densities of the amphiphile and solvent is too low for X-rays, or if X-ray exposure degrades the sample during the experiment. SLS is generally used to probe colloids of hydrodynamic diameter above 300-500 nm. Interestingly, SAXS employed with synchrotron radiation provides access to fast acquisition rates, meaning that many self-assembly processes can be measured in a time-resolved (as low as the millisecond scale) *in situ* approach.[69] SANS provides the so-called "contrast-matching", which allows selective study of specific regions in supramolecular aggregates (e.g., hydrophilic headgroup, hydrophobic core in micelles, bilayers, vesicles…) with sub-nanometer resolution, by controlling the contrast between the aggregate and the solvent. This is practically done by controlling the hydrogen-deuterium (H/D) ratio in the solution. To a much lesser extent, SAXS can be employed in the same way in specific Anomalous-SAXS (A-SAXS) experiments. These probe the enhanced scattering of selected counterions when they are irradiated at their spectral absorption edge.[70,71] SAXS, SANS and SLS are particularly important for their statistical relevance, illustrated by a simple calculation. A typical concentration of surfactant in solution is in the order of 0.5 wt%, which, for an average molecular weight of 500 $gmol^{-1}$, yields a solution of about $10^{-2}$ $molL^{-1}$. The volume explored varies roughly between $10^{-2}$ $mm^3$ and $10^2$ $mm^3$, proving a range of about $10^{13}$ and $10^{17}$ molecules. In the case that the surfactant self-assembles into ellipsoidal nanometer-scale micelles of classical aggregation number (number of molecules in the micelles at equilibrium) of 100, the number of analyzed objects, of which the size and morphology is averaged at once, varies between $10^{11}$ and $10^{15}$. This illustrates the fact that the data collected by SAXS, SANS and SLS are extremely reliable, when the experiments and data treatment are properly done. Nonetheless, these techniques suffer from a number of drawbacks. Firstly, they are indirect techniques, as they probe the Fourier space, thus needing model-dependent and model-independent analyses to extract quantitative data.[68] For many classical situations and experienced users this is not a problem; however, complex structures can make modelling tedious, or even not possible. At the same time, newcomers are not able to exploit and interpret the data on their own, even for the simplest systems (e.g., spherical micelles). For this reason, and to avoid misinterpretation, SAXS, SANS and SLS should always be coupled to electron and/or optical microscopy, depending on the desired scale.[68] Finally, accesses to SAXS and SANS instruments are limited and require access to large-scale facilities like synchrotron light and neutron sources. Additionally, the limited number of SAXS and SLS instruments available at lab scale may not be powerful enough to study supramolecular systems at high dilutions (< ~ 1 wt%).

A particularly popular scattering technique employed to study colloids in solution is dynamic light scattering (DLS), which has the double advantage of being accessible in many laboratories and also easy-to-use.[72] DLS provides information on the hydrodynamic diameter (diameter of the colloid plus its hydration corona) and size distribution and it is very helpful for studies on colloidal stability.[72] However, DLS provides no information on morphology and is characterized by a number of possible biases (e.g., scattering from dust or aggregates), which make it very prone to misinterpretation.[72] DLS should be used with care and as a complement to other scattering and microscopy techniques.

Whenever possible, scattering techniques should systematically be coupled to direct observation. This is not only necessary to avoid misinterpretation but also to correctly attribute those scattering signals, which can be produced by multiple structures and which cannot unambiguously be attributed (e.g., flat lamellar vs. curved vesicular bilayers). Microscopy techniques can be divided between invasive and non-invasive. The former modifies the concentration by drying and the latter can be performed in the parent solution. Non-invasive microscopy can be performed from the nanometre to the millimetre scales. Cryogenic transmission electron microscopy (cryo-TEM) has a nanometer resolution and it is a convenient technique, which preserves hydration and original concentration of the sample in its parent solvent, thus providing access to the nano-to-microscale morphology of the aggregate. For these reasons, it is generally preferred to conventional TEM which requires sample drying.[73] Cryo-TEM is nonetheless employed for fairly diluted samples (< 1-2 wt%) and artefacts due to vitrification, ethane adsorption and poor statistics (like any microscopy technique) could occur.[73,74] For this reason, the analysis should be approached with caution, repeated, and if possible coupled with scattering techniques. Finally, access to a cryo-TEM microscope is limited. Standard TEM microscopes can be employed under cryogenic conditions, however only a reduced number of laboratories is actually equipped and possesses the know-how to perform routine cryo-TEM experiments. Similarly, SEM can be employed in a cryogenic mode, although such instrument is ever rarer. Optical microscopy, despite its lower resolution (micrometre to millimetre scales), can be used in a hydrated environment, constitutes the alternative to cryo-TEM for micron-scale structures and can be coupled to data obtained by SLS. Any microscopy approach requiring drying (standard TEM, SEM or even atomic force microscopy, AFM) should be avoided or at least employed with extreme care. This is due to the amphiphile concentration changes and aggregation or unexpected phase transitions, not reflecting the self-assembled state in solution which can potentially occur. For a robust interpretation, combining SAXS/SANS and cryo-TEM data should be coherent and provide the same information.

*Structure*. The structure of the condensed state (crystalline or liquid crystalline) of self-assembled aggregates formed by surface-active compounds can be accessed with diffraction techniques and polarized light microscopy (PLM).[75] X-ray diffraction (XRD), and in rare cases neutron diffraction (NeD), is very practical, although standard diffractometers in $\theta$-$2\theta$ geometry do not allow a reliable analysis of wet samples and diluted solutions. In this case, one should employ a wide-angle X-ray scattering (WAXS) configuration, which may sometimes be available as a supplementary tool in a SAXS instrument, at a laboratory scale or at a synchrotron facility. Like scattering techniques, XRD or WAXS, are measured in the Fourier space but in this case modelling is not necessary and data interpretation generally occurs on the simple analysis of the peak positions relative one to the other and the general principles of crystallography apply. In practice, the variety of possible crystal systems is generally limited to few recurrent ones (2D oblique lamellar, 3D hexagonal and cubic) compared to crystalline inorganic solids and straightforward interpretations are not uncommon. Neutron diffraction could replace use of X-rays in specific cases of sample instability under X-rays, seeking light-weight atoms like hydrogen, or contrast-matching experiments. PLM has long been used to analyze surfactant mesophases; however, despite the ease of accessibility to polarized light microscopes, image analysis requires a long experience and coupling to X-ray diffraction experiments may be necessary.[76] Circular dichroism (CD) is a spectroscopic technique which is useful to probe chirality in molecular and supramolecular systems in solution. Finally, same as above, any technique which requires sample drying like infrared spectroscopy, FT-IR, and standard TEM should be employed with care.

*Dynamics and interactions.* More advanced studies on the functionality of surface-active compounds may include analysis of the dynamics and intra/inter-molecular interactions in the aggregates formed by these compounds. Probing the elastic constants of lipid membranes, the thermodynamic and kinetic parameters of self-assembly but also the possible existence of raft regions in lipid membranes are important topics in biophysics and quantification of intermolecular forces are all important aspects of advanced characterization of self-assembled systems.[58,77,78] Accessing this class of information requires experience, even if the analytical technique itself is relatively accessible and easy to use. The study of both local (e.g., intra-aggregate) and collective (e.g., membrane fluctuations) dynamics of self-assembled systems can be performed with neutron spin echo (NSE) or NMR spectroscopy.[77,79–82] The advantage of neutron spin-echo is the combination of both the spatial (1-100 Å) and time (~ps) domains, thus discriminating between single-molecule within and collective dynamics of self-assembled aggregates, like membranes. However, NSE is only accessible upon proposal approval in neutron-generating facilities. NMR is not a spatially-resolved technique but its advantage over NSE is certainly the spectroscopic filtration, providing access to the dynamics of specific regions of the molecule (hydrophilic, hydrophobic) or the solvent, and even both simultaneously over time range between the millisecond to the tens of seconds.[83] However, for large, slow-moving aggregates, solution NMR is useless. In the specific case of vesicle bilayers, a number of other techniques can also be used to study their dynamic properties and measure their elastic properties. The review by Monzel summarizes all techniques, their space and time scales as well as their advantages and inconveniences.[77]

Surfactant-solvent, inter-surfactant or surfactant-macromolecule interactions can be explored with isothermal titration calorimetry.[84] This is the preferred technique for quantifying the enthalpy change, association constant (or binding affinity) and stoichiometry, and consequently the Gibbs free energy and entropy changes, between two or more molecules in solution.[85] However, ITC apparatus is not readily accessible, and its usage requires a high level of experimental practice and data analysis. Collective intermolecular forces (electrostatic, steric, hydration, entropic at scales between Å and $10^3$ nm) can be quantified in so-called pressure-distance experiments, consisting in following a structural parameter (e.g., interlamellar distance in lamellar phases) with osmotic pressure. These experiments can be performed with a more accessible X-ray (or neutron, in some cases) diffraction apparatus. Despite the experimental ease, the understanding and analytical treatment of data related to intermolecular forces is still quite complex and reserved to expert users.[86–88] Spectroscopic techniques like UV-Vis, FT-IR, CD and NMR can also provide a qualitative insight on interactions at the scale of chemical bond (< 1 nm). NMR is of particular interest for the broad panel of 2D and 3D homo and heteronuclear experiments based on intramolecular, through-bond, *J*-couplings as well as intermolecular, through-space, dipolar and quadrupolar couplings. By selecting the appropriate pulse sequence, it is possible to visualize direct interactions between homo (e.g. $^1$H-$^1$H) and hetero (e.g. $^1$H-$^{13}$C; $^1$H-$^{15}$N) nuclei belonging to the same molecule or to adjacent molecules. However, if ITC, WAXS, UV-Vis and CD allow to work in solution, FT-IR generally requires sample drying. In the case of large, slow-tumbling, aggregates solution NMR becomes less efficient due to short relaxation times and use of solid-state NMR (ssNMR) may be necessary. However, ssNMR generally requires sample dehydration, which can modify the structure and interactions, thus leading to misinterpreted data. Using ssNMR with wet samples should be utilized when possible, although it always represents a challenge because ssNMR experiments require fast spinning of the sample holder around its axis (frequency ranges from from 0.1 kHz up to $10^2$ kHz), increasing the risk of modifying the actual volume fraction, or even of sample dehydration, to due centrifugation in the sample holder.[89] To overcome this problem one can either employ specific probes and sample holders specifically developed for soft diluted materials like gels or biological samples such as the high resolution magic angle spinning (HRMAS) NMR method.[90] One can also prevent centrifugation effects by freezing the sample solution within the sample holder and use fast sample spinning to gain in spectral resolution. Whichever solution, one either need a specific HRMAS probe or a freezing unit coupled to the NMR spectrometer.

Table 1 – A summary of the most common techniques that can be employed in the study of amphiphile self-assembly in solution, classified in four domains. The techniques are classified by the type of information to which they provide access.

| **Information** | **Technique** | **Size domain** | **Physical state**<br>**D: diluted**<br>**C: concentrated** | **Level of accessibility** | **Level of analysis** |
|---|---|---|---|---|---|
| Morphology & Size | SAXS | 1-500 nm | D/C solution | Limited | Experienced |
| | SANS | 1-500 nm | D/C solution | Very limited | Experienced |
| | SLS | 200-1000 nm | D solution | Limited | Experienced |
| | Cryo-TEM | nm to µm | D solution | Limited | Intermediate |
| | Optical microscopy | ~ 200 nm – mm | D/C solution | Broad | Intermediate |
| | TEM | Å to µm | Powder | Medium | Beginner |
| | AFM | nm to µm | Powder | Medium | Intermediate |
| | SEM | > 100 nm | Powder | Medium | Intermediate |
| | | | | | |
| Size only | DLS | nm to ~ 1 µm | D solution | Broad | Intermediate |
| | Turbidimetry | nm to µm | D solution | Broad | Beginner |
| | | | | | |
| Structure | XRD/NeD | < 5 nm | D/C solution Powder | Broad /Limited | Intermediate |
| | WAXS | < 5 nm | D/C solution Powder | Limited/Very Limited | Intermediate |
| | PLM | ~ 200 nm – mm | D/C solution | Broad | Intermediate |
| | CD | nm to µm | D solution | Medium | Beginner |
| | TEM | Å to µm | Powder | Medium | Beginner |
| | FT-IR | < nm | Powder | Broad | Beginner |
| | | | | | |
| Dynamics | Neutron SE | Å to 100 nm | D/C solution | Very limited | Experienced |
| | NMR | < 1 nm | D solution | Broad | Experienced |
| | | | | | |
| Interactions | ITC | Å | D solution | Limited | Experienced |
| | WAXS | < 5 nm | D/C solution Powder | Limited/Very Limited | Experienced |
| | UV-Vis | Å | D solution | Broad | Experienced |
| | CD | nm to µm | D solution | Medium | Experienced |
| | NMR | < nm | D solution | Broad | Experienced |
| | FT-IR | < nm | Powder | Broad | Experienced |
| | ssNMR | < nm | Powder | Limited | Experienced |

## 2. Biosurfactants in solution

### 2.1 Introduction to surfactants, biosurfactants and microbial biosurfactants

The global surfactants market accounted for 43.7 billion dollars in 2017 and has been projected to reach 66.4 billion dollars by 2025[91] accounting to about 20 million tons of surfactants produced each year. Although this is a huge volume, the market is largely scattered with hundreds of different compounds covering an array of varying functionalities. Looking into the sourcing of surfactants, the surfactant market can roughly be divided into three segments: fossil-based surfactants, partly bio-based- and fully bio-based surfactants (biosurfactants). About 4% of all produced surfactants are 'fully bio-based' meaning all the therein contained carbon is derived from biomass (e.g. derived from plants, animals, algae, etc.) and thus do not contain any fossil derived carbon. These fully bio-based surfactants, or biosurfactants, can be produced through chemical or biological means (Figure 6). Some well-known examples of the first type are sucrose esters, alkyl polyglucosides, fatty acid glucamides, sorbitan esters. Biologically produced biosurfactants can be divided in three types: those extracted from plants, those obtained through biocatalysis and a last type obtained through fermentation, i.e. the so-called microbial biosurfactants.

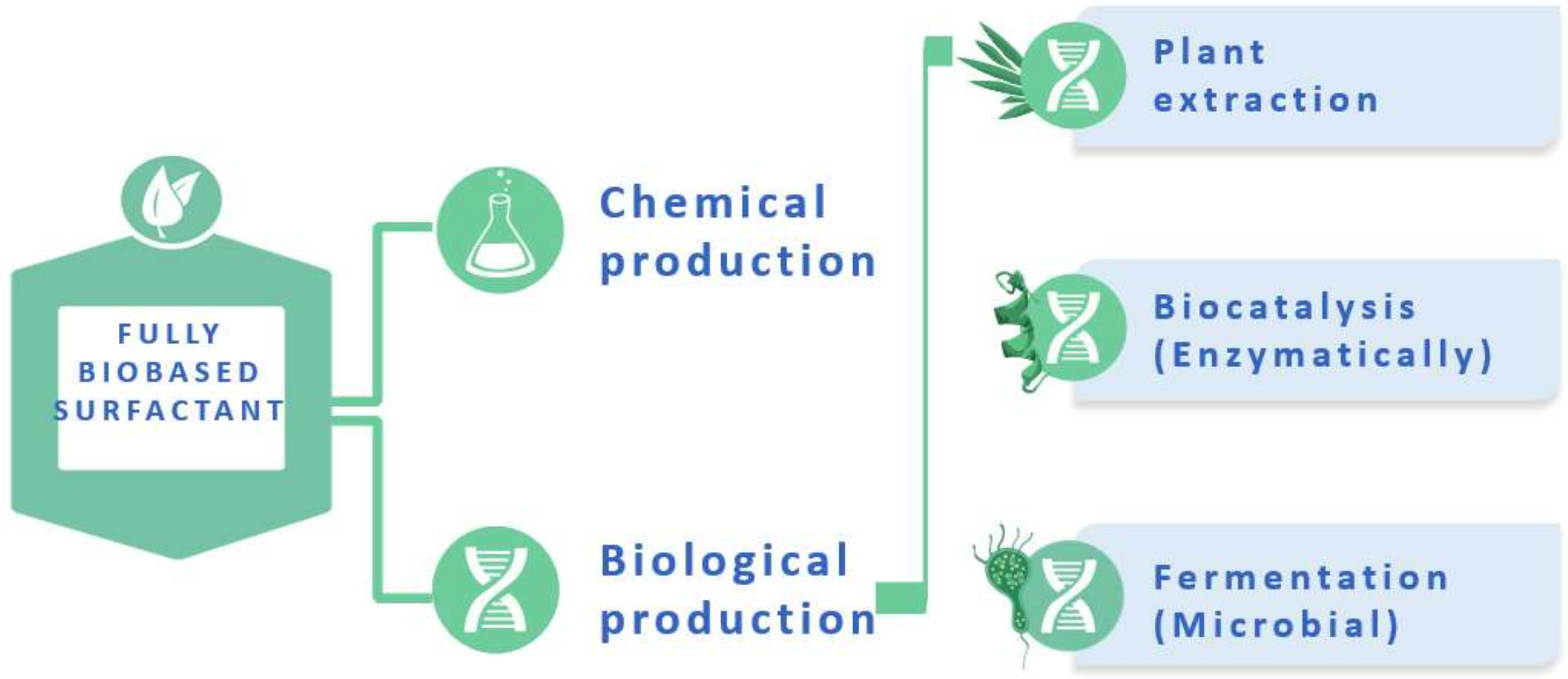

Figure 6 – Known categories of fully bio-based surfactants

In this review, we will address only biosurfactants prepared from microbial fermentation. Within this category, five main groups can be distinguished: glycolipids, lipopeptides, phospholipids, neutral lipids and polymeric microbial biosurfactants. The best known microbial biosurfactants are derived from the group of the glycolipids with commercialized molecules, i.e. wild type sophorolipids (SL),[92] rhamnolipids (RL)[93] and mannosylerythritol lipids (MEL).[94]

SL are already commercialized by companies such as Evonik and Wheatoleo and applied in consumer products by several companies, e.g. Ecover, Givaudan, Saraya Co., Henkel etc. mainly for applications in household detergents and personal care products. Applications in other sectors, such as in the agrochemical, food- and pharmaceutical industry are expected to follow. Rhamnolipids were recently launched on the business-to-consumer (B2C) market by a partnership between Evonik and Unilever and big scale production of the rhamnolipids at a multi thousand-ton scale is expected in the next few years. This will dramatically open up the perspectives for microbial biosurfactants.

Microbial biosurfactants are generally produced by the wild type microorganism from a culture medium rich in glucose and vegetable oil. Use of modified strains or chemical modifications of given biosurfactants are also possible routes to extend the diversification of these molecules and, hence, their properties and application potential. Many reviews describe the origin of biosurfactants, their producing strains and conditions as well as the main widely accepted categories. Giving information on such topic is out of the scope of this document and we address readers to the abundant literature published on this topic since at least three decades.[10,11,17,20–22,94–102]

The number of existing biosurfactants from microbial origin is quite impressive, as well as the number of microorganisms producing them.[103] However, only few can be produced in sufficient amount, with acceptable purity and homogeneity to be satisfactorily studied from a physicochemical point of view. Rhamnolipids (RLs), sophorolipids (SLs), cellobioselipids (CLs), mannosylerythritol lipids (MELs), trehalolipids (TLs) and surfactin are broadly recognized as the most classical ones.[10,11,17,20–22,94–102] However, even in this case, they have not been studied evenly. For instance, we are not aware of any specific study on the self-assembly and interface properties of trehalolipids, despite some non-negligible work that has been done on this family of coumpounds since the mid 50's.[104] For this reason, trehalolipids will only be partially discussed in this review. In the meanwhile, chemical derivatizations of existing biosurfactants,[17] and more recent trends in the production of new biosurfactants from engineered strains,[15–17,21,22,92,103,105–109] constitute promising alternatives to expand the biosurfactant portfolio in the future. The availability of these new compounds has occurred since less than a decade and ready collaboration between researchers across disciplines has made their advanced characterization possible. These recent works are included in this review and the broad, although non-exhaustie, list of compounds is given in Figure 7.

Compounds from **1** to **16** reflect the wide variety of sophorolipids (SL) and their derivatives. SL are generally obtained in a 80:20 acetylated C18:1-*cis* lactone(**3**):acidic(**2**) mixture, although the nonacetylated acidic form (**1**) can be easily obtained by alkaline hydrolysis. Figure 7 separates lactonic SL (**3**) from the general acidic SL structure (**0**), to which derivatives **1**,**2** and **4**-**16** can be compared in terms of structural linearity and bolaform shape. We address the reader to the respective literature cited all along the text to obtain more insight on the synthesis of each compound. Besides SL, cellobioselipids (CL) in their hydrolyzed form (**17**), rhamnolipids (RL), both di- (**18**) and mono- (**19**), surfactin (**20**) and MELs (**21**) are also widely discussed throughout this work. One should note that MELs are commonly divided into four sub-compounds (MEL-A, MEL-B, MEL-C and MEL-D) according to their acetylation degree (**21A-D**, $R^1$=H, Ac; $R^2$= H, Ac, Figure 7).

An important note. Although the compounds in Figure 7 are commonly referred to as '*biosurfactants'*, we will show in the next sections that they often don't show a typical 'surfactant' behavior. The most correct term should probably be "*bioamphiphiles*", in analogy to what is outlined in section 1.1. Considering the presence of two polar sides connected by an apolar part, most of these molecules should actually be referred to as bolaamphiphiles, or even "*biobolamphiphiles*". If one should be aware of such subtilities, for a matter of consistency with the broad literature cited, we will keep the term *biosurfactants* most of the time throughout this review.

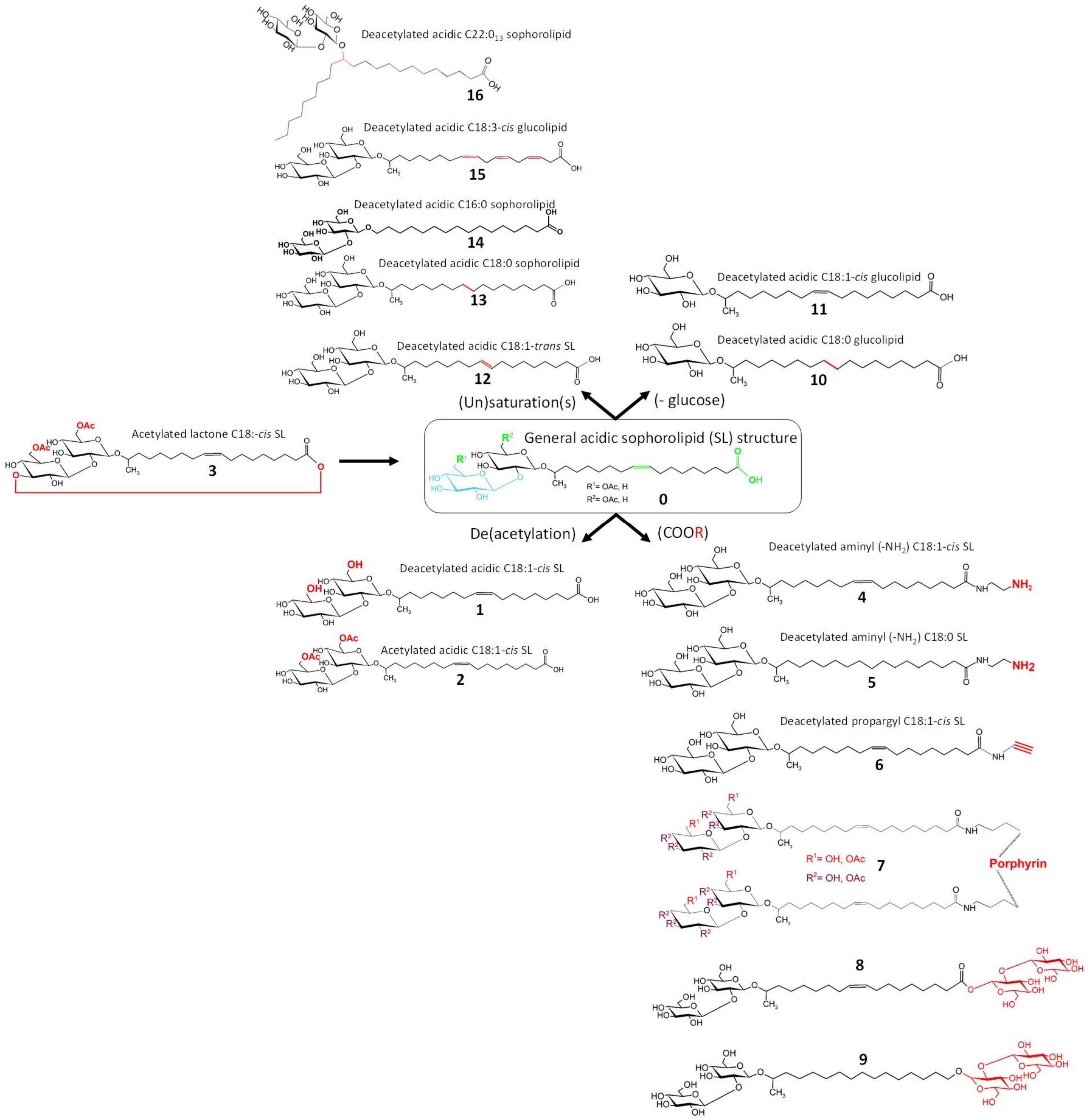
Deacetylated acidic C22:0$_{13}$ sophorolipid
16
Deacetylated acidic C18:3-*cis* glucolipid
15
Deacetylated acidic C16:0 sophorolipid
14
Deacetylated acidic C18:0 sophorolipid
13
Deacetylated acidic C18:1-*cis* glucolipid
11
Deacetylated acidic C18:1-*trans* SL
12
Deacetylated acidic C18:0 glucolipid
10
(Un)saturation(s)
(- glucose)
Acetylated lactone C18:-*cis* SL
3
General acidic sophorolipid (SL) structure
$R^1$= OAc, H
$R^2$= OAc, H
0
De(acetylation)
(COOR)
Deacetylated aminyl (-$NH_2$) C18:1-*cis* SL
4
Deacetylated acidic C18:1-*cis* SL
1
Acetylated acidic C18:1-*cis* SL
2
Deacetylated aminyl (-$NH_2$) C18:0 SL
5
Deacetylated propargyl C18:1-*cis* SL
6
$R^1$= OH, OAc
$R^2$= OH, OAc
7
Porphyrin
8
9

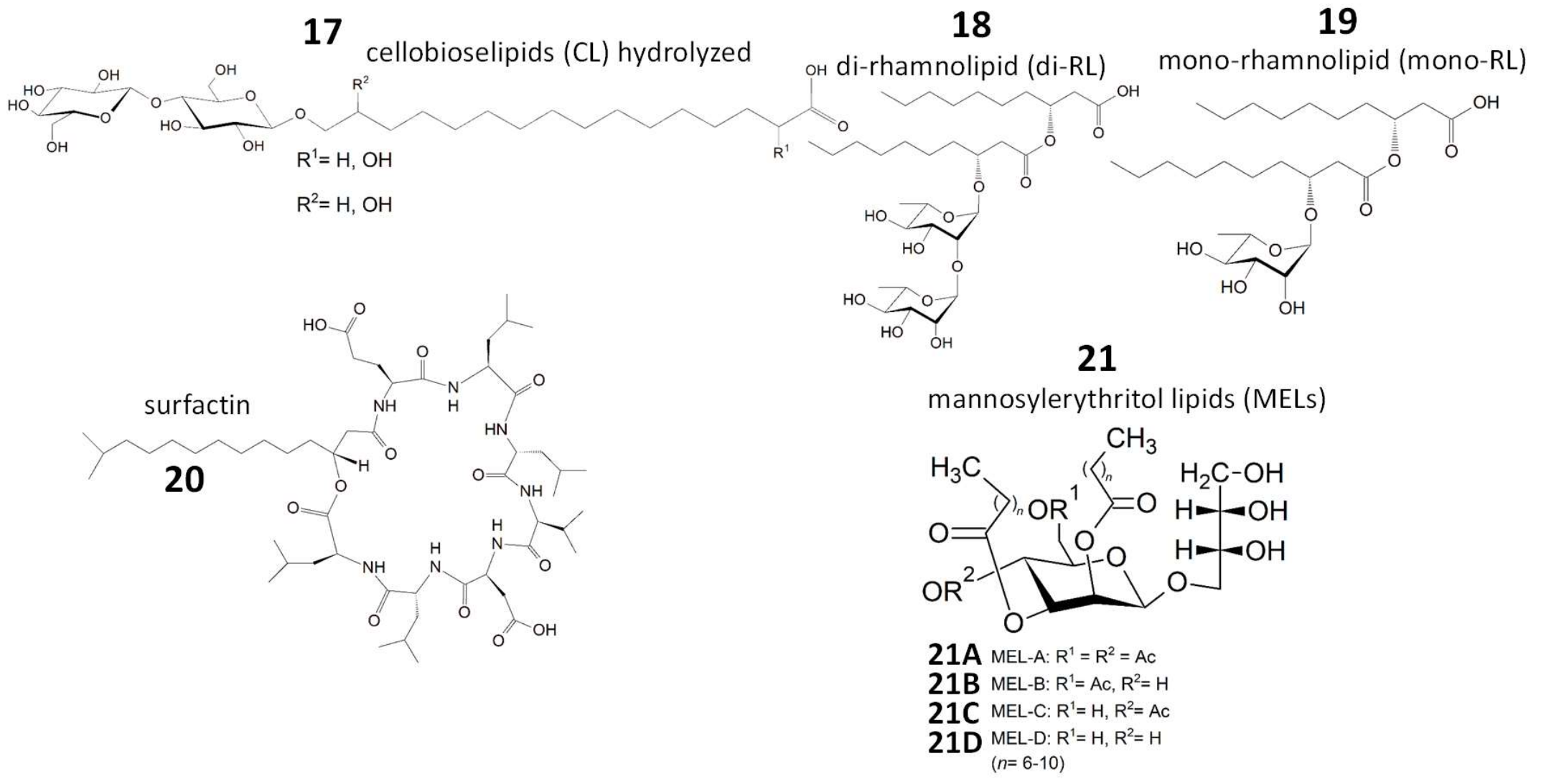

Figure 7 – Chemical structure of the main biosurfactant molecules presented in this review.

Similarly to surfactants, biosurfactants can be primarily characterized in solution through the same generic approaches discussed in section 1.7. In particular, surface tension and CMC are largely referenced in the biosurfactants' literature since decades; however, as for standard surfactants, these parameters are particularly inefficient to understand, or even predict, self-assembly or stimuli-responsivity of biosurfactants. If some reviews covering the self-assembly/physicochemical properties of biosurfactants,[19,25,110] rhamnolipids,[102,111] mannosylerythritol lipids[98] or surfactin[112] exist, they are neither comprehensive nor they critically analyze the biosurfactants' behavior in relationship to the general concepts of surfactants assembly discussed in section 1.5. Furthermore, they do not focus on one of the most interesting characteristics of biosurfactants, namely their complex phase behaviour driven by external stimuli like pH, temperature or ionic strength. Such a rich, dynamic, phase behavior in water at low concentrations is useful for a large panel of applications[113–117] but above all, it is a rich source for fundamental understanding of the self-assembly phenomena of natural compounds, often under nonequilibrium conditions.

This section is structured as follows: we propose a qualitative overview of the classical aspects of surfactant's science (HLB, surface tension and CMC) reported for biosurfactants. Then, we describe the complex self-assembly properties of biosurfactants in water, including stimuli-responsivity, within and beyond the limits of the packing parameter approach. Finally, we will describe the relationship between self-assembly and macroscopic behaviour, like hydrogel formation.

### 2.2 HLB, surface tension, CMC

HLB, surface tension and CMC, given in Table 2, Table 3 and Table 4, are generally estimated or measured for biosurfactants. However, they are limited indicators, because a wide range of CMC and surface tension values have been reported in the literature for the same molecules. Many biosurfactants are not pure and their aggregation state (e.g., micelles vs. vesicles), and even solubility, depends on physicochemical conditions like pH. For this reason, the surface tension and CMC values found in the literature are poorly indicative, if not at all, of their actual aggregation properties. For this reason, although very practical and broadly understood, these parameters should be interpreted with care and carefully used in comparison to petrochemical surfactants. Since these parameters have been extensively reviewed in the past for many biosurfactants,[11,19,20,94] here we only aim at: reporting the typical range, comparing it with petrochemical and glycolipid synthetic surfactants and discussing such limits in a critical way for the biosurfactants family.

Hydrophilic/lipophilic balance (HLB)

HLB, broadly discussed in section 1.2, is an indicator of surfactants' emulsification properties. Few papers discuss the HLB of biosurfactants and one of the latest was published on surfactin in 2005.[118] Table 2 shows the calculated HLB, and expected properties, of the major glycolipid biosurfactants according to the Griffin formula.

Table 2 – HLB values, calculated according to the Griffin formula (HLB= 20 x ($MW_H$/$MW_S$), with $MW_H$ and $MW_S$ respectively being the molecular weight of the hydrophilic part and of the whole surfactant), for biosurfactants (N° refers to Figure 7).

| BS | N° | Properties | HLB | Inferior limit | Superior limit | References |
|---|---|---|---|---|---|---|
| Glycolipids | | | | | | |
| RL | 18, 19 | Wetting agent, o/w, detergent. | 9 – 13 | mono-RL C12:0, C10:0. | di-RL C10:0, C8:0. | This work;[119] |
| SL | 1-3 | o/w, detergent | 12 – 13 | Acidic-SL C18:1. | Acidic di-acetylated-SL C18:1. | This work |
| TL | - | w/o, Wetting agent, o/w | 5 – 11 | TL-dimycolate. | isopropyl myristate–TL–water.[120] | This work;[120] |
| MEL | 21 | w/o, Wetting agent, o/w, detergent | 6 – 13 | Tri-acylated MEL-A, from *P. aphidis MUCL 27852.*[121] | Acylated MEL-A from *P. antarctica T34.*[122] | [121–123] |
| CL | 17 | Detergent, solubilizer. | 15 | | | This work |
| Lipoprotein | | | | | | |
| Surfactin | 20 | Solubilizer | 21 | / | / | [118] |

For many biosurfactants, the properties expected according to the calculation of the HLB are in agreement with the broad range of properties experimentally observed, like o/w emulsification and detergency.[124,125] However, calculated HLB for biosurfactants can be very broad, as in the case of TL, or the expected properties may not correspond to the value of HLB, thus generating confusion and bad expectations. Marqués *et al*. estimate an HLB of 11 for a TL mixture so that o/w emulsion is expected, although w/o emulsion is obtained.[120] Acidic sophorolipids are expected to be o/w emulsifiers, but in fact their bolaform nature make them poor emulsifying agents. To improve their emulsifying character, the hydrophobic character of the tail must be improved by chemical modification.[126] HLB also fails to predict the behavior in mixture of compounds with different HLBs. Some studies provide the HLB for a given biosurfactant, as reported for individual MELs (**21**), but calculated HLB fails to predict and understand the interfacial behavior of a mixture of MELs (**21**), which constitute the actual raw compound.[121] Finally, HLB becomes unsuitable to predict the behavior of polymeric and proteic biosurfactants like surfactin (**20**), because the HLB range expected by surface efficiency of surfactin (**20**) [127,128] is far from the HLB calculated by emulsification method.[118] HLB of surfactin (**20**) is varying with environmental conditions like pH, specific ions condensation and temperature and it is a source of debate.[129–131] These specificities render HLB useless and require more refined understanding of the biosurfactant behavior in solution and in oil and water mixtures.

Minimal surface tension

Early surface tension (ST) experiments on biosurfactants started already in the 60's,[132–134] although thorough ST measurements were only carried out from the 80's onward, when low-molecular weight glycolipid biosurfactants, like rhamnolipids, sophorolipids or trehalolipids, appeared to have a better market potential in view of replacing petroleum-based surfactants.[135–137] In the meanwhile, constant improvements in developing both structural variety and increasing production rates contributed to promote surface tension studies later on.[138,139] Interestingly enough, even if some biosurfactants were discovered in the in the 50's, like cellobioselipids, the study of their interfacial properties only started half a century later.[140,141] ST measurements are nowadays often associated to the study of interfacial properties of biosurfactant derivatives in relationship to the evaluation of the area-per-molecule at the water-air interface. These specific features will be reviewed later in this document (section 3.1).

Biosurfactants have similar concentration-dependent ST profiles, as synthetic surfactants, but the mechanism of surface stabilization depends on their molecular weight. Low-molecular weight (LMW) biosurfactants, like RL (**18**, **19**), SL (**1-3**) or MEL (**21**) (< 1 kDa), follow the classical adsorption/desorption mechanism at the air-water interface considered to be at thermal equilibrium (~kT). High-molecular weight (HMW) biosurfactants, like surfactin (**20**), emulsan or alasan (1 < Mw / kDa < 500), on the contrary, follow a colloidal interfacial adsorption behavior, considered to be irreversible in the range of kT. In the literature, the former are generally referred to as biosurfactants and the latter bioemulsifiers.[18,142]

Table 3 - Minimal surface tension of common biosurfactants (N° refers to Figure 7) in aqueous media.

| **BS** | **N°** | **ST (mN/m)** | **Inferior limit** | **Superior limit** | **Ref.** |
|---|---|---|---|---|---|
| Glycolipids | | | | | |
| RL | 18,19 | 25.0 – 37.4 | Mixture from *P. aeraginosa 44T1*.[143] | Di-RL from *P. aeruginosa* in borax-HCl buffer pH9.[144] | 143–148 |
| SL | 1-3 | 34.2 – 48.0 | Mixture from *S. bombicola NRRL Y-17069*.[125] | Mixture from *C. bombicola ATCC 22214*.[149] | 20,125,149 |
| TL | - | 19.0 – 43 | Succinoyl TL from *Rhodococcus sp. SD-74*.[150] | TL dicorynomycolates from *R. erythropolis DSM43215*.[137] | 137,150,151 |
| MEL | 21 | 24.2 – 33.8 | Purified mono acetylated MEL-C from *P. graminicola CBS10092*.[152] | Purified acylated MEL-A from *P. antarctica T34*.[122] | 122,139,152–154 |
| CL | 17 | 34.7 – 41.9 | Sodium salt of CL mixture from *C. humicola JCM 10251*.[155] | Mixture from *C. humicola JCM 1461*, in 30mM citrate - 100mM phosphate buffer pH4.[156] | 141,155,156 |
| Lipoprotein | | | | | |
| Surfactin | 20 | 27.0 – 30.0 | Surfactin from *B. subtilis IAM 1213* in 100mM $NaHCO_3$.[132] | Industrial Surfactin from *Wako pure chemical Ind, Japan*.[157] | 127,132,158,159 |
| Polymeric | | | | | |
| Emulsan | - | 27 – 46.4 | Emulsan from *A. calcoaceticus PTCC 1318.*[160] | Emulsan from *A. calcoaceticus RAG-1 (ATCC 31012).*[161] | 160,161 |
| Alasan, liposan | - | Good emulsifying capability | | | 11,162–164 |

Table 3 gives the range of minimal surface tension for LMW[165–167] and HMW biosurfactants.[168–170] For all compounds, ST varies between 50 and 20 mN/m, although large disparities can be found for the same compound, as it is the case of trehalolipids, for which a range of 19 to 43 mM/m could be found in the literature. Overall, these values are comparable to the ST of classical surfactants: anionic surfactants, like sodium hexadecyl sulfate (SHS) and sodium dodecylsulfate (SDS), reduce the surface tension to 36 and 38 mN/m, respectively while nonionic surfactants like triton X-100, -114 and -165 have a minimal surface tension of 33, 30 and 39 mN/m.[171] Similar values are also reported for more exotic cationic gemini surfactants, of which the ST in water in the order of 33 – 41 mN/m according the carbon chain length.[172] Even alkyl polyglucosides, synthetic glycosidic amphiphiles, like lauryl glucoside sulfosuccinate and ß-D-octyl, decyl and dodecyl glucoside have a minimal surface tension contained between 30 and 40 mN/m.[173,174] According to the above, one can conclude that biosurfactants display classical values of minimal surface tension and it is then hard to attribute a non-ionic or an ionic character to these molecules on such basis. However, the efficiency of biosurfactants to reduce ST, and in particular their absolute values, should be interpreted with caution, and the key limiting factors will be discussed below.

Critical micelle concentration (CMC)

CMC, described in more detail in sections 1.3 and 1.7, is classically measured for biosurfactants and relative data is abundant in the literature.[11,19,20,94] For this reason, Table 4 reports only the typical range of CMC for each biosurfactant and the reader is encouraged to refer to previous literature for a more extensive list of CMC values.

Table 4 - CMC in aqueous media of common biosurfactants (N° are given in Figure 7) in aqueous media. Literature reports CMC values in both mM and wt:vol units. To allow direct comparison, we convert the reported values using reported or calculated values of the molecular mass, Mw. Superscripts in the Mw column refer to the specific values used to convert concentration units across studies: a) Mw= 503 g/mol (mainly mono RL);[21] b) given in the corresponding article ; c) Mw= 689 g/mol (weighted average);[21] d) Mw= 705 g/mol (main acetylated acidic form);[21] e) Mw= 870 g/mol (weighted average);[120] f) Mw= 2542 g/mol (calculated from Mw of trehalose and mycolic acid); g) Mw= 1354 g/mol;[137] h) Mw= 1212 g/mol (weighted average);[120] i) Mw= 648 g/mol (MEL-A2);[175] j) Mw= 490 g/mol (mono acyl MEL A);[122] k) Calculated after Ref. [156]; l) Mw= 750 g/mol;[141] m) Mw= 780 g/mol (calculated after Ref. [156]); n) Mw= 1036 g/mol (most used).

| **BS** | **N°** | **Mw (g/mol)** | **CMC** | | **Inferior limit** | **Superior limit** | **Ref.** |
|---|---|---|---|---|---|---|---|
| | | | **(mM)** | **(mg/L)** | | | |
| Glycolipids | | | | | | | |
| RL | 17,18 | 475 – 677[176] | 0.004[a] – 0.36[b] | 2[b] – 181.1[a] | Mono-RL from *P. spec. DSM 2874* | Mono-RL from *P. aeruginosa* | 135,143–148,177,178 |
| | | | | | in industrial buffer pH 3 + 0.17M NaCl.[135] | in borax-HCl buffer pH 9.[144] | |
| SL | 1-3 | 621 – 707[176] | 0.009[c] – 0.97[d] | 6[b] – 680[b] | Mixture from *Candida bombicola ATCC 22214* in 3 mM Na-acetate buffer pH6.[179] | Acidic SL from *C. bombicola ATCC 22214.*[158] | 20,125,149,158,179 |
| TL | - | 870[e] – 2500[f;120,151] | 0.0005[g] – 0.039[h] | 0.7[b] – 34[b] | TL dicorynomycolates from *R. erythropolis DSM43215.*[137] | TL tetraester from *R. erythropolis 51T7* in 100mM sodium-citrate buffer pH4+100mM NaCl.[120] | 120,137,150,151 |
| MEL | 21 | 490 – 916[121,122] | 0.0017[b] – 0.36[b] | 1.1[i] – 176.4[j] | Tri-acetylated MEL (MEL-A2) from *P. churashimaensis OK96.*[175] | Mono acylated MEL-A from *P. antarctica T34.*[122] | 122,139,153,154,175 |
| CL | 17 | 750 – 780[k] [141,156] | 0.02[b] – 0.41[b] | 15[l] – 319.8[m] | CL mixture from *C. humicola 9-6* in citrate phosphate buffer pH4.[141] | CL mixture from *C. humicola JCM 1461*, in 100mM phosphate buffer pH7.[156] | 141,155,156 |
| Lipoprotein | | | | | | | |
| Surfactin | 20 | 1022 – 1088[180–182] | 0.009[b] – 0.096[b] | 9.3 – 99.5[n] | Industrial Surfactin from *Wako pure chemical Ind, Japan*.[157] | Surfactin from *B. subtilis YB7.*[131] | 127,131,132,157,158 |
| Polymeric | | | | | | | |
| Emulsan | - | 1000000[133] | $2x10^{-5}$ – $6x10^{-5}$ | 20[b] – 58[b] | Emulsan from *A. venetianus RAG-1 (ATCC 31012).*[169] | Emulsan from *A calcoaceticus RAG-1 (ATCC 31012).*[161] | 160,161,169 |
| Alasan, liposan | - | (700 000 – 900 000[164]), (28 000[163]) | | / | / | / | / |

Most biosurfactants have a bolaform, double hydrophilic, structure and, according to the CMC predictions for bolaamphiphiles by Nagarajan (Sections 1.3 and 1.4),[37] one could expect better solubility and higher CMC than head-tail surfactants. He reports CMCs in the order of 1-10 mM and in the range 0.1 – 10 mM for, respectively, cationic and nonionic bolaamphiphiles. From Table 4, the CMC for biosurfactants are rather in the µM than in the mM range, that is between one and up to three orders of magnitude smaller than what is predicted for bolaamphiphiles, thus confirming the fact that the behavior of biosurfactants in aqueous solution cannot be easily predicted on the sole basis of their gross molecular structure. Interestingly, the CMC of biosurfactants are also smaller, on average, than classical head-tail ionic surfactants and rather in the order of non-ionic surfactants. For instance, the CMC range corresponding to short (x= 6) and long (x= 18) chain cationic $C_xTAB$ is contained between 1008 mM and 0.26 mM,[183] between 136.1 mM (x= 8) and 0.16 mM (x= 18) for the anionic $C_xSO_4Na$,[184,185] and between 10 mM (x= 8) and 0.5 µM (x= 16) for nonionic $C_xE_8$.[36,186] In addition, the CMCs of alkyl polyglucosides (APGs) with an alkyl chain varying between $C_8$ and $C_{14}$ were reported to be in the range 1.7 – 25 mM, 1.2 mM, 0.8 – 2.2 mM, 0.19 – 0.30 mM at RT and 0.27 µM at 50°C for the $C_{14}$ derivative.[174,187]

At a first glance, the CMC of biosurfactants is comparable to the CMC of non-ionic surfactants with long tails rather than to the CMC of ionic surfactants. However, an appropriate comparison is very risky, because the CMC for bisurfactants is extremely variable among different molecules and even for a given molecule. The highest CMC range, between few µM and up to the mM, corresponds to SL (**1-3**), RL (**18, 19**), CL (**17**) and MEL (**21**), while the lowest ranges are reported for TL, surfactin (**20**) and emulsan (below the µM). Furthermore, values between 0.008 and nearly 1 mM are reported for sophorolipids, just to cite one example, but similar variations are reported for rhamnolipids (**18, 19**) or MELs (**21**) (Table 4). If, the effect of pH on biosurfactants is very important and it partially explains different values for the same molecule (from pH 7 to 9 CMCs of mono- and di-RL are respectively 2 and 1.6 times higher),[144] it cannot explain such a systematic, impressively wide, range of CMC for a given molecule. Molecular purity and uniformity is undoubtedly another problem to consider for biosurfactants, as specifically discussed below.

Phase behavior under dilute conditions can be the third explanation for such incoherent values. Biosurfactants have a rich phase behavior, even under dilute conditions. For instance, MEL were not reported to have a CMC but rather a critical aggregation concentration (CAC), because no micellar phase was observed between the free molecular state and the first aggregated structures, found to be vesicles at a first CAC and sponge phase at a second CAC for MEL-A (**21A**).[188–190] In fact, formations of more complex phases than micellar are classically observed for many biosurfactants. Different self-assembled structures can be obtained at low concentrations with SL (**1-3**), RL (**18, 19**) or MEL (**21**) according to pH, as discussed later in this section. As for peptidic biosurfactants, Ishigami *et al*. have shown that surfactin has a specific capability to form ß-sheet structure by self-assembling in aqueous media. ß-sheet formation associates with the high aggregation number, suggesting a rod-shape micelle at basic pH.[157] However, at neutral pH, it has been shown by Shen *et al*. a ball-like structure with remarkably low aggregation number.[182]

Finally, although a crucial parameter, CMC is largely insufficient to study the aggregation behavior of surfactants in general and biosurfactants, in particular. The wide range of CMC values available in the literature for biosurfactants makes this parameter unreliable and of practical poor use.

*Surface tension and CMC data dispersion*

Impurities and poor batch uniformity are certainly the main causes in the broad range of ST and CMC values reported. Impurities are well-known factors influencing the value of ST and CMC in petrochemical surfactants, as the well-known case of dodecanol, a hydrolysis byproduct in SDS formulations.[36] On the other hand, batch uniformity is specific to biosurfactants and it was recently shown to play an important role on the phase behavior of sophorolipids.[191] Batch homogeneity depends on many factors, including biosurfactants production processes, and in particular the kind of microorganism (Table 3, Table 4) but also the carbon source (soybean oil, olive oil, rapeseed oil).[124,149,167,192] In all cases, many congeners can be produced at the same time at different ratios from one process to another, thus influencing the final property. This is known for many systems including RL (**18, 19**), SL (**1-3**),[176] TL,[193] MEL (**21**),[121] and CL (**17**)[194] and often include the number of acetylation, the unsaturation of the tail, or the number of glucosidic moieties. For instance, a large minimal surface tension range is observed with two homologues of purified TL (19.0 – 43 mN/m). In the case of more or less complex batches, synergistic effects, also known for chemical surfactants,[1] can strongly influence both ST and CMC, as shown by Hirata *et al.*, according to who the natural ratio between lactonic and acidic SL provide the lower minimal surface tension.[158]

Physicochemical conditions like ionic strength, type of ions and pH of course play an important role because, if most alkylpolyglycosides are neutral surfactants, biosurfactants have a chargeable chemical group like COOH. Some studies focusing on mono- (**19**) and di-RL (**18**) show no variation of the minimal surface tension with addition of NaCl < 1 M for di-RL (**18**), and a very weak decrease for RL. However, a visible increase in phosphate buffer at the same pH is observed, indicating the stronger influence of hydrogenophosphate ions. Moreover, the weak NaCl-dependence about surface tension is an indication of a non-ionic surfactant.[144,195] From an acidic to a more basic pH, minimal surface tension increase for both mono- (**19**) and di-RL (**18**). Furthermore, the higher CMC at basic pH provide information about the assumption of molecules negatively charged, due to carboxyl group.[144,196]

In summary, the complex structure and dual neutral/charged nature of most biosurfactants push towards a more detailed structural study of the aggregation behavior in water, thus making the study of CMC or surface tension poorly informative. In the next section, we will focus on the advanced self-assembly behavior of biosurfactants.

### 2.3 Self-assembly and phase diagramme

To the best of our knowledge, the first study on biosurfactants' self-assembly can be traced back to 1987, when the aggregation of RL (**18**, **19**) was explored as a function of pH.[159] The first study of surfactin self-assembly is published in 1995, of MEL in 2000,[189] and sophorolipids (SL) and cellobioselipids (CL) in 2004[197] and 2012,[198] respectively. The self-assembly properties of lipids and surfactants are studied since the 50's and their rationalization has occurred in the 70's with the work of Tanford[38] and Israelachvili and coworkers (refer to section 1.4).[39,199] This comparison shows that a gap of at least 10 years, but more realistically 25-30 years, exists between the development of fundamental concepts in surfactants' science and their employment in the field of biosurfactants.

Several reasons explain such gap. In particular, the chemical structure of biosurfactants is more complex than common head-tail surfactants; biosurfactants are often bolaform and in many cases they have ionizable chemical groups like carboxylic acids for most glycolipids or lipopeptides. In this regard, external stimuli like pH, ionic strength or temperature are particularly affecting their phase behavior. This multi-functionality strongly favors additional weak interactions in the self-assembly process, such as hydrogen bond or pi-pi stacking at the same time as ionic, steric, van der Waals and entropic forces. For this reason, straightforward predictions of the morphology of biosurfactants' aggregates and phase behavior in water often fails. The packing parameter theory based on molecular shape can sometimes be used to explain biosurfactants' aggregation, but the classical theory described in section 1.4 is not adapted for this class of functional compounds, for which the notion of clearly distinct hydrophilic and hydrophobic regions is sometimes dim, as clearly stated for surfactin.[182]

Specific considerations about the packing parameter theory associated to biosurfactants will be discussed at the end of this section. This section then focuses on the aqueous self-assembly and phase behaviour of biosurfactants given in Figure 7, making a distinction between dilute and concentrated solutions. Table 5 and Table 6 express in different ways the summary of phase behavior of biosurfactants highlighted in Figure 7. Additionally, we address the unexperienced reader to section 1.7 for a qualitative explanation of the most relevant characterization techniques applied to the study of biosurfactants.

Table 5 – Summary of the self-assembly and phase behaviour in water of biosurfactants shown in Figure 7. The data are organized by diluted and concentrated systems and by type of phase. Note, *: for these BS, clear-cut attribution of SUV, MLV or lamellar phase is not possible on the sole basis of the data provided in the corresponding references; authors of this work provide a tentative attribution after the SAXS/SANS or microscopy data reported. Please, refer to discussion in sections 2.3.3 and 2.3.4 for more information.

| Biosurfactant N° | Biosurfactant name | Conditions | Ref. |
|---|---|---|---|
| **Diluted (> CMC or CAC), < 0.1-1 wt% (often up to 5-10 wt%)**<br>**Major phase** | | | |
| 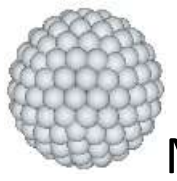 Micellar ($L_1$) (spherical, ellipsoidal) | | | |
| *Sophorolipids and derivatives of SL* | | | |
| 1 | Nonacetyl. acidic sophorolipids (C18:1-*cis*) | 10 < pH < 4 | 200,201 |
| 4 | Aminyl (-$NH_2$) sophorolipids (C18:1-*cis*) | 10 < pH < 4 | 202 |
| 12, 13, 14, 16 | Acidic (C18:1-*trans*, C18:0, C16:0, C22:$0_{13}$) sophorolipids | Basic (pH> 7.5) | 201,203–205 |
| 5 | Aminyl (-$NH_2$) sophorolipids (C18:0) | Acidic (pH< 7.5) | 202 |
| 6 | Alkynyl (-C≡CH) sophorolipids (C18:1-*cis*) | T> ~52°C | 202 |
| 2 | Acetylated acidic sophorolipids (C18:1-*cis*) | 7 < pH < 2 | 200,206 |
| *1i, 1h, 1f, 1k, 1j, 1l, 1g, 2g* in Ref. [207] | Ammonium- and amine oxide derivatives of SL | C < 10 wt% | 207 |
| 8 | Symmetrical bola C18:1-*cis* sophorolipids | C < 10 wt% | 16 |
| 9 | Symmetrical bola C16:0 sophorosides | T> 28°C | 208 |
| *3b, 3c, 3d, 4d, 5c, 5d, 6a, 6d*, in Ref. 207 | Symmetrical amine SL derivatives | Neutral form at basic pH | 207 |
| 7<br>($R^1$=$R^2$=OAc) | Peracetylated porphyrin sophorolipids | T< $T_e$ (~36°C) | 209 |
| | | | |
| *Glucolipids* | | | |
| 11, 10 | Acidic glucolipids (C18:1-*cis*, C18:0) | Basic (pH> 7.5) | 69,201 |
| | | | |
| *Cellobioselipids* | | | |
| 17 | Cellobioselipids (hydrolyzed) | Basic (pH> 7.5) | 201 |
| | | | |
| *Rhamnolipids* | | | |
| 18, 19 | Rhamnolipids (di-, mono-) | Neutral/Basic (pH> 6.8) | 144,159,210,211 |
| | | | |
| *Surfactin* | | | |
| 20 | Surfactin | Basic (pH> 7.5) | 182,212 |
| | | | |
| 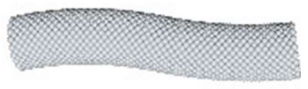 Micellar ($L_1$) (cylindrical, wormlike) | | | |
| *Sophorolipids* | | | |
| 1 | Acidic sophorolipids (C18:1-*cis*) | pH 4.5, 2 < C/wt% < 5 | 213 |
| | | | |
| *Glucolipids* | | | |
| 11 | Acidic glucolipids (C18:1-*cis*) | Neutral (6.5-7.5) | 69,201 |
| | | | |
| *Surfactin* | | | |
| 20 | Surfactin | Neutral (6.5-7.5) | 212 |
| | | | |
| 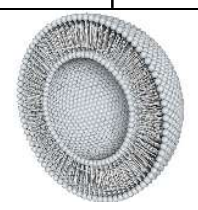 Unilamellar vesicle ($L_4$, *V*, *Ves*, *SUV* or $L_{\alpha 1}$) | | | |
| *Sophorolipids* | | | |
| 15 | Acidic sophorolipids (C18:3-*cis*) | Neutral pH | 214 |
| 3 | Lactonic sophorolipids (C18:1-*cis*) | Neutral (1 < C/mM < 5) | 200 |
| | | | |
| *Glucolipids* | | | |
| 11 | Acidic glucolipids (C18:1-*cis*) | Acidic (7 < pH < 4.5) | 69,201 |
| 10 | Acidic glucolipids (C18:0) | Acidic (7 < pH < 4.5), T> $T_m$ | 69,201 |
| | | | |
| *Mannosylerythritol lipids* | | | |
| 21A ($R^1=R^2=Ac$) | MEL-A | Neutral, < CAC2 ($2 \times 10^{-5}$ M) | 188–190 |
| 21B ($R^1$= Ac, $R^2$=H) | MEL-B | Neutral, < CAC ($6 \times 10^{-6}$ M) | 188–190,215 |
| 21C ($R^1$= H, $R^2$=Ac) | MEL-C | Neutral | 189 |
| | | | |
| Surfactin | | | |
| 20 | Surfactin* | Acidic (< 6.5)* | 182,212* |
| | | | |
| 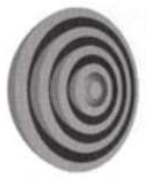 Multilamellar vesicle (*MLV*) | | | |
| *Sophorolipids* | | | |
| 16 | Acidic sophorolipids ($C22:0_{13}$) | Neutral/acidic (pH < ~7) | 205 |
| | | | |
| *Glucolipids* | | | |
| 11 | Acidic glucolipids (C18:1-*cis*) | (pH 3 →) pH 6 (*Lam* to *MLV* phase change) | 216 |
| | | | |
| *Mannosylerythritol lipids* | | | |
| 21B ($R^1$= Ac, $R^2$=H) | MEL-B | Neutral, > CAC (6 x $10^{-6}$ M) | 188–190,215 |
| 21C ($R^1$= H, $R^2$=Ac) | MEL-C | Neutral | 189 |
| 21D ($R^1$= H, $R^2$=H) | MEL-D | Neutral | 215 |
| | | | |
| *Rhamnolipids* | | | |
| 18, 19 | Rhamnolipids (di-, mono-, di/mono) * | Acidic (pH < ~7) * | 159,178,195,217* |
| 18, 19 | Rhamnolipids (mono-, di-/mono-)* | Basic pH 9, C> 20-40 mM* | 144* |
| | | | |
| *Surfactin* | | | |
| 20 | Surfactin* | $Ba^{2+}$ (pH 7.5)* | 212* |
| | | | |
| 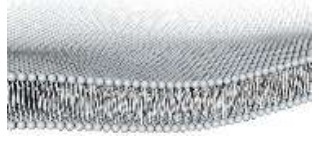 Flat lamellar | | | |
| *Glucolipids* | | | |
| 10 | Acidic glucolipids (C18:0) | Neutral/Acidic (pH < 7.8) | 69,201,218 |
| 11 | Acidic glucolipids (C18:1-*cis*) | Acidic (7 < pH < 4.5), $T< T_m$ | 69,201 |
| | | | |
| *Rhamnolipids* | | | |
| 18, 19 | Rhamnolipids (mono-, mono-/di-)* | pH < 6* | 210,211* |
| | | | |
| *Surfactin* | | | |
| 20 | Surfactin* | Acidic (pH < 5.5) * | 212* |
| | | | |
| 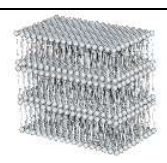 Flat lamellar (condensed) | | | |
| *Sophorolipids* | | | |
| 16 | Acidic sophorolipids ($C22:0_{13}$) | pH < 4 | 205 |
| | | | |
| *Glucolipids* | | | |
| 11, 10 | Acidic glucolipids (C18:1-*cis*, C18:0) | pH < 4 | 69,201 |
| | | | |
| *Surfactin* | | | |
| 20 | Surfactin* | $Ba^{2+}$ (pH 7.5)* | 212* |
| | | | |
| 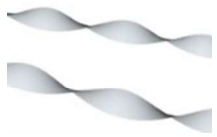 Fiber | | | |
| *Sophorolipids* | | | |
| 12, 13, 14 | Acidic sophorolipids (C18:0, C18:1-*trans*, C16:0) | Neutral/acidic 7.5 < pH < 3 | 201,203,204,219–221 |
| 5 | Aminyl ($-NH_2$) sophorolipids (C18:0) | Neutral/basic 10 < pH < 7.5 | 202 |
| 1/13 | Acidic sophorolipids (C18:1-*cis*/C18:0 mixtures) | Neutral/Acidic, pH < ~7 | 191,197 |
| 6 | Alkynyl sophorolipids (C18:1-*cis*) | T< ~52°C | 202 |
| 9 | Symmetrical bola C16:0 sophorosides | T< 28°C | 208 |
| | | | |
| *Cellobioselipids* | | | |
| 17 | Cellobioselipids (hydrolyzed) | Neutral/Acidic, pH < ~7 | 201 |
| | | | |
| 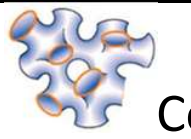 Coacervate, $L_3$ | | | |
| *Mannosylerythritol lipids* | | | |
| 21A ($R^1$= $R^2$=Ac) | MEL-A | Neutral, > CAC2 (2 x $10^{-5}$ M) | 188–190 |
| | | | |
| 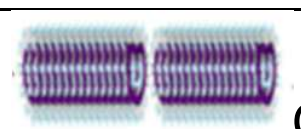 Columnar | | | |
| *Sophorolipids* | | | |
| 7 ($R^1$= OAc, $R^2$=OH) ($R^1$=$R^2$=OH) | Di-/de-acetylated porphyrin sophorolipids | T< $T_e$ (~36°C) | 209 |
| | | | |
| **Diluted (Minor phase)** | | | |
| | | | |
| 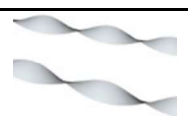 Fiber | | | |
| *Sophorolipids* | | | |
| 12 | Acidic sophorolipids (C18:1-*trans*) | Basic (pH> 8, time> 1h) | 203 |
| | | | |
| 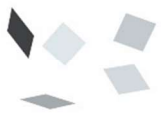 Nanoplatelets | | | |
| *Sophorolipids* | | | |
| 1, 13 | Acidic sophorolipids (C18:1-*cis*, C18:0) | Neutral/basic (pH> 8) | 220 |
| | | | |
| Filaments | | | |
| *Cellobioselipids* | | | |
| 17 | Cellobioselipids (hydrolyzed) | Neutral/basic (pH> 8) | 201 |
| | | | |
| Bilayer fragments | | | |
| *Glucolipids* | | | |
| 11, 10 | Acidic glucolipids (C18:1-*cis*, C18:0) | Neutral/basic (pH> 8) | 201 |
| | | | |
| Ill-defined | | | |
| *Sophorolipids* | | | |
| 3 | Diacetylated lactonic sophorolipids (C18:1-*cis*) | Neutral (C> 7 mM) | 200 |
| 16 | Acidic sophorolipids ($C22:0_{13}$) | pH> ~7 | 205 |
| 2 | Acetylated acidic sophorolipids (C18:1-*cis*) | Neutral/acidic pH | 109,200 |
| 5 | Aminyl (-$NH_2$) sophorolipids (C18:0) | Acidic (pH< 7.5) | 202 |
| | | | |
| *Rhamnolipids* | | | |
| 18, 19 | Rhamnolipids (mono-, di-) | pH 6.2, NaCl> 500 mM | 222 |
| | | | |
| **Concentrated, > 10 wt%** | | | |
| | | | |
| 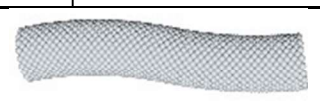 Micellar ($L_1$) (cylindrical, wormlike) | | | |
| *Sophorolipids* | | | |
| 1 | Acidic sophorolipids (C18:1-*cis*) | pH 4.5, C< 20 wt% | 213 |
| | | | |
| 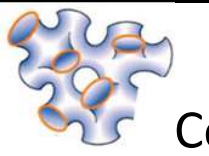 Coacervate, $L_3$ | | | |
| *Mannosylerythritol lipids* | | | |
| 21A ($R^1$=$R^2$=Ac) | MEL-A | Neutral, C< ~ 55 wt% | 223 |
| | | | |
| 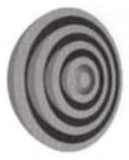 MLV | | | |
| *Mannosylerythritol lipids* | | | |
| 21B (R[1]= Ac, R[2]=H) | MEL-B (*R*-, *S*-diastereoisomers) | Neutral, C< ~ 60 wt% | 215 |
| 21D (R[1]= H, R[2]=H) | MEL-D (*R*-, *S*-diastereoisomers) | Neutral, C< ~ 60 wt% | 215 |
| | | | |
| 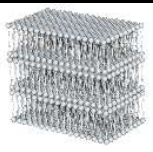 *Lamellar, $L_\alpha$* | | | |
| *Mannosylerythritol lipids* | | | |
| 21A (R[1]=R[2]=Ac) | MEL-A | Neutral, C> ~ 65 wt% | 223 |
| 21B (R[1]= Ac, R[2]=H) | MEL-B | Neutral, C> ~ 60 wt% | 190,215,224,225 |
| 21C (R[1]= H, R[2]=Ac) | MEL-C | Neutral (C not defined) | 152,226,227 |
| 21D (R[1]= H, R[2]=H) | MEL-D | Neutral, C> ~ 60 wt% | 225,228 |
| | | | |
| *Cubic, $V_2$* | | | |
| *Mannosylerythritol lipids* | | | |
| 21A (R[1]=R[2]=Ac) | MEL-A | Neutral, 55 < C/wt% < 65 | 223 |

Table 6 – Condensed overview of Table 5, sorted out by type of biosurfactant. Unless otherwise stated, temperature is at room temperature. Note, *: for these BS, clear-cut attribution of SUV, MLV or lamellar phase is not possible on the sole basis of the data provided in the corresponding references; authors of this work provide a tentative attribution after the SAXS/SANS or microscopy data reported. Please, refer to discussion in sections 2.3.3 and 2.3.4 for more information. #: 1i, 1h, 1f, 1k, 1j, 1l, 1g, 2g (Ammonium- and amine oxide derivatives of SL) and 3b, 3c, 3d, 4d, 5c, 5d, 6a, 6d (Symmetrical amine SL derivatives) in Ref. [207]

| BS | Name | N° | Micellar ($L_1$) | | Vesicle | | Lamellar | | Fiber | Coacervate ($L_3$) | Columnar | Cubic ($V_2$) | Minor phases | | | Ref. |
|---|---|---|---|---|---|---|---|---|---|---|---|---|---|---|---|---|
| | | | sphere | cylinder | SUV | MLV | Flat | Condensed | | | | | Platelets | Filaments | Ill-defined | |
| | | | 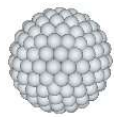 | 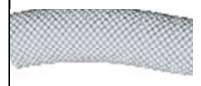 | 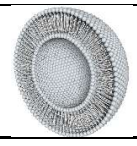 | 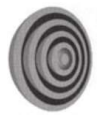 | 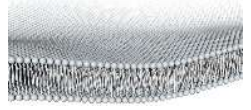 | 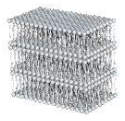 | 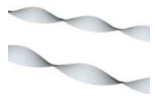 | 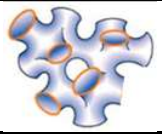 | 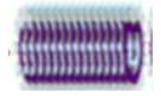 | | 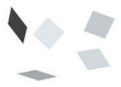 | | | |
| Sophorolipids | Nonacetyl. acidic (C18:1-cis) | 1 | 10 < pH < 4, C< 5 wt% | pH 4.5, 5 < C/wt% < 20 | | | | | Mix with N° 13 pH < ~7 | | | | pH> 8 | | | 191,197,200,201,213,220 |
| | Acetylated acidic (C18:1-cis) | 2 | 7 < pH < 2 | | | | | | | | | | | | pH< ~7 | 109,200 |
| | Lactonic (C18:1-cis) | 3 | | | Neutral (1 < C/mM < 5) | | | | | | | | | | Neutral (C> 7 mM) | 200 |
| | Aminyl (-$NH_2$) (C18:1-cis) | 4 | 10 < pH < 4 | | | | | | | | | | | | | 202 |
| | Aminyl (-$NH_2$) (C18:0) | 5 | pH< 7.5 | | | | | | 10 < pH < 7.5 | | | | | | pH< 7.5 | 202 |
| | Alkynyl (-C≡CH) (C18:1-cis) | 6 | T> ~52°C | | | | | | T< ~52°C | | | | | | | 202 |
| | Peracetylated porphyrin sophorolipids | 7 | ($R^1$=$R^2$=OAc) T< $T_e$ (~36°C) | | | | | | | | ($R^1$= OAc, $R^2$=OH) ($R^1$=$R^2$=OH) T< $T_e$ (~36°C) | | | | | 209 |
| | Symmetrical bola C18:1-cis sophorolipids | 8 | C < 10 wt% | | | | | | | | | | | | | 16 |
| | Symmetrical bola C16:0 sophorosides | 9 | T> 28°C | | | | | | T< 28°C | | | | | | | 208 |
| | Acidic (C18:1-trans) | 12 | pH> 7.5 | | | | | | 7.5 < pH < 3;<br><br>pH> 8, time> 1h (minor phase) | | | | | | | 203,219 |
| | Acidic (C18:0) | 13 | pH> 7.5 | | | | | | 7.5 < pH < 3 | | | | pH> 8 | | | 201,203,219–221 |
| | Acidic (C16:0) | 14 | pH> 7.5 | | | | | | 7.5 < pH < 3 | | | | | | | 204 |
| | Acidic (C18:3-cis) | 15 | | | Neutral pH | | | | | | | | | | | 214 |
| | Acidic ($C22:0_{13}$) | 16 | pH> 7.5 | | | pH < ~7 | | pH < 4 | | | | | | | pH> ~7 | 205 |
| **Other sophorolipids#** | - | - | C < 10 wt%<br><br>Neutral form at basic pH | | | | | | | | | | | | | 207 |
| | | | | | | | | | | | | | | | | |
| **Glucolipids** | Acidic (C18:0) | 10 | pH> 7.5 | | 7 < pH < 4.5) T> $T_m$ | | Neutral/Acidic (pH < 7.8) | pH < 4 | | | | | | | pH> 8 | 69,201,216,218 |
| | Acidic (C18:1-cis) | 11 | pH> 7.5 | 6.5< pH < 7.5 | 7 < pH < 4.5 | (pH 3 →) pH 6 (Lam to MLV phase change) | 7 < pH < 4.5, T< $T_m$ | pH < 4 | | | | | | | pH> 8 | 69,201 |
| | | | | | | | | | | | | | | | | |
| **Cellobioselipids** | Hydrolyzed | 17 | pH> 7.5 | | | | | | pH < ~7 | | | | | pH> 8 ( | | 201 |
| | | | | | | | | | | | | | | | | |
| **Rhamnolipids** | di- | 18 | pH> 6.8 | | | pH < ~7 and pH 9 ( C> 20-40 mM) | pH < 6* | | | | | | | | pH 6.2, NaCl> 500 mM | 144,159,178,195,210,211,217,222 |
| | Mono- | 19 | pH> 6.8 | | | pH < ~7 and pH 9 ( C> 20-40 mM)* | pH < 6* | | | | | | | | pH 6.2, NaCl> 500 mM | 144,159,178,195,210,211,217,222 |
| | | | | | | | | | | | | | | | | |
| **Surfactin** | Cyclic | 20 | pH> 7.5 | 6.5< pH < 7.5 | pH < 6.5 | $Ba^{2+}$ (pH 7.5) * | pH < 5.5* | $Ba^{2+}$ (pH 7.5) * | | | | | | | | 182,212 |
| | | | | | | | | | | | | | | | | |
| **Mannosylerythritol lipids** | MEL-A | 21A | | | Neutral, < CAC2 (2 x $10^{-5}$ M) | | $L_\alpha$ Neutral, C> ~ 65 wt% | | | Neutral, CAC2 (2 x $10^{-5}$ M) < C < ~ 55 wt% | | 55 < C/wt% < 65 | | | | 188–190,223 |
| | MEL-B | 21B | | | Neutral, < CAC (6 x $10^{-6}$ M) | Neutral pH, CAC (6 x $10^{-6}$ M) < C< ~ 60 wt% | $L_\alpha$ Neutral, C> ~ 60 wt% | | | | | | | | | 188–190,215,224,225 |
| | MEL-C | 21C | | | Neutral | Neutral pH | $L_\alpha$ Neutral (C not defined) | | | | | | | | | 152,189,226,227 |
| | MEL-D | 21D | | | | Neutral pH | $L_\alpha$ | | | | | | | | | 215,225,228 |
| | | | | | | C< ~ 60 wt% | Neutral, C> ~ 60 wt% | | | | | | | | | |

Diluted systems (CAC,CMC < C < 5-10 wt%)

Most of the work related to the self-assembly properties of biosurfactants has been done for dilute solutions, generally between the CMC/CAC and about 5-10 wt%. For this reason, we make a distinction between diluted (this subsection) and concentrated (next section) biosurfactant systems. We organize this subsection by type of phase (micellar, vesicle, lamellar, etc…) and for each phase we report the data corresponding to specific biosurfactants. Such organization corresponds to Table 5. On the contrary, for an overview of the phase behavior associated to a given family of biosurfactants, one cas refer to Table 6.

### 2.3.1 Micellar phase ($L_1$)

Micelles are certainly the most common self-assembled morphology observed for classical head-tail surfactants under dilute conditions. The corresponding packing parameter for spherical micelles is 0 < *PP* < 0.33 and for elongated, rod- until worm-like micelles, is 0.33 < *PP* < 0.5 (section 1.4, Figure 2b). Micelles are very well characterized by SAXS/SANS (spherical micelles: lack of q-dependency followed by -4 q-dependency of the intensity; cylindrical micelles: -1/-4 I(q) dependency, always in log-log scale) but important morphological information could be obtained by cryogenic TEM (acronyms and notions of the analytical tools are given in section 1.7 for non-experts).

***Rhamnolipids*** (**mono-** (**19**), **di-** (**18**))

Aggregation behavior of rhamnolipids was firstly studied using natural mixture, generally produced by *Pseudomonas aeruginosa* as a not far from 50:50 mixture of mono- and di-rhamnolipids. Each rhamnose is monoacetylated and the length of alkyl chain may vary between 8 – 14 carbons. However, various publications provide data on both mixtures and mono- (**19**) or di-RL (**18**), purified by chromatography.

The first work approaching the self-assembly of rhamnolipids is published in 1987 by Ishigami *et al.*[159] A micellar behavior under neutral/basic conditions (pH > 6.8) is deduced from dynamic light scattering (DLS) and tensiometry while bilayer structures are observed by electron microscopy at lower pH. Due to high potential of weakly charged micelles of natural RLs to oil dispersion, discovered in 1992,[229] the self-assembly as a function of pH was confirmed with cryo-transmission electron microscopy (cryo-TEM) by Champion *et al.*,[211] who described a micellar phase at pH 8. Others confirm these results on raw mixtures (pH 13.2)[230] and an artificial 1:1 (w/w) mixture of mono-RL and di-RL (pH 7-8).[210] SANS experiments probing individual congener solutions have shown that at pH 9 di-RL prefer a micellar environment in a concentration range up to 100 mM while mono-RL prefer a micellar phase below 50 mM (Figure 8). For intermediate mono- and di-RL mixtures, the phase behaviour is entirely micellar up to 40% in mono-RL and mixed micellar-flat membrane at higher mono-RL fractions for concentrations above 20 mM.[144] According to the best fit of SANS data, the inner core and outer radius of di-RL micelles are 12 and 15 Å, respectively, in good agreement with a gyration radius of 17-18 Å, previously reported in water at pH 13.2 in presence of $Na^+$ and $K^+$ for a RLs mixture.[230] Complementary SAXS data recorded on a commercial 1:0.35 mono-/di-RL mixture at pH 7 has shown slightly different values. Authors found that RL micelles form prolate ellipsoids of revolution with a core radius of 7.7 Å, a shell thickness of 12 Å and an aspect ratio of 2.4.[231] Differences between the two studies were also found in the aggregation numbers, being of 11 in Ref. [231] and 37 in a mono-/di- RL mixture.[144] Possible discrepancies could be due to the exact mono-/di-RL content, use of $D_2O$ against $H_2O$, respectively in SANS[144] and SAXS,[231] or different overall RLs concentrations. It must be nonetheless said that SAXS data show a full oscillation of the form factor in the mid-q/high-q range, thus making modelling probably more reliable than in SANS data.

An interesting finding was proposed by Zhong *et al.*,[232] who explored the pre-micellar regime of di-RL using cryo-TEM and DLS. They find round-shaped aggregates at both pH 6 and 8 with hydrodynamic diameters ranging from roughly 50 nm and 10 nm.[232] If pre-micellar aggregation is not uncommon, and it was actually observed for head-tail and synthetic gemini surfactants,[172,233] aggregates of similar shaper, size and contrast as reported by Zhong *et al.* are often observed in cryo-TEM experiments and could be related to freezing artifacts. Nonetheless, pre-micellar aggregates for mono-RL were reported by others,[234] but the typical size is rather in the order of 2.5 nm, corresponding to few RL monomers, that is, more than one order of magnitude smaller than the type of aggregate reported by Zhong *et al*. Whatever the case, the nature of pre-micellar RLs aggregates probably requires to be confirmed by further studies combining microscopy and scattering.

*Comment on experiments performed at basic pH*. RL are held by ester bonds between rhamnose and the hydroxyl fatty acids as well as between the fatty acids themselves. One should then consider the fact that highly basic pH (at least pH> 9) values could hydrolyze, if not all, at least a fraction of the RL into a mixture of hydroxyl fatty acids, RL and free rhamnose. To the best of our knowledge, there is no evidence that such a mechanism occurs in the cited literature. However, it must be said that none of the papers involving RL experiments at high pH specifically investigates the stability of RL. In this regard, the corresponding phase behavior could be potentially biased.

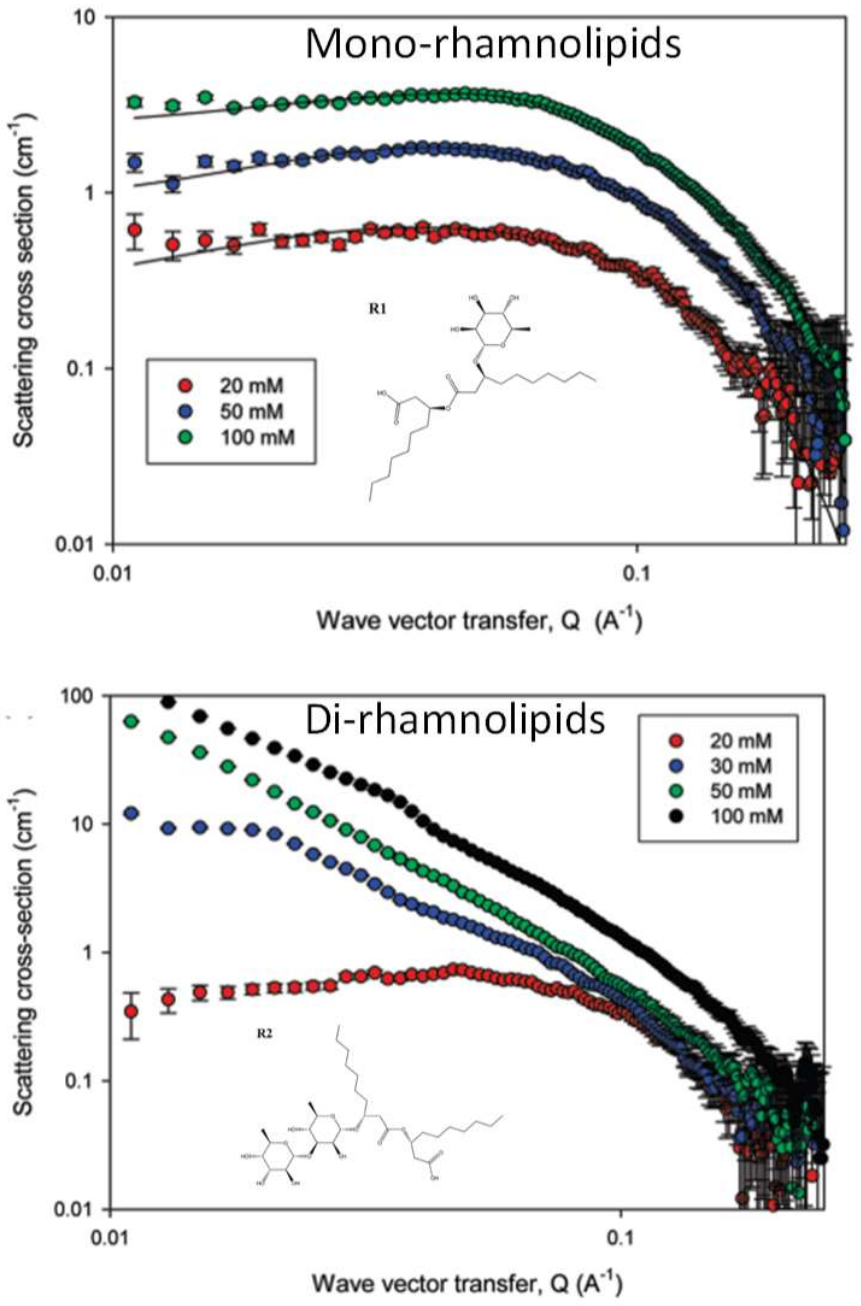

Figure 8 – Concentration-dependent SANS experiments recorded for mono- and di-RL at pH 9. Reprinted with permission from [144]. Copyright (2010) American Chemical Society.

## ***Sophorolipids*** (**1-3**)

The general structure of sophorolipids (**0**) is given in Figure 7. Lactonic (**3**) and acidic (**1, 2**) form of sophorolipids generally compose a typical SL batch in a 80:20 molar ratio. The alkyl chain, generally C18, may be saturated, mono- or di- unsaturated (*cis* or *trans*), while the sophorose group can be either nonacetylated or mono/diacetylated but also, in some specific cases, peracetylated. Speaking of the self-assembly behavior of SL is then imprecise, given the fact that phase is generally associated to a specific congener or to a mixture of congeners. In the following, we try to describe the main results reported for SL by evoking the type of SL congener at the beginning of each dedicated paragraph.

*Nonacetylated C18:1-cis acidic* (**1**). Zhou *et al*. reported an angle-dependent DLS study of the micellar aggregation of (**1**) at pH 7.8 and 1.4 mg/mL, supposing anisotropic aggregates characterized by a bimodal (~3 nm and ~50 nm) apparent radius of gyration at high angles and a monomodal radius at low angles (~150 nm). The same authors reported crystalline ribbon structures at lower pH.[197] Micelle formation of the same congener, and in particular insight on micellar structure, charge and interactions are reported by Baccile *et al.*,[213,235–237] combining X-ray and neutron scattering and cryo-TEM, and by Penfold *et al*.,[200] employing neutron scattering.

An early evidence of micelle formation through SANS was given in 2010,[213] where authors show a transition from spherical to cylindrical micelles between 0.05 wt% and 5% wt% at the equilibrium pH of about 4.5, together with an effect of pH, with the appearance of a broad correlation peak, typical of repulsive interactions, at pH 6.5. Penfold *et al*.[200] confirmed these data in 2011 while working on four different partially purified acidic sophorolipids. The sample referred to S3 and containing 93.2% of nonacetylated acidic SL and 6.8% of lactonic SL is the closest to a single-component nonacetylated acidic SL. However, the corresponding structural parameters (aggregation number of 28 ± 2, ellipsoidal shape of radii 12 and 15 Å) were given for its mixture with 20% content of lactonic SL (LSL). To the best of our knowledge, no quantitative information were given for the sample S3 as such. In that work, the SANS data were modelled with a core-shell ellipsoid of revolution model, classical for many surfactant micelles and which nicely interprets the strongly hydrophilic sophorose (shell) and hydrophobic oleic acid (core) regions.

Baccile *et al.* have further employed the same core-shell ellipsoid of revolution model to study the effects of pH and counter ions with SANS[235,236] but also the intimate structure of the sophorolipid micelles,[237] by combining SAXS, anomalous SAXS, SANS and molecular dynamics simulations. If the use of SANS is generally enough to estimate the shape and the size of the micelles, the strong contrast between the solvent ($D_2O$), the hydrophilic core and hydrophobic shell produces a flattened scattering signal, where the oscillation of the form factor is generally not observed. The consequent fit of SANS data is then possible with a broader set of structural parameters. On the contrary, SAXS data show a well-defined oscillation of the form factor, which imposes a strong constraint in the fitting process, thus being more selective towards the choice of the structural parameters. pH-resolved experiments have shown that acidic nonacetylated sophorolipids form a micellar phase in a wide pH range, from 10 to 3.[69,235]

The combination of SANS and SAXS experiments at pH ~5 (low ionization degree) agree on the formation of non-ionic spheroidal micelles, of which the cross-sectional radius (core + shell) is in the order of 20 Å. The degree of anisotropy (core + shell) depends on concentration and it was found an aspect ratio between 1.5 and 1.8 at concentration of 0.5 wt%.[69] Increasing pH, that is the $COO^-$/COOH ratio, contributes to increase the negative charge density on sophorolipids micelles, thus introducing repulsive electrostatic interactions in the pH range between 6 and 9. This is demonstrated by the appearance of a broad peak in both SANS/SAXS and centered between 0.07 and 0.1 $Å^{-1}$.[213,235–237] The electrostatic nature of these interactions is demonstrated by the progressive disappearance of the correlation peak with increasing ionic strength, and in particular with addition of few mM of divalent cations, like $Ca^{2+}$.[236]

Upon ionization of the COOH group, micelles become negatively charged and the scattering signal is dominated by two concomitant phases, one related to spherical micelles above 0.02 $Å^{-1}$ and one related to large-scale objects below 0.02 $Å^{-1}$.[220,235] In the micellar phase, the size is strongly reduced at more alkaline pH values: the hard-sphere radius declines below 30 Å[236] while the cross-sectional radius (core + shell) falls below 20 Å and even below 15 Å at pH above 10.[69,236] The reduction of the micellar size can be explained by the strong density of negative charges, which strongly increase the micellar curvature due to electrostatic repulsions.

The structure of C18:1-*cis* micelles and the distribution of the carboxylate group was specifically studied by a combination of SAXS, anomalous SAXS, SANS and molecular dynamics simulation, concluding to an atypical micellar model structure defined as "*coffee bean*" (Figure 9a). Three regions of different polarity were identified: a globular aliphatic core, a mixed aliphatic/COOH/sophorose/water ellipsoid and a sophorose/water shell.[237] Sophorolipids were supposed to adopt both a bent configuration at the polar regions of the ellipsoid but they also cross the micelle at its equatorial region, whereas the cross-sectional diameter is compatible with the size of a single sophorolipid molecule. Anomalous SAXS experiments, probing the counterion distribution around the micelles, seemed to indicate that the carboxylate groups are not specifically located at the micelle-water palisade, as in classical cationic or anionic micelles, but in a more diffuse region. It goes without saying that the "*coffee bean*" model is a crude simplification and not a real picture, but it is in quite good agreement with the self-assembly results obtained by molecular dynamics simulations.

These studies above indicate that nonacetylted acidic C18:1-*cis* SL (**1**) form a micellar phase in a broad pH range, from acidic to basic. This is in contrast with ribbon formation at acidic pH reported in [197]. Such discrepancy will be discussed in detail in the fiber phase (section 2.3.2).

*Acetylated acidic* (**2**). The effect of acetylation was studied by Penfold[200] and Baccile,[109] although both works are carried out on mixtures of congeners, due to the complexity of having single-components batches of mono- or di-acetylated SL. According to SANS, lactone-free samples with 47.5% acetylation (mono- and di-), corresponding to sample S4, exhibit globular core-shell micelles with radius of 2.4 – 2.7 nm for a bulk concentration *C* between 0.5 – 50 mM (Figure 9b). In the meanwhile, the sample having the highest degree of acetylation (89.4% diacetylated), but a 8.4% of lactone form (sample S1), shows a micellar/bilayer mixed phase (Figure 9c). If upon increase of the lactone congener fraction the micellar phase seems the most stable one (up to a total concentration of 30 mM) and 90% volume fraction of lactone (Figure 9c), it is still delicate to separate the lactone from the acetylation contributions to self-assembly.

Lactone-free acidic C18:1-*cis* SL batch having a global (mono- and di-) acetylation degree of 88% could be obtained by the engineered *S. bombicola* lactone esterase knockout strain (*Δsble*). SAXS of this sample shows a global tendency to form globular core-shell micelles at equilibrium acidic pH (4.8) under diluted conditions up to about 5 wt%, whereas coexistence with larger, ill-defined, structures were observed at pH 2.[109] The body of existing data suggests that, as long as the main SL compound is the nonacetylated acidic C18:1-*cis* SL, variations in the ratio between the acetylated and lactonic congeners, in ionic strength and type of salt or pH do not perturb the micelle formation but they only have an impact on the aggregation number and structural micellar parameters (core and shell radii).[109,200,237]

In summary, acidic C18:1-cis sophorolipids (**1**) are easy to obtain through alkaline hydrolysis of a raw SL mixture. Although the product is never 100% uniform and subject to variation in the congener content from batch to batch, it is generally water-soluble in a broad pH range[69] and it forms a micellar phase, at least up to 5-10 wt%. The effect of pH mainly plays a role in the length of the hydrophobic core and in the aspect ratio of the hydrophilic shell. At acidic pH, micelles are more ellipsoidal and the hydrophilic contribution is more prominent in the equatorial than in the axial direction of the ellipsoid. At basic pH, the micellar hydrophobic core becomes more spherical, while the hydrophilic contribution is more prominent in the axial direction than in the equatorial one. Finally, due to the risk of hydrolysis, acetylated (**2**) and lactone (**3**) SL cannot be evaluated at high pH in a reliable manner.

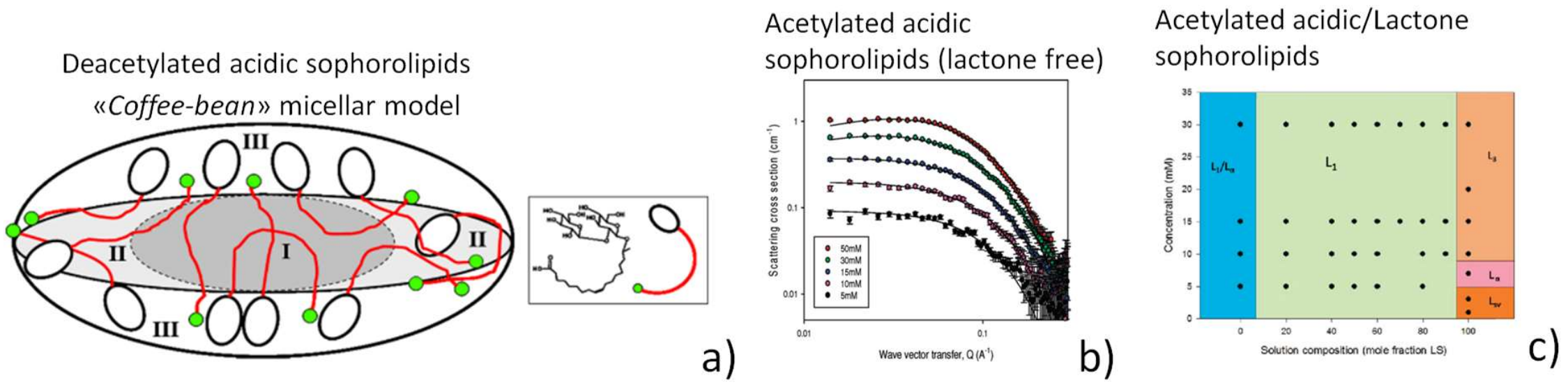

Figure 9 – a) "*Coffee bean*"-like model used to describe the micellar structure of deaetylated acidic C18:1-*cis* (**1**) sophorolipids.[237] Adapted with permission from [237]. Copyright (2015) American Chemical Society. b) SANS profiles showing a micellar behavior of partially acetylated acidic sophorolipids (sample S4, 47.5% acetylation) (**1**,**2**).[200] c) Concentration-dependent phase diagramme for acetylated acidic (sample S1)/lactonic sophorolipid mixture (**1**,**3**). b) and c) are adapted with permission from [200]. Copyright (2011) American Chemical Society.

*Other acidic nonacetylated SL (C18:0* (**13**)*, C18:1-trans* (**12**)*,* $C22:0_{13}$ (**16**))

Saturated C18:0 SL exhibit a micellar behavior at pH>7.4.[69] Formation of a micellar phase at basic pH seems to be a common feature for other less common sophorolipids, the C18:1-*trans*[203] and $C22:0_{13}$[205] derivatives, both showing a predominant micellar phase under alkaline conditions.

*Symmetrical bolaform sophorolipids* (**8**)

Symmetrical C18:1-*cis* sophorolipids were shown to form stable spherical micelles, of which the size (radius of 1.79 nm, determined by SAXS) does not evolve with concentration, which is quite an atypical behavior for surfactants.[16] On the contrary, similar C16:0 symmetrical sophorosides (SS) were only shown to form micelles above a critical temperature of about 28°C[208], while synthetic, neutral, symmetrical SL with amine derivatives form micelles at room temperature.[207]

*Sophorolipid derivatives.* If a number of sophorolipid derivatives has been prepared so far,[17] only a limited, although not negligible, share has been studied in terms of their self-assembly. Koh *et al.*[238] have studied SL ester derivatives by diffusion ordered nuclear magnetic resonance (DOSY-NMR). The carboxylic group of both di- and nonacetylated C18:1-*cis* SL (**2**,**1**) is modified by an alkyl chain of length between 1 and 10 carbon atoms. DOSY-NMR, probing the diffusion coefficient, combined with DLS demonstrate the presence of spherical micelles, of which the size slightly increases with the length of the alkyl chain until C8, above which the length is three times larger. The CMC was shown to decrease with the number of alkyl chain length for both di- and nonacetylated SL, as expected (see also section 1.3),[36] while surface tension increases for the di- and decreases for the nonacetylated SL as a function of chain length. The higher CMC (a factor ten) and inversed CMC trend for the diacetylated SL (**3**) was explained by the less hydrophilic character of the sophorose headgroup, reducing the surface activity of this family of compounds.

A large number of ammonium and amine oxide derivatives of SL were recently synthesized to develop antibacterial and gene transfection applications.[239–243] In these samples, the functional groups were located in place of the *cis* double bond after its ozonolysis, rather than on the COOH group. This approach generates a series of SL with shorter chain and interesting divalent and Y shaped morphologies. Interestingly, only those samples with long alkyl chain substitutes on the ammonium headgroup show the formation of micelles, of which the size is proportional to the chain length. All other derivatives having small-sized substituents show poor self-assembly properties.[207]

In a recent study, amine derivatives of C18:1-*cis* (**4**) and C18:0 SL (**5**) were prepared by grafting ethylene diamine on the COOH group.[202] Despite the expected charge inversion, from neutral/negative for COOH/$COO^-$ to neureal/positive for $NH_2/NH_3^+$, the self-assembly behaviour was shown to be exactly the same as for acidic SL but at inverted pH. Charged (whether positive of negative) C18:1-*cis* SL form small spheroidal micelles with an enhanced hydrophilic region along the polar direction, while the corresponding neutral (whether COOH or $NH_2$) molecule forms ellipsoidal micelles with an enhanced hydrophilic region on the radial direction of the ellipsoid. Similarly, the positively-charged C18:0 derivative[202] forms spheroidal micelles, as its negatively-charged counterpart.[201]

The functionalization potential of SL is practically infinite, especially if the COOH group is modified with either alkyne or azide groups, which can undergo the well-known copper-catalyzed cycloaddition "click" reaction. Azide-modified C18:1-*cis* SL were recently functionalized with zinc porphyrin complexes their interesting self-assembly were studied.[209] In particular, the per-acetylated derivative was supposed to form, above an onset temperature, micellar aggregates, of which the size and structure was however not fully characterized. Alkynyl SL (**6**) were also recently synthesized and a micellar phase was shown to form above 52°C. The micellar radius obtained by the Guinier region of the corresponding SANS pattern is 3.25 ± 0.03 nm, compatible with the size of the modified SL.[202]

***Other glycolipids*** *(cellobioselipids* (**17**)*, glucolipids* (**10**, **11**)*).*

The micellar phase is quite common for a number of other glycolipid systems, including cellobioselipids and glucolipids at basic pH.[69,201] In the case of glucolipids, the micellar phase is obtained at basic pH and it is described by a classical core-shell ellipsoid morphology, according to SAXS data analysis.[201]

***Surfactin*** (**20**)

Surfactin is affected by pH and ionic strength due to the presence of glutamic and aspartic acids in his hydrophilic ring, negatively charged at basic pH. pKa of surfactin is determined at 5.8 by alkaline titration.[157] Ishigami *et al.*[157] have first studied surfactin solutions (2 mM) in a $NaHCO_3$ buffer at pH 8.7. Knowing the molecular mass of the surfactin (1050 Da),[132] static light scattering measurements (SLS) determine the molecular weight of the aggregate in solution, so the aggregation number (173). Asymmetry of intensity ratio at 50° and 130° assumes a cylindrical shape, pushed by a β-sheet signal evidenced by circular dichroism (CD). Shen *et al.*[182,212] have studied surfactin at 1 mM in solution at pH 7.5 in perdeuterated water by SANS (Figure 10), neutron reflectivity and freeze fracture transmission electron microscopy (FF-TEM). Measurements reveal a micellar phase with an aggregation number around 20, a total micelle diameter of 5 nm and a spherical core of 2.2 nm while the monolayer at the air/water interface is found to be 1.4 nm according neutron reflectivity (NR), in agreement with the micellar size. For higher pH until 9.5, a stable micellar phase is deduced from SANS (Figure 10) while a phase transition between pH 7.5 and 6.5 is clearly shown by the stronger scattering at low-q. The scattering profile at pH 6.5 is attributed to a rod like micelle with a rod radius of 1.9 nm while the one at pH 5.5 indicates formation of a bilayer. Addition of monovalent ions at pH 7.5 does not affect the micellar phase while addition of divalent ions promote the formation of a flat lamellar structure, attributed to the higher binding affinity towards $COO^-$. Similar results are also reported by Zou *et al.*[244] in phosphate buffer at pH 7.4, where a dimensional analysis of SANS spectra at the low-q ( < 0.02 Å) region shows a power law dependence of the intensity as $\sim q^{-3}$, interpreted by a fractal-like surface region with a dense core. In addition to SANS, freeze fracture-TEM experiments show a coexistence between flat lamellar structure and rod-like micelle. Spherical-to-rod micelle transition induced by salt and ion binding affinity are common and well-known for synthetic surfactant like alkyl sulfate by increasing salt concentration.[245–247]

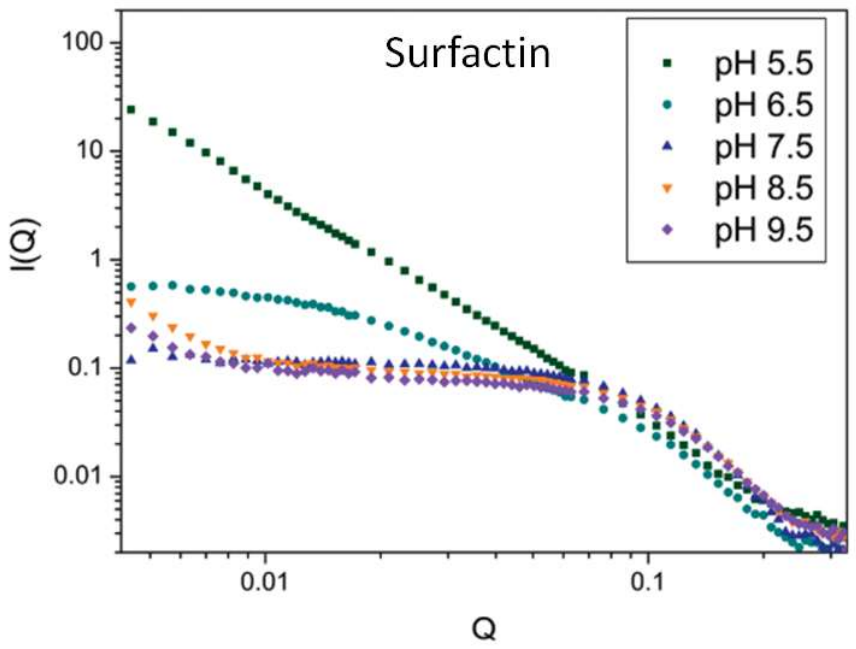

Figure 10 – pH-dependent SANS profiles recorded for surfactin (2 mM). Reprinted with permission from [212]. Copyright (2011) American Chemical Society.

### *General comments*

Table 7 summarizes the structural parameters of a number of biosurfactants' micellar systems prepared at acidic and basic pH. One can find the following general trends:

- SL with a neutralized end-group (-COOH at acidic pH or $–NH_2$ at basic pH) form ellipsoidal micelles with an axial radius of about two to three times the equatorial radius.
- When micelles are charged (may the charge be positive or negative), the shape becomes more spherical and the equatorial shell is thinner than the axial shell, $T_{a\text{-}s} > T_{e\text{-}s}$ (image in Table 7)
- SANS experiments (superscripts: *, §, °) on both SL and RL seem to underestimate the hydrophilic shell contribution with respect to SAXS (superscript #, ^). However, this could be explained by the complex electron density distribution between the shell and core regions along the equatorial and axial directions of the micelles: SANS data are fitted with a core-(homogeneous)shell model while SAXS data are fitted with core-(inhomogeneous)shell model, which better suits the SAXS form factor. Another explanation could be that SAXS data recorded on biosurfactants generally show the first oscillation of the form factor, of which the fitting is probably more reliable when more complex models are employed.
- When micelles are neutral, the equatorial shell is thicker than the axial shell, $T_{a\text{-}s} < T_{e\text{-}s}$.
- Despite the bulkiness of the peptide group, surfactin is reported to have an astonishingly thin shell thickness.

Table 7 – Biosurfactants (N° given in Figure 7) micelles in solution observed at room temperature at concentration below 5 wt%. Micellar structural parameters are extracted from fitting SAXS (#, ^) or SANS (*, §, °) data using core-shell spheroidal models. The typical cartoon of the model is reported below: $R_{e\text{-}c}$= Equatorial radius of the hydrophobic core, $T_{e\text{-}s}$ = Equatorial thickness of the hydrophilic shell, $R_{a\text{-}c}$= Axial radius of the hydrophobic core, $T_{a\text{-}s}$ = Axial thickness of the hydrophilic shell. Total equatorial radius= $R_{e\text{-}c} + T_{e\text{-}s}$, Total axial radius= $R_{a\text{-}c} + T_{a\text{-}s}$. All compounds are in their open acidic form and nonacetylated, unless specified otherwise. In §, * and °, micellar radii were adapted from Ref. [200] and [144] as follows: $R_{e\text{-}c} \equiv R_1$, $T_{e\text{-}s} \equiv T_{a\text{-}s} \equiv (R_2 - R_1)$ and $R_{a\text{-}c} \equiv (R_1 \times ee)$, with $ee$ = elliptical axial ratio. Values of $R_1$, $R_2$ and $ee$ are averaged between: 5 mM and 50 mM in * (Table 3 in ref [200]), 3 mM and 10 mM (Table 3 in ref [200]) in ° and 20 mM and 50 mM (Table 6 in Ref. [144]) in §. For the systems indicated by #, we have extracted the corresponding values from Figure 5 and Figure S4 in Ref. [69].

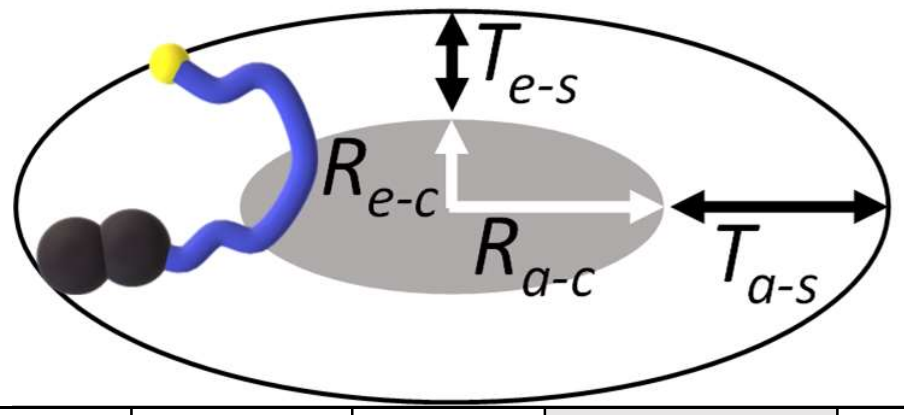

| BS | Type | N° | pH | $R_{e-c}$ | $T_{e-s}$ | *Total equatorial radius* ($R_{e-c}$ + $T_{e-s}$) | $R_{a-c}$ | $T_{a-s}$ | *Total axial radius* ($R_{a-c}$ + $T_{a-s}$) | Ref. |
|---|---|---|---|---|---|---|---|---|---|---|
| **Units** | | | | Å | | | | | | |
| RL§ | Di-RL | 18 | 9 | 11.7 | 3.3 | 15 | 30.3 | 3.3 | 33.6 | 144 |
| RL | Mono-/di-RL | 19,18 | 13.2 | - | - | 17.5 | - | - | - | 230 |
| RL^ | Mono-/di-RL | 19,18 | 7 | 7.7 | 12.0 | 19.7 | 18.5 | 12.0 | 30.5 | 231 |
| SL* | C18:1-*cis* (acetyl.) | 2 | < 6 | 15.3* | 5.3 | 20.6 | 45.0 | 5.3 | 50.3 | 200 Sample S1 |
| SL° | C18:1-*cis* (acetyl.) | 1,2 | < 6 | 11.4 | 2.8 | 14.2 | 31.7 | 2.8 | 34.5 | 200 Sample S4 |
| SL# | C18:1-*cis* | 1 | 5 | 13 | 7 | 20 | 52 | 2 | 54 | 69 |
| SL# | C18:1-*cis* | 1 | 9 | 4 | 9 | 13 | 7 | 31 | 38 | 69 |
| SL | Aminyl (-$NH_2$) C18:1-*cis* | 4 | 10.6 | 7 | 1 | 8 | 31 | 0 | 31 | 202 |
| SL | Aminyl (-$NH_2$) C18:1-*cis* | 4 | 4.6 | 7 | 11 | 18 | 5 | 19 | 25 | 202 |
| SL# | C18:0 | 13 | 9 | 9 | 5 | 14 | 14 | 5 | 19 | 69 |
| SL | Aminyl (-$NH_2$) C18:0 | 5 | 4.6 | 7 | 6 | 13 | 11 | 21 | 32 | 202 |
| SL | $C22:0_{13}$ | 16 | 9 | 9 | 10 | 19 | 9 | 10 | 19 | 205 |
| GL# | C18:1-*cis* | 11 | 9 | 6 | 9 | 15 | 11 | 9 | 20 | 69 |
| GL# | C18:0 | 10 | 9 | 7 | 10 | 17 | 19 | 10 | 29 | 69 |
| Surf | Cyclic | 20 | 7.5 | 22 | 3.0 | 25 | 47.0 | 3.0 | 50.0 | 182 |
| Surf | Cyclic | 20 | 7 | Hydrodynamic radius (by DLS): 34 | | | | | | 129 |

### 2.3.2 Fiber phase

Fibrillation of amphiphiles and proteins is a well-known crystallization phenomenon occurring in soft matter and responsible for many living processes, like bacterial motility through actin[248] or neuronal degeneration through the protein Tau. Fibrillation is not uncommon for low-molecular weight bolaamphiphile[27,249,250] and peptides-based[251] surfactants, and several reports show similar processes in biosurfactants. In terms of packing parameter, one could be tempted to attribute a unitary *PP* for flat fibers or contained between 0.3 and 0.5 for cylindrical fibers. However, the packing parameter description cannot be associated to crystalline objects, because the theory behind the PP contains the hypothesis of liquid hydrocarbon core, which is not the case for crystalline objects (see also section 1.4).

In terms of characterization, flat and cylindrical fibers respectively provide a $q^{-2}$ and $q^{-1}$ (log-log) dependence of the scattered intensity, I(q), in SAXS/SANS experiments. However, polydispersity, fiber aggregation and formation of spherulites generally provide a non-integer exponent, to be rather interpreted as a fractal dimension. In any case, optical microscopy and/or cryo-TEM should be employed for a clear-cut attribution of the shape and type of twist (ribbon, helical). In the case of crystalline flat fibers, a diffraction peak should be observed at distances in the order of the size of the molecule.

***Sophorolipids*** (**1**)

Fiber formation by SL is a disputed topic, which illustrates the complexity of biosurfactants self-assembly properties in relationship to their congener mixture. The first work on fiber formation by SL was reported within the context of nonacetylated acidic C18:1-*cis* (**1**) in 2004, when this compound was shown to spontaneously self-assemble into giant ribbons at acidic pH on the basis of optical microscopy and XRD arguments.[197] Ribbon formation of the same compound was also reported some years later by another group under comparable conditions.[219] However, concomitant studies have claimed the formation of micelles in a wide pH range, from acidic to basic, for the same compound (please refer to the sophorolipid paragraph in section 2.3.1).[69,200,235] To better understand the origin of this apparent discrepancy, a thorough investigation was done on a series of nonacetylated acidic C18:1-*cis* SL batches, including pure (100%) standards and mixtures of C18:1-*cis* (**1**) and C18:0 (**13**) congeners. It was possible to show that the nature and amount of congener drives the self-assembly of acidic nonacetylated C18:1-*cis* SL (**1**) into micelles or twisted ribbons. SL from batches containing above 85% of the C18:1-*cis* congener (**1**), classically obtained after alkaline hydrolysis, systematically self-assembles into micelles at all pH values. Mixtures of acidic C18:1-*cis* SL (**1**) with its acetylated (mono- or di-) (**2**) and/or lactonic (**3**) congeners were also shown to form a micellar phase up to about 85%-90% of the acidic congener (**1**), in agreement with literature data mentioned above.[200] However, if the batch specifically contains an excess (> 10%-15%) of the C18:0 congener (**13**), ribbons are systematically obtained at acidic pH.[191] This complex behaviour is explained by the fact that the C18:0 conger (**13**), under the same conditions of use (room temperature and acidic pH), self-assembles into twisted nanoribbons.[221,252] However, it was also found that the 100% pure C18:1-*cis* congener (**1**) also crystallizes in water at room temperature,[191] thus suggesting that the self-assembly of acidic C18:1-*cis* (**1**) strongly depend on the batch purity.

For a more pragmatic approach in view of reproducible experiments, one should retain that micellar solutions can be obtained from a nonacetylated, lactone-free, congener mixture (the C18:1-*cis* (**1**) being > 85%-90%) of acidic SL obtained by the alkaline hydrolysis of a raw SL sample (**0**); fibers in the morphology of twisted nanoribbons are systematically obtained from C18:0 congener (**13**).

Saturated acidic C18:0 SL (**13**) are insoluble in water at room temperature. If this compound is dispersed in an excess of water, it forms large planar structures upon sonication,[191,219] while it forms twisted nanoribbons though a pH jump method, according to cryo-TEM, DLS, CD, WAXS, SANS, SAXS and NMR data (Figure 11a). Cryo-TEM images at pH 6 put in evidence an average cross-section of the ribbons contained between 10 and 15 nm, while SAXS/SANS show a repeated interlayer distance of 2.65 nm, corresponding to the intermolecular stacking of SL within the ribbon plane.[69,191,221,252] The broad diffraction peak suggests that the lipid order occurs along the cross-section of the ribbons rather than along the ribbon main axis. Interestingly, the interlayer distance reported for pure C18:0 SL (**13**),[221] C18:0:C18:1-*cis* [13%:77%][191] and fiber-forming C18:1-*cis* (C18:0/C18:1-*cis* ratio not given)[197] are, respectively, 2.65 nm (in solution), 2.70 nm (in solution) and 2.78 (dry powder). This comparison, which is not strictly rigorous due to the different state of the powder (dry vs. wet), still seems to suggest that the presence of fraction of C18:1-*cis* (**1**) slightly modifies the angle of the SL packing within the ribbon.[253] In terms of mechanism, pH-resolved *in situ* SAXS shows a well-defined transition from micelles to ribbons,[69] and not an isostructural process, meaning that micelles may play a role of reservoir of matter, rather than fibrillation nuclei.

Saturated acidic palmitic acid C16:0 (**14**) sophorolipids is also insoluble in water and it fibrillates upon pH jump from basic to acidic, below about 6.[204] However, differently than its stearic acid counterpart (**13**), pH-resolved *in situ* SAXS experiments show a direct micelle-to-fiber transition. In terms of the inter-fiber packing, C16:0 SL (**14**) seems to show two polymorphs, a main one with a typical distance of 3.31 nm and a minor one at 2.62 nm, the latter probably indicating a tilt of the molecules. We will discuss in section 2.7 that both (**13**) and (**14**) spontaneously form fibrillar hydrogels above about 1 wt%. The difference in fibrillation mechanism between (**13**) (nucleation and growth) and (**14**) (continuous micelle-to-fiber) has a direct impact on the homogeneity of the fibers and, consequently, on the macroscopic mechanical properties of the respective hydrogels. (**13**) is more sensitive to spherulite formation, providing hydrogels with poor elastic properties, while (**14**) forms highly homogeneous fibers and hydrogels with very good elastic properties.

The *trans* derivative of nonacetylated acidic C18:1 SL (C18:1-*trans* (**12**)) also forms twisted ribbons in water at acidic pH (Figure 11b) with a repeating distance of the intra-ribbon lipid arrangement of 3.05 nm,[219] that is more than 10% higher than for the C18:0 SL. Again, the reason for this discrepancy could be due to the different tilt angle of the lipid stacking within the ribbon. The most interesting feature of (**12**) is certainly their resilience in forming twisted ribbons, even at basic pH, and in complete contradiction with the C18:0 SL (**13**). pH- and time-resolved *in situ* SAXS experiments have shown that C18:1-*trans* SL (**12**) initially form a micellar phase above pH 8, as found for C18:0 SL (**13**); however, the micellar phase slowly evolves with time (~hours) towards a crystalline fibrous precipitate,[203] with similar morphological characteristics and interlipid distance (2.06 nm) as found at acidic pH.[191] These data are quite surprising and still partially hard to rationalize because they show that non-covalent weak interactions, especially H-bonding and most likely pi-pi stacking at the *trans* double bond level play an important role in the kinetics of fiber formation and in the self-assembly process in general. It could be also surprising that stable ribbons can be obtained with negatively charged lipids, which should otherwise stabilize higher curvatures. If it is not uncommon that small charged molecule containing strongly hydrophic moieties, as in FMOC peptide derivatives, fibrillate in their negatively-charged form,[254,255] it could be interesting to verify the charge of ribbons composed of (**12**) in future work.

***Other glycolipids*** *(cellobioselipids* (**17**)**,** *glucolipids* (**10**)**,** *symmetrical sophorolipids* (**9**)*)*.

Other microbial glycolipids have been reported to form a fiber phase. Acidic cellobioselipids (**17**), which also contain two glucosidic groups (cellobiose) show twisted, helical ribbons and flat fibers in water[155,201] and specific organic solvents.[198] The behaviour of this compound is then similar to the fibrillation of saturated (**13**) and *trans* SL (**12**), despite the fact that acidic cellobioselipids have side hydroxyl groups, of which the influence on fibrillation does not seem highly relevant. Ribbon formation has also been observed for the aminyl derivative of saturated SL (C18:0-$NH_2$) (**5**) in their neutral form at high pH (Figure 11c), thus exhibiting a similar behaviour as saturated COOH-containing C18:0 SL (**13**) in their neutral form at acidic pH.[202] This shows that, up to a certain extent, the chemical nature of the end-group is not of paramount importance, as long as it shows lower solubility and, possibly, the tendency to form hydrogen bonds. Nonetheless, the question whether the bolaform shape and hydrogen bonds drive fibrillation is still open, as shown by the fibrillation at room temperature of alkynyl C18:1-*cis* SL (**6**).[202] In this case, the COOH group is replaced by an alkyne function, thus making this compound look like a more classical head-tail surfactant rather than a bolaform amphiphile.

Symmetrical bolaform sophorolipids were also shown to form fibers at room temperature, irrespective of the synthetic of natural origin.[207,208] C16:0 symmetrical sophorosides (SS) bolas (**9**), obtained by the fermentation of the modified *S. bombicola* strain *Δat Δsble Δfao1*, form twisted nanoribbons having a interlipid distance within the ribbon of 2.48 nm, that is sensitively smaller than what the expected full length of this molecule and even smaller than what is found for acidic sophorolipid fibers commented above. This strongly suggests a tilted geometry of the given lipid. Similar conclusions arise from synthetic symmetric SL bolas having a charged ammonium group in the middle of the connecting aliphatic chain.[207] In this case as well, one find a repeating distance 2.73 nm, more in agreement with previous SL fibrous systems, but still smaller than the fully extended length of the bolaamphiphile, thus suggesting a tilted arrangement of the lipid.

As a general comment, most systems show a strong polydispersity both in the ribbon length and cross section.[201,203,219,221] It was shown that salt may have an influence on the homogeneity although a clear explanation was not given.[252] In fact, in the worst case scenario, cryo-TEM images show a mixture of long and short fibers, polydisperse cross sections and even spherulitic networks, similarly to what it has been described for fibrillated low-molecular weight gelators (LMWG).[256] Similarly to temperature-driven fibrillation of LMWG, it was recently pointed out the importance of kinetics in the fibrillation of C18:0 acidic SL (**13**), and in particular in the rate of pH variation. Although the influence of salt was not specifically pointed out, the most homogeneous fiber networks free of spherulites were obtained in the absence of extra salt and at low pH-change rates.[257] In this regard, the chemical nature of the lipid seems to influence the kinetics of self-assembly and, consequently, the homogeneity of the structures (e.g., spherulites vs. homogeneous fibers), as discussed for (**13**) and (**14**) in the sophorolipid paragraph of this section. This may have direct impact on the macroscale properties of the corresponding soft material (e.g., hydrogel, section 2.7).

Influence of kinetics, ionic strength and type of salt on the crystallization of biosurfactants is still underestimated but it is definitely a topic to be explored further in the future to control the fibrillation process for these complex systems.

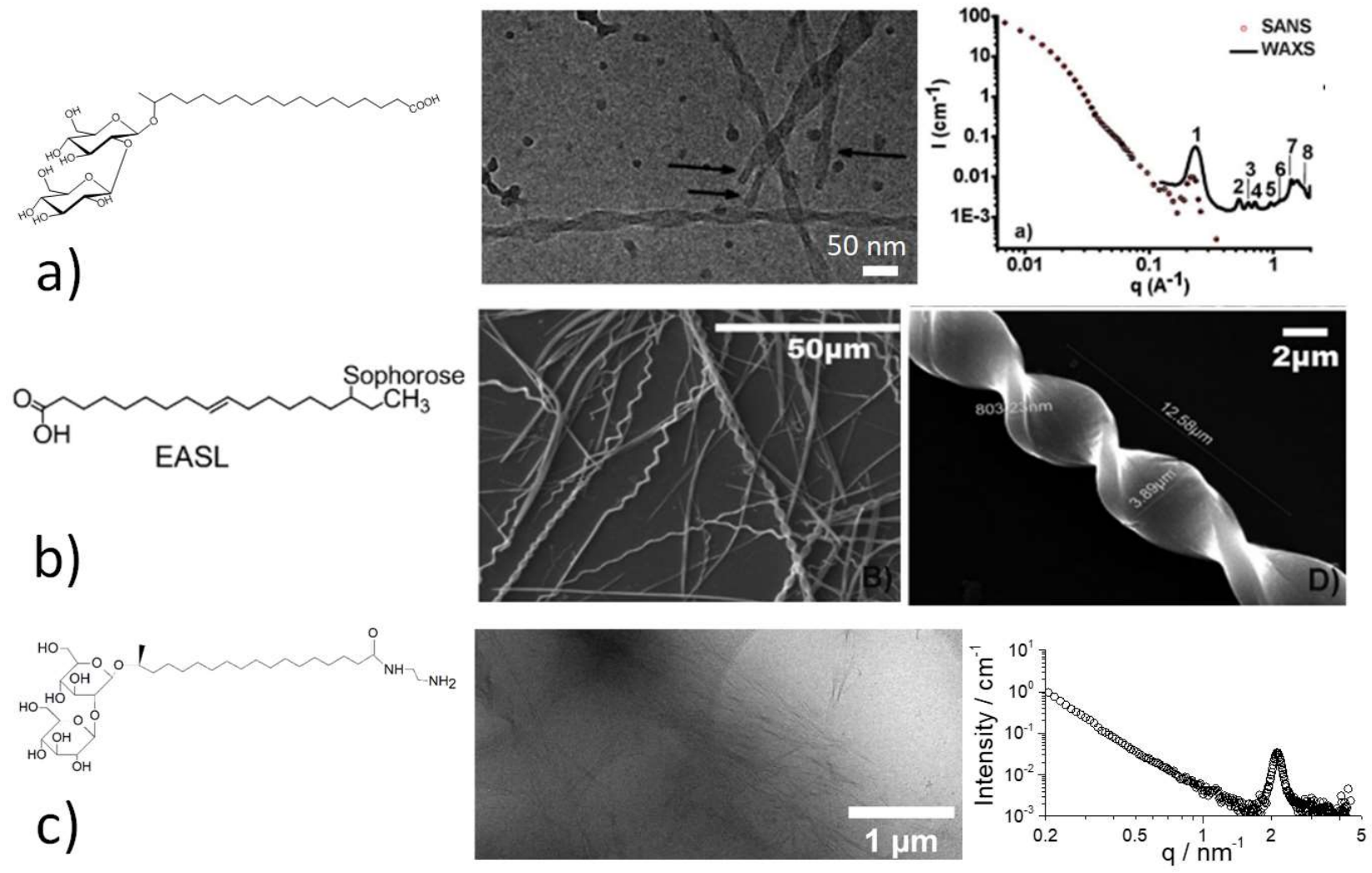

Figure 11 – Twisted ribbon structures observed for nonacetylated a) acidic C18:0 (**13**),[221] b) acidic C18:1-*trans* (**12**) (adapted from Ref. [219]. Copyright John Wiley & Sons) both at acidic pH and c) (-$NH_2$) C18:0 SL (**5**) at basic pH.[202] a) and c) are reproduced from Ref. [221] and [202] with permission from The Royal Society of Chemistry.

### 2.3.3 Vesicle ($L_4$, *V*, *Ves*, *SUV*, $L_{\alpha 1}$) and multilamellar vesicle (*MLV*) phase

Vesicles, generally a metastable phase, are unique objects having the capability to encapsulate, transport and deliver a cargo (hydrophobic drugs, macromolecules, nanoparticles) with strong benefit in many domains from medicine to cosmetics. The packing parameter for vesicles is $0.5 < PP < 1$. In the literature, single-wall, or unilamellar, vesicles are referenced as $L_4$, *V*, *Ves*, *SUV*[258] or $L_{\alpha 1}$[259] phase, while multilamellar vesicles are generally referenced to as *MLV* phase.[260,261]

A comment on phase notation. Some studies reported below attribute the notation $L_\alpha$ to a "vesicle" phase although this is not precise, as the classical attribution for the $L_\alpha$ notation is (flat) lamellar liquid crystal, characterized by an "oily streaks" texture under crossed polarizers[258] and a diffraction peak corresponding to d= 0.45 nm.[262,263] If the confusion may come from the fact *MLV* have a lamellar order in SAXS, $L_\alpha$ and *MLV* phases are commonly separated in the lipid and colloid science community[260] (please refer to P. 1294 and 1324 onwards in Ref. [258]). However, exceptions to such notations are not uncommon and some authors refer to *MLV* as being part of flat lamellar phases ($L_\alpha$, $L_\beta$, $P_\beta$), although in this case *MLV* are not specifically associated to the $L_\alpha$ phase only.[261] This subsection then reports all communications in which the authors of this document could identify specific vesicle phase. Whenever possible, we will try to distinguish between unilamellar ($L_4$, *V*, *Ves*, *SUV* or $L_{\alpha 1}$) and multilamellar (*MLV*) vesicle phases on the basis of SAXS, XRD and/or microscopy arguments and we will warn the reader of the imprecise use of the $L_\alpha$ notation.

Vesicles are constituted by a lipid mono or bilayer, which is characterized by a typical $q^{-2}$ dependence of the scattered intensity, I(q), in SAXS/SANS experiments, whereas small vesicles are also characterized by a plateau at low scattering vectors. In the case of multilamellar vesicles, at least two diffraction peaks (at a 1:2 ratio in position) characterize the SAXS pattern. However, in the absence of a plateau, ambiguity with flat lamellar systems is always possible, as the latter display exactly the same small angle scattering profile. Optical, polarized light, fluorescence or cryo-TEM microscopy are then necessary to discriminate spherical and flat morphologies (refer to section 1.7 for more information on the characterization techniques). This is not always possible for some of the systems reported below (mainly RLs and surfactin) and some ambiguity in terms of phase attribution may still exist.

***Rhamnolipids*** (**mono-** (**19**), **di-** (**18**))

Rhamnolipids self-assemble into bilayers, as shown by SAXS and SANS experiments, [144,230,264] in the shape of vesicles under acidic pH conditions (~4 < pH < ~6.8), as shown by fluorescence microscopy and cryogenic TEM.[159,210,211] The first report of vesicle formation of RL is attributed to Ishigami *et al.* in 1987 (Figure 12a) and later confirmed by Champion *et al.* in 1995 and Shin *et al*. in 2008. Interestingly, the surface activity (surface tension, ST) of RLs is maximal (lowest ST) for a pH close to the vesicle-to-micelle transition (∼6.8) while it is reduced for the micellar phase $L_1$ and drastically reduced in the vesicle region (highest ST).[229]

Some interesting works[265,266] show that the bimodal distribution (60 – 80 nm and 350 – 400 nm) for a crude RL extract could be attributed to the almost monomodal distributions of the corresponding mono- and dirhamnolipids. However, if DLS experiments taken by themselves are always subject to a large uncertainty, they definitely corroborate, in relationship with the studies cited above, the presence of vesicular objects having a multimodal size distribution for RL aqueous solutions.[178,265–267] Nice vesicular objects have been observed with confocal scanning laser microscopy for a crude mixture of RL in water with 875 mM NaCl and pH 6.2.[222] In this work, the size distribution is broad and it is shown that addition of NaCl increases the size distribution from 100-200 nm to several microns for crude RL mixture but also for mono- and di-RL. However, in the case of mono- and di-RL at 875 mM NaCl, authors show the presence of dense spheroidal aggregates of several microns in size and of which the phase is not reported, although it may deserve future investigation.

Chen *et al*. have contributed to this field by showing that the vesicle phase can be maintained even at basic (pH 9) but only for the mono-RL, while di-RL mainly form a micellar phase. In this regard, the phase diagram at room temperature of mono-RL/di-RL systems can be quite complex, with coexistence of micelles and vesicles together (pH 9) according to the proportion of mono-di-RL and overall RL concentration.[144] Below 20 mM, RL are always in the micellar phase, while above 20 mM the vesicle phase appears roughly above 60% of mono-RL.

A qualitative overview of the body of existing data agrees on the vesicle formation of RL mixtures at acidic pH and mono-RL at pH 9 above 20 mM. However, whether RL form a vesicle or *MLV* phase, or even a flat lamellar phase, is still unclear, because scattering (SAXS/SANS) studies are not associated to microscopy on the same set of samples. This aspect is commented below. SAXS and SANS experiments by several authors[144,230,264,265] support the evidence of a broad correlation peaks, generally at low-q (< 0.5 $nm^{-1}$) and superimposed to a $q^{-2}$ dependence of the intensity. Such a scattering profile is typical of a *MLV*,[268] rather than a single unilamellar vesicle (SUV) phase. However, in the absence of microscopy support, such signal could also be attributed to a swollen flat lamellar phase.[269] If the majority of cryo-TEM or fluorescence microscopy data support the existence of a *MLV* phase[159,210,211,222,265] and exclude a *SUV* phase, some other data, on the basis of scattering only without microscopy support (see also Table 5 and Table 6), do not exclude the existence of flat lamellae at pH below 6.[210,211] To date, one could safely consider that mono-RL and/or di-RL form a *MLV* at acidic pH. However, formation of a poorly ordered flat lamellar phase under specific conditions of pH, concentration or ionic strength at room temperature should not be excluded.

Finally, the little amount of quantitative data concerning the thickness of RL membranes and obtained from SAXS/SANS show a certain degree of discrepancy (Table 8). Values of 32 Å[265], 27 Å[264] or 15 Å[230] have been reported for mono-/di-RL or di-RL vesicle solutions at acidic pH. Since periodicity in multilamellar membranes is always given by the sum of the membrane thickness and the intermembrane water layer, the disparity of these values could then be possibly explained by an inaccurate estimation of the intermembrane water layer.

***Sophorolipids.***

Non- (**1**) and di-acetylated (**2**) acidic C18:1-*cis* SL form a micellar phase, as discussed above. A single study reports on vesicle formation by nonacetylated C18:1-*cis* SL (**1**) under basic pH conditions; however, this is only supported by single-object freeze-fracture TEM image and it lacks statistically-relevant techniques like scattering, cryo-TEM or fluorescence microscopy.[270] At the present state of the art, the body of existing data indicates that this compound forms a major micellar phase under alkaline conditions,[69,197,235,236] although coexistance with nanoplatelets (discussed later) has been reported.[220]

In order to observe a vesicle phase with acidic sophorolipids, one must vary the nature of the tail. Dhasaiyan *et al.*[214] reported vesicles for linolenic (tri-unsaturated, C18:3-*cis* (**15**)) SL prepared under acidic pH conditions (Figure 12c). If vesicle formation is also supported by molecular dynamics simulations, the multilamellar wall structure should be verified in the future by complementary SAXS/WAXS analysis collected on the wet sample, rather than powder XRD. Baccile *et al.*[205] reported the *MLV* and vesicle-in-vesicle (called glucosomes, Figure 12e, and referring to glucose-rich vesosomes[271,272]) structures for a behenic acid derivative of sophorolipids (**16**), prepared from *Pseudohyphozyma bogoriensis* and confirmed by a combination of SAXS and cryo-TEM. A vesicle phase could also be reasonably hypothesized by Penfold *et al.*[200] using SANS data recorded on an acetylated lactone C18:1-*cis* SL (**3**) sample in water (Figure 12b). They reported the coexistence of micelles and nanovesicles under dilute conditions, nanovesicles around 3 mM, large unilamellar vesicles below 10 mM and an undetermined phase, possibly tubular or disordered vesicle, above 10 mM. If all hypotheses are reasonable on the basis of the rigorous SANS data analysis, future microscopy data will be needed to confirm them.

***Glucolipids*** (**10**, **11**)

Single-glucose head lipids, glucolipids, refer to microbial glycolipids obtained by fermentation of the modified strain *S. bombicola ΔugtB1*. C18:1-*cis* glucolipids (**11**) exhibit a *SUV* vesicle phase at acidic pH [6.2 – 4], as shown by the combination of SAXS and cryo-TEM (Figure 12f).[69,201] pH-resolved *in situ* SAXS that vesicles form from a micelle-to-rod-to-vesicle transition between alkaline and acidic pH. Recent unpublished in-house experiments show that vesicles can actually form spontaneously by dispersion (or sonication) of the compound in water. Vesicles are characterized by a single-wall membrane of approximate size of 3 nm, compatible with an interdigitated organization of the glucolipid. Interestingly, a vesicle-to-lamellar transition is experienced when pH is lowered to 4 and below. Upon increasing the pH again above 4-5, a *MLV* phase is obtained instead of a *SUV*, with better encapsulation potential.[216]

Saturated acidic C18:0 glucolipids (**10**), derived from the C18:1-*cis* compound, were shown to form vesicles at 70°C, above the $T_m$ of the lipid.[201]

***Mannosylerythritol lipids*** (**21**)

The self-assembly of MEL has deeply been investigated by D. Kitamoto and his co-workers in a series of papers from 2000 to 2012[19,188–190,215,223,224,228] and reviewed several times.[19,98] MEL-x corresponds to a general family of MEL containing various homologues (variation of the alkyl chain length, number of unsaturation or acetylation). Unless stated otherwise, all studies on MELs have been performed on the *S*- isomer shown in Figure 7 (**21**). Derivation of the *R*- isomer is provided in Ref. [225].

*MEL-A* (**21A**). Giant *SUV* and *MLV* vesicles (~10 µm) were observed by optical microcopy at room temperature in water at pH 7 using the classical hydration process of a condensed lamellar deposition of the lipid onto a glass slide and for a final MEL-A concentration of about 2 wt% (10 mg in 0.5 mL).[189] Interestingly, several years later, the same authors proposed a new phase attribution (coacervate, or $L_3$, phase) for MEL-A at equivalent concentrations (2 mM) using freeze fracture TEM as evidence.[188] In a more consistent work published in 2006, authors clearly show an *SUV* to $L_3$ transition between critical aggregation concentration 1 (CAC1, 4 x $10^{-6}$ M) to and above CAC2 (2 x $10^{-5}$ M).[190]

*MEL-B* (**21B**). Studied at the same time as MEL-A, MEL-B shows a different phase behaviour under dilute conditions. Initially, giant *SUV* and *MLV* vesicles (~10 µm) were observed by optical microcopy at room temperature in water at pH 7.[188,189] Later, this behaviour was rationalized against concentration and authors found a single CAC (6 x $10^{-6}$ M), below and above which it forms, respectively, *SUV* and *MLV* phases,[190] as confirmed by a combination of freeze fracture and SAXS. Compared to MEL-A, MEL-B has a hydroxyl group in $R^2$ position (Figure 7), thus showing, for this family of biosurfactants, the importance of acetylation on the self-assembly properties. In a later study, the same authors compared the self-assembly at 2 wt% of the *S*- and *R*- diastereoisomers of MEL-B[215] and showed that both form a *MLV* phase, with the *R*- isomer forming larger (> 1 µm) objects than the *S*- isomer (~500 nm). [188,189190215]

*MEL-C* (**21C**). Giant *SUV* and *MLV* vesicles (~10 µm) were also observed for MEL-C at room temperature in water at pH 7.[189]

*MEL-D* (**21D**). MEL-D is produced by enzymatic hydrolysis from MEL-B and it corresponds to the fully hydroxylated form of the latter. MEL-D has been produced in *S*- and *R*- diastereoisomers their self-assembly behaviour at 2 wt% in water shows the formation of *MLV*, as shown by a combination of DLS, DIC and polarized light microscopy (Figure 12d),[215] the latter showing a typical maltese cross, characteristics for *MLV* . The main difference between the *R*- and *S*- isomers seems to consist in the size: *R*- isomer form large (> 1 µm) objects while the *S*- isomer form sub-micron (~500 nm) objects.

***Surfactin*** (**20**)

Possible vesicle formation was proposed by Arutchelvi *et al.*[131] for surfactin solutions at pH 8.6 at about 0.2 mM using differential interference contrast microscopy and standard TEM and observed strong coalescence upon addition of divalent cations. However, these data are not corroborated by other studies, showing that surfactin forms a stable micellar phase at pH above 7.[129,157,182,212,244] Such discrepancy could be explained from the improper use of TEM in [131], which requires sample dehydration, thus modifying its solution volume fraction.

SANS experiments at pH 5.5 suggest the presence of a bilayer structure with an estimated thickness around 2.4 nm and, potentially, a vesicle phase.[212] Authors of the mentioned work evoke the existence of a lamellar phase, although its shape, vesicular or flat, should be verified. Additional experiments should be performed to clarify this issue. In the same study, it is shown the effect of adding ions. Divalent ions ($Ca^{2+}$ and $Ba^{2+}$) suggest the formation of a bilayer structure, which could possibly have a vesicular shape, although, in the absence of microscopy data, flat lamellae cannot be excluded. Addition of $Ba^{2+}$ seems to promote formation of multilamellar structures (structure peak at about 0.25 Å$^{-1}$), may them be in a MLV or flat condensed lamellar phase (see also comments on analytical techniques in section 1.7).[212]

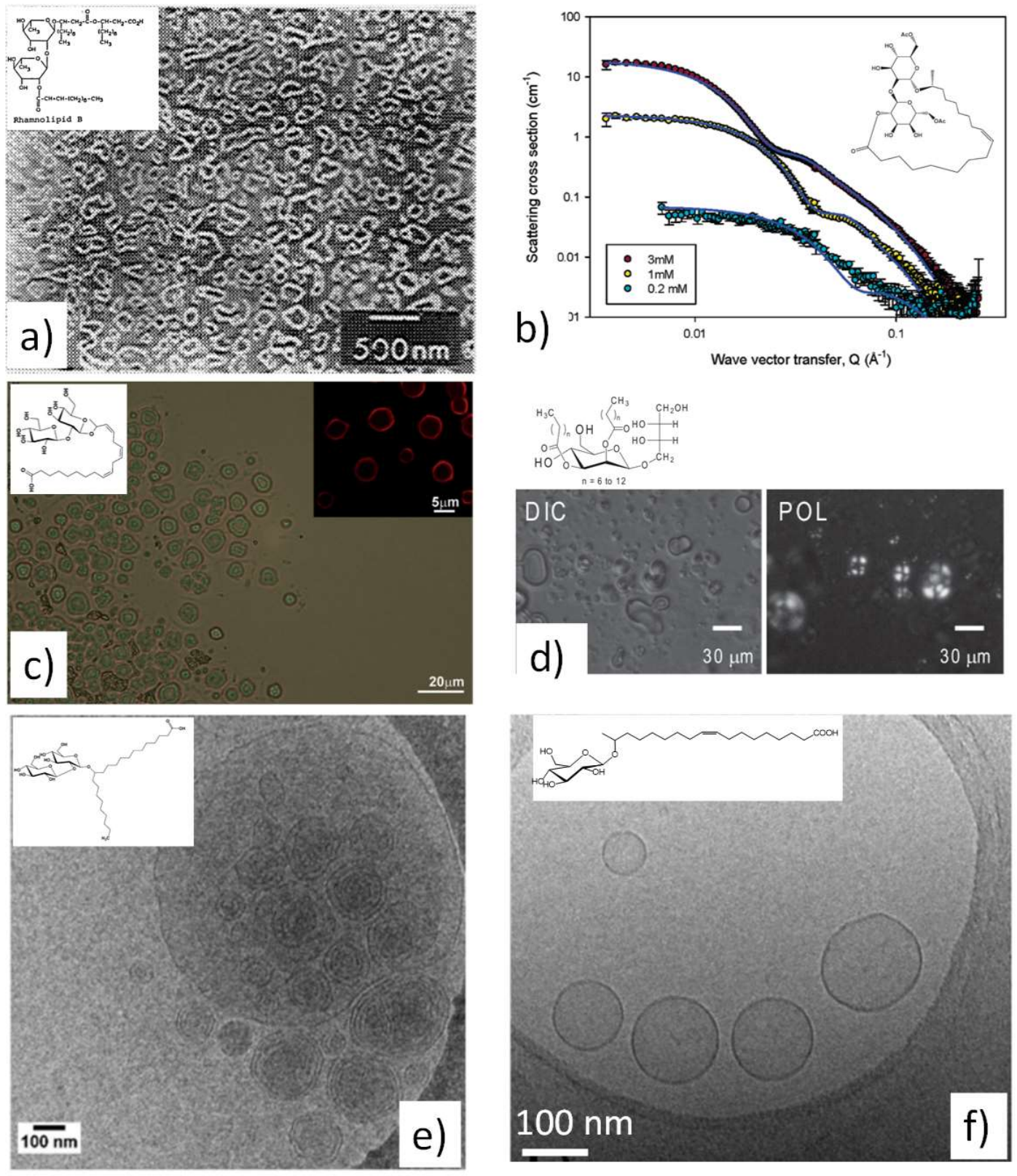

Figure 12 – Vesicle structures reported at room temperature under acidic pH conditions for diluted biosurfactants in water: a) di-rhamnolipids (**18**) (reprinted with permission from [159], copyright Chemical Society of Japan), b) acetylated lactonic sophorolipids (**3**) (adapted with permission from [200], copyright (2011) American Chemical Society), c) acidic C18:3-*cis* sophorolipids (**15**) (adapted from Ref. [214], copyright John Wiley & Sons), d) mannosylerythritol lipid-D (MEL-D) (**21D**) (reprinted with permission from [215], copyright Japan Oil Chemical Society), e) acidic $C22:0_{13}$ sophorolipids (**16**)[205] and f) acidic C18:1-*cis* glucolipids (**11**) (adapted with permission from [201], copyright (2016) American Chemical Society).

### 2.3.4 Flat lamellar phase.

Lamellar phases are of major importance in living organisms. Phospholipids, not discussed in this review, are the major component of biological membranes. Lamellar phases have different notations according to their order, hydration or lipid tilt. $L_\alpha$ refers to a (flat) lamellar liquid crystalline phase, being the most hydrated and less ordered (generally found below the lipid melting temperature, $T_m$), abundant in living organisms. The $L_\beta$, gel, phase is less hydrated, more ordered and characterized by perpendicular, or tilted $L_{\beta'}$, lipid chains with respect to the surface.[273] The ripple phase is, on the contrary, a periodically undulated bilayer, with the notation $P_\beta$, or $P_{\beta'}$ in the case of tilted lipids.[258,274] Each phase is generally characterized by a typical intramembrane lipid distance, typical of a more or less ordered state.[262,263]

These phases are an important topic of research in biophysics, for the better understanding of living organisms. The corresponding packing parameter for this phase is 1 (Figure 2). As a general rule, lamellar phases are characterized by a typical $q^{-2}$ dependence of the scattered intensity in SAXS/SANS experiments and by at least two diffraction peaks, one being the harmonics of the other (1:2 ratio in position). In this section we maily consider diluted lamellar phases with poor long-range order, or swollen systems, which make exception to these rules. If the $q^{-2}$ dependence is generally always observed, the diffraction peaks may either be broad or even not appear at all and confusion with the scattering signal of giant unilamellar vesicles and *MLV* is always possible. As for the vesicle phase, clear-cut attribution is not always possible for most BS systems discussed below.

Polarized light microscopy, or cryo electron microscopy on diluted samples, should be generally employed as complementary techniques for a more clear-cut attribution. For instance, $L_\alpha$ phases are generally characterized by an "oily streaks" texture under crossed polarizers and a diffraction peak corresponding to d= 0.45 nm.[258,262,263]

***Rhamnolipids*** (**mono-** (**19**), **di-** (**18**))

If RL have been shown to form bilayers in a *MLV* phase at acidic pH for both mono-RL and di-RL, and at basic pH for mono-RL only (**19**) (within the concentration range explored and at room temperature), their role in the stabilization of a flat lamellar (non-vesicular morphology) phase is less clear for pH well below 6 or for other specific physicochemical conditions.[210,211] We have outlined this possible ambiguity at the end of the RL paragraph in section 2.3.3, to which we address the reader for more details.

***Sophorolipids.***

Diacetylated lactonic C18:1-*cis* SL (**3**) were shown to form a bilayer, attributed to a vesicle phase, at 7 mM. However, the lack of a plateau at low-q could arise the question whether the morphology is constituted by flat lamellae or large unilamellar vesicles, the latter interpretation being preferred by the authors.[200]

***Glucolipids*** (**10, 11**)

Saturated acidic C18:0 glucolipids (**10**) can be derived from the corresponding C18:1-*cis* glucolipids (**11**), prepared from the modified strain *S. bombicola ΔugtB1*, and forming a vesicle phase below pH 6.2. Saturation of the fatty acid backbone induces a rigidity in the aliphatic part and of which the effect is to generate more rigid membranes. A combination of *ex situ*, *in situ* SAXS and cryo-TEM recorded at room temperature, in the excess of water, at different pH for glucolipid C18:0 demonstrates the presence of lamellar phase in the form of flat floating objects (Figure 13a,b).[69,201] The low–q region of SAXS profiles (< 0.04 Å) is characterized by a $q^{-2}$ dependence of the scattered intensity, typical for flat structures, while cryo-TEM shows the presence of infinitely flat membranes, rather than vesicles (the latter being observed for (**10**) at 70°C, above the $T_m$ of the lipid). Modelling of the SAXS data shows a membrane having a thickness of about 3.6 nm (± 10%), in agreement with the lipid length, thus suggesting an interdigitated monolayer rather than a classical bilayer. Such behavior is maintained up to about 10 wt%, where the flat structure persists. The interlamellar distance decreases from about 20 nm to less than 1 nm, both with concentration and/or addition of salt, thus demonstrating the existence of long-range interlamellar interactions of electrostatic repulsive nature (Figure 13b).[218]

Condensed lamellar phases are also supported for various compounds (C18:1-*cis* (**11**) and C18:0 glucolipids (**10**),[69,201] behenic acid derivatives of sophorolipids (**16**)[205]) at pH below 4, when all carboxylic groups are supposed to be protonated, or at high ionic strength.[218]

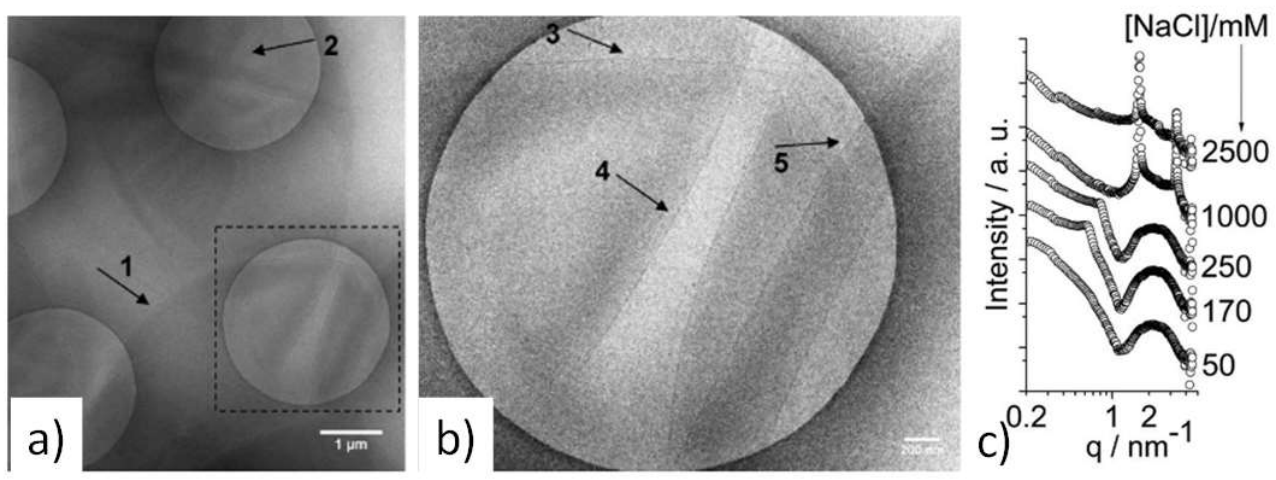

Figure 13 – a,b) Cryo-TEM of flat lamellar structures recorded for acidic glucolipid C18:0 (**10**) (0.5 wt%) at pH 7.8 and room temperature. Adapted with permission from [201]. Copyright (2016) American Chemical Society. c) NaCl-dependent evolution of the SAXS pattern recorded for a 5 wt% of the same compound (pH 7, RT). The high-q evolution of the scattering/diffraction peak with increasing NaCl indicates the electrostatic nature of the repulsive interlamellar forces. Reproduced from Ref. [218] with permission from The Royal Society of Chemistry.

***Surfactin*** (**20**)

A lamellar structure was claimed for a diluted solution (2 mM) of surfactin at pH 5.5 and at pH 7.5 in the presence of divalent ions ($Ca^{2+}$ and $Ba^{2+}$) on the basis of SANS. However, in the absence of complementary microscopy data, no clear-cut statement can be done about its vesicular of flat morphology.[212] Addition of $Ba^{2+}$ seems to promote formation of multilamellar structures (structure peak at about 0.25 Å$^{-1}$), may them be in a *MLV* or flat condensed lamellar phase.

***General comments.***

Most biosurfactants forming bent (vesicles) or planar (lamellae) membranes display a membrane thickness, $T_m$, between 15 Å and 36 Å, as summarized in Table 8. The length of the hydrophobic core, $L_c$, is often comparable between amphiphiles and it settles around 10 Å, while the thickness of the hydrophilic shell, $T_s$, is more variable, roughly between 6 Å and 14 Å, according to the biosurfactants. These values, compared to size of the amphiphiles, indicate that biosurfactants are in an interdigitated configuration, rather than a classical bilayer. In some cases, like RL, the data suggest both single- (15 Å)[230] and bilayers (32 Å)[265], probably due to the erroneous estimation of the interlamellar water layer. Most systems studied are in the shape of unilamellar vesicles, from nanosized with relatively low polydispersity[200] to large, highly polydispersed, systems.[69,205,265] However, some exception[214] exist, especially when it comes to RL[159,210,211,222,265] and MELs,[188–190,215] which seem to stabilize *MLV* instead.

Flat lamellae, instead of vesicles, are observed without any doubt only for the saturated form of a C18:1 glucolipid (**10**), and this is most likely explained its melting temperature higher than RT.[69,218] However, as commented in sections 2.3.3 and 2.3.4, this phase cannot be fully excluded for RLs and surfactin under specific physicochemical conditions (Table 5 and Table 6). This fact should be verified in future work.

†: note on the $L_\alpha$ phase. Several studies attribute optical microscopy or scattering data to a $L_\alpha$ phase. We prefer not to report such attribution in the *Phase* column. $L_\alpha$ is characterized by a liquid-like order of the acyl chain in the membrane and it could be probably supposed for vesicular morphologies. However, $L_\alpha$ is generally intended as (flat) lamellar liquid crystalline phase,[258,273] characterized by diffraction peaks at 1.47 Å$^{-1}$ [262,263,275] and an "oily streak" texture under cross polarizers[258,273] (please refer to P. 1294 and 1324 onwards in Ref. [258]). None of these is provided in the studies associated to the BS reported in Table 8.

Table 8 – Biosurfactants (N° given in Figure 7) vesicles and lamellar phases in solution observed at room temperature at concentration below 5 wt%. Structural parameters are extracted from fitting SAXS or SANS data using adapted vesicle or bilayer models. The typical cartoon of the model is reported below: $T_s$= Hydrophilic shell

thickness, $L_c$ = Hydrophobic core length, $T_m$= $2T_s + L_c$= Total membrane thickness, $R$= External radius. Notes: in *, $T_m$ is calculated by the difference between the external ($R_2$) and internal ($R_1$) radii given in Table 2 of Ref. [200]. To be noted, vesicle morphology is not confirmed by complementary microscopy. In #, we have extracted the corresponding values from Figure S4 in Ref. [69]. In °, membrane thickness is estimated from molecular dynamics (MD) simulations (Figure 4 of Ref. [214]), considered to be more reliable than SAXS arguments.

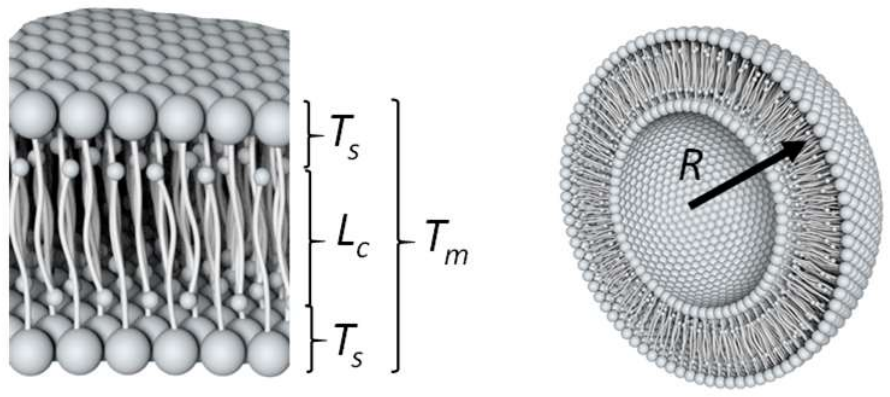

| BS | Type | N° | pH | $L_c$ / Å | $T_s$ / Å | $T_m$ / Å | R / Å | Phase† | Technique | Morphology | Ref. |
|---|---|---|---|---|---|---|---|---|---|---|---|
| RL | di-RL | 18 | 7.4 | 10 | 6 | 32 | $< 10^4$ | | SAXS | *MLV* | 265 |
| RL | Mono-/di- | 18, 19 | 5.5, 6.5 | - | - | 15 | < 600 | | SANS | *MLV* | 230 |
| RL | Mono-/di- | 18, 19 | < 7 | - | - | 27 | - | | SAXS | *MLV* | 264 |
| SL* | Lactone | 3 | - | - | - | 15-24 | 37-104 | | SANS | Vesicle* | 200 |
| SL | $C22:0_{13}$ | 16 | 5 | 8 | 9 | 26 | $< 10^4$ | | SAXS | Vesicle | 205 |
| SL | $C22:0_{13}$ | 16 | <5 | - | - | - | - | *Multilamellar* | SAXS | Not verified | 205 |
| SL° | C18:3 | 15 | < 7 | - | - | ~ 20 | $< 10^4$ | | MD | Vesicle | 214 |
| GL# | C18:1-*cis* | 11 | 6 | 11 | 9 | 29 | $< 10^4$ | | SAXS | Vesicle | 69 |
| GL# | C18:1-*cis* | 11 | <4 | - | - | - | - | *Multilamellar* | SAXS | Not verified | 69 |
| GL# | C18:0 | 10 | 6 | 8 | 14 | 36 | ∞ | $P_\beta$ | SAXS | Lamellae | 69,218 |
| GL# | C18:0 | 10 | <4 | - | - | - | - | *Multilamellar* | SAXS | Lamellae | 69,218 |

### 2.3.5 Others phases

*Cylindrical/wormlike micelles*

This phase has been rarely observed for biosurfactants. It's clear-cut characterization can only occur by SANS/SAXS experiments presenting a -1, or -1.6, q-dependence of the intensity in a log-log scale. Preferentially, cryo-TEM experiments should be used to confirm scattering data. To date, the only reports showing a cylindrical/wormlike phase at room temperature concern nonacetylated acidic C18:1-*cis* SL (**1**) in water at acidic pH and concentrations above 10 wt% (SANS),[213] dilute solutions (< 1 wt%) of acidic C18:1-*cis* glucolipids (**11**) in a transitory pH region (7.5 – 6.5) between the (spherical) micellar and vesicle phases (SAXS, cryo-TEM),[201] diluted solutions (2 mM) of surfactin at pH 6.5 (SANS).[212] Reports claiming cylindrical structures and based on single-angle scattering DLS experiments alone should be taken with care, because this technique is not reliable for anisotropic objects. In this sense, angle-dependent DLS experiments have shown strong anisotropy of micellar objects prepared from nonacetylated acidic C18:1-*cis* SL (**1**) in water (1.40 mg/mL) at pH 7.8.[197]

*Ill-defined large-scale structures*

The phases presented above are intended as major phases, although in some specific cases they coexist with other much less abundant, and often hard-to-define, phases. Such phases are often detected by an important contribution to SAXS/SANS scattering in the low-q region but which generally represents less than 5-10% of the scattering signal.

The important low-q scattering observed in SANS experiments collected on diacetylated acidic C18:1-*cis* (**2**) was interpreted by Penfold *et al*.[200] as the proof of large-scale structures coexisting with micelles and attributed to residual vesicles, although further microscopy data will be needed to support this claim. Similar results were shown by Baccile *et al*.[109] on an acetylated-rich C18:1-*cis* (**2**) sample at pH 2, and cryo-TEM images have shown that the large-scale structure was more compatible with ill-defined aggregates. Similar observations have been done on the nonacetylated acidic C18:1-*cis* (**1**) and C18:0 (**13**) SL at basic pH, of which the SAXS, SANS and SLS scattering profiles are characterized by a strong low-q signal,[235] which was later on attributed to nanoplatelets.[220]

A strong low-q scattering signal with -4 dependence of the intensity was reported in rhamonlipid solutions (**18**, **19**) at pH 13.2 and attributed to ill-defined large-scale structures. However, it is unclear to which chemical species the signal could be attributed to, because RL may undergo alkaline hydrolysis at such pH values.[230]

*Nanoplatelets*

Nonacetylated acidic C18:1-*cis* SL (**1**) form nanoplatelets under basic pH conditions, characterized by a strong low-q scattering signal in SAXS, SANS and SLS. After an initial erroneous attribution to ill-defined structures on the basis of cryo-TEM and actually corresponding to ethane adsorption, a classical artifact in cryo-TEM, additional high-resolution cryo-TEM experiments have shown the spurious existence of fractal aggregates constituted by aggregated nanoscale platelets, of which the composition in terms of ionic/neutral SL and structure is still not fully elucidated.[220]

Similar platelets/filamentous structures are not uncommon in other systems and generally at basic pH values. SAXS and/or cryo TEM data have been recorded for cellobioselipids (**17**), glucolipids (**11**)[201] and behenic acid sophorolipids (**16**).[205] In many cases, these structures recall a bicelle phase, sometimes observed in phospholipid systems.[276] If their exact nature is biosurfactant-dependent and unclear in many systems, they have several common features: 1) they are often observed at basic pH, when BS are in their $COO^-$ form; 2) they only represent a small fraction with respect to the main phase, the reason for which they are generally detected only in the low-q portion of SAXS/SANS profiles; 3) they tend to aggregate into fractal aggregates, and this constitutes the main reason why the can be easily observable by scattering but hard to detect by cryo microscopy techniques.

*Columnar phase*

A new set of SL derivatives containing porphyrin (**7**) has recently been synthesized by the group of Gross.[209,277,278] Porphyrins are well-known macrocycles acting as light-harvesting antennae in photosynthetic systems and concentrating a number of weak interactions, like π-π stacking and metal-ligand coordination. Its conjugation to SL will provide additional interactions such as hydrogen bonding and hydrophobic effects. Authors have synthesized a broad range of SL-porphyrin derivatives (**7**), including monounsaturated, saturated, mono-, di-and peracetylated compounds, as well as short-chain derivatives and mono-, di- and tetra-substituted porphyrin. They show that excitonically coupled J-type columnar aggregates are mainly obtained for di- and nonacetylated diporphyrin C18:1-*cis* SL derivatives below a transition temperature ranging around 34-37°C, following a cooperative nucleation and growth mechanism. On the contrary, peracetylation of the sophorose groups disrupts the hydrogen-bonding network and it induces the formation of uncoupled micellar aggregates, instead.[209] Interestingly, shortening the sophorose-to-phorphyrin linker (17 to 10 carbons) in fully hydrogenated SL-diporphyrin derivatives induces a chiral inversion in the coupled columnar aggregates, from mainly left-handed (long derivative) to mainly right-handed (short derivative).[277]

*Coacervate ($L_3$) phase*

MEL-A (**21A**) is shown to form a coacervate, or $L_3$, phase (whereas single-molecule coacervation, especially at such low concentrations, is quite rare in lipid science[279]) at low concentrations (2 mM) using freeze fracture TEM, optical microscopy, polarized light microscopy and SAXS as evidence.[188,190] In particular, authors find that the $L_3$ phase is observed above MEL-A's (**21A**) CAC 2 of 2 x $10^{-5}$ M. A structural variant of classical MEL-B (**21B**) has been reported and studied. The new MEL-B presents only an additional hydroxyl group on the main alkyl chain[280] and its self-assembly properties in solution measured at 0.34 wt% (about 5 mM) show the formation of dense liquid-like, non-birefringent, droplets instead of the expected vesicle phase, found for standard MEL-B (**21B**). Authors attributed this phase to a $L_3$, coacervate, in analogy with the data reported for MEL-A (**21A**).[188–190] However, they acknowledge that additional microscopy and X-ray diffraction data are needed to confirm this hypothesis.

## 2.4 Concentrated systems

Concentration is a common parameter in evaluating the phase diagramme of surfactants. Most studies related to biosurfactants, discussed above, have been performed under dilute to semi dilute conditions, rarely above 10 wt%. One of the main reasons is probably the lack of large amounts of matter having an acceptable molecular uniformity. If the phase diagramme may be relatively influenced by impurities at low volume fractions, it will certainly be at higher concentrations. However, few studies do report on the self-assembly properties of biosurfactants at large volume fractions in water.

***Rhamnolipids*** (**mono-** (**19**), **di-** (**18**))

A series of work has dedicated some attention to the phase behaviour of mono and di-RL, as well as the influence of mono-di- ratio, pH and ionic strength upon concentration.[178,195,196] However, the method employed by the authors, consisting in drying a 10 wt% RL solution on a glass slide and using polarized light microscopy to identify the phase should not be considered more than a qualitative screening of the RL phase behaviour. The optical polarized light microscopy shows the presence of mesophases, but the volume fraction of RL is not controlled and it varies all along the experiment with evaporation. Additional XRD should provided to support the hypothesis drawn form microscopy.

Long-chain (C14-C14) dirhamnolipids obtained from *Burkholderia* (*Pseudomonas*) *plantarii* were shown to have a complex phase behaviour at 5wt% and 25wt%, if compared to standard RL extracted from *P. aeruginosa* (C10-C10).[281] On the basis of X-ray diffraction, authors show the existence of a (not yet indexed) cubic phase in HEPES at pH 7 at only 5 wt%. A lamellar order is on the contrary observed at 25 wt%. If this work shows some interesting features, the body of data should be confirmed by additional studies for this specific compound: some of the peaks could not be attributed, others were hardly connected to the size of the RL and the cubic phase was not fully indexed.

***Sophorolipids.***

Nonacetylated acidic C18:1-*cis* SL (**1**) were studied above 10 wt% in water at equilibrium pH (< 5). Authors show a transition from spherical to wormlike micelles between 0.05 wt% and 17% wt% and room temperature, whereas the assumption of wormlike morphology comes from the $q^{-1.6}$ dependence of the intensity in the SANS experiments, and classical for wormlike systems. Additional microscopy experiments should nonetheless confirm this claim.[213]

***Mannosylerythritol lipids*** (**21**)

*MEL-A* (**21A**). Imura *et al.*[223] have studied the phase behaviour of MEL-A by polarized optical microscopy, SAXS and DSC. The sponge phase ($L_3$), detailed earlier, spreads from 0 to 55 wt% and from 20 to 65°C.[188,190] At room temperature, the sponge phase is stable between the CAC2[190] and up to 56 wt%.[223] Between 56 wt% and 65 wt%, the system forms a transparent phase, highly viscous, and attributed to a reverse cubic phase $V_2$ of space group *Ia3d*. At C > 65 wt%, the solution becomes turbid but less viscous than $V_2$ phase identifying a lamellar phase with a d-spacing of 3.6 nm. The nature of the lamellar phase is correctly identified as being $L_\alpha$ on the basis of WAXS arguments (s= 2.2 $nm^{-1}$ ≡ d= 0.45 nm), although typical "oily streaks" were not shown in PLM (refer to sections 2.3.3 and 2.3.4).[262,263] An isotropic fluid phase is finally observed at concentrations higher than 85% and up to 80°C (Figure 14c). In terms of temperature, the $L_3$ phase is stable up to at least 60°C.

The qualitative phase behaviour of a new form of MEL-A, referred to triacetylated MEL-A2 and where the third acetylation is carried by the C2 of the mannosyl group, has been reported.[175] On the basis of polarized light microscopy combined to water penetration experiments, authors suppose the formation of an isotropic phase followed by a lamellar phase and at higher concentration a neat surfactant phase. If these data should be confirmed further by x-ray diffraction at low and high angles to confirm the claimed $L_\alpha$ phase, it seems that MEL-A2 has a different phase diagram than it congener MEL-A, thus proving the impact of the substitution on the C2 atom of mannose.

*MEL-B* (**21B**). The phase behaviour of MEL-B has been qualitatively studied in 2006[190] by water penetration experiments and SAXS/WAXS and more deeply in 2008.[224] A 25°C, it was already shown how MEL-B forms myelin figures followed by a lamellar phase and neat surfactant phase. The lamellar phase is attributed to a $L_\alpha$ on the basis of WAXS arguments, showing a peak at 2.2 $nm^{-1}$ and corresponding to an interlipid distance of about 0.45 nm, classical for $L_\alpha$.[262,263] A more quantitative study of the phase diagram between 0 – 100 wt% is performed in a second work.[224] Optical microscopy performed on a sample at 10 wt% shows the presence of vesicular aggregates with a thick wall. This is largely consistent with both XRD (lamellar peaks) and polarized light microscopy (maltese crosses) and observed until 60 wt%. Authors attribute this phase to a biphasic $L_\alpha$/water system, although a more correct attribution would be *MLV* phase (please refer to the comment on *MLV* and $L_\alpha$ at the beginning of the vesicle and lamellar paragraphs in section 2.3.3 and 2.3.4). In terms of interlamellar distance, MEL-B lamellar phase shows a constant d-spacing of 4.7 nm until 60 wt%, while for higher concentration d-spacing linearly decreases until 3.1 nm up to 100 wt%. Such behavior is typical of a lamellar systems due to the increased volume fraction and water expulsion from the interlamellar space.[224] Above 60% MEL-B forms a condensed lamellar phase, attributed to $L_\alpha$. Authors also show that both the *MLV* and lamellar phases are stable between 5°C and 95°C and up to 80 wt%.

*S*- and *D*- isomers of MEL-B have similar phase diagrams, MLV below about 55 wt% and $L_\alpha$ lamellar above, whereas the *S*- isomer has a slightly higher concentration for the $MLV/L_\alpha$ transition.[225]

*MEL-C* (**21C**). A series of MEL-C compounds has been identified from the fermentation of various *Pseudozyma* microorganisms.[152,226,227] For all molecules, authors reported a qualitative water penetration study associated to polarized light microscopy. Interestingly, all molecules show a fairly extended $L_\alpha$ lamellar domain. However, according to the production organism, they observe an isotropic-hexagonal-lamellar phase transition,[227] water-myelin-lamellar transition[226] and direct water-lamellar transition.[152] If the origin of such discrepancy is unclear, one could guess that these data should be confirmed with more accurate diffraction experiments, and in any case it is also possible that each MEL-C, being obtained by a different microorganism, contains an uneven distribution of fatty acids.[152]

*MEL-D* (**21D**). This compound corresponds to the nonacetylated version of MEL-B (**21B**), obtained by enzymatic catalysis. MEL-D displays a similar phase behaviour as MEL-B (**21B**) is detected: between 0 and 50 wt%, authors find a *MLV* phase (shown by both XRD and microscopy, but labelled $L_\alpha$ + w), while above 50 wt%, condensation into a dense lamellar phase is most likely obtained (Figure 14d).[228] The higher capacity for lamellar phase of MEL-D to retain water is attributed to the force of the nonacetylated hydrophilic group of MEL-D contrary to MEL–B (**21B**) which have one acetyl group.[19] Finally, *S*- and *D*- diastereomers on the mannose group of MEL–B (**21B**) and MEL–D show differences in terms of surface activity and CMC but equal phase behaviour, except for the phase boundary between vesicle and lamellar phases (Figure 14b):[225] *S*- isomers systematically display a slightly higher concentration for the $MLV/L_\alpha$ transition.

***Surfactin*** (**20**)

Imura *et al.*[282] have determined the phase behavior of natural cyclic and linear surfactin, the latter chemically derived from the cyclic congener, above 50 wt% in water. They found that cyclic surfactin forms a lamellar structure between 50 wt% and 70 wt%, determined by polarized light microscopy and SAXS, while linear surfactin displays no birefringence, with a corresponding SAXS pattern attributed to a bicontinuous cubic ($V_2$) phase of *Pn3m* symmetry at 50 and 60 wt% (Figure 14a). At 70 wt%, the cubic phase coexists with a reverse hexagonal phase $H_2$. The repeating distance in the lamellar phase varies from from 3.9 nm to 3.2 nm between 45 wt% and 85 wt%. According these data, the bilayer thickness is estimated at 3.01 nm. Considering the estimated thickness of a surfactin double layer to be of 4.32 nm, SAXS data suggest the existence of an interdigitated layer. Interestingly, it is discussed that the lactonic ring of surfactin tends to reduce the negative curved close to 0, although this is somewhat unexpected in terms of packing parameter, if one considers the lactone to be bulkier. Considerations on the packing parameters will be done in the next section.

***Trehaloselipids.***

To the best of our knowledge, the phase behavior of trehalolipids was never reported, although its effects on different phospholipidic membranes have been studied.[120,283,284] This topic falls out of the scope of this review.

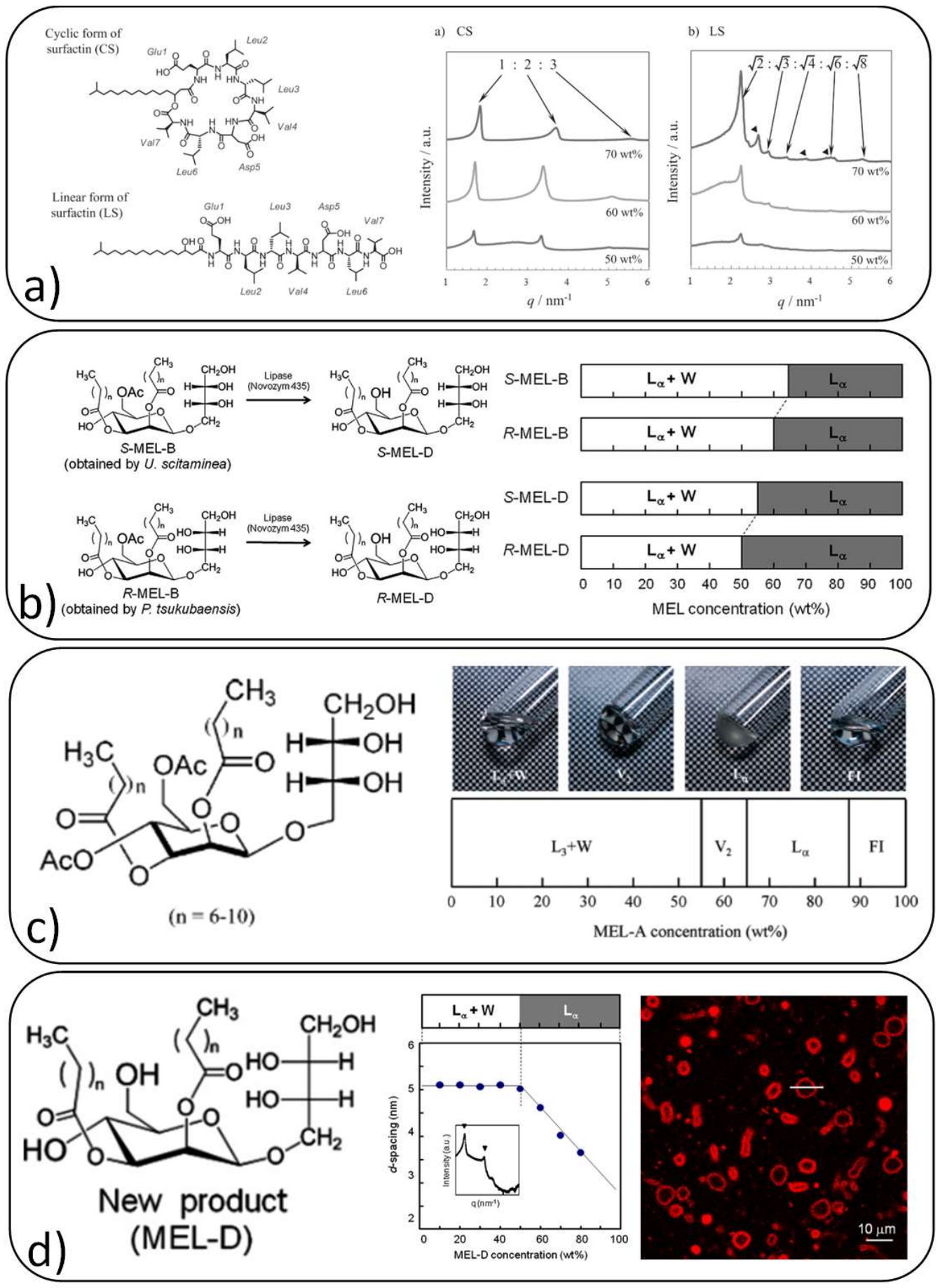

Figure 14 – Concentration-dependent phase diagrammes for a) surfactin (cyclic and linear) (**20**) (adapted from Ref. [282], copyright Japan Oil Chemical Society), b) *R*- and *S*- MEL-B (**21B**) and MEL-D (**21D**) (reprinted from Ref. [225] with permission from Elsevier), c) MEL-A (**21A**) (adapted with permission from Ref. [223], copyright (2007) American Chemical Society), and d) MEL-D (**21D**) (reprinted from Ref. [228] with permission from Elsevier). The $L_\alpha$+W in b) and d) should rather be interpreted as a *MLV* phase. Please refer to sections 2.3.3 and 2.3.4 for more explanation.

## 2.5 The notion of Packing Parameter associated to biosurfactants

The packing parameter, *PP*, concept, recalled in section 1.4, has been discussed by several authors in relationship to the self-assembly of biosurfactants. It was employed to discuss, understand and tentatively explain the self-assembly properties of rhamnolipids,[144,178,195,217] sophorolipids,[200,201] glucolipids,[201] and surfactin.[182,217,282] However, a deeper look at the existing data shows that this approach, classically used for standard head-tail surfactants, does not satisfactorily explain the experimental evidence and in this regard it cannot be employed in a straightforward manner to predict the self-assembly of biosurfactants. This is illustrated in Table 9, which gives the calculated (*calc* columns) and experimental (*Exp* columns) values of *PP*, as well as the expected and observed morphologies, for a number of biosurfactants. In about 50% and 25% of the examples, the *PP* theory either fails or it is only partially verified, while it is verified for only 25% of the examples reported.

In clear, the *PP* theory partially works, with exceptions, for many biosurfactants under alkaline conditions (noted B, for basic, in Table 9). It generally fails for saturated systems under acidic conditions (noted A, in Table 9), while the discussion is open for surfactin, of which the role of the peptidic part, may it be linear or cyclic, is an open question. It is then spontaneous to question the reason why the *PP* theory is unreliable to explain and predict the self-assembly of most biosurfactants. Please note that *PP* for MELs is not reported in Table 9 because we could not find any evaluation.

Table 9 – Calculated and experimental packing parameter, *PP*, found for a number of biosurfactants (N° in Figure 7) in water at room temperature and under dilute conditions, generally below 5 wt%. Unless otherwise stated, SL, GL and CL are in their open acidic form and nonacetylated. Acronyms: BS : Biosurfactant ; Calc : calculated ; Exp : experimental ; *PP* : Packing Parameter; A, B, N: acid, basic, neutral. Symbols: *= biphasic system where the major phase is micellar; §= fiber morphology is not discussed in the *P* theory (refer to sections 1.4 and 2.3.2), however, in all BS systems fibers are flat objects and the best approximation would be a P of 1, here indicated in brackets; #= C> 50 wt%

| **BS** | **Type** | **N°** | **pH** | ***Calc PP*** | ***Calc shape*** | ***Exp PP*** | ***Exp shape*** | ***PP* theory** | **Ref.** |
|---|---|---|---|---|---|---|---|---|---|
| RL | Mono- | 19 | B | 0.67 | Vesicle | > 0.5 (C> 20 mM)<br>< 0.33 (C< 20 mM) | Planar<br>Micellar | Verified<br>Fails | 144 |
| RL | Di- | 18 | B | 0.5 | Cylinder/vesicle | < 0.33 | Sphere | Fails | 144 |
| RL | Mono- | 19 | A | 0.24, 0.62 | Micelle/vesicle | > 0.5 | Vesicle | Partly Verified | 178,195 |
| RL | Di- | 18 | A | 0.27, 0.73 | Micelle/vesicle | > 0.5 | Vesicle | Partly Verified | 178,195 |
| RL | Di/Mono | 18,19 | A | > 0.5 | Cylinder/vesicle | >0.5 | Vesicle | Verified | 159 |
| RL | Di/Mono | 18,19 | B | > 0.5 | Cylinder/vesicle | < 0.3 | Sphere | Fails | 159 |
| RL | Di/Mono | 18,19 | N? | 0.544 | Vesicle | - | - | - | 217 |
| SL | Lactonic Part acetyl | 3 | - | 0.65 | Vesicle | > 0.5 | Planar | Verified | 200 |
| SL | C18:1-*cis* Part acetyl | 2 | A | 0.22 | Sphere | <0.33 | Sphere | Verified | 200 |
| SL | C18:1-*cis* | 1 | A,N,B | 0.36 | Sphere/cylinder | < 0.33 | Sphere | Partly Verified* | 201 |
| SL | $-NH_2$ C18:1-*cis* | 4 | A,N,B | 0.36 | Sphere/cylinder | < 0.33 | Sphere | Partly Verified* | 202 |
| SL | C18:1-*trans* | 12 | B | 0.28 | Sphere | / (1)§ | Sphere/Fiber | Fails | 203,219 |
| SL | C18:1-*trans* | 12 | A | 0.28 | Sphere | / (1)§ | Fiber | Fails | 203,219 |
| SL | C18:0 | 13 | B | 0.28 | Sphere | < 0.33 | Sphere | Partly Verified* | 201 |
| SL | C18:0 | 13 | A | 0.28 | Sphere | / (1)§ | Fiber | Fails | 201,219,221 |
| SL | $-NH_2$ C18:0 | 5 | B | 0.28 | Sphere | / (1)§ | Fiber | Fails | 202 |
| SL | $-NH_2$ C18:0 | 5 | A | 0.28 | Sphere | < 0.33 | Sphere | Partly Verified* | 202 |
| GL | C18:1-*cis* | 11 | B | 0.72 | Vesicle | < 0.33 | Sphere | Fails | 201 |
| GL | C18:1-*cis* | 11 | A | 0.72 | Vesicle | > 0.5 | Vesicle | Partly Verified* | 201 |
| GL | C18:0 | 10 | B | 0.56 | Vesicle | < 0.33 | Sphere | Fails | 201 |
| GL | C18:0 | 10 | A | 0.56 | Vesicle | 1 | Lamellae | Fails | 201 |
| CL | hydrolyzed | 17 | B | 0.28 | Filaments | < 0.33 | Sphere | Fails | 201 |
| CL | hydrolyzed | 17 | A | 0.28 | Sphere | / (1)§ | Fiber | Fails | 201 |
| Surf | Cyclic | 20 | N | 0.21, 0.65 | Sphere, Vesicle | < 0.33 | Sphere | Partly verified | 182,212 |
| Surf | Cyclic | 20 | B | 0.21, 0.65 | Sphere, Vesicle | < 0.33 | Sphere | Partly verified | 182,212 |
| Surf | Cyclic | 20 | A | 0.21, 0.65 | Sphere, Vesicle | > 0.5 | Bilayer | Partly verified | 182,212 |
| Surf | Cyclic | 20 | N? | 0.176 | Sphere | - | - | - | 217 |
| Surf# | Cyclic | 20 | N? | - | - | 1 | Lamellae | - | 282 |
| Surf# | Linear | - | N? | - | - | < 0.33 | Sphere (cubic) | - | 282 |

Most BS fall in the category of bolaamphiphiles. The general outlines of the *PP* theory applied to head-tail surfactants is presented in Section 1.4, while Nagarajan generalizes it to bolaamphiphiles in the late '80s.[37] According to his work, calculation of *PP* for bolaamphiphiles relies on twice the headgroup surface area and half the chain length per molecule. Despite such a discrepancy, the upper and lower end values of *PP* defining the shape of the aggregate do not change between one-headed and bolaamphiphiles. This fact rules out possible errors between the calculated *PP* and experimental morphology observed for bolaamphiphiles structures. The poor agreement between the *PP* theory and biosurfactants self-assembly must then be found elsewhere. Below, we list the most plausible flaws.

a) Flaws in the theory.

*Hydrophobic tail.* The main hypothesis of the *PP* theory is based on the consideration that the hydrophobic region of the micelles is liquid and of homogeneous density, the latter being similar to hydrocarbon liquids. This strong hypothesis holds for most one-headed amphiphiles because their backbone is constituted by an aliphatic chain, which is generally liquid at room temperature. However, the backbone of most biosurfactants, glycolipids in particular, is constituted by fatty acids, which are not necessarily liquid at RT. This is particularly true for biosurfactants at acidic pH having a saturated fatty acid, like C18:0 sophoro- (**13**) and glucolipids (**10**) or cellobioselipids (**17**). Saturated SL (**13**) and CL (**17**) form semicrystalline fibers upon pH jump,[201,252] or plaquettes if dispersed in water.[219] C18:0 glucolipids (**10**) stabilize flat membranes, which are more rigid than vesicles, the preferred morphology of the corresponding monounsaturated glucolipid.[201] Such behavior is directly related to the tendency of the saturated fatty acid to crystallize at room temperature. Control experiments of these compounds show flat-to-vesicle or a vesicle-to-flat membrane transitions respectively upon heating and cooling processes.[201] This is also in agreement with the phase diagrams of fatty acids in water and which were shown to depend on temperature with respect to the melting temperature of the fatty acid.[285,286]

*Hydrophilic headgroup*. The case of surfactin (**20**) is particularly interesting, because its self-assembly behavior with respect to the *PP* theory is specifically discussed by Shen *et al*.[182], commented in Déjugnat *et al.*[129] and partially studied by Imura *et al*. [282]. With respect to glycolipids, surfactin has a classical aliphatic tail and it's not bolaform; however, definition of the hydrophilic headgroup, intended in the sense of the *PP* theory, is certainly a challenge. In fact, part of the surfactin peptidic ring could be integrated in the hydrophobic palisade, the notion of headgroup equilibrium area is certainly influenced by the peptide conformation, onto which pH has a crucial effect. This is clearly shown by Imura *et al*.,[282] who put in evidence the critical impact of the linearity vs. cyclicity of the peptidic headgroup on the phase diagram of surfactin above 50 wt%, the former providing a lamellar phase and the latter a major cubic phase.

b) Flaws in the calculation of *PP* for biosurfactants

The molecular structure of BS is often complex and the choice of the proper values of the tail length, volume and surface area of the headgroup may be cumbersome, as discussed by Zdziennicka *et al*.[217] and Helvacı *et al*.[195] The general approach consists in using the Tanford formula for the tail length ($L$= 1.54 + 1.265$n_c$) and volume ($V$= 27.4 + 26.9$n_c$, with $n_c$ being the number of methylene groups in the acyl chain)[38] in combination either with a gross estimation of the headgroup area on the basis of its size[201] or with the value measured at the air-water interface using techniques like surface tension, pressure-distance or neutron reflectivity.[144,195,217,287] Each method is unfortunately fallacious: the Tanford approach is not the only one,[288,289] the relationship between the headgroup size and its area is not straightforward because it depends on the physicochemical parameters and, finally, the packing at a flat air-water interface is not necessarily the same at a curved micelle/water interface. The result of these uncertainties can be observed in the large variety of calculated *P* values for rhamnolipids (may them be di- (**18**), mono- (**19**) or mixtures), and going from, roughly, 0.25 to 0.75 (Table 9).[178,195,196,217] Similarly for surfactin (**20**), calculation of *PP* is not reliable, as shown by the values given by Zdziennicka *et al.*[217] (*PP*= 0.176) and Shen *et al.*[182] (0.21 to 0.65). The large variation in the latter case is explained by the authors as follows: according to neutron scattering, the leucine groups in the peptide ring are hydrophobic and could be part of the hydrophobic region, providing *PP*= 0.65, whereas if leucines are considered as a part of the hydrophilic peptide ring, where they actually are, then *PP*= 0.21. Experimentally, authors find micellar aggregates, which are more compatible with the latter value. Interestingly, to the best of our knowledge, no study mentions the calculated *PP* of MELs (**21**).

c) Effect of pH: ionic, neutral and mixture of ionic and neutral species

All biosurfactants, may them be glycolipids and lipopeptides, have free carboxylic acids. The effect of pH is then important and this was shown by many authors since 1987, the first work on rhamnolipids.[159] According to pH, biosurfactants are charged, neutral or a mixture of neutral and charged species. In respect to the *PP* theory, the headgroup repulsive component in the free energy of micellization does not have the same magnitude and the equilibrium area per headgroup strongly varies between the ionic and neutral forms (more information on the *PP* in section 1.4, note 1, and sections 1.5 and 1.6). Upon full deprotonation at high pH, electrostatic repulsions between carboxylates dominate and the curvature will necessarily be higher. At low pH, when all carboxylic acids are protonated, only the steric component appears in the free energy and the area per headgroup will be smaller. This is systematically observed experimentally, where practically all biosurfactants considered here tend to form spheroidal micelles at high pH and low-curvature membranes at low pH. For intermediate pH values, the system is always a mixture, the neutral and ionic forms, of which the proportion differs at each pH value. The self-assembly behavior is then even harder to predict, although it generally follows a logical change in curvature, as observed for C18:1-*cis* glucolipids (**11**), of which the self-assembled structures varies between spheres, cylinders, worms, flat membranes and vesicles.[69] In other cases, like for acidic C18:0 SL (**13**), a true gap exists between the charged (micelles) and neutral (fibers) species, with no morphological gap.[69] This is explained by point a), that is the poor solubility of the neutral species at room temperature.

d) Hydrogen bonding

Biosurfactants contain several functional chemical groups which can undergo hydrogen bonding, like OH or amides. This specific interaction is not considered in the classical *PP* theory, although it can have a non-negligible impact on the overall driving force towards specific morphologies, like fibers, for instance. Although hydrogen bonding has been shown before to drive the final structures in glycolipid bolaamphiphiles, and in particular in association to the orientation of the OH groups in the sugar moiety,[62] a clear-cut proof of the impact of hydrogen bonding on the structure of biosurfactants was never really provided. Several authors tried to elucidate the effect of hydrogen bonding by studying the differences in terms of self-assembly between acetylated (**2**) and nonacetylated (**1**) sophorolipids,[109,200] but the results were not absolutely clear, mainly due to the coexistence with lactonic and/or nonacetylated SL in non-negligible amounts. A recent work on porphyrin derivatives of SL (**7**) has indeed shown the importance of hydrogen bonding on the type of self-assembled structure,[209] whereas the fully acetylated compound tends to form micellar aggregates while the de- and partially acetylated compounds rather form columnar aggregates. The most striking impact of hydrogen bonding is probably shown in the phase behavior of MELs (**21**), whereas partial, or total, nonacetylation from MEL-A (**21A**) to MEL-*X* (**21*X***) (*X*= B, C, D) induces a phase transition from a coacervate, $L_3$, phase to a MLV phase below about 60 wt% and RT.[188–190,223,225,228] The OH groups were ascribed to play a somewhat, although still not understood, important role in the positive Gaussian and mean curvatures (notions about the curvature are recalled in Section 1.4, Figure 3) of the *MLV* phase.

e) Effect of impurities

Studies on highly purified and homogeneous biosurfactants are extremely rare and, in fact, of little interest, because all of them are never obtained with a high degree of homogeneity. Extreme purification can only be performed with analytical chromatography, which generates only mg-scale amounts. This is obviously not compatible with real-life applications. Preparative column chromatography could be of a certain interest, but in the case of biosurfactants, it does not guarantee full separation, because many congeners have very similar elution times. It was for instance shown that an undesired higher concentration of saturated C18:0 SL (**13**) in a C18:1-*cis* SL (**1**) batch after preparative chromatography was responsible for the fiber formation of the latter.[191]

A more practical approach consists in removing acetylated (**2**) and, when existing, lactonic (**3**) congeners through a relatively simple alkaline hydrolysis step.[290] This method is performing, although it does not remove the, often large, number of congeners presenting variations in the number of unsaturation and tail length. Fortunately, the total fraction of such molecules is generally below the 10%, an amount which is not exceedingly high, but which could still influence the self-assembly properties.[191]

Acetylation is an interesting way to study the effect of hydrogen bonding on the self-assembly properties.[190,209] However, controlling selective and full acetylation is not an obvious task. Penfold *et al*. chose chromatography to prepare acetylated SL-rich (**2**) batches, although still containing a mixture of mono-, di-lactonic (**3**) and in some cases even nonacetylated compounds (**1**).[200] Baccile *et al*. preferred to employ genetic engineering, but acetylation was also partial.[109] Peters *et al*. chose enzymatic catalysis to achieve peracetylation, but that was done on a SL derivative with quite a complex self-assembly behavior *per se*. The most striking effect of acetylation probably occurs on the self-assembly of MELs (**21**). Introduction of one acetyl group in the 4' position of the mannosyl ring (MEL-A, **21A**) promotes the formation of a coacervate *L3* phase, while its absence and replacement with a OH group (MEL-B, **21B**) drives the assembly into a *MLV* phase.[188]

It then goes without saying that use of the *PP* theory for systems which always contain at least a 5% to 10% of congeners, each having it own self-assembly behavior cannot be fully reliable. Table 9 shows that nonacetylated acidic C18:1-*cis* SL (**1**) form a stable micellar phase from acidic to basic pH, in good agreement with the calculated *PP* and in disagreement with many other biosurfactants, which show different properties between acidic and basic pH. However, this is probably a coincidence, because it was shown that the 100% pure compound, isolated by column chromatography, is actually water-insoluble at acidic pH.[191] It then most likely seems that the *PP* theory actually fails for this compound, although it is valid within the context of its common batch heterogeneity.

f) Multi-phases

Coexistence between multiple phases is another hint that the *PP* theory is not adapted for biosurfactants. Many glycolipids presented in Table 9 show one main and a second minority phase. Acidic nonacetylated C18:1-*cis* (**1**),[220,235] C18:0 (**13**),[220] behenic sophorolipids (**16**),[205] cellobioselipids (**17**), C18:1-*cis* (**11**) and C18:0 (**10**) glucolipids[201] at basic pH, diacetylated sophorolipids (**2**)[109,200] and surfactin (**20**)[129] at neutral/acidic pH present a low-q scattering signal in SAXS/SANS experiments, often attributed to a minor amount of large-scale structures. In most cases, these structures are constituted by ill-defined flat geometries, may them be platelets or membranes, and which can sometimes be observed flocculating by naked eye in the sample solution. The most striking case is certainly constituted by the time-dependent micelle-to-fiber evolution in nonacetylated acidic C18:1-*trans* SL (**12**) at basic pH, where the fiber fraction becomes more and more important and easy to observe and isolate.[203,219]

None of these structures is compatible with the *PP* theory, especially at basic pH, where all lipids are supposed to be negatively charged and are expected to be in a high-curvature environment. In fact, the actual origin of these structures is not clear, yet. It is not excluded that these structures are rich in the common impurities existing in all biosurfactant batches, as discussed above in point d), and in this regard, use of the *PP* theory would actually be meaningless. Unfortunately, These structures are often stable in a narrow pH range, as shown by the evolution of the fractal dimension in pH-resolved *in situ* SAXS experiments[69] and they represent less than 10% of the phase mixture, if one tries to model the SAXS data. Their isolation and analysis then becomes a true challenge.

g) Fast kinetics vs. thermodynamic equilibrium

The stimuli-responsive character of biosurfactants raises a fundamental question in terms of their behavior. Are their self-assembly properties the same under fast kinetic conditions with respect to thermodynamic equilibrium? To the best of our knowledge, nobody has specifically studied in depth the effect of kinetics *vs*. thermodynamics for these molecules. Many glycolipids have been studied under fast pH-change conditions, which certainly can play a role in the final equilibrium phase.[69] This seems to be the case for nonacetylated acidic C18:0 SL (**13**) at acidic pH, which form large crystalline plates when they are dispersed in water,[219] while they systematically form nanoribbons through a fast pH jump process.[221] Similarly, nonacetylated acidic C18:1-*trans* SL (**12**) at basic pH form micelles upon dispersion in water; however, twisted fibers form after a longer equilibration time (~ hour).[203]

A certain consensus is reached on surfactin (**20**), which has often been studied under equilibrium conditions at a given pH and shown to form vesicular and micellar structures, respectively under acidic and neutral/basic conditions.[129,212,265,282] However, its properties were not specifically evaluated under fast kinetics, to the best of our knowledge. In other systems, the impact of kinetics seems to be moderated. We have reported the formation of vesicles and flat lamellae, respectively for acidic C18:1-*cis* (**11**) and C18:0 (**10**) glucolipids when pH is varied from basic to acidic.[69,201] However, we have recently observed that both vesicle (unpublished results) and lamellar[218] phases can be obtained by simple dispersion of the respective compound in water at acidic pH after sonication.

h) Physicochemical parameters

Like all ionic surfactants, the self-assembly of biosurfactants is influenced by classical physicochemical parameters like temperature, ionic strength and nature of counterions, which play a role in the effective surface area of the hydrophilic group (please refer to Section 1.4 for the notion of effective area). In addition to these, pH plays a crucial role, as already discussed throughout this section. If several authors have reported the effect of ionic strength and counterions on the self-assembly of biosurfactants,[129,195,212,230,236,265,291] there is no complete study to date. Nonetheless, the general trend seems to confirm what is known for classical surfactants, that is: divalent cations bind strongly onto the carboxylate groups and can induce phase transitions, generally from highly curved to less curved morphologies.

Temperature also plays an obvious role, because it has a strong impact on the melting of the hydrophobic tail and, in this sense, on the elasticity of the corresponding membranes. If the effect of temperature on biosurfactants phase transitions has been reported,[198,201,208,209] no general study can be noted to date. In all cases, the effect of physicochemical parameters on self-assembly of surfactants and the impact on the effective headgroup area and chain volume are well-known, and in this regard there is no specific reason to expect special behaviors for biosurfactants. If the effect of physicochemical parameters on the phase can be partially forcecast in light of the *PP* theory (e.g., high ionic strength and/or strong binding affinity decrease the effective equilibrium area of the hydrophilic headgroup and induce a high-to-low curvature transition, section 1.4, notes 1 and 2, and sections 1.5 and 1.6), empirical studies must always be performed, due to the often unexpected behavior of counterions in water (e.g., Hofmeister *vs.* anti-Hofmeister series). This is particularly important, especially because the medium which is often used in the literature is not the same from one study to the other. Some authors employ pure water while others employ buffer solutions, a difference which could also contribute to observe different self-assembly behaviours.

## 2.6 Additional aspect on biosurfactants self-assembly

*Chirality*. Both glycolipid and peptidic biosurfactants have chiral centers and their corresponding self-assembled structures could be driven by chirality. Nonacetylated acidic C18:0 SL (**13**), but also (**14**), (**9**) and cellobioselipids (**17**) form twisted or helical ribbons, two chiral fibrillar structures.[201,252] In the case of SL, it was shown that both right- and left-handed fibers are observed, indicating that no specific sense is predominant. Similar results, although with a slighter excess in either left- or right-handed chiral structures, are reported for porphyrin-modified SL (**7**).[209,277] Also in this case, both structures are observed, although the length of porphyrin-to-SL spacer seems to have an influence on the sense of chirality. If no systematic studies associate the molecular chirality of biosurfactants to the sense of chirality of their corresponding self-assembled structure, this preliminary set of data strongly suggests that molecular chirality is not the driving force of structural chirality in biosurfactants self-assembled fibers. If this was the case, one would certainly observe nearly one rotatory sense. In other fibrous amphiphile-based systems, the sense of chirality was driven by the chirality of the counterion,[54] which cannot obviously be the case in these systems. Although counterions seems to have an effect, that is mainly related to the fibrillation process and fiber homogeneity.[252]

Fiber formation of low-molecular weight amphiphile is far from being reserved to biosurfactants and it has been reported for a large number of molecules.[292] In many cases, twisted ribbons are an intermediate between micellar aggregates and helical fibers, where the equilibrium structure is generally a lipid nanotube. In the specific case of biosurfactants, nanotubes were never reported and the twisted ribbon phase seems to be the stable phase even at long incubation time. Although the nanotube phase is often observed as the equilibrium phase for many systems, including synthetic glycolipids bolaamphiphiles with a COOH end-group[293] and strongly similar to sophoro and cellobioselipids, stable ribbons phase is in fact not uncommon in many amphiphilic systems.[292] If structural chirality can be driven by a chiral center on the amphiphile or by a chiral counterion, it can also be driven by a steric effect and/or by side-by-side hydrogen bonding interactions providing a non-zero tilt-angle between adjacent molecules.[292,294,295] To date, steric effect between adjacent disaccharide headgroups (Figure 15), possibly associated to hydrogen bonding, and presence of a fatty acid with high melting temperature are the most plausible explanations for twisted and helical fiber formation in sophoro- and cellobioselipids at acidic pH. Furthermore, the specific carbohydrate-COOH asymmetry in sophoro- and cellobioselipids is also excluded from the possible driving forces, because chiral fibers are also observed for symmetric sophorolipids (**9**),[208] C18:0 aminyl SL (**5**) at basic pH (neutral form) and alkynyl SL (**6**).[202]

In all cases, it seems that the fiber phase occurs for the neutral form of the biosurfactant, thus excluding the role of the negative (or positive, for aminyl derivatives) charge. This could also explain the reason why alkynyl SL (**6**) also form fibers. However, we should also note the exception constituted by the nonacetylated acidic C18:1-*trans* SL (**12**), which form stable ribbons even at basic pH.[203,219] In the latter case, although not impossible, it is however unclear whether or not the ribbons are constituted by negatively charged SL or not. The question of the ribbon composition and their actual charge is still open.

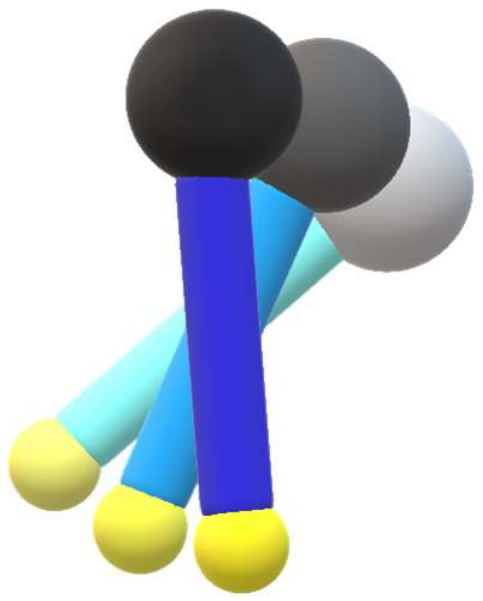

Figure 15 – Chiral molecular assembly depending on the steric hindrance of adjacent molecules packed at a nonzero angle one with respect to the other. Image redrawn by authors, adapted from Ref. [295].

*Stimuli-induced phase transitions*. All biosurfactants considered in this work, except MELs, undergo a phase transition in response to pH. Often, the structures are studied *ex situ* at a given pH, but advanced pH-resolved *in situ* SAXS studies have been performed, revealing the precise mechanism of the morphological transition.

Table 10 summarizes the main data about the pH-driven phases reported for biosurfactants while Figure 16 highlights some selected examples concerning acidic C18:1-*cis* (**1**), C18:0 sophoro- (**13**) and glucolipids (**10**, **11**), rhamnolipids (**18**, **19**), surfactin (**20**) but also porphyrin-derived (**7**) sophorolipids. When the phase is given in italic in Table 10, it refers to a minor fraction, while a phase given in brackets indicates the most plausible morphology. Table 10 does not report the work of Zhou *et al.*[197] and Imura *et al.*,[270] both concerning onacetylated acidic C18:1-*cis* SL (**1**). If the former, consistently with other reports, shows a micellar phase at pH 7.8, it proposes the formation of ribbons at acidic pH. However, as commented in the sophorolipid discussion in section 2.3.2, ribbons are observed for C18:0 (**13**) and C18:1-*cis*/C18:0 (**1**, **13**) mixtures. We then prone for a congener effect to explain ribbon formation in Ref. [197]. This issue is supposed to be settled in Ref. [191], although further work would not be unwelcome. The second work proposes the formation of a vesicle phase at high pH values. However, this is the only report showing on vesicle formation from (**1**) and it relies on a single freeze-fracture image. Vesicles were reported within the context of SL but only for their lactonic form using statistically-relevant SANS arguments.[200] Similarly, SL-based vesicles were also reported but for linolenic, C18:3, derivatives of SL (**15**) at acidic pH.[214]

Surfactin (**20**) shows micellar formation at neutral-basic pH and bilayer formation at acidic pH, with an intermediate cylindrical phase in between. Nonetheless, the morphology, flat or vesicular, of the bilayer is not clear to date.[212] Finally, one should note that

Table 10 does not report the self-assembly of lactonic and acetylated systems, because high pH values would result in a potentially partial nonacetylation process. Similar considerations could be done for RL (**18**, **19**), of which the chemical structure is based on ester bonds, highly sensitive to hydrolysis at basic pH. However, most works on RL are carried out at pH not higher than 9, where hydrolysis should still be minimized.

The effect of temperature is also important, although it was so far studied for a limited number of biosurfactants. Table 11 reports the main biosurfactant systems where temperature-induced transitions have been studied. Finally, an example of light-responsive SL has been described in the literature,[296] however, the chemical stability of SL after irradiation as well as the final phase and mechanism of fluorescence emission are still unclear. For these reasons, this example will not be discussed further.

Table 10 – pH-driven phase transitions in biosurfactants (N° given in Figure 7). *: in brackets it is given the most plausible morphology; §: in italic it is given the minority phase. †: The micelle phase has been reported by [200,201,235] and it is favored by a spurious (< 5%) amount of C18:2, C16:0, C18:0 sophorolipid congeners[191] while the ribbon phase has been reported by [197,252] and it is favored by an excess amount (> 10-15%) of C18:0 SL congeners in the batch.[191]

| BS | Type | N° | Structure | | | | Method | Ref |
|---|---|---|---|---|---|---|---|---|
| | | | **Basic pH > 8-7** | **Neutral 8 < pH < 7** | **Acidic 7 < pH < 4** | **Strongly acidic pH< 4** | | |
| RL | Mono | 19 | Planar (vesicle) (> 20 mM) Micelles (< 20 mM) | Unknown | *Bilayer* (vesicle)* | - | *Ex situ* | 144,159,211,267 |
| RL | Di | 18 | Micelles | Unknown | *Bilayer* (vesicle)* | - | *Ex situ* | 144,159,211,267 |
| SL | C18:1-*cis* | 1 | Micelles /*Platelets*§ | Micelles | Micelles | Micelles/*Ribbons*† | *In situ* | 197,200,201,219,220,235 |
| SL | -$NH_2$ C18:1-*cis* | 4 | Micelles | Micelles | Micelles /*Platelets*§ | - | *Ex situ* | 202 |
| SL | C18:1-*trans* | 12 | Micelles/*Ribbons*§ | Micelles | Ribbons | Ribbons | *In situ* | 203,219 |
| SL | C18:0 | 13 | Micelles/*Platelets*§ | None | Ribbons | Ribbons | *In situ* | 201,221 |
| SL | -$NH_2$ C18:0 | 5 | Ribbons | None | Micelles /*Platelets*§ | - | *Ex situ* | 202 |
| SL | C18:3 | 15 | - | Unknown | Vesicle | - | *Ex situ* | 214 |
| SL | $C22:0_{13}$ | 16 | Micelles/*Vesicles*§ | Undetermined | Vesicle | multilamellar | *In situ* | 205 |
| GL | C18:1-*cis* | 11 | Micelles/*Lamellar*§ | Cylinders-wormlike | Vesicle/*MLV* | multilamellar | *In situ* | 69,201,216 |
| GL | C18:0 | 10 | Micelles/*Lamellar*§ | None | $P_\beta$ (Lamellar) | multilamellar | *In situ* | 69,201 |
| CL | Hydrolyzed | 17 | Filaments | Unknown | Ribbons | - | *Ex situ* | 201 |
| Surf | Cyclic | 20 | Micelles/*Other* | Unknown | Bilayer (Vesicle) | - | *Ex situ* | 129,182,212,265,282 |

Table 11 - Temperature-driven phase transitions in biosurfactants (N° given in Figure 7).

| BS | Type | N° | Structure | | Transition Temperature | Ref |
|---|---|---|---|---|---|---|
| | | | Low T | High T | | |
| SS | Symmetric C16:0 | 9 | Ribbons | Micelles | 28°C | 201 |
| SL | Phorphyrin-derivatives | 7 | Columnar or micelles | Monomers | 34-37°C | 209 |
| GL | C18:1-*cis* | 11 | Lamellar | Vesicle | Below RT | 201 |
| GL | C18:0 | 10 | Lamellar | Vesicle | Above RT | 201 |
| MEL | Mel-A (<~57 wt%) | 21A | Coacervate ($L_3$) | Isotropic | ~63°C | 223 |
| MEL | MEL-B (<~60 wt%) | 21B | *MLV* | *MLV* | - | 224 |

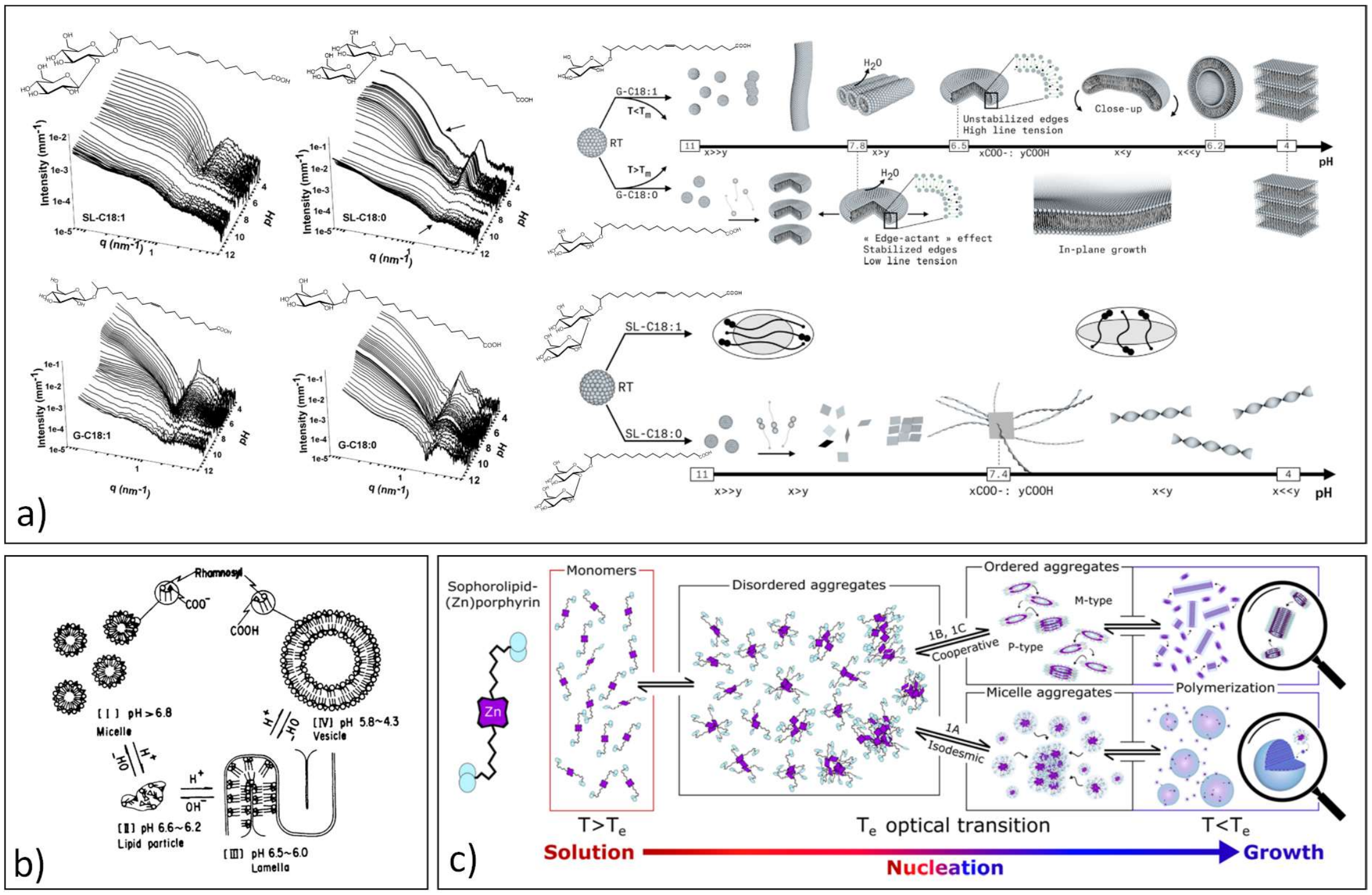

Figure 16 – pH-dependent phase transitions for a) nonacetylated acidic C18:1-*cis* (**1**), C18:0 sophoro (**13**) and glucolipids (**10**)[69] (adapted with permission from [69]. Copyright (2016) American Chemical Society) and b) rhamnolipids (**18**, **19**) (adapted with permission from [159], copyright Japan Chemical Society). c) Temperature-dependent phase transition for porphyrin-derivative of sophorolipids (**7**).[209] (adapted with permission from [209]. Copyright (2019) American Chemical Society).

## 2.7 Macroscale properties and possible applications of self-assembled biosurfactants

A number of more or less recent reviews present the physicochemical properties of biosurfactants[19,25,98,110] in relationship to several examples of application as antimicrobial, antiviral, gene delivery[297] molecules or as components in cosmetic formulations.[19,94,166]

However, most of the knowledge generated from the study of their self-assembly properties has not yet much been explored further. Precise control over the self-assembly opens an infinite number of opportunities in the fields of physics, chemistry, colloids and materials science. Amphiphiles can stabilize interfaces, thus giving opportunities of developing bio-based emulsions; lamellar objects can be employed to encapsulate and release colloids and molecules; fibers can be used to modulate viscosity but also to template chiral inorganic materials for enantioselective catalysis, and so on. In this short section, we want to show how the knowledge of the self-assembly properties of biosurfactants can lead to new systems, thus broadening the application potential of these compounds.[298] We stress the fact that we will only address biosurfactant systems, thus excluding formulations with other surfactants or lipids. A specific comment on this topic will be given at the end of this section.

*Hydrogelling*. Hydrogels can be obtained by colloids, polymers and low molecular weight gelators (LMWG). In this paragraph, we will considers biosurfactants as LMWG and we will exclude those studies in which BS are used as adjuvant to other gelling systems, like polymers.[299–303] LMWG are a well-known class of molecules, which, upon cooling, self-assemble into fibers and stabilize a solvent into a gel. If the solvent is organic, they are referred to organogelators, while in the case of water, they are referred to as hydrogelators.[298,304] Fiber-forming BS are given in section 2.3.2.

The first reports in this field were probably proposed by Imura *et al*., who found that cellobioselipids (**17**) can act as gelators of organic solvents[198] and water, in their ionic form.[155] Hydrogels are prepared by the classical heating (to 80°C) and cooling (to RT) approach and oscillatory rheology experiments displayed moderate-low elastic properties, with the elastic modulus, G′, in the order of 10 to 30 Pa at about 1 Hz and 2 wt%. The temperature dependent G′/ G′′ show a gel-to-sol transition at about 40°C (Figure 17a).

Two more recent reports extensively studied sophorolipids-based hydrogels. Symmetrical C16:0 sophorosides (**9**) spontaneously form twisted ribbons at room temperature and undergo a fiber-to-micelle phase transition at about 28°C, as shown by a combination of SAXS, SANS, cryo-TEM and rheology experiments. Upon dispersion in water, ribbons entangle and retain water in a gel state.[208] The evolution of G′ with concentration after an equilibration time of 15h displays a classical scale law with an exponent of 2.6, suggesting fibrillar entanglement as found in polymers (Figure 17b). Hydrogels are quite strong as they reach a G′ in the order of 1 kPa at 3 wt% and RT and it was shown that the sol-gel transition is highly reproducible over several hours and up to 20 heating-cooling cycles. Similarly to other temperature-driven LMW gels described in the literature,[256,305,306] the elastic properties of the sophoroside gels can be controlled by a slow decrease in temperature, a process which limits supersaturation and, consequently, fractal and spherulite formation, generally considered to be responsible for weak gels within the context of self-assembled fiber network (SAFiN) gels.

Sophorolipid-based SAFiN hydrogels were also found for nonacetylated acidic C18:0 SL (**13**), which are known to self-assemble into semicrystalline twisted ribbons. In this case, hydrogels are controlled by pH variation (Figure 17c)[257] rather than temperature, as found for LMWG peptide amphiphiles.[304,307,308] The values of G′ are shown to be impressively high, up to 10 kPa at about 3-4 wt%, when the gels are prepared using a homogeneous gelation process, rather than a manually uncontrolled pH variation.[257] Interestingly, and quite astonishingly, fast pH variations lead to precipitation of spherulites. In most of pH-dependent SAFiN, pH controls the micelle-to-fiber transition and in peptide-based amphiphiles it has been found that the final pH controls the strength of the hydrogels, thus explaining variations in the elastic properties of the SAFiN hydrogels.[309] However, within the context of SL, it was found that the final pH has little influence on the elastic properties, but it is rather found that the rate of pH variation has a huge impact, instead. This has been explained, and demonstrated by solution NMR, by the same supersaturation effect, known for temperature-driven, but never reported for pH-driven, SAFiN. It also seems that this specific system has a poor tolerance to salt, the higher the ionic strength, the poorer the elastic properties, in agreement with previous reports, which showed better ribbon homogeneity upon dialysis.[252]

SAFiN hydrogels with thyxotropic properties are also formed by palmitic acid C16:0 sophorolipids (**14**) upon pH jump form basic to acidic (details in section 2.3.2).[204] When pH approaches about 6, the solution becomes a hydrogel at already 0.5 wt% of the compound in solution at G' in the order of 50 Pa. The elastic modulus can increase above 10 kPa above 2.5 wt%, making C16:0 SL hydrogels very interesting. They also show a thixotropic behaviour with fast recovery rates, varying from 80% to 100% between 20 s and 5 min. Interestingly, the gel strenght depends on the final pH, as found for other peptide hydrogels,[309] rather than on the pH variation rate, as found for the C18:0 SL (**13**) hydrogels. The reason of this difference is still unclear but it may be related to the different fibrillation mechanism of the two molecules: pH-resolved in situ SAXS shows that C16:0 SL (**14**) undergoes a continuous micelle-to-fiber transition and show very homogeneous fibers, while C18:0 SL (**13**) undergoes an abrupt micelle-to-fiber transition and it can precipitate into spherulites at fast pH change rates, as expected in difusion-limited crystallization process.

Lamellar hydrogels are a class of gels, which are much rarer to obtain with respect to fibrillar, micellar or vesicle hydrogels.[268,304] Obtained for the first time in 1996,[310] they can only be obtained from defective lipid membranes, where defects are introduced by mean of polymer-grafted phospholipids[310] or surfactants.[311] It has been recently shown that nonacetylated acidic C18:0 glucolipids (**10**), known to undergo a pH-induced micellar-to-lamellar transition (refer to section 2.3.4),[69,201] spontaneously form lamellar hydrogels above 1 wt% in water below pH 7 (Figure 17c).[218,312] The lamellar character of the hydrogels has been shown by *ex situ* SAXS/SANS, *in situ* rheo SAXS[312] and polarized light microscopy experiments, as well as by the strong dependency of the interlamellar distance upon ionic strength, a typical feature and electrostatic-stabilized lamellar phases (Figure 13c). Differently than what was currently, although rarely, shown, glucolipids are able to form lamellar hydrogels without any additive, the natural $COOH/COO^-$ mixture being enough to provide flat stabilized lamellae, with an interlamellar distance varying between 25 nm and 10 nm, according to the ionic strength. The latter is shown to play a critical role in the gel strength: at low ionic strength, hydrogels are rather highly viscous solutions, while in the range of 50-100 mM and above NaCl, hydrogels become stronger, with elastic moduli reaching 10 to 100 kPa at room temperature at concentration in the order of 3 wt%. If pH and time have also shown to play an important role, these gels are always out of equilibrium and their elastic properties seem to be strongly variable. Lamellar hydrogels are generally described as a defectuous single lamellar phase. In the present case, a typical glucolipid gelled solution can be described by a heterogeneous network of large (hundred of microns) and highly defective lamellar domains.[218]

The results above are summarized on Table 12.

Table 12 – Additive-free biosurfactants (N° given in Figure 7) hydrogels and related stimuli-responsitivity and mechanical performances. *= Mechanical properties strongly vary with pH change rate; the value reported is obtained with homogeneous pH variation with glucono-δ-lactone. §= Elastic properties depend on rate of temperature variation. Slow temperature variation could improve G´of a factor 10. #= At RT, elastic properties strongly depend on a combination of pH, ionic strength and time. For a given time, pH and temperature, ionic strength strongly improves the elastic properties.

| BS | Type | N° | Gel | G´ / kPa ν= 1 Hz linear domain | C / wt% | pH | T / °C | Stimuli | Ref |
|---|---|---|---|---|---|---|---|---|---|
| SL* | C18:0 | 13 | Fibrous | Up to 10 | 3 | < 7 | 25 | pH | 257 |
| SL | C16:0 | 14 | Fibrous | ~4 | 3 | 5 | 20 | pH | 204 |
| SS§ | C16:0 symmetric | 9 | Fibrous | 1 | 3 | neutral | 25 | T | 208 |
| CL | Sodium salt | 17 | Fibrous | 0.01-0.03 | 2 | neutral | 25 | T | 198 |
| GL# | C18:0 | 10 | Lamellar | 10-100 | 3 | < 7 | 25 | pH, T, ionic strength, time | 218,312 |

*Solid foams.* Processing of the biosurfactant hydrogels into soft condensed materials has been shown on the fibrillar and lamellar sophoro- (**13**) and glucolipid (**10**) hydrogels. Through the freeze-casting, also known as ice-templating, technique, it is possible to prepare macroporous materials through the directional freezing of water, which is eventually removed by freeze-drying.[313–316] Hydrogels of similar elastic properties, but different self-assembled structures (fibrillar against lamellar), were frozen at rates up to 10°C/min without losing their nanostructure, as confirmed by temperature-responsive *in situ* SAXS and electron microscopy experiments.[317] After drying, the fibrous foams obtained from (**13**) display a preferential orientation of the macropores along the freezing axis, with expected anisotropic mechanical properties: the axial Young modulus is in the order of 20 times the transversal modulus. On the contrary, the lamellar foam from C18:0 glucolipids (**10**) shows Young moduli in the order of 20-30 kPa in both the axial and transversal directions, explained by the more isotropic orientation of the macropores within the material and they can withstand up to 1000 times their own weight, while fibrillar foams barely reach a factor 100 (Figure 17d). This result is particularly unexpected considering the strong anisotropic growth of ice and it displayed superior properties of the self-assembled lamellar structures compared to the fibrous one.

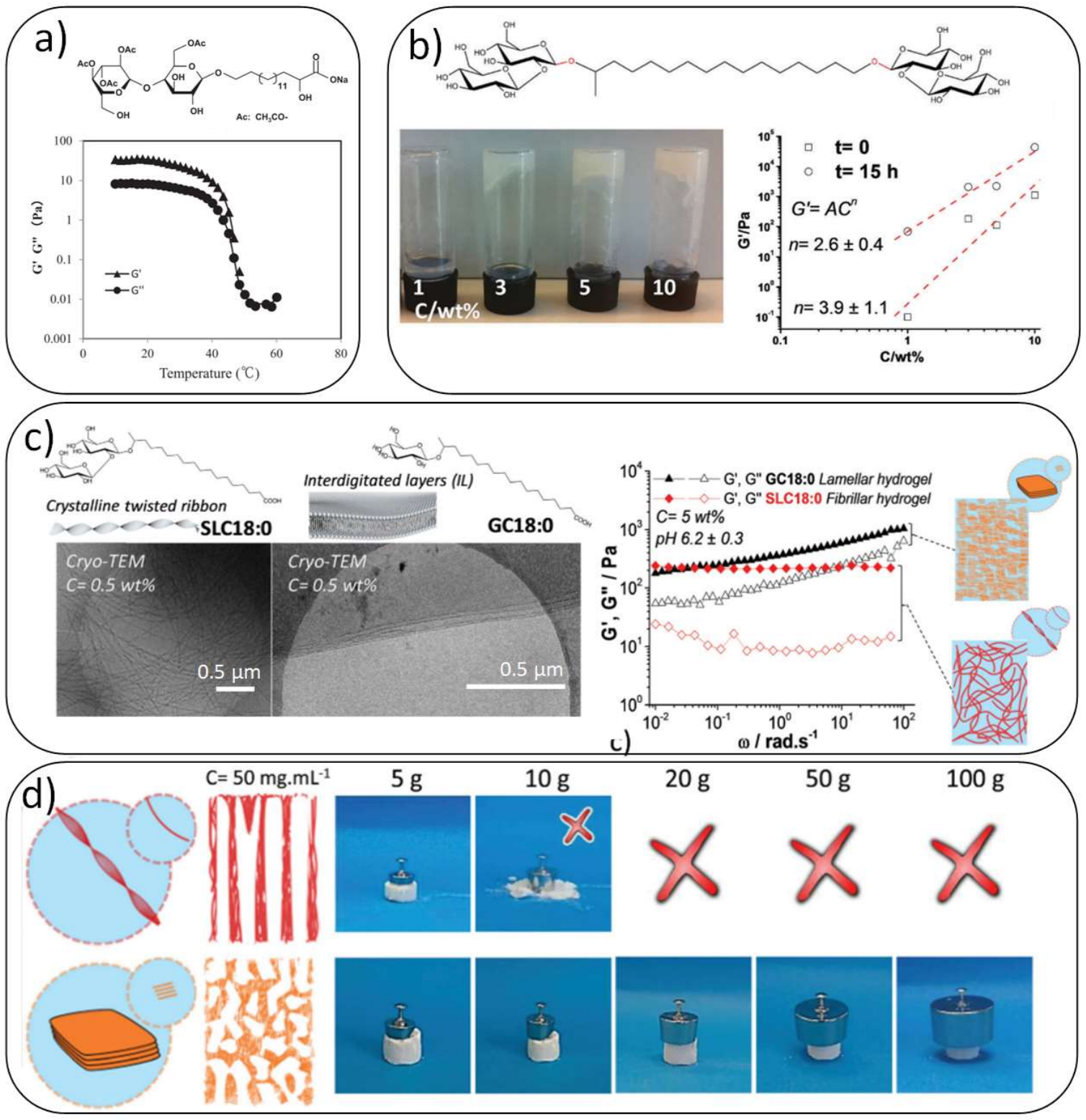

Figure 17 –Low-molecular weight biosurfactant-only gelators in water. a) Temperature evolution of G′, G′′ for cellobioselipids (**17**) hydrogels (reproduced from Ref. [155], copyright Japan Oil Chemical Society). b) Concentration evolution of G′, G′′ for C16:0 sophoroside (**9**) hydrogels (reproduced from Ref. [208] with permission from The Royal Society of Chemistry). c) Comparison of frequency-dependent G′, G′′ between fibrillar and lamellar hydrogels respectively obtained from C18:0 sophoro (**13**) and glucolipids (**10**) at acidic pH.[218,257,317] d) Comparison of the axial compression applied to solid foams prepared from fibrillar and lamellar hydrogels using the ice-templating process (c) and d) arereproduced from Ref. [317] with permission from The Royal Society of Chemistry).

*Fundamental biophysics*. Self-assembled lamellar structures associated to hydrogels revealed to be particularly useful as indirect probe for measuring the endogenic lateral pressures generated between ice columns during the ice-templating process.[318]

Lipid lamellar phases are, since long time, interesting tools to measure colloidal forces within and out of the context of DLVO theory. Pressure-distance profiles recorded on both supported and bulk lamellar phases are used to estimate interlamellar attractive Van de Waals and repulsive hydration, electrostatic or undulation forces.[58,87,319,320] Colloidal forces were measured for both mono-/di- (di-/mono- 1.2) rhamnolipids (**18**, **19**)[321–323] and C18:0 glucolipids (**10**).[324] In the former, authors used a Scheludko-Exerowa cell to measure the thickness of a thin rhamnolipid self-assembled membrane at the air-water interface at constant capillary pressure and explored the long-range (> 2-3 nm) electrostatic regime (discussed in section 3.1.1). In the glucolipid system, authors rather explored short-range (< 2-3 nm) primary and secondary hydration forces on glucolipid multilamellar layer deposited on silica and thermalized in a humidity pressure chamber, using neutron diffraction to measure the interlamellar distance (discussed in section 3.1.3).

The pressure-distance profiles recorded for C18:0 glucolipids (**10**) have been used as calibration curves for the first quantitative estimation of the interstitial pressure exerted by ice during the ice-templating process applied to the same C18:0 lamellar system, and for which the interlamellar distance is recorded with temperature-driven *in situ* SAXS experiments. Except for fast freezing rates and close to the cold finger, authors found average lateral pressure values between 1 and 2 kbar after water freezing and for temperature between -20°C and -60°C.[318]

*Template for porous oxides*. Self-assembled amphiphilic structures are known as soft endo-templates to prepare inorganic oxides with controlled porosity.[325–327] By mean of the evaporation-induced self-assembled process, nonacetylated acidic C18:1-*cis* SL (**1**) micelles have been used to template silica thin mesoporous films.[213,328,329] A combination of SAXS/SANS and TEM experiments show a correlation between the micellar size in solution, ~ 3 nm, and pore size within the silica network, ~ 2.4 nm. As proved by concomitant and later studies on micellar solutions of SL, the effect of pH reveals to be important also during the templating process. The porous structure strongly varies in size and morphology when templating occurs between acidic and basic pH, with the possibility to form hierarchical porous structures at pH above neutrality.[213,329]

Interestingly, biosurfactants having some antimicrobial and antiviral properties, it has been shown that SL in particular, and BS in general, could have an interesting double role of soft templates and drug, to be released from the mineral scaffold under controlled conditions after the templating process.[328]

*Encapsulation*. Encapsulation and release of hydrophobic compounds within amphiphilic carriers is a common, although highly demanded, application in medicine and it is essentially based on the self-assembly properties of the amphiphile in solution. Within the context of biosurfactants, several tests have been reported in the fields of drug and gene delivery, although in formulations with phospholipid liposomes, a case which is commented separately. When the lipid/biosurfactant formulation approach is excluded, one must acknowledge the fact that encapsulation within biosurfactants-only devices is quite a virgin field, probably because the phase diagrams of biosurfactants are not as well-known and controlled as for the case of phospholipids, which benefit of decades of studies worldwide.

One of the first reports about the encapsulation of hydrophobic compounds by a pure BS system is reported by Pornsunthorntawee *et al.*, who encapsulate the hydrophobic dye Sudan III into vesicles prepared from a batch containing mainly mono-RL (**19**) and other congeners. They found an encapsulation efficiency (amount of encapsulated dye over the total dye concentration) varying between 5-10% in PBS or PBS-NaCl medium and 30% in PBS-ethanol medium. Despite those preliminary and encouraging results, the potential impact of ethanol on the RL phase diagram and consequent encapsulation was not studied.[267]

A recent work has shown the dynamic encapsulation properties of vesicles prepared from nonacetylated acidic C18:1-*cis* glucolipids (**11**). This compound undergoes a micelle-to-vesicle phase transition from basic to acidic pH at RT and it was exploited to encapsulate various types of colloids: uncoated magnetic iron oxide (Figure 18a) and hydrophobic luminescent up-converting nanoparticles (Figure 18b), both unstable in water, but also stable ferritin nanocages (Figure 18c).[216] Interestingly, the micelle-to-vesicle transition was not suitable to encapsulate the colloids, which are systematically kept out of the vesicle lumen. However, a lamellar-to-*MLV* phase transition, where a condensed lamellar phase is generally obtained at pH below 4 for this compound, revealed to be the key to trap the colloids within the lumen with an average efficiency of at least 50% to 70% and in spite of their intrinsic colloidal stability in the absence of the vesicles (Figure 18d). After encapsulation, the colloids show a good stability over the range of hours upon both sponteneous and forced (magnetic or centrifuged) decantation, while stability in the absence of vesicle and upon forced decantation is only in the order of seconds. If the exact mechanism of encapsulation is still poorly understood, the lamellar-to-*MLV* transition is the key step, as also demonstrated by the multilamellar structure of the vesicles, differently than their classical single-wall structure obtained for the same molecule (Table 5 and Table *6*).

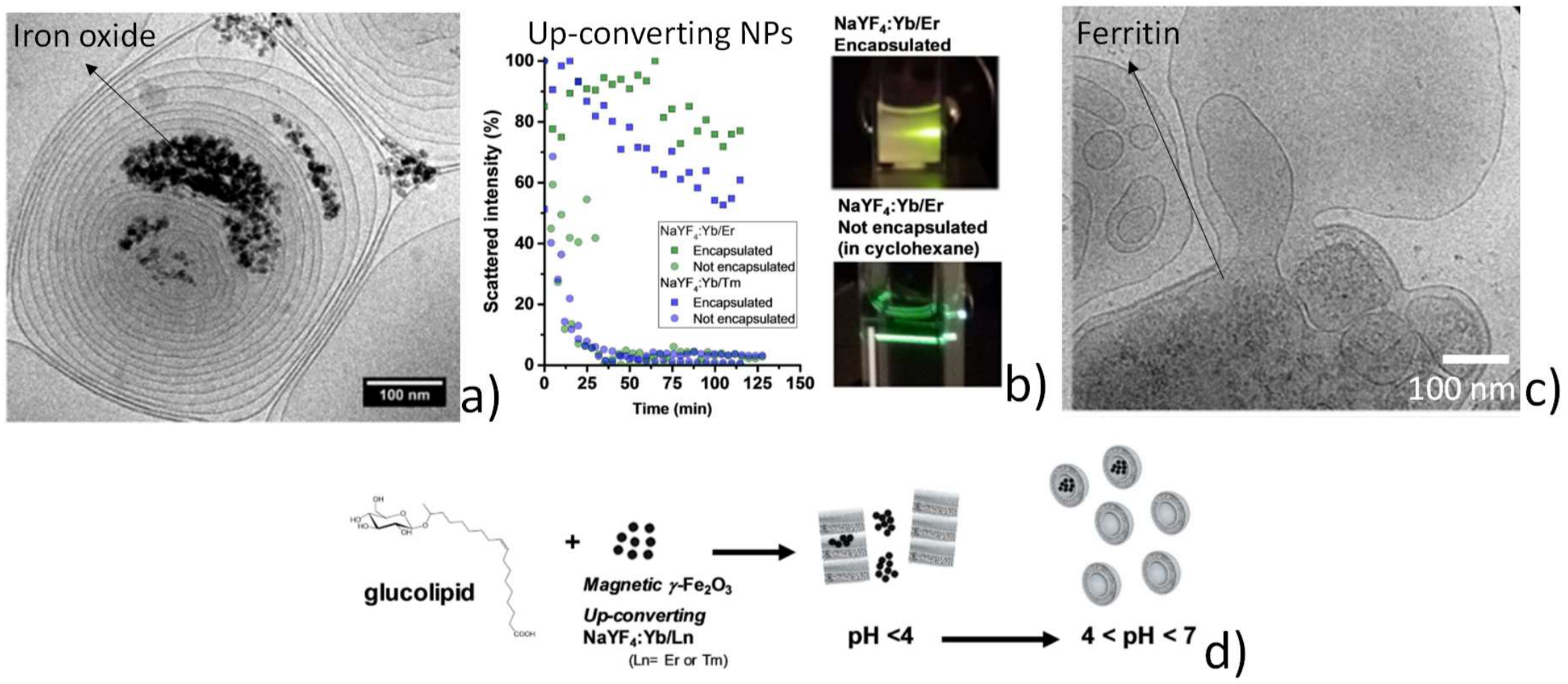

Figure 18 – Encapsulation of a,b) unstable (uncoated iron oxide, hydrophobic $NaYF_4$:Yb/Er) and c) stable (ferritin) nanocolloids within multilamellar vesicles prepared through a d) phase-change (lamellar-to-vesicle) process using nonacetylated acidic C18:1-*cis* glucolipids (**11**) (adapted from Ref. [216]. Copyright John Wiley & Sons).

Micelles are known and used for their solubilization potential of hydrophobic molecules in water (direct micellization) or hydrophilic molecules in organic solvents (reverse micellization). Within the context of solubilizing molecules of biomedical interest, antibiotics or antioxidants, a series of work has shown the interest of employing sophorolipids. Nonacetylated acidic C18:1-*cis* SL (**1**) were used to solubilize curcumin, an antioxidant with interesting effects in cancer therapy, in water. A curcumin/SL solution is sonicated to improve curcumin solubility and its impact on breast cancer cell lines, MCF-7 and MDA-MB- 231, is shown to be significant above 6.66 µg/mL of SL-curcumin mixture, while non-cancerous cell line HEK 293 is not affected.[330] Similarly, the same authors have shown uptake of SL-curcumin systems within *E. coli* and *S. aureus*, thus proposing these systems as biomarkers *in vitro*, and have shown antibiofilm effects against *P. aeruginosa*.[331] Uptake of SL-curcumin in combination of gold nanoparticles, synthesized by the same SL-curcumin complex as capping and reducing agent, by rats was also explored for potential *in vivo* use of these systems.[332] If this body of works presents a certain interest in the solubilization of curcumin, some questions are still open, and in particular concerning the nature of the SL self-assembled form. Authors generally speak of curcumin-containing micelles, of which the size is supposed to be in the order few nanometers, but they use SEM arguments to show that in one case SL form twisted ribbons and fibers,[330] in another they use TEM arguments to show large micellar aggregates of about 100 nm in size[331] and DLS arguments to prove SL-curcumin nanoparticles of 275 nm in diameter and SL-curcumin gold nanoparticles of 5 nm in diameter.[332] It was also unclear the reason why no clear spectroscopic FT-IR signature of curcumin was observed in the SL-curcumin systems. Further studies will certainly clarify these structural and spectroscopic details.

In a similar approach, SL (**0**) have been used to solubilize antibiotics.[333,334] Chuo *et al.* employed an unspecified form of SL to extract erythromycin and amoxicillin from aqueous solutions and transfer them into isooctane.[333] These molecules are insoluble in the organic medium and both forward and backward extraction methods were shown to improve solubility of the antibiotics in the organic phase, where SL are supposed to form an inverse micellar phase, although no specific data were supposed to support such a mypothesis. pH has an important role in the extraction, and particular efficiency was found for pH< 9.5 and above >4.5, respectively for erythromycin and amoxicillin. In the meanwhile, Joshi-Navare *et al.* have explored the solubilization potential and the synergistic effect of SL towards selected cefaclor and tetracycline.[334] Enhanced efficiency of antibiotics was noticed when co-administered with SL in both typical gram-positive and gram-negative bacteria, even if they differ in their cell membrane structure and mechanism of action against antibiotics. Tetracycline (15 µg/mL) is unable to totally inhibit the microorganism even after 6 h of exposure and a concentration of 300 µg/mL of SL alone efficiently inhibits *S. aureus* within 4 h. Upon combination of both agents, ~22% more inhibition was estimated at 2 h of exposure. The authors similarly explored the conjugative effect of SL (500 µg/mL) and Cefaclor (50 µg/mL) against *E. coli*. The microorganism was almost totally inhibited by Cefaclor alone after 6 h exposure. SL alone could not achieve a total inhibition of the bacterial growth, but co-administration of both agents provided equivalent results to Cefaclor alone in a reduced time: ~98% killing was reported within 4 h, against 6 h without SL.

Similarly, several bio-based and synthetic rhamnolipid molecules have been used to encapsulate drugs, namely Nile red, dexamethasone and tacrolimus nanoparticles with up to 30% drug loading (w/w). Authors report two methods of preparation and refer to the RL-drug system as "nanoparticle", whereas the phase of RL (**18**, **19**) is not specified, although the RL-drug interactions seem to convey towards colloidal systems with potential application in dermal drug delivery. Authors used both human dermal fibroblasts and human skin and showed efficient penetration of the model drug both into the stratum corneum and, although to a lesser extent, into the lower epidermis.[335]

*Interactions with other amphiphiles*. Interaction of biosurfactants with other surfactants and lipids is a topic which generates some peculiar interest. However, interactions with other amphiphiles generate a level of complexity which goes beyond the understanding of biosurfactants' self-assembly, the actual goal of this review. Nonetheless, we cannot ignore these systems, which are then only briefly mentioned here. In short, mixed BS/surfactants and BS/lipid systems are studied either to develop formulations for greener detergents[200,291,336] or more efficient gene transfection and drug carriers[98,239,283,284,297,337–347] but also to study the impact of BS on lipid membranes, for a better understanding of their interaction with living organisms.[283,284,338–342] We address the reader to the cited studies for more information. In particular, interactions between BS and lipid membranes have been recently reviewed by Otzen.[340]

## 3. Biosurfactants at interfaces

In this section we will describe the self-assembly of surfactants at interfaces, a crucial topic in many fields like food processing, health, chemical, agricultural and cosmetic industries.[348] In general, some examples of classical low and high molecular weight surface active compounds are: sodium lauryl sulfate, a anionic surfactant employed in cleansing or in the cosmetic industry;[349] polymers that allows stabilization of suspensions and nanoemulsions for pharmaceutical products, such as the non-ionic hydroxypropyl methylcellulose;[350] lecithin, used as an emulsifier in sauces like mayonnaise in food industry.[351] These amphiphilic molecules can then be employed at liquid/liquid, liquid/air, liquid/solid and solid/solid interfaces.

The complex structure of biosurfactants, their double headgroup as well as their chemical asymmetry make this class of molecules particularly interesting for studying their interfacial properties in view of their real-life applications. As in the rest of the review, only biosurfactants from microbial origin are considered in this section, which is organized in two main parts: 1) flat interfaces, generally looked at as model systems and which include air/liquid, liquid/liquid, solid/liquid and solid/solid interfaces; 2) curved interfaces, mainly involved in foaming, emulsion or stabilization of nanoparticles.

### 3.1 Flat interfaces

#### 3.1.1 Air-Liquid

In this section, we discuss the lowering of the surface tension at the air-liquid interface, and in particular air-water interface. Decrease in surface tension depends on the surfactant packing and in particular on the orientation of both the hydrocarbon chains and hydrophilic headgroup with respect to the air/liquid plane. The methods to measure surface tension analysis were reviewed in section 1.7, while the values of surface tension itself for various families of biosurfactants were shown and commented in section 2.2 (Table 3). For these reasons, here we will focus on the structural characterization of biosurfactants, including packing density through the area per molecule, studied by surface tension, pressure-area profiles using Langmuir trough or neutron reflectivity.

*Rhamnolipids (mono- (**19**), di- (**18**))*

The pH dependence of the surface and the interfacial behavior of mono- and di-RL was reported by Özdemir *et al.*,[196] who observed that CMC increases one order of magnitude from pH 5 to 6.8 for both compounds. The area-to-molecule ratio at the air/water interface varied from 0.60 nm$^2$ to 1.3 nm$^2$ by assuming a Gibbs prefactors of 1 (noionic at lower pH) and 2 (anionic at higher pH), in line with the increasing charge of the COOH group. Similarly, Helvaci *et al.*[195] have shown the expected opposite trend, a decrease of CMC and area-to-molecule ratio, while adding an electrolyte, known to screen the electrical charges.

In 2010, Chen *et al.*[144] used surface tension and introduced neutron reflectivity (NR) to study the adsorption behavior and structure of the mono- and di-RL and their mixtures at the air-water interface. Measurements were made at pH 7 and 9 in the presence of 0.5 M NaCl and found, as a general trend, larger surface area per headgroup at pH 9 (77 Å$^2$ for mono-RL and 80 Å$^2$ for di-RL) than at pH 7 (66 Å$^2$ for mono-RL and 77 Å$^2$ for di-RL), whereas the di-RL has an overall larger area than mono-RL. This is consistent with the bulkier dirhamnose headgroup of di-RL, inhibiting closer packing at the surface. The thickness of RL at the interface is about 20-23 Å, a range which reveals to be quite independent of the bulk conditions. The structure of each RL is also evaluated, with a denser packing for the di-RL and a major difference only in terms of the solvent distribution within the RL layer. Ö

Cohen and Exerowa published a series of papers on the biophysical study of thin liquid films, or planar film foams (Figure 19c), constituted of mono-RL,[322,352] di-RL[352] or mono-/di-RL mixtures,[321,323] but also RL and pulmonary surfactants.[353] Their goal was to characterize a RL thin film of RL and to determine the relationship between the type of film and the intermolecular forces responsible for its stability, within, and out of, the framework of the DLVO theory. A specific apparatus, the Scheludko-Exerowa measuring cell (Figure 19a), allows the formation of a microscopic surfactant/water film/air interface (Figure 19c), observed with a metallographic microscope. The evolution of its thickness, measured by interferometry (Figure 19b), against electrolyte concentration (Figure 19d) in solution is associated to the disjoining pressure evolution, measured by Exerowa-Kolarov-Khristov "thin liquid film" pressure balance technique. We recall that the obtained films can be either 'thick' common films (CF, appearing white to gray in reflected light) or black films (BF, virtually reflect no light). BF can be common (CBF, they comprise a water core between the adsorbed layers at its air/solution interfaces) or Newton (NBF, two adsorbed on each other monolayers of the surface-active substance). Authors found that NBF (film thickness < 6 nm) form around 1 M NaCl while CBF (film thickness < 20 nm) form at around 0.1 M NaCl (Figure 19d). From these data, they calculate pressure-distance curves, from which they can evaluate the intermolecular forces in the BS film.

They discuss that CBF are stabilized by an equilibrium between attractive Van de Waals and repulsive electrostatic forces, classical in the DLVO theory. However, the stability of thinner NBF films is explained by the possible presence of non-DLVO forces, namely repulsive hydration or, as proposed by the authors, steric contributions. Authors argue that hydration forces are very important for thickness below 1 nm. However, they did not consider secondary hydration forces, which are observed at longer distances (2-3 nm) in the presence of electrolytes.[88] These forces are reported in a recent study on multilamellar C18:0 glucolipids (**10**) at the solid/air interface,[324] a study detailed in section 2.7 (fundamental biophisics paragraph) and section 3.1.3 (glucolipid paragraph).

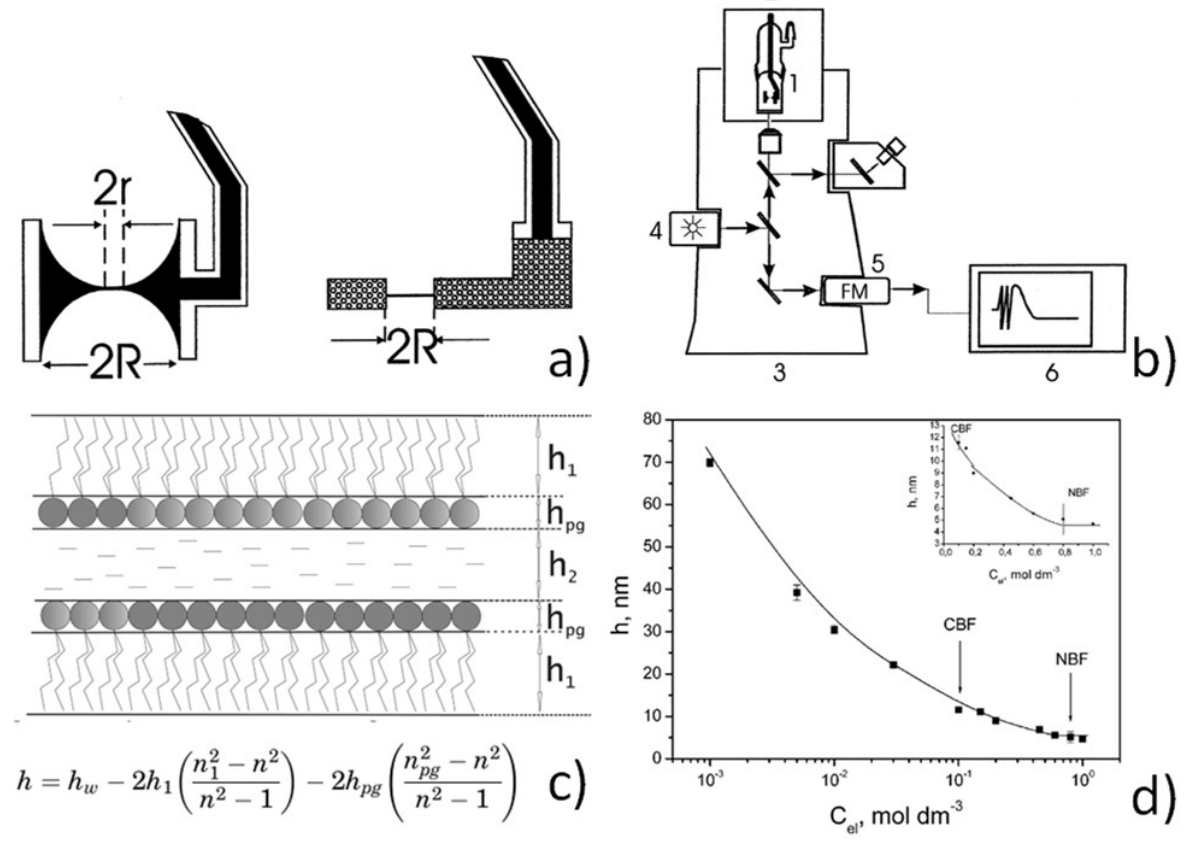

Figure 19 – a) Scheludko-Exerowa (left) and Exerowa-Scheludko porous plate cells; b) scheme of the interferometer; c) structure of flat foam layer and relationship between the thickness and the structural parameters of the foam film (*n* are the refractive indexes); d) typical thickness, *h*, vs. electrolyte concentration, $C_{el}$, with limits indicating formation of CBF (common black films) and NBF (Newton black films). Image adapted from Ref. [322], reprinted by permission of the Taylor & Francis Ltd.

In an interesting work, Zdziennicka *et al.*[171] have measured the thermodynamic parameters (standard free energy, enthalpy and entropy of adsorption) of rhamnolipids (and surfactin) at the air-water interface and compared their standard free energy of adsorption ($\Delta G^0_{ads}$) to the values of other known surfactants. At 293 K (20°C), -43 < $\Delta G^0_{ads}$ / kJ/mol < -55 and -48 < $\Delta G^0_{ads}$ / kJ/mol < -66, respectively for RL and surfactin. They find that the efficiency of adsorption of surfactin is higher than the one of RL and other common surfactants. The only drawback of this work seems to be the nature of the RL, a commercial source of which the mono-di-RL ratio is not specified.

*Sophorolipids*

Chen *et al*. reported the NR study of acetylated acidic C18:1-*cis* sophorolipids (**2**) containing lactone SL (**3**) at the air/water interface.[287] The cmc of lactonic C18:1-*cis* SL (**3**) is 10 times lower than the corresponding acidic SL and the former show stronger absorption at the air/water interface, that is a more pronounced surface active effect. The surface area per molecule is larger for acidic (0.85-0.90 $nm^2$) than for lactonic SL (0.70 $nm^2$). The surface tension also decreases to 38.5 and 36 mN/m for acidic and lactonic SL respectively, which still remain values significantly higher than the ones reported for rhamnolipids, being less surface active.[354] The adsorption behavior in the mixtures predominantly depends on the packing constraints associated to the geometries of the surfactant. The structural information is obtained by the kinematic approximation, which provides data of the volume fraction distributions. The plot of these values reflects the width distribution of lactonic (12 Å) and acidic SL (16 Å). Despite having the same headgroup, they show a different distribution, which correlates with the difference in alkyl chain structure. If the fraction of lactonic SL is more pronounced at the interface, the lactonic/acidic system is described as being close to an ideal mixture.

Imura *et al*. have challenged the same question using pressure-area profiles collected with a Langmuir-Blodgett trough for various mono-/di-acetylated acidic-lactonic SL mixtures (**2**).[355] In the medium compression range (20 mN/m), they find values between 0.6 and 0.8 $nm^2$ per molecule, in good agreement with Chen *et al*., while at high compression (25 mN/m) they find values between 0.3 and 0.5 $nm^2$ per molecule. The range corresponds to a complete acidic/lactonic mixed composition and, interestingly, they systematically observe an excess area per molecule at the acidic/lactonic ratio of 0.3, thus showing that the mixing between the acidic and lactonic forms is, first, not ideal, and, secondly, optimized at this ratio, in disagreement with Chen *et al*.

A study by the group of Gross[238] reports the minimum surface area per molecule of a series of ester derivatives of acetylated (**2**) and nonacetylated C18:1-*cis* SL (**1**) at 298 K. They find values decreasing from 0.75 $nm^2$ for the nonacetylated SL-methyl ester to 0.08 $nm^2$ for the SL-octyl (Table 13), indicating that derivatives having longer chains pack more efficiently at the interface. Interestingly, the diacetylated counterparts show the same trend and the same order of magnitude, from 0.86 $nm^2$ for SL-methyl to 0.06 $nm^2$ for SL-octyl (SL-decyl with 0.40 $nm^2$ is an exception), indicating that the presumed bulkier headgroup does not actually play a significant role.

*Glucolipids*

The acetylated form of acidic C18:1-*cis* glucolipids (**11**) can be derived from the corresponding sophorolipid through enzymatic catalysis.[356] Imura *et al*. have evaluated the surface area per molecule of both acetylated and nonacetylated glucolipids using surface tension and found values of 0.47 $nm^2$ and 0.35 $nm^2$ respectively for acetylated and nonacetylated GL. These values are smaller than the area per molecule values measured for sophorolipids under the same conditions (0.60 $nm^2$ and 0.44 $nm^2$ respectively for diacetylated (**2**) and nonacetylated (**1**) SL). These data show that the presence of a single sugar and lack of acetylation reduce the surface area per molecule at the air-water interface, as one would reasonably expect. Interestingly, these results support the formation of both curved (vesicles) and flat (lamellae) interdigitated membranes in water from this family of compounds (**10**, **11**) at acidic pH (Table 5, Table 6, Table 8, Table 9, Table 10).

*Surfactin* (**20**)

Early studies of surfactin at the air-water interface were performed by Maget-Dana *et al.*[170] in 1992 and Ishigami *et al*. in 1995.[157] Using pressure-area experiments, (Figure 20a), the former have shown that the limiting molecular area ($A_0$, low compression) of surfactin at 30°C varies between 2.20 nm$^2$ at pH 3.03 and 2.83 nm$^2$ at pH 9.08, while the transition area ($A_t$, medium compression) varies between 1.44 nm$^2$ and 1.80 nm$^2$, whereas the larger area at basic pH depends on the electrostatic repulsion due ionization of the COOH groups. At even higher compression, the molecular area, $A_c$, falls below 0.5 nm$^2$. Authors have interpreted these data as follows: at low compression, the peptide cycle lies at the interface; at mild compression, the peptide lies perpendicular to the interface while at high compression the peptide is submerged in the aqueous phase and the tail lies perpendicular to the interphase. Ishigami *et al*. have also studied the conformation of surfactin through pressure-area experiments. They find the same trend with compression, but lower absolute values of area per molecule at $A_0$ and $A_t$, respectively in the order of 1.8 nm$^2$ and 0.8 nm$^2$. To account for these smaller values compared to Maget-Dana *et al.*,[170] they propose a dimerization of the acyl chains at the interface for low compression (Figure 21a-1) and perpendicular orientation of the acyl chain, with the peptide in the aqueous phase, at mild compression (Figure 21a-2).

Finally, Maget-Dana *et al.*[357] also found that surfactin promotes a synergistic effect in the hemolysis activity of Iturin A. Iturin A is another lipopeptide with similar structure to surfactin, well-known for its activity as antifungal and hemolytic agent. This synergy shows an improvement in the hemolytic behavior of Iturin A as well as its antibiotic activity. The adsorption of Iturin A into surfactin monolayers occurs when surfactin is in its ionized state at physiological pH.

Several years later, Shen *et al.*[182] studied the adsorption and conformation of surfactin at air/water and solid/water interfaces. Results suggest that surfactin adopts a ball-like structure, with a thickness of 14 Å and an area per molecule of 1.47 nm$^2$. This value is higher than the corresponding $A_t$ proposed by others[170,358] (Table 13) and for this reason, authors suggest another interfacial structure of surfactin at the air-water interface: 1) the hydrocarbon chain folds back into the structure, probably associating itself with the hydrophobic amino acid residues (Figure 21b); 2) the thickness of 14 Å suggests that peptide ring is always aligned parallel with the surface; 3) the separation of the hydrophile is 5.5 Å from its relatively large hydrophobic neighbor (center-to-center distance). In 2011, they further analyzed the effect of pH and cations.[212]

Also, they employed NR in combination with selective isotopic labelling of surfactin to study its structure at both air-water and hydrophobic solid-water interfaces. At the air/water interface, surfactin shows a persistent ball-like structure, although the surface area is slightly affected by pH (from 1.40 nm$^2$ t pH 6.5 to 1.50 nm$^2$ at pH 8.5), in agreement with the same trend reported by Maget–Dana between pH 3 and 9 several years before.[170,359] The effect of cation also follows the results presented by Maget–Dana, with an average area in the order of 1.60 nm$^2$ for $Ca^{2+}$ and $Ba^{2+}$ and 1.56 nm$^2$, 1.64 nm$^2$ respectively for $Li^+$ and $K^+$. Authors suggest that divalent cations bind more strongly, thus have a stronger screening effect, than monovalent ones, as expected for divalent cations in the presence of carboxylate groups. To better understand the distribution of surfactin, these experiments were carried out with perdeuterated, perhydrogenated surfactin and both samples containing hydrogenated or deuterated leucines Figure 20b.

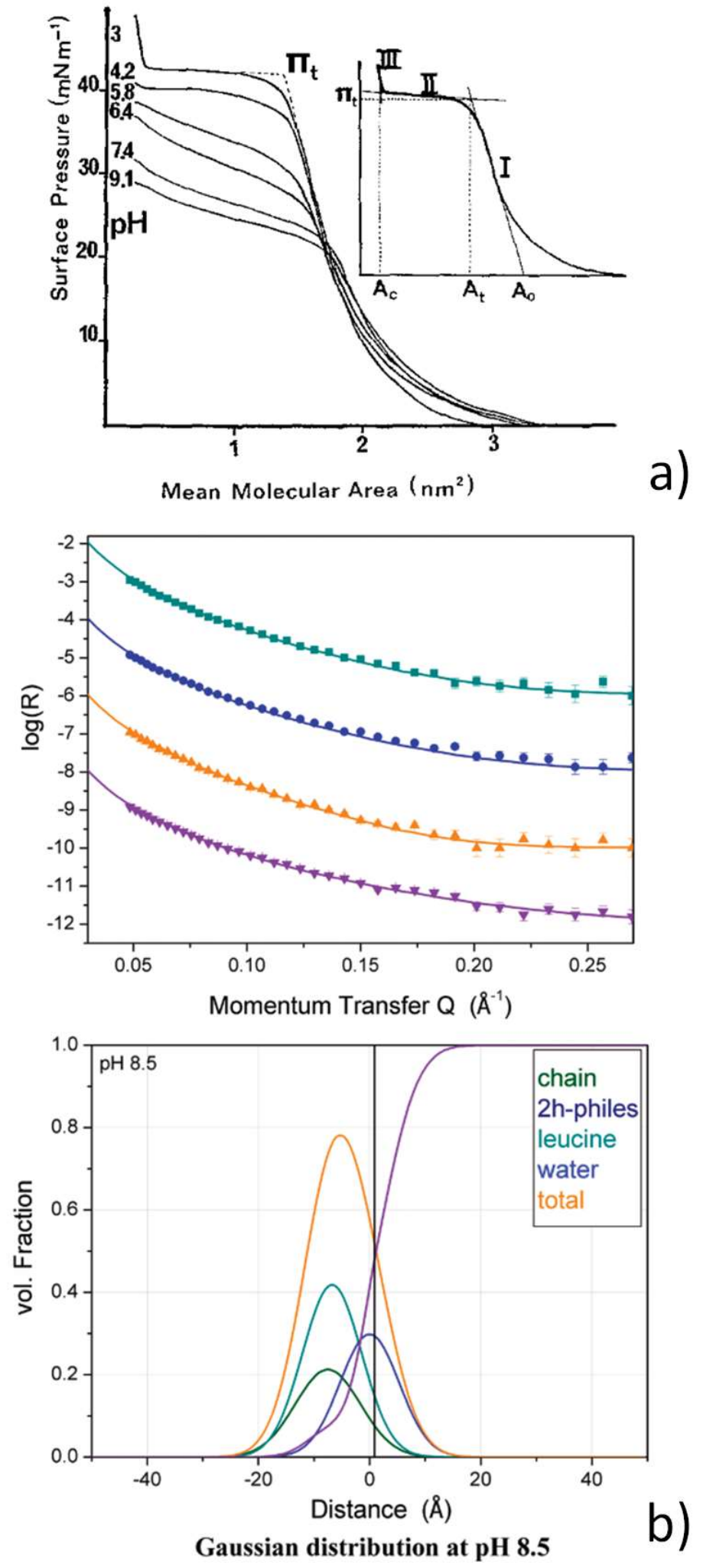

Figure 20 – a) Pressure-area curves recorded at various pH values for surfactin (**20**). The limit areas, $A_0$, $A_t$ and $A_c$ are shown in the inset (reprinted from [170], Copyright (1992), Elsevier). b) Neutron reflectivity (NR) profiles (pH 8.5 in $D_2O$) and corresponding distribution profiles of adsorbed surfactin. Color code: green= H-surfactin, blue: D-surfactin(H-leucines), orange: H-surfactin(D-leucines), purple= H-surfactin (reprinted from [212], copyright (2011) American Chemical Society). H- and D- refer to isotopic labelling of surfactin and leucine groups.

Onaizi *et al.*[360] characterized the surfactin behavior at the air-water (buffered) interface containing relatively high concentration of surfactin. They confirmed the possibility to correlate the surface tension measurements with theoretical calculations to obtain further details in the biosurfactant system behavior. They combined the Gibbs equation with the Langmuir adsorption isotherm to estimate the surfactin surface excess, being at saturation ($\Gamma_\infty$) 1.05 ± 0.05 µmol $m^{-2}$. The adsorption equilibrium constant determined (K= (1.5 ± 0.6) × $10^6$ $M^{-1}$) provides information about its high affinity for the interface. The Gibbs elasticity ($E_G$) was reported to be 183 mN $m^{-1}$, a value comparable to other biosurfactant systems.

In an effort to further estimate the maximum adsorption density ($\Gamma_\infty$) and understand the interaction between adsorbed surfactin molecules at the interface, they reported in 2016 the use of surface tension and the Frumkin model to obtain quantitative data.[361] Results showed that the lateral interaction between surfactin molecules is attractive despite the negative charges, at pH 8, of the glutamic (Glu) and aspartic (Asp) acids, suggesting that this is due to the proximity of the hydrophobic moieties of surfactin, and that sodium counter-ions would decrease the Debye length. In the same contribution, they reported that the value of the maximum adsorption density is 1.16 µmol $m^{-2}$, which corresponds to an area per surfactin molecule of 1.43 $nm^2$, in good agreement with the value found by Shen *et al.*[182] They also compared the surfactin values with two synthetic surfactants, an anionic sodium dodecylbenzenesulphonate (SDBS), and nonionic, octaethylene glycol monotetradecyl ether ($C_{14}E_8$), being the surfactin value 3- and 2.5-times higher, respectively.

In 2018, Onaizi *et al.*[362] investigated the adsorption mechanism of surfactin at the air-water interface, and also compared it with the anionic SDBS and nonionic $C_{14}E_8$ synthetic surfactants. This mechanism can be determined as a pure molecular diffusion, pure barrier or a mixed version of diffusion-barrier mechanism. They also studied the dynamic surface tension from the Ward-Torday approximations [363] for short ($\gamma_{t\rightarrow 0}$) and long time approximations ($\gamma_{t\rightarrow \infty}$). The values obtained for effective diffusivity ($D_{eff}$) are eventually compared to the diffusion coefficient ($D$): for $D_{eff}$ ~$D$, adsorption is governed by a pure diffusion-controlled mechanism. On the contrary, if $D_{eff} < D$, one expects either the existence of an energy barrier or a mixed mechanism involving an energy barrier coupled to a diffusion control. Typical energy barriers include repulsive electrostatic interactions, steric hindrance and rearrangements of the molecules at the interface.

The conditions of the solutions were prepared in 20 mM sodium phosphate buffer containing 38.5 mM of sodium ions, and at pH 8, where surfactin remains anionic. Results showed that surfactin follows a pure diffusion mechanism at initial adsorption, without being influenced by the presence or absence of $Na^+$ in solution, which may provide an energy barrier due to electrostatic interactions. However, at longer times when surface tension is close to equilibrium ($\gamma_{t\rightarrow \infty}$), it changes to a mixed diffusion-barrier mechanism, being the energy per surfactin molecule at least 1.8-9.5 kJ, to be able to overcome such barrier. Interestingly, surfactin behaves more likely the nonionic $C_{14}E_8$ than like the anionic SDBS synthetic surfactants.

If most studies in the literature concerning the adsorption and aggregation behavior of surfactin have been done in aqueous solutions (discussed above in the RL paragraph),[171] recent works explored surfacing behavior in organic media, which can act as a co-solvent or as a co-surfactant.[364,365] Rekiel *et al.*[366] have recently evaluated the addition of ethanol to the aqueous solutions of surfactin and analyzed the effect on their properties at the liquid-air interface. Measurements were performed at the surfactin range concentration from 0 to 40 mg/L and ethanol content was evaluated from 0 to 100%. The surface tension was measured by the ring detachment method and also predicted by the equations of Miller *et al.*,[367,368] Connors *et al.*[369,370] and by independent adsorption methods. Results show that the Gibbs free energy of surfactin adsorption in the absence of ethanol can be evaluated between -47 and -48 kJ/mol and upon adding ethanol data suggest that there is a mutual influence in the adsorption properties at the solution-air interface, which depends on the concentration of both in solution.

General comments

Table 13 summarizes the major literature values of the area per molecule for a series of biosurfactants. If the surface area can induce some interpretation of the molecular conformation at the interface in a sparse and dense regime, direct interpretation should be performed with care in the absence of additional experiments. Area/structure relationship has been particularly discussed for surfactin, with disparities in terms for interpretation among authors. In the medium density regime, which some authors refer to the transition area, $A_t$ (Figure 20a), it has been proposed that surfactin lays parallel to the liquid air interface ($A_t$> 1.5 $nm^2$),[170] or it lays perpendicular ($A_t$< 1 $nm^2$, Figure 21a-2)[157] or it undergoes internal folding of the hydrophobic acyl chain into the more hydrophobic regions of the peptide ring ($A_t$< 1.5 $nm^2$, Figure 21b).[182]

In terms of consistency, and beyond specific pH, temperature, mixing or ion effects discussed in the corresponding papers, the values of surface area (average from Table 13 for each family of BS) settle at about 0.76 ± 0.07 $nm^2$ (10% rel. err.) for RL, 0.72 ± 0.15 $nm^2$ (21% rel. err.) $nm^2$ for SL and 1.36 ± 0.34 $nm^2$ (25% rel. err.) for surfactin ($A_t$ values taken only). Interestingly, given the crude approach employed here, the values given in the literature are quite consistent within maximum a 25% relative error in the case of surfactin and 10% relative error for RL. Considering variety of the methods employed in the literature spanning from surface tension, Langmuir trough and neutron reflectivity, the overall result is quite good. Within the errors, RL and SL occupy the same surface area, in line with their similar molecular structures, while surfactin occupies a larger area, about twice their value, in good agreement with the larger peptidic ring.

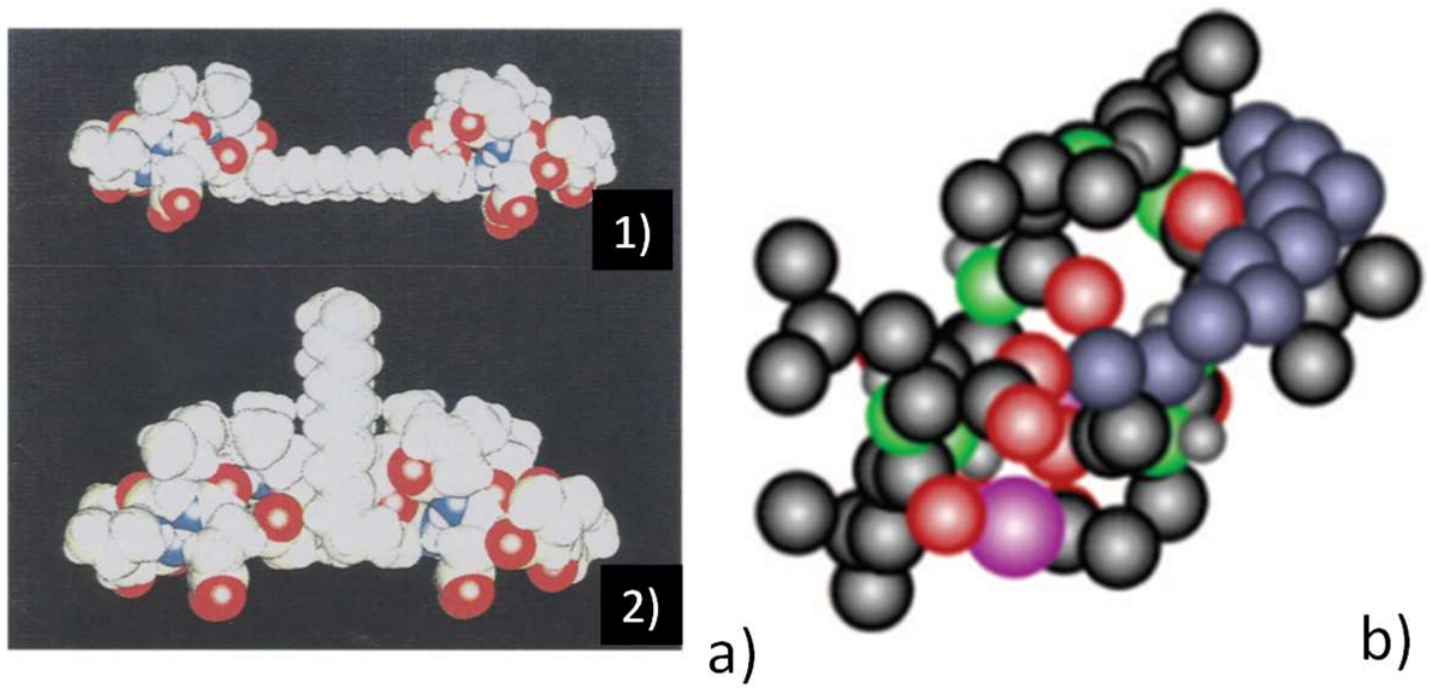

a) b)

Figure 21 – Structural interpretation of surfactin (**20**) at the air-water interface. a) Surfactin forms a dimer at mild surface pressures (1) while it flips perpendicular to the interface at higher compression ($A_t$< 1 $nm^2$). Image adapted (reprinted from [157], copyright (1995), with permission from Elsevier). b) Surfactin undergoes an internal folding, where the acyl chains interact with the hydrophobic regions of the peptide ($A_t$< 1.5 $nm^2$) (reprinted from [182], copyright (2009), American Chemical Society). In Reference [170], authors propose, on the contrary, that a single surfactin molecule lays parallel to the interface ($A_t$> 1.5 $nm^2$).

Table 13 – Values of surface area per molecule of various biosurfactants (N° are given in Figure 7). The temperature range is between 20°C and 30°C. In pressure-area experiments, the area values $A_c$ refers to high compression, $A_t$, the transition area, to medium compression and $A_0$, the limiting area, to low compression (refer to Figure 20a). Techniques are: ST: surface tension; NR: neutron reflectivity; LT: Langmuir trough.

| BS | Type | N° | pH | A / nm$^2$ | Technique | Ref |
|---|---|---|---|---|---|---|
| RL | mono | 19 | 9 | 0.77 | ST | [144] |
| RL | di | 18 | 9 | 0.80 | ST | [144] |
| RL | mono | 19 | 7 | 0.66 | ST | [144] |
| RL | di | 18 | 7 | 0.77 | ST | [144] |
| RL | mono (0.30) di (0.70) | 18, 19 | 9 | 0.83 (1 mM) | ST | [144] |
| RL | mono (0.70) di (0.30) | 18, 19 | 9 | 0.76 (1mM) | ST | [144] |
| RL | mono | 19 | 9 | (1 mM RL) 0.63 ($Ca^{2+}$) | NR | [291] |
| RL | di | 18 | 9 | (1 mM RL) 0.83 ($Ca^{2+}$) | NR | [291] |
| SL | acidic C18:1-*cis* | 1 | Not given | 0.93 (2 mM) 2.60 (1 µM) | NR | [287] (Table S4) |
| SL | lactone C18:1-cis | 3 | Not given | 0.72 (1 mM) 2.03 (5 µM) | NR | [287] (Table 4) |
| SL | d-Lactone (0.15) h-Acidic (0.85) | 3, 1 | Not given | 0.86 (1 mM) | NR | [287] (Table 5) |
| SL | d-Lactone (0.15) h-Acidic (0.85) | 3, 1 | Not given | 3.73 (1 mM) | NR | [287] (Table 5) |
| SL | (mono-/diacetylated) Acidic/lactone ratio | 1, 3 | Not given | (P= 20 mN/m) ~0.62 ($X_{AS}$=0) ~0.80 ($X_{AS}$=0.3) ~0.60 ($X_{AS}$=1) | LT | [355] (Figure 4 a) |
| SL | Diacetylated C18:1-*cis* | 2 | Not given | 0.60 | ST | [356] (Table 3) |
| SL | Nonacetylated C18:1-*cis* | 1 | Not given | 0.44 | ST | [356] (Table 3) |
| SL | C18:1-*cis* Ester derivatives | - | Not given | 0.75 (methyl) 0.65 (ethyl) 0.59 (propyl) 0.41 (butyl) 0.13 (hexyl) 0.09 (octyl) 0.08 (decyl) | ST | [238] (Table S2, 298 K) |
| SL | Diacetylated C18:1-*cis* Ester derivatives | - | Not given | 0.86 (methyl) 0.60 (ethyl) 0.31 (propyl) 0.19 (butyl) 0.14 (hexyl) 0.06 (octyl) 0.40 (decyl) | ST | [238] (Table S2, 298 K) |
| GL | Acetylated C18:1-*cis* | - | Not given | 0.47 | ST | [356] (Table 3) |
| GL | Nonacetylated C18:1-*cis* | 11 | Not given | 0.35 | ST | [356] (Table 3) |
| surfactin | cyclic | 20 | 3 | 0.27 ($A_c$)<br>1.44 ($A_t$)<br>2.20 ($A_0$) | LT | [170] |
| surfactin | cyclic | 20 | 9 | 1.80 ($A_t$)<br>2.83 ($A_0$) | LT | [170] |
| surfactin | cyclic | 20 | 4.2 | 0.89 ($A_t$)<br>1.84 ($A_0$) | LT | [157] |
| surfactin | cyclic | 20 | 4.8 | 0.81 ($A_t$)<br>1.82 ($A_0$) | LT | [157] |
| surfactin | cyclic | 20 | 5.4 | 0.79 ($A_t$)<br>2.02 ($A_0$) | LT | [157] |
| surfactin | cyclic | 20 | 7.5 | 1.47 (6 µM, 10 µM) (equivalent to $A_t$) | NR | [182,212] |
| surfactin | cyclic | 20 | 6.5 | 1.40 (10 µM) | NR | [212] |
| surfactin | cyclic | 20 | 8.5 | 1.50 (10 µM) | NR | [212] |
| surfactin | cyclic | 20 | 7.5 | (10 µM surfactin)<br>1.60 ($Ca^{2+}$, $Ba^{2+}$)<br>1.56 ($Li^{+}$)<br>1.64 ($Li^{+}$) | NR | [212] |
| surfactin | cyclic | 20 | 8 | 1.43 | ST | [361] |

### 3.1.2 Liquid-solid

*Rhamnolipids (di- (**18**), mono- (**19**))*

Costa *et al.*[371] described the wetting properties of rhamnolipids solutions over hydrophobic polymer surfaces (PCL, PVC, PET, and PVC–PCL blend). At low RL concentration, the contact angle is larger than at higher concentrations, with similar results observed for SDS. At increasing concentration, the contact angle decreases to 13° on PET, 15.7° on PVC–PCL blend, 19.1° on PVC and 40.8° on the PCL surfaces. To explain these data, authors made the hypothesis that RLs adsorb in a random manner and horizontally to the hydrophobic surface at low concentration, while their configuration changes from horizontal to an ortogonal arrangement when concentration increases. In the latter, RLs expose the hydrophilic carbohydrate towards the air, while their aliphatic chain eventually interacts with the hydrophobic surfaces. Finally, experiments performed on more hydrophilic glass surfaces show that the contact angle remains constant for both RL (6.5°) and SDS (15°), being RL a better wetting agent on a glass surface.

Zdziennicka *et al.*[372] performed water contact angle measurements of RL (unclear the mono-/di- ratio) and surfactin solutions on several surfaces: polytetrafluoroethylene (PTFE), polyethylene (PE), polymethyl methacrylate (PMMA), nylon 6 and quartz. At the same time, they theoretically predicted the contact angle values on the hydrophobic surfaces by knowing the Lifshitz-van der Waals component of the water surface tension, the surface tension of the solids and that of the solution. They found that the calculated contact angle values for both RL and surfactin solutions match the measured ones for the air-water and PTFE interfaces, meaning that at a given biosurfactant concentration the orientation of its molecules and their density at both interfaces are the same. A more complex behavior is found for other surfaces, which suggest that, for instance, in the case of PMMA and nylon-6, the structure of both biosurfactants at the solid-water interface probably changes, with the hydrophobic chain being directed towards the water solution.

Terziyski *et al*. studied the thickness of di-RL films with and without NaCl at the liquid-(hydrophilic) silica interface.[352] They found a maximum contact angle at 20 µM for both solutions but a minimum thickness of about 43 nm only for the aqueous solution, while in the presence of NaCl, the film thickness was concentration-independent and settled at about 30 nm. Authors argue that saturation of the film surface by di-RL could explain the lowering of the film thickness with concentration to the minimum, where RL could adsorb with the hydrophilic rhamnose and carboxylate groups pointing toward the silica substrate leaving the alkyl chains in solution. At higher concentration, after the minimum, increase in the thickness could be explained by the formation of bilamellar patches, where the carboxylate group could point towards the solution, a fact which could also explain the decrease in the contact angle. These hypotheses are to be verified further.

*Surfactin* (**20**)

Onaizi *et al*. explored the surfactin activity at the liquid-solid interface by surface plasmon resonance (SPR). A surfactin nanolayer was analyzed on two different surfaces, one hydrophobic (octadecanethiol, ODT) and another hydrophilic (β-mercaptoethanol, BME) over 45 nm gold film.[373] This system was also complemented with hydrophilic air-liquid interface data obtained from dynamic surface tension measurements through drop shape analysis (DSA). Surfaces were equilibrated with 20 mM sodium phosphate buffer at pH 8. Surfactin adsorption was strongly influenced by the hydrophobicity of the surfaces, being its affinity 3 times higher than for hydrophilic surface. For this analysis, the SPR shift angles were correlated with Langmuir isotherm calculations. Interestingly, they observed no significant difference in the packing of surfactin of each surface studied.

Enhanced oil recovery (EOR) from reservoirs is an important strategy to improve the efficiency of industrial processes and biosurfactants have demonstrated their capability to replace chemical surfactants in this field. Park *et al.*[374] described a microbial enhanced oil recovery (MEOR) from the bacteria *Bacillus subtilis*, producing surfactin, under variable environmental conditions such as pH, salinity, pressure (~10 MPa) or temperature (35−45 °C). To confirm the production of surfactin, they observed the dodecane-brine interfacial tension (IFT) being reduced from ~50 to ~10 mN/m and wettability by contact angle of dodecane-brine-quartz surface obtaining an intermediate water-wet condition (θ = ~45−50°) to a strong water-wet condition (θ = ~20−25°). They concluded that MEOR could be an interesting alternative for the recovery of oil.

Adsorption of surfactin within the context of heterogeneous surfaces has been commented above in the RL paragraph.[372]

*MELs* (**21**)

Immunoglobulins are a well-known and useful class of glycoproteins. Their carbohydrate-recognition motifs are widely studied to extend their use in immunodiagnostics, epitope mapping and therapeutic applications. Many efforts were devoted to the development of new, stable, affinity ligands for the proteins, especially immunoglobulins, to replace the expensive protein-A, or –G, actually in use. High affinity of gangliosides, among other glycolipids, toward immunoglobulins make these molecules a potential alternative, but their development is mainly restricted by their low availability. Yeast glycolipid biosurfactants thus gained attention, and several studies identified MELs as an alternative ligand for immunoglobulins after investigation of their binding behavior to immunoglobulins. In the field of self-assembled monolayers (SAMs), Kitamoto and coworkers[375–377] developed a classical SAM employing an alkane thiol primer onto a gold substrate and MEL-A (**21A**) on top. They studied the binding affinity of the mannosyl headgroup of MEL-A (**21A**), MEL-B (**21B**) and MEL-C (**21C**) towards various classes of immunoglobulins and lectins using surface plasmon resonance (SPR), surface pressure measurements and AFM. They found that MEL-A (**21A**), with two acetyl groups, have the highest affinity towards a number of immunoglobulins, while MEL-B (**21B**) and MEL-C (**21C**) have poor affinity in general. In particular, MEL-A (**21B**) gave good affinity (binding constants in the order of $10^{-18}$ M.mol.cm$^{-2}$) towards immunoglobulin from humans (HIgG, HIgA and HIgM) and other species (bovine, cat, sheep, goat, horse, mouse, pig, rabbit, and rat).[376] This affinity is directly related to the packing density of the assembly, which can also be modulated by changing the temperature, and to the multivalency of the carbohydrate-rich SAM. In more specific works, they studied the surface structure of those immunoglobulins (HIgG and HIgM) having the highest affinity (respectively, $K_a = 9.4 \cdot 10^6$ M$^{-1}$ for HIgG and $5.4 \cdot 10^6$ M$^{-1}$ for HIgM) towards MEL-A (**21A**)[377] and the effect of free fatty acid impurities in the MEL-A (**21A**) batch.[375] They found that HIgG and HIgM adsorb through their F(ab')$_2$ and Fab fragments and have an overall globular structure (Figure 22) and that the content of unsaturated fatty acid impurities (MEL-A (**21A**) derived from olive oil) favor the affinity.

These results were expanded by Konishi and co-workers, who investigated the binding of mannose specific ConA and sialic acid specific *Maackia amurensis* lectin-I (MAL-I) towards MEL-A (**21A**) using surface plasmon resonance (SPR) spectroscopy.[378] Both lectins showed high affinity towards self-assembled monolayers (SAMs) of MEL-A (**21A**). Affinity constants towards ConA and MAL-I were respectively found to be in the order of $1{\cdot}10^7$ and $3{\cdot}10^6$ M$^{-1}$, which are comparable with affinity constants of the ConA's specific probe Manα1–6(Manα1–3)Man. ConA binding was further investigated on MEL-B (**21B**) and MEL-C (**21C**) SAMs and, very interestingly, no affinity was detected. These results induced the authors to suggest that the binding is favored, or maybe even conditioned, by the two O-acetyl groups on the mannose moiety of MEL-A (**21A**). The question of inhibiting the ConA-MEL-A (**21A**) interaction was addressed using a-methyl-D-mannopyranoside, which however did not have any inhibiting effect, showing the strength of the interaction between ConA and the SAMs formed by MEL-A. If the lectin- and immunoglobulin-binding data are promising, further work should still be performed to clarify some incoherencies: the poor binding affinity of MEL-B (**21B**) and MEL-C (**21C**) towards immunoglobulins and SAMs composed of ConA, but their good binding affinity towards ConA in their vesicular form.[189] Similarly, ConA displays the inverse behaviour, that is good affinity in a SAMs but no affinity in a vesicle configuration in solution.[189]

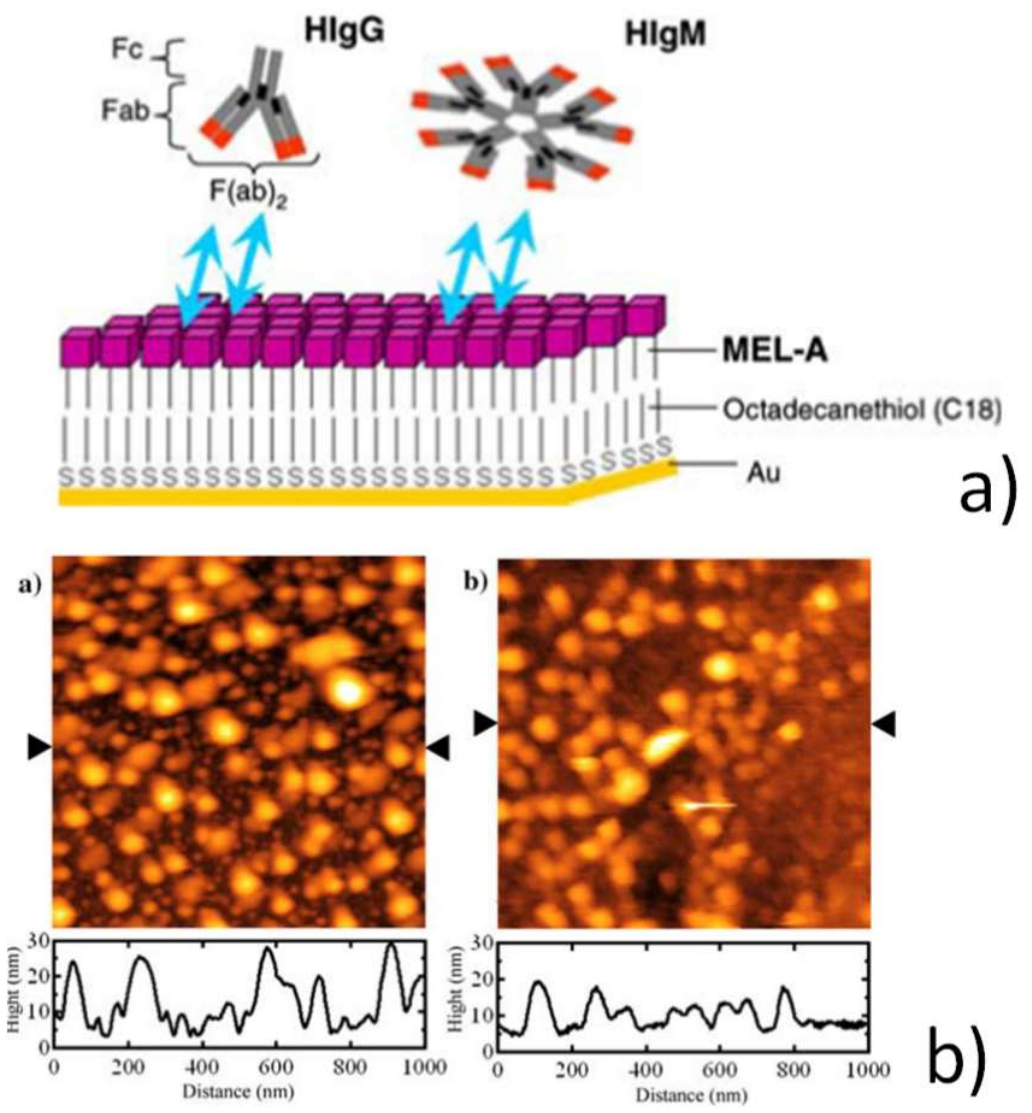

Figure 22 – a) MEL-A (**21A**) self-assembled monolayers (SAMs) deposited on an alkanethiol primer deposited on a flat gold substrate. Possible interaction with the F(ab)$_2$ region of human immunoglobulins HIgG and HIgM is sketched. b) Corresponding AFM images of MEL-A SAMs with HIgG and HIgM. Reprinted by permission from Springer Nature Customer Service Centre GmbH: Springer, freom Ref. [377].

### 3.1.3 Solid-air

The presence of bacteria in different scenarios such as food industry and health care is a challenging fight to avoid the proliferation of infections and diseases in humans. Many bacterial pathogens attach to biomaterials end up forming biofilms, which are directly related to the resistance to antibiotics.[379,380]

Biosurfactants have exhibited their role as antimicrobial and antiadhesive agents against several pathogens. In the first case, the antimicrobial activity involves the disruption of the bacterial membrane and therefore the bacterial growth decreases. Concerning the antiadhesive effect, biosurfactants modify the possible interactions between solid surfaces and bacteria which can be: specific (i.e. adhesin–ligand interactions) and/or non-specific (i.e., hydrophobic effect).[381] Another important advantage of these coated surfaces consists in not spreading antimicrobial compounds in the environment, and in this manner, the possible development of resistant strains is limited.

Some research articles on the antimicrobial and antiadhesive activity of biosurfactants have shown their effectiveness by directly measuring their effect on different bacterial strains.[382] However, little information is given on how these molecules interact with the microbial membrane, if and how are incorporated into the biofilms, therefore disrupting and solubilizing them. In this subsection, we will focus on the limited number of research examples that have shed light into the mechanism and specific properties of biosurfactants at the solid-air interfaces, and why they are becoming an interesting strategy against bacterial pathogens.

*Sophorolipids*

Valotteau *et al*.[383] first reported a glycosylated surface with biocidal properties, in which they employed either nonacetylated acidic C18:1-*cis* (**1**) or C18:0 SL (**13**), covalently grafted to a model flat gold surface via short thiolamines primer coupled to the COOH group. We stress the fact that, differently than the strategies developed by Kitamoto and coworkers,[375–377] SL are anchored to the gold surface through an amide coupling between the COOH and amine of the primer. The antimicrobial activity was tested against non-pathogenic Gram+ bacteria (*Listeria ivanovii*) by atomic force microscopy (AFM), scanning electron microscopy with a field emission gun (SEM-FEG), fluorescent staining and bacterial growth, which also allowed to see the disruption of the bacteria membrane (Figure 23c). A bacterial decrease (approximately 45%) was observed on those surfaces grafted with the C18:1-*cis* (**1**) derivative, while no effect was observed on surfaces grafted with the C18:0 (**13**) derivative. The wettability between samples is also different, the C18:1-*cis* (**1**) grafted surface being more hydrophilic, probably suggesting that the difference in lipid chain orientation promotes exposition of the C-OH group towards the solution.

The simultaneous presence of sophorose unit and double bond seems to promote the biocidal effect on bacteria. Later studies[384,385] by the same group have shown 1) a comparable efficiency on Gram+ bacteria and a less performant (about 20%) efficiency on Gram- bacterial strains, the latter being challenging because of their additional outer lipid membrane and lipopolysaccharide layer; 2) similar antibacterial properties are shown by other microbial glycolipids, of which the efficiency strongly depends on the number of sugars (two-sugar biosurfactants are more efficient) and backbone structure (saturated acyl chains are less efficient). The advantages of this grafting strategy rely on providing knowledge on the role of each part of the glycolipids (sugar *vs*. backbone), which is difficult to determine in solution. Consequently, these studies established the action of disaccharide as membrane-disrupting, that cannot be considered as a "surfactant effect" due to the anchoring -COOH to the gold surface. In addition, the effect on the orientation of biosurfactant determines its biocidal efficiency on non-pathogenic Gram + bacteria. The different wettability and antibacterial properties of like SL molecules indicate the crucial importance of sugar orientation, thus going in the sense of the different affinity of MELs towards ConA discussed above.

Valotteau *et al*.[386] have also studied the bacterial adhesion at short times (100 ms) on nonacetylated acidic C18:1-*cis* (**1**) grafted substrates, whereas the interactions can be specific (i.e. adhesin–ligand interactions) or non-specific (i.e. hydrophobic). They studied modified gold surfaces (-$CH_3$, -OH, -SL) using two different bacteria: Gram+ *S. aureus* and Gram– uropathogenic *E. coli*. Single-cell force spectroscopy shows that bacterial cell surface proteins are engaged in strong hydrophobic interactions with surfaces, while they also modestly contribute to hydrophilic adhesion (Figure 23a,b). In addition to that, SL (**1**) in solution contributes in avoiding the adhesion of the bacteria onto surfaces by their surfactant and detergent behavior. Considering these results and the work developed in ref. [383–385], they proposed a 2 steps mechanism for the SL (**1**) activity onto surfaces and in solution at the same time. At early seconds, the anti-adhesive effect predominates,[386] while at longer times the biocidal role of the glycolipids becomes more relevant.

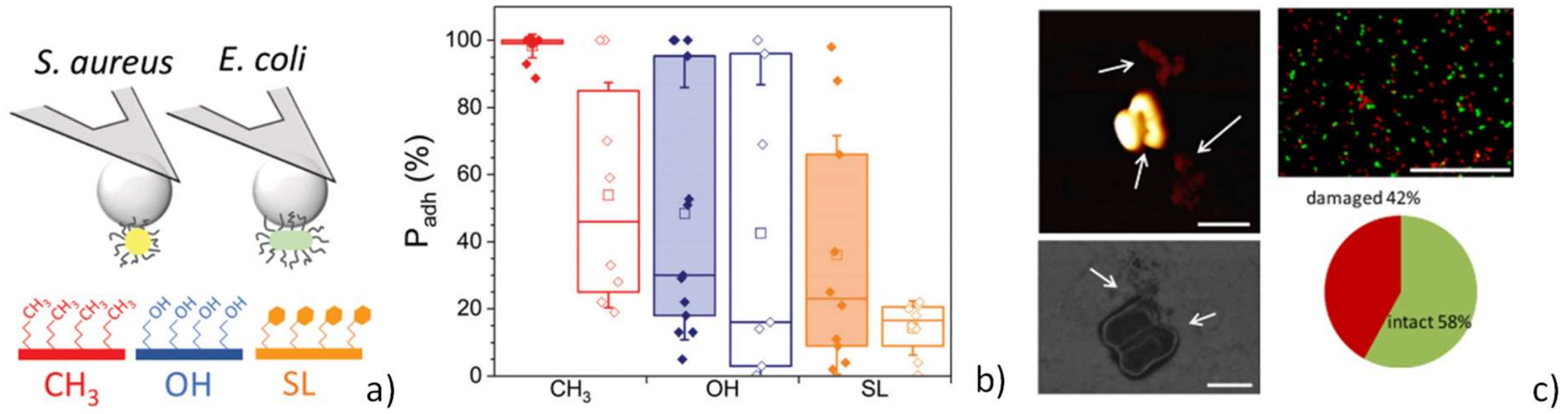

Figure 23 – a,b) Single-molecule AFM experiments showing the anti-adhesive properties of C18:1-*cis* (**1**) sophorolipid-grafted gold substrate compared to methylene and hydroxyl-modified controls. Microorganism: *S. aureus* Newman WT (filled bars) *vs*. the Newman *ΔsrtA* mutant (open bars) (reproduced from [386], copyright The Royal Society of Chemistry). These experiments are performed at short time-scales (100 ms). c) Antibacterial properties of SL-grafted gold substrates against *L. ivanovii* at long time scales (3 h) (Reprinted (adapted) with permission from [383], copyright American Chemical Society).

To study the structure of physisorbed SL, Imura *et al.*[355] observed the structure of (mono-/di)acetylated acidic (**2**)-lactonic (**3**) mixtures onto mica surfaces by AFM. They found spheroidal structures of several hundred nm in diameter but only about 3 nm in height. These were found to be more and more frequent with the increasing fraction of acidic SL (**2**) and their thickness was attributed to a single SL layer.

A similar approach was adopted by Peyre *et al.*[387] in view of developing surfaces coated with (physisorbed) deacetyalted acidic C18:1-*cis* SL (**1**). They explore, qualitatively and quantitatively, the surface self-assembly of SL (**1**) by the dip-coating method onto 3 different surfaces ( i.e. gold, silicon (111) and $TiO_2$ anatase). The coated surfaces were analyzed by AFM for the local structure of the SL films and SEM for spatial homogeneity. The surface assembly was put through different pHs and correlated depending of their surface tension of each material substrate. Remarkably, three different surface self-assembled structures at three different pHs are found on $SiO_2$. An entangled network of needles forms at pH 4, small and large circular aggregates of thickness above 50-100 nm form at pH 6 while a thin (< 3 nm) SL (**1**) layer forms at pH 11. Compared to $SiO_2$, $TiO_2$ substrates showed a much less marked pH dependency and no dependency at all was found for Au surfaces. However, at constant solution pH, the nature of the surface played an important role on the surface structures. Authors explained such differences using surface energy arguments between the various surfaces.

*Rhamnolipids (mono-* (**19**), *di-*(**18**))

The structure and phase behavior of mono- and di-RL is generally studied in solution. However, a series of works has reported their behavior at the surface of a glass slide using optical microscopy under polarized light.[178,195,196] Although authors pretend associating the surface phase behavior to the phase behavior in solution and in relationship to the intrinsic RL packing parameter, it should be said that this should be done with care, because a number of critical parameters, like volume fraction, vary during drying, thus affecting the phase behavior. The nature of the surface and the interactions can also modify the type of RL aggregates, as found for SL.[387] Nonetheless, authors give an interesting qualitative trend of what surface aggregates of mono- and di-RL look like. A transition from *MLV* drops, characterized by a typical Maltese cross, to an isotropic phase is observed with increasing NaCl concentration.[195] The *MLV* drop phase seems to be stable against pH and RL structure at both pH 5 and 6.8,[196] in agreement with the corresponding solution phase, which is known to be lamellar/vesicular at room temperature,[211] and mono-/di-RL ratio.[178]

Rhamnolipids were tested onto different surface materials for applications such as food and health industry. One important application concerns their activity as antimicrobial and antibiofilm behavior.[388] However, we must note that it is challenging to compare the systems because of the degree of purification of the biosurfactant, or the direct use of the raw mixture of RL.

In 2005, Irie *et al.*[389] showed the disruption of the respiratory pathogen *B. bronchiseptica* biofilms by rhamnolipids secreted from *Pseudomonas aeruginosa*. Remarkably, they observed that the purified rhamnolipid was less effective than the one secreted from *P. aeruginosa*, contemplating in this way that it is possible that other secreted molecules could contribute to the antimicrobial activity such as pyocyanin.

In 2006 Rodrigues *et al*.[390] described the effects of adhesion of different bacterial strains on the surface of silicone rubbers with and without an adsorbed rhamnolipid layer. Silicone rubbers have special interest because they are employed in voice prostheses, urinary catheters and contact lens materials among others. In particular, voice prostheses have a limited life-time of 3-4 months due to the formation of bacterial biofilms once inserted. In order to avoid this frequent replacement, they tested the effect of covering them with an adsorbed layer of rhamnolipid biosurfactant in a parallel-plate flow chamber. Results demonstrate that the bacterial adhesion onto the silicone rubber treated with rhamnolipid was significantly reduced in comparison to the untreated one. Interestingly, a low concentration of the rhamnolipid is needed to considerably reduce the surface tension value, but a high concentration was needed to adsorb the rhamnolipid layer in the silicone rubber. Considering the water contact angles values, they believed that this fact is a consequence of the washing out of the rhamnolipid as it is bound to the silicone surface by the relatively weak Lifshitz–van der Waals forces.

Disrupting marine biofilms in industrial systems is also an important concern. Dusane et *al.*[391] studied surfaces exposed to seawater and identified *B. pumilus*, of which biofilms were formed more efficiently in surfaces such as polysterene microtitres plates and glass plates than *P. aureofaciens* and *B. licheniformis*. Rhamnolipids demonstrated to act as an effective antiadhesive and disrupting biofilm agent with a concentration below the MIC (minimum inhibitory concentration). Also, they showed by SEM that rhamnolipids added in solution are able to remove the exopolymeric substances (EPS) and micro-colonies of pre-formed biofilms.

Sodagari *et al.* [392] employed rhamnolipid biosurfactants to study their effect on the attachment of bacteria on hydrophilic glass and hydrophobic octadecyltrichlorosilane (OTS) modified glass under continuous-flow conditions. Results showed a reduced effect on the attachment of different Gram-positive and Gram-negative species, and they explored different experiments to explain its mechanism (e.g. inhibition of microbial growth, change of cell surface hydrophobicity, easier detachment of cells already attached to substratum, and modification of substratum surface properties). However, these experiments did not provide enough evidence to elucidate the mechanism involved.

Hajfarajollah *et al*.[393] studied the antimicrobial and antiadhesive capability of rhamnolipids adsorbed on polypropylene (PoP) surfaces. This material is of great utility because it is one of the most common polymers in food industry packaging. The adsorption of rhamnolipids on the PoP plasma surface decreased considerably the hydrophobicity of the surface measured by water contact angle. Then, the different surface topologies and morphologies were analyzed by AFM. The parameters considered here are: average roughness ($R_a$), root mean roughness ($R_q$), skewness ($R_{sk}$) and kurtosis ($R_{ku}$). In general, the air and oxygen plasma treated surfaces increased their roughness by an increase in the values of $R_a$, as also observed from AFM. Then, the air-treated ones showed an irregular surface, that after adsorption of the rhamnolipid became smoother. The chemical analysis of the surface was studied first by FTIR, but as it did not give information about the changes in the surface chemistry of the plasma film, XPS is performed in combination with a labeling technique for the elemental analysis. A good rhamnolipid adsorption was obtained in the plasma films with antimicrobial and antiadhesive activity and film stability, which may be a proper candidate for food and pharmaceutical industry materials.

The effective anti-adhesive potential of rhamnolipids (crude extract) and surfactin on polystyrene surfaces against *L. monocytogenes* strains was demonstrated by Araujo *et al.*[394] (Figure 24). They found that both surfactin and RL reduced the adhesion against specific strains (ATCC7644, ATCC15313) of more than 80%, with values as low as 53% on ATCC19112. In comparison, SDS reduced the adhesion in the order of 20% for the same strains. After these results, they investigated in 2016 the relationship between this anti-adhesive activity and the physicochemical modifications on different surfaces (i.e. polystyrene and stainless steel AISI 304).[395] The parameters considered are: total surface energy, Lifshitz-van der Waals, Lewis acid-basic, electron donor and electron acceptor components. In the early stages of the biofilm formation, the non-polar Lifshitz-van der Waals interactions predominate at a distance between microorganism and surface above 50 nm. The surface hydrophobicity parameter is crucial in the second stage (10-20 nm from the surface) of the biofilm formation due to electrostatic repulsion. Lewis acid-base interactions occur at distances smaller than 1.5 nm being responsible of polar interactions. And finally, the degree of hydrophobicity of bacteria is related to the capability of behaving as strong (high hydrophobicity) or weak electron donor (low hydrophobicity). These parameters indicated the significant decrease of biofilm formation of both Gram-positive and Gram-negative microorganisms evaluated at different stages for both rhamnolipid and surfactin. This latter biosurfactant showed no increase in hydrophilicity compared to rhamnolipid.

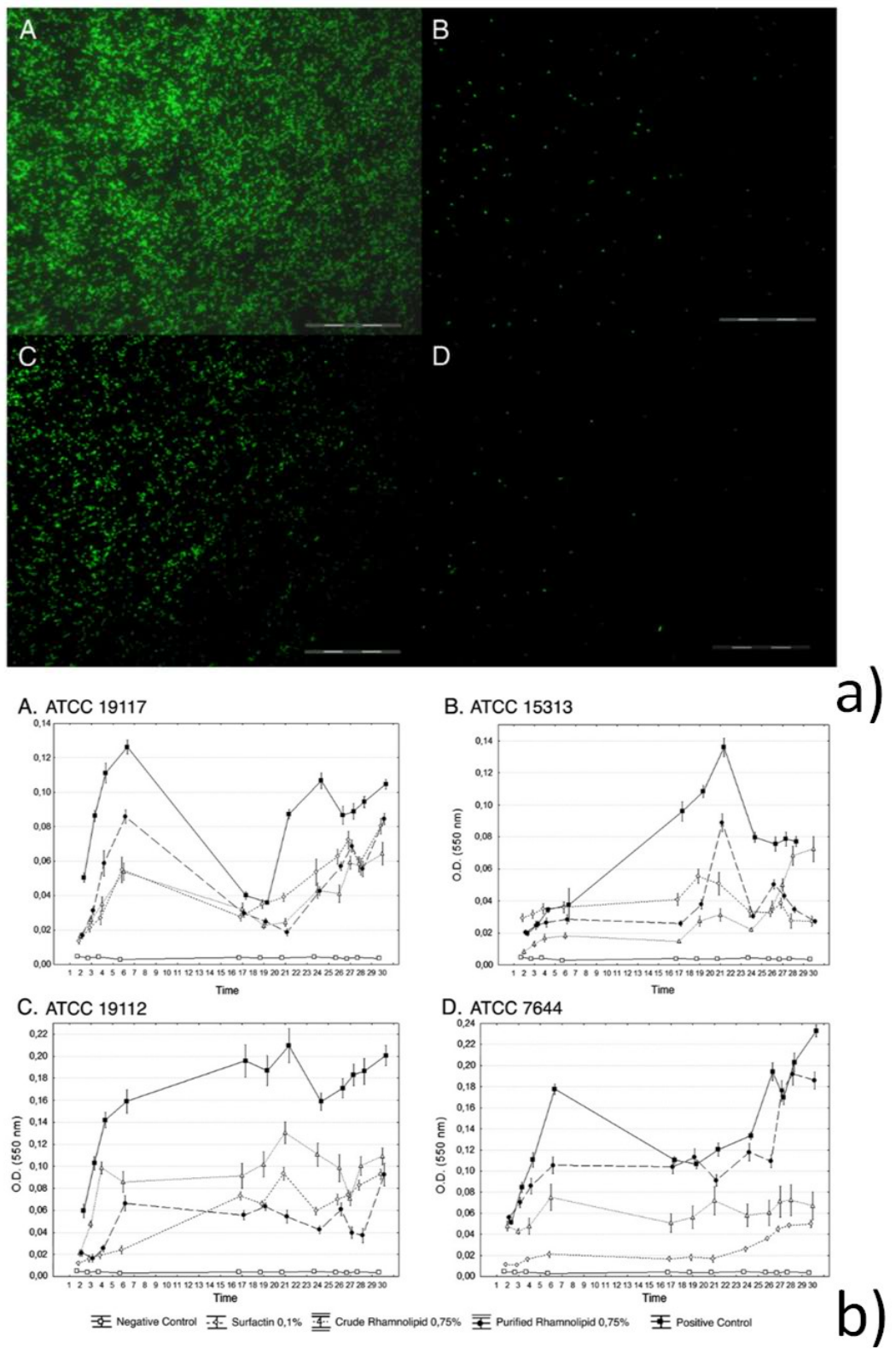

Figure 24 – a) Antiadhesive properties against *Listeria monocytogenes* of biosurfactants (A, B: Rhamnolipids (**18**, **19**); C,D: surfactin (**20**)) dispersed on polystyrene surfaces. b) Adhesion kinetics of various strains of *Listeria monocytogenes* on polystyrene surfaces in the presence of biosurfactants. Reprinted from [394]¸ Copyright (2011), with permission from Elsevier.

Another material that suffers from the action of some bacterial strains is carbon steel (API 5LX), a material employed in gas and petrochemical industry. It has good mechanical strength and anticorrosive capability, but this latter one is compromised in the presence of corrosive bacteria strains. Other surfactants have been previously studied for avoiding this problem, but Parthipan *et al.*[396] recently proposed the use of rhamnolipids. Weight loss and corrosion rates were observed to be reduced in the presence of rhamnolipids, with an inhibition of around 80-87%. These results were corroborated by potentiodynamic polarization and impedance measurements. Also, XRD results demonstrate that the formation of $Fe_2O_3$ from $Fe^{2+}$ was diminished in the presence of the biosurfactant, confirming that it is a good alternative candidate in carbon steel anticorrosive strategy.

*Glucolipids* (**10**)

Studies on intermolecular forces have been performed on multilamellar coating of the acidic C18:0 glucolipid through osmotic stress experiments.[324] A 1 wt% glucolipid lamellar gel (presented in section 2.7) at pH 6.2 and ionic strength varying between 16 and 100 mM is drop-cast on a silicon wafer and introduced in an adiabatic desiccation chamber. After drying, the relative humidity is adjusted from 10 % to 98% while the corresponding neutron diffraction pattern, after equilibration, is recorded for each value of the relative humidity. The interlamellar aqueous layer is found to be below 2 nm, thus excluding repulsive electrostatic and entropic contributions, which are generally observed at much larger distances. The pressure-distance profile below 2 nm are then nicely modeled by two hydration forces with decay lengths of 0.4 ± 0.1 nm and 2.0 ± 0.8 nm, respectively corresponding to short and long-range repulsive hydration forces, counterbalancing the attractive Van der Waals force. If the primary hydration force is poorly affected by the presence of salt, as expected, the secondary hydration force is directly related to the presence of the salt, as found in other similar systems[87] but never reported for biosurfactants.

The spontaneous tendency of glucolipid C18:0 (**10**) to form flat lamellae (Table 5, Table 6 and Table 8) was exploited to prepare single-layer coatings of high homogeneity on silicon and gold substrates. Baccile *et al.*[397] used dip-coating as a novel tool to prepare lipid membranes. This deposition technique, generally employed to prepare thin layers of metal oxides of controlled thickness, was well-adapted to deposit the colloidal lamellar structures of (**10**) from an aqueous solution. Controlled dipping rate in the order of 0.1 mm.s$^{-1}$ allow the the deposition of a highly homogeneous, defect-free, layer of (**10**), with a thickness of about 3 nm, equivalent to a single, interdigitated, molecular layer over a cm$^2$-scale surface. Homogeneity was controlled by crossing large-scale (ellipsometry) and nanoscale (AFM and infrared nanospectroscopy) analytical tools. Since dip-coating, differently than other coating technique, can be applied to morphologies of virtually any size and shape, this work shows the possibility to prepare homogeneous coatings of bioamphiphiles for, possibly, antifouling or antibacterial applications.

*Mannosylerythritol lipids* (**21**)

As above mentioned, hydrophobicity on solid surfaces favors the adhesion and growth of several bacteria. For this reason, the use of the amphiphilic property of different biosurfactants provides an interesting strategy for various industrial and medical applications. But, another way to take advantage of these molecules, is their use in agriculture as pesticides against pythopathogenic fungi on plant surfaces. Yoshida *et al.*[398] reported the use of various biosurfactants of the family of mannosylerythritol lipids, MELs (MEL-A (**21A**), MEL-B (**21B**), MEL-C (**21C**) and isoMEL-B) as good candidates to significantly decrease their water contact angle values and therefore the hydrophobicity of surfaces, with comparable results as other commercial surfactants such as Tween 20 and Brij 35. After, they confirmed that such change in hydrophobicity showed an influence in the germination of several phytopathogenic fungal conidia on the Gelbong surfaces.

*Trehalolipids*

This glycolipid has demonstrated its antiadhesive and antimicrobial over polystyrene and silicone urethral catheters towards a large number of strains. Authors found excellent antiadhesive activity against *C. albicans* (90-95%) and good properties (60-70%) against *Escherichia coli*, *Enterococcus hirae*, *Enterococcus faecalis* and *Proteus mirabilis*. Authors coupled these studies to calculations of the interaction energy between trehalolipids and both silicone and polystyrene, with respective values of -11.1 and -7.5 kcal.mol$^{-1}$, indicating a strong mutual interaction. The better interaction with silicone is explained by the strong hydrogen bonds (1.71 Å) between the trehalose lipid and polysilicon molecule. If authors eventually propose two adsorption mechanisms between trehalose lipid and the investigated surfaces being monolayer adsorption below the CMC and a bilayer or micelle adsorption above CMC, they do not provide any experimental proof about the effective thickness and structure of the adsorbed trehalolipid layer.[399]

*Surfactin* (**20**)

As previously described along with rhamnolipids, surfactin has been described to have antiadhesive and antimicrobial activity on different surfaces, but not reaching the efficiency of rhamnolipid.[395] Surfactin was additionally studied at the hydrophilic and hydrophobic (modified with octadecyl trichlorosilane, OTS) silica and sapphire interface with air.[182,212] If no adsorption was found on the hydrophilic surface, authors found irreversible adsorption independent of pH on the OTS surface with an area per surfactin of 1.45 $nm^2$ and a thickness of 15 Å.[182] Sapphire was chosen for its point of zero charge being around 8. Differently than silica, which is negatively-charged above pH 2, sapphire is positively-charged below pH 8. Surfactin was then adsorbed at pH 7.5 and below (5.5 and 6.5) and in the presence of $Ca^{2+}$ at pH 7.5. Authors found no adsorption of surfactin at pH 7.5, except in the presence of $Ca^{2+}$ (50% coverage) which confirms charge screening and interaction with surfactin, although the surfactin structure was not obvious to model. At lower pH, authors found an incomplete coverage at pH 6.5 and almost full coverage at pH 5.5. The thickness is slightly higher at pH 5.5 (20 Å) than at pH 6.5 (13 Å), thus suggesting a double surfactin layer, confirming the strong adsorption. Authors correctly explain that at pH 5.5 the adsorption is mainly driven by electrostatic attraction between negatively-charged surfactin and positive sapphire, while at higher pH, when electrostatic attraction is weaker, or not existing, they find that the hydrophobic effect is not enough to drive surfactin adsorption.

### 3.2 Other interfaces

In this section, we consider those studies involving biosurfactants at non-model curved interfaces. At the liquid-air interfaces, we include the formation of foams; at liquid-liquid, all the processes involving emulsification, and finally at the liquid-solid interfaces, various systems involving both natural materials, like hair, and synthetic materials, like nanoparticles. In the latter, biosurfactants are certainly very promising surface capping compounds providing good dispersibility and stability in water and, probably, good biocompatibility for biological applications. However, a much larger effort must be undertaken to reduce particle size distribution in one-step processes, to prove grafting of a BS monolayer on the NP surface and to purify the NPs from the excess of BS. Specific comments on these aspects are given at the end of this section in view of improving the quality of such promising nanosystems.

#### 3.2.1 Liquid-air

Foaming is a critical property of surfactants, highly demanded for industrial applications. Strong foaming ability determines the application potential in the house-hold and personal-care fields: low-foaming compounds are more interesting for dish- and laundry-washing while high-foaming surfactants are more interesting for detergency or teeth-care. One of the precursors, although highly fundamental, study on the foaming potential of BS has been performed by Cohen and coworkers several years ago. This set of papers, discussed in more detail in the previous section on flat liquid-air interfaces, determined the intermolecular forces and disjoining pressures of common rhamnolipid films.[321–323,353] Study of the foaming properties of BS presented below are, on the contrary, more empiric.

*Sophorolipids*

The foaming properties of acidic (**1**, **2**), lactonic (**3**) C18:1-*cis* SL and their mixtures (acetylation is partial, although not fully specified) in pure and hard water (calculated as ppm of $CaCO_3$) at pH 8.94 and 40°C has been reported by Hirata *et al*.[158] They find that pure acidic SL and acidic SL-rich solutions display the best foaming properties in terms of foam height (120 mm in pure and 80 mm in hard water) immediately after draining and after 5 minutes. Foaming properties are practically not existing for lactone rich and pure lactone SL, in agreement with the lactone poorer surface active properties and poor tendency to form micelles (Table 5 and Table *6*) compared to the acidic SL. The same group performed a comparative foaming study between lactone-rich SL (7-to-1 with respect to acidic SL), surfactin (**20**) and other commercial petrochemical surfactants and found that the foaming properties of this specific SL mixture is among the lowest, all conditions of time and water hardness tested.[158] The foam height settles at about 20 mm, while surfactin settles between 120 and 150 mm, depending on conditions. SDS is the surfactant having the highest foaming performance (> 200 mm foam height) while foaming of SL is rather comparable to pluronic block copolymers (< 30 mm foam height).

*Sophorolipids derivatives*

A series of ester derivatives of (**1**), from SL-methyl (SL-ME) to SL-decyl (SL-DE), were studied in terms of foam formation and compared to a natural 1:1 acidic and lactonic mixture as well as to an acidic SL (**1**) and tween80 (Figure 25).[238,400] In a first study,[238] authors found a comparable initial foam height, except for the ethyl ester derivative, which shows a foam height of almost 1.5 to 2 times higher compared to the others. However, after 10 min and 1h, the foam height collapsed by a factor three, or more, for all derivatives. In a second more specific study,[400] authors found that the initial foam height (1 min) sets around 12-16 cm for all compounds but the SL-DE derivative shows longer stability after 60 min, with a comparable height, while all other compounds show heights falling below 2 cm. In the presence of paraffin oil, the SL sample constituted by a mixture of (**1**, **2** and **3**, named L+A) shows poor foaming at any time, while ester derivatives show good stability after 1 min but foam disruption after 60 min.

Qualitative foaming properties were also evaluated for a naturally-produced acetylated acidic C18:1-*cis* SL in comparison with sodium laureth ether sulfate (SLES).[109] After shaking, the foam height was 71% of the SLES foam height, while after a rest of 10 min it was 82% of the fast-collapsing, resting SLES foam height.

Nonacetylated neutral symmetrical bolaform C18:1-cis SL underwent a qualitative comparative foaming test and showed poor foaming capacity, in the order of 10 mm, comparable to acidic SL, commercial SL mixture and alkylpolyglucosides (foam height between 10 mm and 20 mm) but 7 times less than SLES (70 mm).[16]

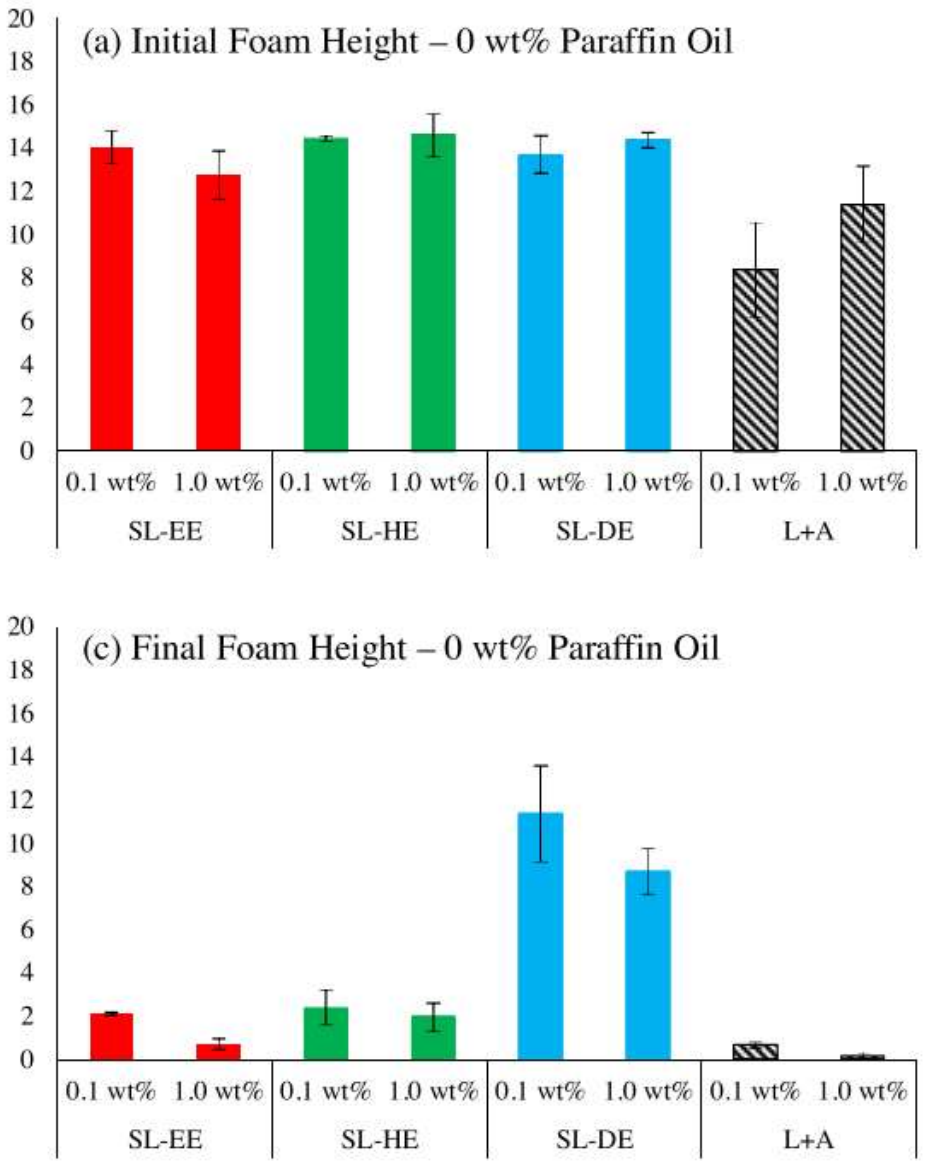

Figure 25 – Foaming properties of ester derivatives of (**1**) sophorolipids. SL-EE: ethyl ester SL, SL-HE: hexyl ester SL, SL-DE: decyl ester SL, L+A: 1:1 unmodified lactonic(**3**):acidic(**1**) SL mixture (Reprinted from [400], Copyright (2016), with permission from Elsevier). The final foaming is measured after 60 min.

*Rhamnolipids (mono- (**19**), di-(**18**))*

Mono- and di-RL foams at pH 5.0 and 6.8 have been reported by Özdemir *et al*.[196]. If they find that higher pH impacts the maximum foam volume, similar for both RL (about 240 $cm^3$), a clear correlation with stability could not be determined, although lower pH seem to promote longer times.

Foaming properties have been employed by Mulligan *et al*. for depollution purposes.[401,402] Pentachlorophenol (PCP)[402] and metal ions[401] were removed by soil using the foaming properties of RL. In the first study, the rhamnolipid's foam stability (time required for half of the liquid in the foam to drain out) was very good and the quality (defined as the ratio of total gas volume per total volume of the foam) was 99%. Upon injection into PCP (1000 mg/kg) contaminated soil, 60 and 61% of the PCP could be removed from a fine sand and sandy-silt soil, respectively. In the second work, authors evaluated the removal of Cd (1710 ppm) and Ni (2010 ppm) from a sandy soil. The removal efficiency using a RL foam was 73.2% for Cd and 68.1% for Ni. These values overrated Triton X-100 foams, which removed 64.7% Cd and 57.3% Ni, and distilled water, which removed only 18% of both Cd and Ni. Authors did not observe specific concentration (0.5, 1.0 and 1.5%) and pH (6.8, 8 and 10) effects, probably because RLs are in their micellar phase above pH 6.8 (Table 10).

*Surfactin* (**20**)

Foaming properties of surfactin at pH 8 (micellar phase, Table 10) were evaluated by Razafindralambo et al. in 1996.[403] Authors found comparable foaming volumes (about 1 mL) for surfactin and SDS, but superior residual volumes (94% for 0.2 mg/mL surfactin and 0% for 0.2 mg/mL SDS). The half-life time of the liquid in foam goes up to 141 s for surfactin while it is only 76 s for SDS.

### 3.2.2 Liquid-liquid

Emulsions, micro and nanoemulsions have an extensive use in commercially available products like in food, beverage, cosmetics, personal care, and pharmaceutical industries, just to cite some.[404] The role of surfactants in emulsion formulations is critical and it relies on the easier formation and rapid absorption to the oil surface and reducing the interface tension, thus stabilizing two otherwise immiscible liquid phases. Another benefit of the use of surfactants is the protection provided from aggregation during storage thanks to repulsive interactions between them. Recently, there has been an emerging need to replace synthetic emulsifiers by more sustainable alternatives. Therefore the use of biosurfactants has become an interesting field of study in emulsions formation,[351] particularly in the field of oil recovery and this since the mid 1980's.[23,405,406]

The field of emulsion formation from biosurfactants is quite wide and it has been reviewed by several authors across the years, including recently, within a number of fields (food science, environmental science, cosmetics).[10,19,20,22,24,95,96,98,102,110,351,407] The importance of this field, compared to the others discussed in this review, is shown by the large number of patens involving BS for emulsification purposes.[406] In addition, emulsion properties are heterogeneous across biosurfactants and for a given biosurfactant. This is explained by the type of medium, oil polarity, emulsion processing, physicochemical conditions of the aqueous phase (pH, hardness…) and methods employed to evaluate emulsification and stability. It is then quite harsh to find a common denominator throughout the literature results. For this reason, we decide not to present a complete summary of the emulsion properties of biosurfactants. One can refer to the previous reviews[10,19,20,22,24,95,96,98,102,110,351,407] and patent collection[406] and to some selected examples, illustrated below.

*Sophorolipids*

Koh *et al*.[126,238,400,408] have published a series of papers in which they study the emulsification properties of a series of ester (from methyl to decyl) derivatives of (**1**) in the presence of paraffin and synthetic crude oil,[400] almond oil[126] and lemon oil (Figure 26).[408] As a general trend, authors find a droplet size between 1 µm and 5 µm for most systems under study with good stability over time, up to a week. Size below 1 µm can be found at 1 wt% in lemon and almond oil concentration. In terms of efficiency, esters with long chain length (hexyl and decyl) appear to be better stabilizers of almond and lemon oils, while such a trend is not observed for paraffin and crude oil, for which the 1:1 acidic:lactonic mixture seem to be an even better stabilizer. These works are of particular interest because they quantify and compare the poor interfacial tension properties at the w/o interface of nonacetylated acidic C18:1-*cis* SL (**1**) compared to its ester derivatives and to its mixture with lactonic SL (**3**), which have good interfacial properties. These data also suggest that, for this specific property, the crude SL (**1**-**3**) extract is superior and much more interesting than the individual lactonic (**2**) and acidic (**1**) components taken individually.

Emulsification was tested on sophorolipid batch containing a major share (not quantified) of the C18:0 congener (**13**), obtain by a direct solid-state fermentation approach.[409] The average droplet size settled at about 3.5 µm and was stable over a week time. No precipitation of SL was observed, although authors report a creaming effect. If the results are encouraging, considering the complex composition of the batch (containing C18:1-*cis* (**1**) and C18:0 (**13**) derivatives at various acetylation and lactonization degrees) it was unclear to determine to which compound the stabilization can be attributed to. However, considering the fibrillation behavior of pure nonacetylated acidic C18:0 SL (**13**) at pH below 7.4 (Table 10),[221] it is unlikely that this specific compound could be responsible for good emulsification properties, although it could contribute to a synergistic effect with the other ones. In this regard, a SL batch containing a major share (> 80%) of acetylated acidic C18:1-*cis* SL (**2**) did not show significant emulsification properties.[109]

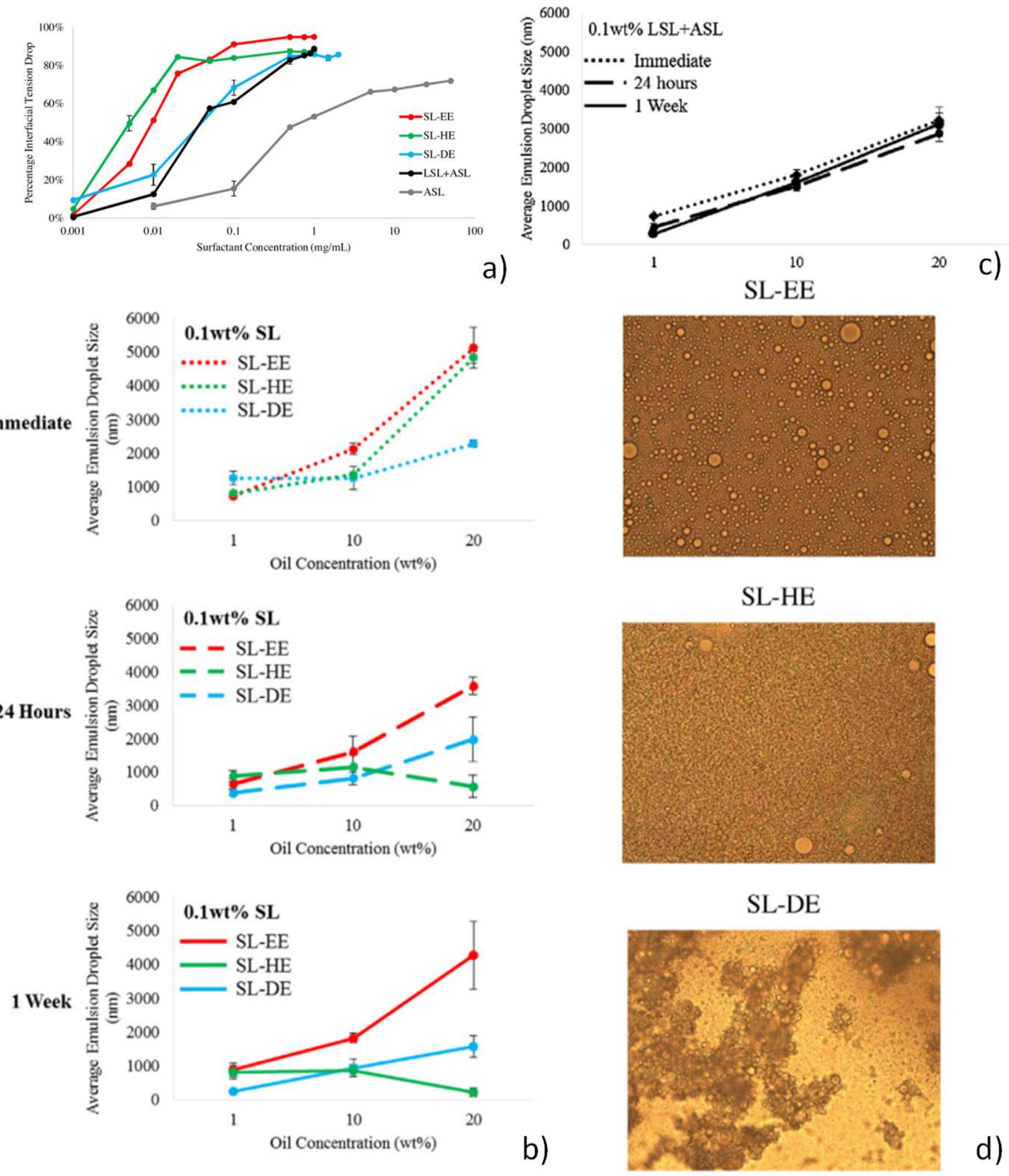

Figure 26 – O/W emulsion properties (oil: lemon oil) of ester derivatives of (**1**). SL-EE: ethyl ester SL, SL-HE: hexyl ester SL, SL-DE: decyl ester SL, ASL: acidic SL (**1**), LSL(**3**)+ASL(**1**): 1:1 unmodified lactonic:acidic SL mixture. a) Concentration dependent interfacial tension. b-c) Emulsion droplet size evolution with time in the presence of b) SL-EE, SL-HE, SL-DE and c) ASL-LSL. d) Emulsion droplet size observed my microscopy (scale bar not provided). Reprinted from Ref. [408], Copyright (2016), with permission from Elsevier.

*Rhamnolipids (mono-* (**19**), *di-*(**18**))

Xie *et al*.[410] have shown the formation of both emulsions and microemulsions between rhamnolipids (mono-/di- ratio not specified) and n-heptane in pseudo-ternary phase diagrams with various mid- and long-chain alcohols (n-propanol, n-butanol, n-pentanol, n-hexanol, n-heptanol, and n-octanol) at 25°C (Figure 27). Authors show that if RL can form emulsions with n-heptane, they can only form single-phase microemulsions by adding an alcohol. In this regard, they find that n-butanol (Figure 27b) has the most favorable influence, whereas the region 3 (single-phase microemulsion) is the widest.

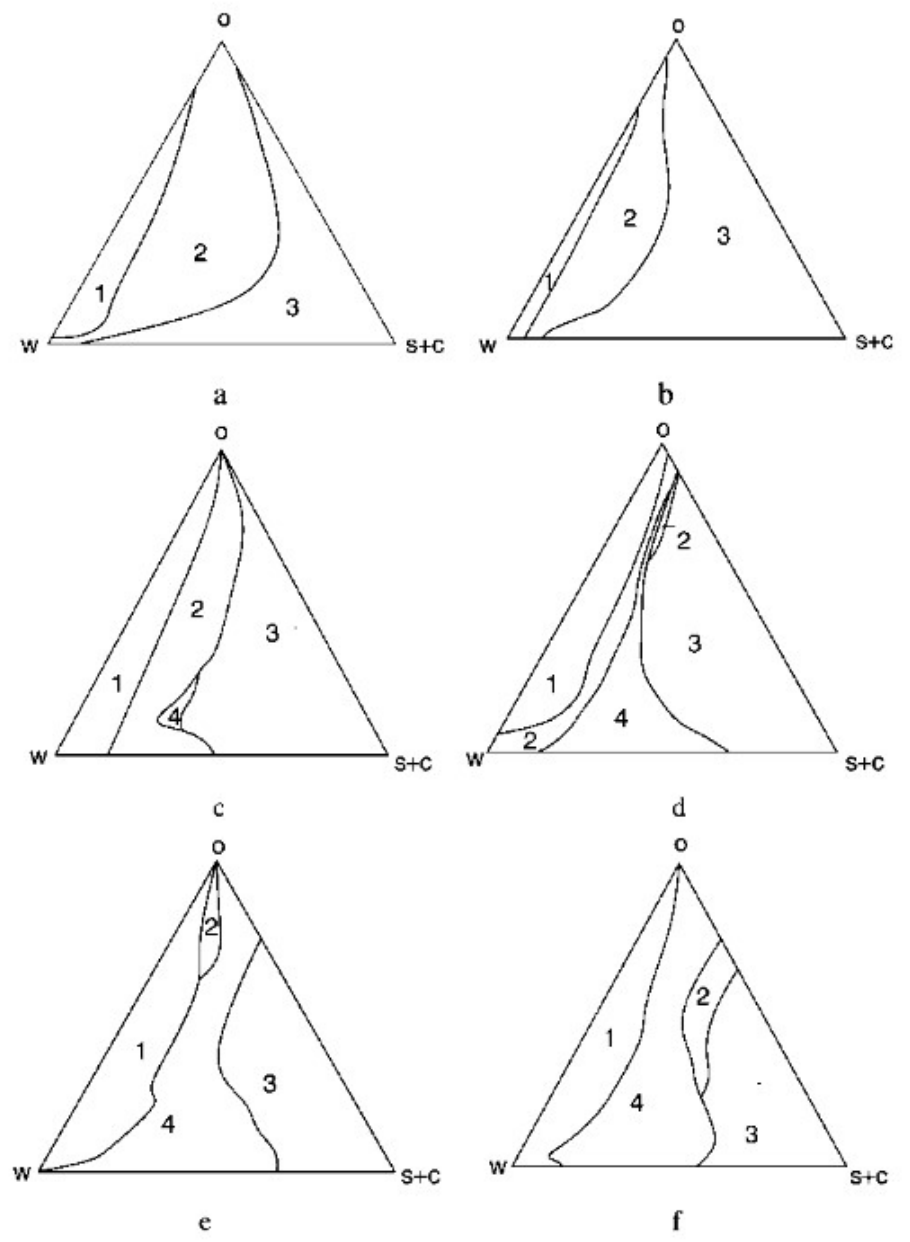

Figure 27 – Phase behaviour of RL/n-heptane with various alcohols: (a) n-propanol; (b) n-butanol; (c) n-pentanol; (d) n-hexanol; (e) n-heptanol; (f) n-octanol. The types of the areas are: (1) emulsion; (2) two-phase micro-emulsion; (3) single-phase microemulsion; (4) liquid crystal. Image adapted from [410] , reprinted by permission of Taylor & Francis Ltd.

Qualitative evaluation of RL (extract, unknown composition) as emulsifier is shown by Nogueira Mendes *et al.*[411], who show good emulsification in the following decreasing order: methyl methacrylate > castor oil > n-heptane > toluene > hexadecane > octane.

Good O/W (oil phase: triglycerides) emulsifying properties of RL was also reported by Li *et al.*[412] In this work, emulsion preparation is standardized by using ultra-turrax combined with a microfluidizer. They find an emulsion stability at least up to 2h for RL concentration between 0.5 wt% and 3 wt% at RT, a negligible effect of temperature on the emulsion particle size between 30°C and 90°C, an important pH and ionic strength effect on the particle size: the size increases from few hundred nm (good stability) to µm (poor stability) at low pH and at high ionic strength, in good agreement with the body of data showing that RL tend to better stabilize interfaces in their ionic form (micelle phase at basic pH, Table 10).

In 2016 Bai *et al.*[413] studied the interfacial properties of rhamnolipids with medium chain triglyceride (MCT) oil-water interfaces, as well as the formation of nanoemulsions. The resulting nanodroplets with a diameter < 150 nm at low surfactant-to-oil ratio, SOR < 1:10, were covered with rhamnolipid, which exhibited a negatively charged surface at high pH values. This is justified by the ionizable -COOH groups in rhamnolipids. In addition, these promising nanoemulsions were stable at different temperatures (30–90 °C), salt concentrations (<100 mM NaCl), and pH conditions (pH 5–9).

A nice application example of the use of rhamnolipids (and sophorolipids) is shown by Drakontis *et al.*, who employ biosurfactants as emulsifiers in complex silicon oil-water formulation containing Aerosil silica particles, for the development of lip gloss.[414] Aerosil improves the rheological properties and the biosurfactant (RL or SL) is mainly used as emulsifier, whereas no particular effect on the rheological properties has been observed.

*Surfactin* (**20**)

Mukherjee[415] has shown the emulsification potential of surfactin against vegetable oils (sunflower, soybean, coconut, musared oils) with an emulsification index (=height of emulsion layer/total height of mixture) between 70% and 80%.

He *et al.*[416] introduced surfactin in the formation of microemulsions with docosahexaenoic acid (DHA), a significant nutrient of the family of omega-3 polyunsaturated fatty acids (PUFAs), for the improvement in their stability. The DHA single cell oil (DHACO) microemulsions showed an increase of their oil/water regions with a decrease on the particle size up to 15 nm when 0.2 mM surfactin was added. Thanks to this addition of the biosurfactant, the physical and antioxidant properties were enhanced considerably even at storage at 37 °C during 60 days.

*Pickering emulsions using biosurfactants*.

Pickering stabilization of emulsions refers to those emulsions of which the o/w or w/o interface is stabilized by solid, generally inorganic (often silica) particles, but also soft, like polymer, colloids.[417–419] Pickering emulsions are known, and sometimes preferred, to conventional surfactant's or polymer's based emulsions for their stability and improved elastic properties. Biosurfactants *per se*, by definition, cannot provide Pickering stability, but they can either be associated to silica nanoparticles,[420] to improve the emulsion stability, or to polyamines. In the latter case, nonacetylated acidic C18:1-*cis* SL (**1**) form complex coacervate stable colloids in water by association with polyamines.[421] More details on this system are given in section 4.1. In turn, SL-polyamine complex coacervates can be used as soft complex Pickering stabilizers for o/w emulsions (oil phase: dodecane). These systems can stabilize an oil fraction up to 70 vol/vol% without creaming and with a virtually "infinite" stability in time. Authors could also show the pH-controlled reversibility of the emulsions, which separate into an oil and water phase when pH is switched out of the stability range of the complex coacervates. Emulsion forms again when pH is changed again in the stability region of the coacervates.[422]

Interestingly, similar results were produced at the same time using rhamnolipids-protein (soy or whey) mixed structures. In these works, authors stabilize either foams or emulsions. However, authors do not mention neither any complex coacervation phenomenon nor the existence of any specific Pickering effect at the water/oil or water/air interfaces.[423,424]

### 3.2.3 Liquid-solid

Biosurfactants represent an interesting tool to stabilize solid surfaces, from simple spherical shapes (e.g., nanoparticles) to complex morphologies (e.g., human hair). In all cases, when the adsorption occurs from the carboxylic end-group (in glycolipids), they provide hydrophilicity, water dispersibility, colloidal stability and biocompatibility. These characteristics are all highly required for the development of green nanotechnology. In particular, biosurfactants can play an important role in the processes of synthesis of the nanomaterials.[425] We highlight this aspect in this section by addressing some examples of biosurfactants as exo-templates, or external stabilizers. A fair number of studies exists on the use of biosurfactants as capping agents for various type of nanoparticle, metal, metal oxide and polymeric. This topic has been already partly reviewed in existing works[17,425–427] and we address to them for a complete picture of the field. In the following, we only report few major examples, selected on the basis of the type of biosurfactant and nature of the core. We conclude this section with a critical discussion and some perspectives on the development of biosurfactants-stabilized nanoparticles.

Endo-templating is another important topic where liquid-solid interfaces could be concerned. However, this is voluntarily kept out of this section. For some examples of endo-templating towards nanostructured oxides, one can refer section 2.7, while some examples of solid lipid nanoparticles with a dense lactone SL core and stabilized by Pluronics and Tween surfactants are proposed in Ref. [428,429].

*Sophorolipids* (**1**)

Most of the studies cited below employ nonacetylayted acidic C18:1-*cis* SL (**1**). We discuss each system by their difference in terms of core chemistry.

*Metal core.* Prasad and coworkers[430–435] and others[436] reported a number of publications dealing with the synthesis of sophorolipid-coated metal nanoparticles, with cobalt, gold and silver cores. In 2007, they introduced the synthesis of water-stable and redispersable cobalt nanoparticles by employing acidic sophorolipid as capping agent.[430] The cobalt nanoparticles were synthesized by reduction of cobalt with $NaBH_4$ using the sophorolipid and characterized by TEM, obtaining polydisperse particles with an average size of 50 nm, FTIR, showing the binding of SL to the surface of Co nanoparticles and room-temperature magnetization checking its magnetic property.

In 2009, they also took advantage of the reducing and capping properties of sophorolipids simulteneously to describe a direct method to prepare SL-AgNPs as above mentioned.[431] Silver nanoparticles are well-known antimicrobial agents and in that work they have also demonstrated their antibacterial effect against *Bacillus subtilis*, *Staphylococcus aureus* and *Pseudomonas aeruginosa.* This effect is comparable to the one exerted by oleic acid capped AgNPs, but in this case the SL-AgNPs are stable and water-dispersible, which makes it more suitable for their use in biological media. In addition to that, they observed a difference with Gram negative bacteria, which has a thinner layer of peptidoglycan than Gram-positive ones, and that may consist in a better interaction between SL from the SL-AgNPs and the bacteria, making it more vulnerable to the release of reactive oxygen species (ROS) generated. Similarly, Shikha *et al*.[436] reported the synthesis of SL-stabilized gold NPs, using SL as both reducing and capping agent. Gold is generally not considered as an efficient antimicrobial element, however authors have shown that SL-capped gold NPs display good antimicrobial properties against Gram+ and, in particular, against Gram- pathogens, such as *V. Cholerae* and *E. coli*. For instance, at 50 µg/mL of SL-capped NPs, the viability of *V. cholera* is reduced by 100%, while SL alone reduces the viability of only 90%. If these results are interesting, the mechanism of action is however still unclear, authors state.

A particular attention was addressed to the antibiofilm properties. Both silver NPs, attached onto porous polyelethylene scaffolds through the mediation of SL,[435] and gold NPs, capped by SL, were evaluated against antibiofilm formation.[436] In both cases, either the action of the surface-stabilized silver, or the membrane-disruption performance of SL-capped gold NPs, show good efficiency against *B. subtilis*, *P. aeruginosa*, *E. coli* or *S. aureus*, incubated between 6h and 12h, but also forming and mature *S. aureus* and *V. cholera* biofilms.

SL-capped silver nanoparticles are also reported to be synthesized by a continuous flow method using two acidic sophorolipids (C18:1-*cis* and C18:0).[432] C18:0 SL (**13**) performed faster reactions than C18:1-*cis* SL (**1**) and higher residence time permits the full reaction obtaining spheric, small and monodisperse nanoparticles. Two methods were also tried, stainless steel tubular microreactor and a spiral millireactor. And this last spiral millireactor lead to the formation of spherical and monodispersed nanoparticles of less than 10 nm in size at long residence time. Following a similar synthetic protocol as in [431], they reported the preparation of both gold and silver nanoparticles stabilized with SL **1**.[433] These functionalized nanoparticles were tested for their cytotoxicity and genotoxocity on human hepatic cells. Results demonstrated their biocompatibility up to concentrations of 100 µM being AuNPs slightly more viable than AgNPs ones. Concerning their DNA damage checked by comet assay, results on DNA damage correspond to the similarities found in cell viability in which AgNPs showed higher damage.

*Metal oxide core*. Baccile *et al*.[437] described the use of acidic C18:1-*cis* sophorolipids as surface stabilizer for magnetic iron oxide nanoparticles in aqueous solutions. Iron oxide nanoparticles are important nanomaterials employed in magnetic resonance imaging (MRI) and also in magnetic hyperthermia applications, which some of these systems indeed have been approved for clinical trials.[438,439] The reaction to obtain the nanoparticles followed a classical co-precipitation protocol in either one- or two-steps synthesis. The one-step synthesis provided poor crystallinity of ferrihydrite nanoparticles while the two-steps, where the sophorolipid was added in the last moment after ammonia addition, obtained maghemite structure both at room temperature and 80 °C. The interaction between the inorganic nanoparticle and the sophorolipid was analyzed *via* FTIR and DLS. FTIR results demonstrate that -COOH is the functional group interacting with the iron oxide nanoparticles, and DLS measured in a mixture of ethanol/water (sophorose has low solubility in ethanol) showed the formation of aggregates, then concluding that the sophorose moiety faces solvents instead of nanoparticle surface, a relevant fact for its biocompatibility. Noteworthy, the colloidal stability was very good, a key aspect for its use in biomedical applications, remaining stable for over a year and ethanol was required to force the precipitation by centrifugation.

Later in 2019, Lassenberger *et al.*[440] developed a two-step method to prepare monodisperse oleic acid coated iron oxides nanoparticles (IONPs) by a classical thermal decomposition synthesis (Figure 28). IONPs were further stabilized with a nitrodopamine-derivative of sophorolipid (**1**) (NDA-SL) through a classical ligand exchange step. This method was tested in view of having SL-coated monodisperse iron oxide NPs, otherwise a hard task to obtain with a direct synthesis in water. The use of a NDA, instead of the classical COOH, end-group lies in the better affinity of catechols for iron oxide. TEM and cryo-TEM exposed the coexistence of SL-nanoparticles and NDA-SL aggregates, difficult to separate but a close to pure SL-IONP was obtained with a core of 4.6 nm. The combination of SANS and SAXS, respectively sensitive to both core-shell and core only in water, demonstrated the presence of a single SL coating layer, having a thickness of 2.5 nm. The SL-IONP cytotoxicity in human monocytes (U937) and human breast cancer cells (MCF7) was checked, demonstrating their viability.

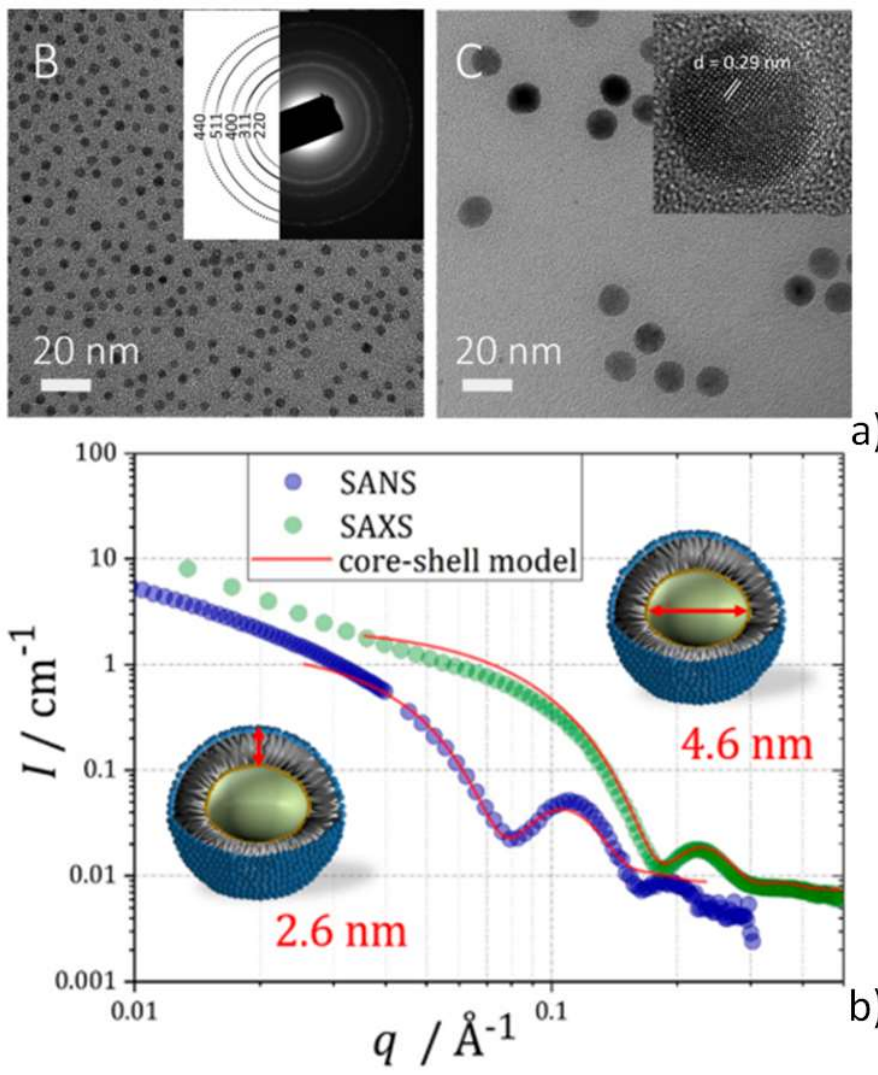

Figure 28 – C18:1-*cis* sophorolipid (**1**)-stabilized FeOx iron oxide nanoparticles (IONPs). a) Typical TEM and high-resolution TEM images of oleic-acid nanoparticles prior to SL stabilization. Oleic acid is exchanged with NDA-modified SL (**1**). NDA stands for nitrodopamine and it is grafted on the COOH group of SL because NDA-FeOx interactions are favoured over COOH-FeOx interactions. b) SAXS-SANS experiment performed on the same SL-NDA/FeOx colloidally-stable sample. SAXS is sensitive to the FeOx core only, while SANS is sensitive to the organic (SL-NDA) shell and FeOx core. Shift between the two profiles yields the SL-NDA thickness. Reprinted with permission from Ref. [440], copyright (2019) American Chemical Society.

*Semiconducting core*. Quantum dots (QDs) are semiconductors nanocrystals having outstanding optical and electronic properties with potential application in fluorescent-biolabelling and imaging thanks to their size dependent fluorescence, but also in cancer therapy.[441] However, the cytotoxicity of these QDs limits its applicability in these theranostic approaches. For this reason, there is a need to find strategies to improve their biocompatibility to be able to exploit their capabilities.

The group of Prabhune described in 2013 the functionalization of cadmium teluride (CdTe) QDs with the acidic form of C18:1-*cis* SL (**1**) conjugated *via a* cross-linking reaction.[442] With the addition of SL, the photoluminescence property remained the same but it improved the water-solubility and chemical stability of QDs. Most importantly, no cytotoxicity was exhibited on NIH3T3 (Mouse Embryonic fibroblast cell line) with 100% viability up to 10 µg $mL^{-1}$, and for MCF-7 (human breast adeno carcinoma cell line) shows a lower viability (around 65%) what would result in a tool for diagnosis and therapeutic as it is able to kill a percentage of the cancer cells. Remarkably, the improvement of this biocompatibility was proven by comparison with the mix of QDs and free SL, what showed a high cytotoxic activity in both cell lines.

*Rhamnolipids (mono-* (**19**), *di-*(**18**))

*Metal core*. Rhamnolipids were also proved to mediate the synthesis of silver nanoparticles.[443,444] Xie *et al*. have demonstrated the possibility to use a crude mixture of RL in a water-in-oil emulsion approach to prepare silver NPs within reverse micelles of RL, whereas NP formation is obtained by a classical reduction by $NaBH_4$. According to TEM and AFM, they find NPs of size in the order of 6 nm. In this case, however, the role of RL seems to be limited to the stabilization of the O/W interface, as authors do not provide any proof of RL as possible capping agents. Kumar *et al*.[444] repeated the same type of experiment using purified rhamnolipids, instead. However, their size distribution seemed to be poorer than the in the work of Xie and the role of RL is not clarified either, may it be a simple nanoreactor of a capping agent. Neither works show specific studies on the colloidal stability and analysis of the NPs surface. Kumar *et al.* show an antibacterial activity of their NP batch for both Gram + and Gram – bacteria and *Candida albicans*. However, it was not clear whether or not the antibacterial properties could be related to the NPs, to the RL or to their synergetic effect.

Gold nanoparticles have been shown to be templated by RL, the latter being the reducing and capping agent in the shape of microtubules (refer to Table 5 and Table 6 for the known phase diagram of RLs).[445] If this work seems interesting, several aspects probably need to be elucidated, namely the origin of the tubule phase, as this was never reported for RL, and the templating and capping mechanism, as NPs seem to lay on the fiber surface.

*Semiconducting core*. Hazra *et al*.[446] in 2013, in order to decrease the amount of toxic wastes in the synthetic protocols, and in particular in the synthesis of ZnS QDs, employed biosurfactants as an alternative eco-friendly protocol to obtain these QDs in aqueous media. The purified rhamnolipid was used 1.2-fold cmc concentration in the reaction. The crystallinity of the QDs was checked by XRD, and SEM and TEM showed their spherical shape, keeping the photoluminescence emission property. This nanomaterial exhibited elevated absorbance and a blue shift to 340 nm if compared with conventional bulk ZnS. In addition, it showed an improved biocompatibility and less effect in the generation of reactive oxygen species (ROS), lipid peroxidation and perturbance of the glutathione balance. In the same work, they checked the photocatalyst role of ZnS in photodegrading the diazo dye direct brown MR, within a context of possible treatment of contaminated waste water effluents. They could show similar percentages of photodegradation, 96 and 94%, for ZnS with SDS as capping ligand and ZnS with rhamnolipid respectively, under UV irradiation (lamp of 125 W) for 180 minutes.

*Metal oxide core*. Biswas *et al*.[447] in 2008, characterized the electrokinetic and rheological behavior of nanozirconia particles with the adsorption of rhamnolipid as coating. Rhamnolipids (230 ppm) strongly adsorbed to the surface of zirconia providing a more electronegative surface. Also their stability was checked by sedimentation studies. With conventional zirconia, the viscosity increases while increasing the concentration. On the contrary, when measuring the rheology of zirconia coated with rhamnolipids, it resulted in a reduced viscosity, including at higher loads. This means that the presence of rhamnolipids in the nanoparticles formulation would imply a significant profit for industrial processing, what would aim at maximizing the content in their processes.

Tentative use of RL having the double role of external capping agent of preformed magnetic iron oxide NPs and primer for the growth of bismuth oxybromide was proposed by Wang *et al*.[448]. If authors show the coexistence of an iron oxide and BiOBr phase through XRD, the role of RL is still unclear, for the combined use of CTAB in their synthesis protocol and lack of a clear-cut FTIR RL signal.

Palanisamy *et al*. employed a microemulsion approach involving RL to prepare NiO NPs, an interesting material for battery cathode and electrochromic devices.[449] Authors prepare $Ni(OH)_2$ at RT and eventually NiO after calcination. If authors do not invoke a capping agent role of RL, as often done for similar synthesis, they do suggest a possible pH dependency of the $Ni(OH)_2$ and NiO particle size, probably in relationship to the self-assembled structures of RL in solution. SEM and TEM show that micrometer-sized particles are obtained by the aggregation of nm-sized particles of less than 100 nm in diameter. These assumptions, which could be reinforced by experiments at acidic pH, are not outrageous.

*Polymer core*. Rhamnolipids were involved in the formulation of polysaccharide nanoparticles, boosting their antimicrobial activity. Marangon *et al.*[450] added rhamnolipids to chitosan NPs, condensed through the use of TPP (a polyphosphate). Stable positively-charged colloidal dispersions, without precipitation during 6 months of storage at 4°C, composed of monodisperse nanoparticles of less than 300 nm are obtained. In addition to the morphological benefits tested with the introduction of rhamnolipids in the chitosan nanoparticles, it was also observed the impressive increase in their antimicrobial and antibiofilm activities against *staphylococci*. This higher antimicrobial activity is due to a higher local concentration of both chitosan and rhamnolipid in the bacteria cell surface, therefore allowing the direct release and targeting of the *Gram positive* bacteria. Remarkably, they were active also against sessile bacteria as well as having low cytotoxicity in human umbilical vein endothelial cells (HUVEC).

In order to prepare biocompatible monodisperse polystyrene (PS) NPs, Hazra *et al*. [451] chose rhamnolipid and surfactin among all biosurfactants to provide innovative and eco-friendly templates for emulsion polymerization of PS. They developed biodegradable and biocompatible polystyrene/biosurfactant bionanocomposites with tunable size ($d_n \approx$ 50–190 nm). According to TEM and FTIR arguments, authors estimated that RL are possibly part of the PS shell, a fact which should be confirmed with additional experiments. An interesting point is the correlation between the PS polymerization mechanism and final morphology with the possible self-assembled structure of RL in solution, vesicular at acidic pH and micellar at basic pH, as discussed in section 2.3 (Table 10). If the data presented by the authors would need more insight, it is of extreme importance to correlate the self-assembled form of the lipid and the final material's structure, as well as the possible impact of the synthesis conditions and precursors on the evolution of the lipids' structure. Author further determined their biodegradation and biocompatibility and compared the profiles with a SDS-based PS and without any surfactant. They found that RL and surfactin derivatives are both more biodegradable (36-38%, compared to 8% SDS-based) and biocompatible ($IC_{50}$, the half maximal inhibitory concentration, ranges from about 55-75 mg.mL$^{-1}$ for RL- and surfactin-based PS to about 25-50 mg.mL$^{-1}$ for SDS-based PS).

The same team further explored the potential of similar core-shell bionanocomposites by employing poly(methylmethacrylate) (PMMA) as polymeric core and rhamnolipid, surfactin or trehalose lipids as amphiphilic shells.[452] They claimed the synthesis of PMMA nanoparticles with nPMMA (core)–biosurfactant (shell) structure on the basis of FTIR, AFM and TEM images. Although the presence of a BS shell is possible, the data are not fully convincing and should be corroborated by additional data, the problem being that AFM and TEM are not sensitive to a poorly-contrasted nm-scale lipid layer and the typical FTIR signals of BS cannot be clearly attributed. For instance, the FTIR profile of surfactin seems to be very similar to the one of RL and TL, although surfactin is a cyclic amphiphilic peptide while RL and TL are glycolipids. If the biocompatibility, cytotoxicity and antibacterial properties of the PMMA-based materials seem to be interesting, it is unclear on how to correlate it to the presence of BS.

*Surfactin* (**20**)

*Metal core*. Reddy *et al*.[453] synthesized silver nanoparticles in the presence of surfactin at different pH values (pH 5, 7 and 9) and temperature (4°C and RT) and was stable for 2 months. The size of the nanoparticles was pH dependent, obtaining a decreasing size with increase of pH from 5 to 9 (17.8 ± 9.8 to 4.3 ± 1.1 nm) at 4 °C. But, on the contrary, at RT the size increased with the increase of pH (4.9 ± 1.4 to 9.7 ± 4.3 nm from pH 5 to 9). The same group applied the previous strategy to the synthesis of gold nanoparticles.[454] In this case, the NPs synthesized at pH 7 and 9 were stable for 2 months, while at pH 5 it was observed the formation of aggregates after just 24 h. At 4°C, the size and shape were uniform for pH 7, while for pH 5 and 9 were polydisperse and anisotropic. And the UV–vis measurements showed a blue shift when changing the pH from 5 to 9 (from 528 to 566 nm) at both temperatures.

*Metal oxide core*. Later in 2011, Reddy *et al.*[455] demonstrated the use of surfactin in the synthesis of ZnO rose-like structures, and its influence in the morphology of the petals, which consequently aimed to a different photodegradability activity of methylene blue molecules. The thickness of the petals varied from 10 to 13 nm, and the prepared nanoparticles with less dense petals (loosely-spread) showed the highest photodegration activity.

The surface of minerals was modified by the adsorption of biosurfactants to combat biological processes like the bio-corrosion and bio-oxidation produces by microorganisms but also to provide new properties, useful in flotation and flocculation processes, typically employed for the separation of minerals on an industrial scale. Didyk-Mucha *et al.*[456] described the adsorption of biosurfactants (surfactin and biosurfactants from *Streptomyces sp*, surfactin analogues) on the minerals serpentinite, magnesite and silica after being activated with nickel (II), and how they altered their electrokinetic behaviour. The role of nickel is to provide a more positively charged surface and therefore improvement of adsorption. The adsorption, described by Somasundaran-Fuerstenau isotherm, is found to be strong between the anionic surfactants and the minerals, although with differences between minerals. For instance, adsorption is found to be stronger on serpentite and magnesite than on silica and this is explained by the larger content of cations and positively-charged metal species on the latter. Favorable interaction is then interpreted to occur between surfactin and the metal centers. These assumptions are verified by the evolution of zeta potential of the mineral particles with pH and authors found that surfactin shifts the isolectric point to lower pH values. Authors also found that the effect of surfactin is stronger than the effect of the SDS control. However, these results must be interpreted with care, because some differences can occur in terms of the nature of the assembled structures of each amphiphile in solution (Table 10) and by their possible excess, which could have a high impact on the zeta potential measurements.

*Semiconducting core*. Singh *et al*.[457] used surfactin from *Bacillus amyloliquifaciens* strain KSU-109 for the synthesis of cadmium sulfide (CdS) nanoparticles, stable up to 120 days. Although surfactin may play a role in the stabilization of the NPs, authors suppose surfactin acting as a capping agent, a fact which is still unclear from the presented data. On the basis of TEM images, authors propose an average size of 3-4 nm, although this hypothesis was not corroborated by additional scattering data.

*Emulsan*

*Metal oxide core*. Amani in 2017 used the biosurfactant emulsan as a strategy for microbial enhanced oil recovery (MEOR) in a synergistic effect with $SiO_2$ nanoparticles, forming a nanofluid.[160] This nanofluid mixture afforded a 90% reduction of interfacial tension in contrast with the biosurfactant alone. They developed a micro model displacement experiment to achieve this aim. The efficiency of this method allowed to recover around 10-20 % of oil emulsions after water flooding. The phase behavior was analyzed by the study of both kerosene and emulsan at the cmc value. In the absence of nanoparticles, it forms oil in water microemulsions, and middle-phase microemulsions in the presence of nanoparticles.

*Mannosylerythritol lipids* (**21**)

Two recent publications claim the synthesis of Ag, ZnO, Ag-ZnO and Au NPs stabilized by MELs (structure not clearly identified by NMR) and studied the antitumor, bactericidal and antidiabetic performances of such systems.[458,459] If the synthesis of the corresponding inorganic materials seem to be confirmed by XRD and electron microscopy, the presence of MELs is not clearly confirmed, neither as a coating nor as a matrix. The FTIR spectra of the nanoparticle systems do not show the C=O resonance above 1700 $cm^{-1}$, characteristics of the ester bonds in MELs; the signal of the mannose ring, between 1000 $cm^{-2}$ and 1400 $cm^{-1}$, is also hard to observe in the MEL/NPs materials. The possible reason of this discrepancy is probably hidden in the conditions of synthesis of the NPs. Authors employ an alkaline aqueous medium (KOH 0.1 M) at T= 80°C, a classical environment under which ester bonds can be irreversibly hydrolyzed. In this case, one expects a conversion from MELs to the corresponding free fatty acids and mannose. If these aspects should be considered further, the bioactivity experiments performed in these works are to be taken with caution.

MELs' amphiphilic structure reminds the one of ceramide and they biochemically act as glycosphingolipids. Previous studies suggesting that MELs would potentially ensure the role of ceramide in cosmetic formulations, the present one thus focuses on the hair care properties of MELs using damaged hair. The authors treated damage hair with 0.5% MEL-A (**21A**), MEL-B (**21B**) or a natural ceramide dissolved in 4.5% lauryl glucoside and have reported for the first time the repairing effect of MELs, supporting the preliminary hypotheses.[460] Regarding the mechanism for hair treatment by MELs, the lamella formation ability of MELs result in their easy adsorption on the surface of damaged hair, providing efficient repair of the hair fiber instead of usual ingredients. Electron microscopic observations witness a significant recover of the damaged hair applying MEL-A (**21A**) and MEL-B (**21B**). The tensile strength of the damaged hair increases by treatment with MEL- A (**21A**) (122.0 ± 13.5 gf/p), MEL-B (**21B**) (119.4 ± 7.6 gf/p) and ceramide (100.7 ± 15.9 gf/p) compared to hear only treated with lauryl glucoside (96.7 ± 12.7 gf/p). Its average friction coefficient was maintained after treatment with MEL-A (**21A**) (0.108 ± 0.002), MEL-B (**21B**) (0.107 ± 0.003) and the ceramide (0.111 ± 0.003), whereas it increased when treated only with lauryl glucoside (0.126 ± 0.003) (the reference value is 0.113 ± 0.005). The increase of bending rigidity by treatment with lauryl glucoside (0.204 ± 0.002) is contained by treatment with MEL-A (**21A**) (0.129 ± 0.002), MEL-B (**21B**) (0.176 ± 0.003) and the ceramide (0.164 ± 0.002), MELs were established not only to repair damaged hair but also to provide the smooth and flexible properties expected for the hair.

The studies reported below focus on interactions between MELs and immunoglobulins at non-planar interfaces, whereas planar surfaces were discussed earlier in the MEL paragraph, section 3.1.2 .[375,377] Im *et al.*[461] demonstrated that MEL-A, attached onto poly (2-hydroxyethyl methacrylate) (polyHEMA) beads, exhibits a significant binding affinity towards the natural human immunoglobulin G (HIgG). They evaluate the binding constant ($1.4 \cdot 10^6$ $M^{-1}$) to be 4-fold higher than that of protein A ($3.7 \cdot 10^5$ $M^{-1}$) and the binding capacity of the two ligands to be similar (17.0 mg HIgG per g).[462] Binding between gangliosides and glycoproteins may be enhanced by a "multivalent or cluster effect" defined as a "simultaneous association of two or more ligands and receptors", driven by the density, orientation and conformation of the saccharide moieties of gangliosides. The hypothesis of a "multivalent effect" is reinforced by the estimated binding molar ratio 1:70 between HIgG and MEL-A (**21A**).

Indeed, based on these results, the same team employed three MEL-polyHEMA composites and investigated their binding affinity towards HIgG.[463] The one reported for the MEL-A (**21**) based composite was the most satisfying. The highest the concentration, the highest the binding yield of HIgG and which was shown to be in the order of 81%, calculated for the maximum concentration of 106 mg HIgG/g of composite. The applied concentration was found to drive the binding according to two different mechanisms, namely Langmuir or Freundlich. In the meanwhile, the binding amount of human serum albumin to the composite was much smaller than that of HIgG, although preliminary binding of human serum albumin was not found to interfere with binding of HIgG, proving the presence of two distinct specific binding sites on the composite. Elution using a phosphate buffer solution at pH 7 allowed to recover 90% of the bound HIgG.

### 3.2.4 Improving the synthesis of biosurfactants-stabilized nanoparticles

The number of studies reporting the synthesis of biosurfactant-mediated NPs is growing more and more. Biosurfactants are either employed in reverse emulsification processes or directly in an aqueous environment. The goal is: to prepare, possibly in a one-step approach, monodisperse NPs capped by a single layer of biosurfactant, dispersible in water and stable for an "infinitely-long" time. The biosurfactant's structure is perfectly suited for such a task as it combines surface active properties and presence of a fatty acid, whereas ionic surfactants and fatty acids are independently used in classical stabilization processes of NPs respectively in water and organic media.

However, despite its obviousness, the proofs that this goal is achieved in the field of BS-NPs are scarce. At the present state of the art, no study can clearly demonstrate to have achieved a colloidally-stable dispersion of BS-capped NPs in water in a single step. In the following, we highlight the main flaws existing in this specific literature, we provide the main problems, specific to the use of biosurfactants, and we suggest some possible directions to address them, having in mind that the task is not easy.

*Current flaws*.

In terms of characterization, many authors claim to synthesize NPs on the simple basis of TEM images. TEM is certainly an important brick in the characterization building of NPs, but many existing studies are based on few TEM images of poor quality, where NPs are often aggregated due to drying and in some worst cases, one cannot distinguish between NPs and impurities. The corresponding XRD patterns, when existing, often show sharp diffraction peaks, which are generally not compatible, from obvious Scherrer (formula) arguments, with the nanoscale size estimated by TEM data. FTIR, used to demonstrate the presence of the BS ligand, is often ambiguous and it is rarely critically discussed, resulting in a lack of clear-cut proof for the ligand. More critically, in the case when FTIR shows the signal of the BS ligand in the NP sample, that is not a proof that the BS acts as a ligand. Since almost none of the systems presented in the BS-NP literature are thoroughly washed before FTIR, an FTIR signal of the BS could simply show that NPs are embedded in a BS medium. This is actually sometimes proved by the spurious TGA analyses which are provided by some authors and which tend to show an excess of the BS in the NP sample. Few are the ones who quantify the organic-to-inorganic ratio, which would show whether or not BS are in excess compared to the available surface area of the NPs. The colloidal stability of the BS-stabilized NPs is rarely studied, although this is one of the most important properties required for NPs. For some specific metal NPs systems, BS are claimed to be both reductive and capping agents. However, this interesting double role is never fully proven. In particular, NMR is not given to support modification of the BS structure, the ratio between the oxidized and initial forms of the BS is not provided nor it is provided the nature of the capping agent, whether it is constituted by the oxidized BS or by the initial one.

Several authors report synthesis conditions of temperature, pH or ionic strength which may not be compatible with the stability of BS or with their expected phase behavior (Table 5, Table 6 and Table 10). In this regard, it is also hard to know the type of molecular structure of the biosurfactant, e.g., mono- or di- for RL or even the natural mixture.

*General problems in using BS for the synthesis of NPs.*

Biosurfactants are undeniably complex molecular systems. If their use as NPs stabilizers has been shown and it is definitely of high potential, they have a number of drawbacks with respect to classical capping agents like ionic surfactants (for aqueous solutions) or fatty acids (for organic media).

*Stability*. Glycolipid biosurfactants, very much used in this field, have the intrinsic problem of thermal and pH stability. Temperatures above about 80°C/90°C and very low pH (< 2) may contribute to their dehydration and hydrolysis, if applied for an extended period of time. Acid-catalyzed and temperature-driven dehydration of carbohydrates is better known as the Maillard reaction and leads to formation of furanes. Please note that strongly alkaline pH, possibly coupled to high temperatures, could also hydrolyze the ester bonds (in lactone SL or RL, for instance).

*Composition*. The problem of purity and homogeneity among BS batches is a known problem. Sensitive differences in composition can lead to an unexpected phase behavior[191] in the synthesis medium, thus providing NPs of various shape and size.

*Phase behavior*. BS are known for their complex phase behavior (Table 5, Table 6), especially with pH (Table 10) and temperature (Table 11). In addition, any precursor, co-solvent or co-reactant employed in the medium during the synthesis of a given family of NPs could potentially influence the BS phase behaviour, thus having a critical impact on the NPs phase, morphology and size.

*Reactivity*. BS have reactive chemical groups, generally COOH. Carboxylic acids may not be neutral in the process of NPs synthesis. Whether in the COOH or $COO^-$ form, they can interact with cations, either through a charge-neutralization process or through metal-ligand coordination. pH then plays an important, multiple, role as it influences not only the biosurfactant's phase behavior but also its ionization state, hence its interaction with metal ions. This process has several concomitant, hard-to-control, consequences: new phase behavior due to the metal-BS interaction, sequestration of metal ions, hence change in the solution stoichiometry, clustering of metal ions, anisotropic phenomena, etc…

*Advises in using BS for the synthesis of NPs.*

Here, we provide some simple advises, which can lead to a reliable claim to a successful synthesis of BS-stabilized NPs. It goes without saying that the following does not need to be performed within the framework of the same study, especially in a proof of concept approach. However, a given claim should be backed up by the corresponding set of experiments.

*Type of Biosurfactants*. BS are never a 100% pure and homogeneous molecule, and in fact it would be useless to use such a degree of purity for the synthesis of NPs. However, one should always use a possibly known composition of which, this is important, the phase behavior is known, at least in water at the conditions of pH and temperature which will be employed for the synthesis of NPs. Stable micellar phases are an interesting and probably the wisest choice, although micelles are tedious to remove from water. If the phase behavior of a given biosurfactant is not known under the conditions of NP synthesis, it should be studied (please refer to section 2.3).

*Synthesis process of NPs*. Many classical synthetic process to obtain monodispersed NPs are performed in organic media, or water, above a temperature threshold which could induce the degradation of BS. These conditions should be avoided and replaced by a milder environment. If this is not possible, one should always perform a control experiment employing the synthesis conditions on the BS alone, without other precursors and prefer solution NMR over FTIR as an analytical technique to control the integrity of the BS before and after.

*Purification of NPs*. For a general synthesis of NP, removal of the ligand excess is just as important as the synthesis itself. The excess BS should then be removed at best using, or sometimes combining, classical approaches like centrifugation, dialysis or phase change processes. Efficient removal of the excess should be controlled by TGA. This step is particularly important if NPs are meant to be used for biomedical applications or antimicrobial purposes, because excess ligand could lead to false positive results.

*Characterization of NPs*.

*Size*. TEM is certainly the preferred technique, although a statistical analysis on at least 100 individual nicely-contrasted NPs should be performed. If size distribution could be double-checked with DLS, SAXS, if applicable within the size windows, should always be the preferred technique, both for its statistical relevance and information on shape and polydispersity. A note on the most suitable analytical techniques is presented in section 1.7 and Table 1.

*Core structure*. XRD (or WAXS) combined to high resolution TEM and electron diffraction are certainly the best options, better if combined together. For monodisperse systems, a quantitative analysis should correlate the NP size to the width of the XRD peak through the Scherrer formula. In the case of small NPs and wide XRD peaks, more advanced approaches, like pair distribution function (PDF) calculation, could be used.

*Shell*. Demonstration of a BS shell of a given thickness is not an easy task. Simplistic approaches based on TEM or FTIR alone should be avoided. In the first case, a single BS layer is not contrasted enough and in the second case excess BS could provide a false positive. A generally-accepted approach involves a preliminary washing step followed by TGA and FTIR. The amount of residual organic should be compatible with the theoretical number of adsorption sites, which depend on both the NP size and crystal structure. The resulting values should be compared to the literature for similar NPs and critically discussed in respect to the BS size. A specific analysis of the IR bands of functional groups should be performed to show possible anchoring of a given functional group. This approach could be coupled to DLS and/or NMR techniques, the latter employing DOSY for the spectroscopical analysis of the diffusion coefficient of colloids in solution. If the experimental environment allows it, coupling of SAXS (sensitive to the dense core) and SANS (sensitive to both core and shell nuclear contrast) should provide a direct proof of the organic shell as well as its thickness.

*Colloidal stability*. Stability studies over time should be performed in water and in any other relevant medium (e.g., high ionic strength). Several techniques can be used in this case, from DLS to UV-Vis to SAXS, etc…, as long as they show a constant and continuous signal clearly attributable to the NPs all along the time span of the study. Experts in the field also suggest to perform multiple filtering steps to show lack of aggregation and reinforced stability.[464]

## 4. Biosurfactants with polymers and biopolymers

Surfactants are generally never used as such in commercial products but they are commonly formulated with other surfactants and/or polymers. This topic is gathering more and more interest worldwide and several studies do report the interaction and mutual effects between biosurfactants and amphiphiles, surfactants[200,291,336] or lipids.[239,283,284,297,341–346] In the latter case, the research heads either towards the integration of biosurfactants in phospholipid liposomes to develop gene transfection carriers[239,297,337,343–347] or towards the study of biosurfactant-membrane interactions to better understand the detergent effect of biosurfactants.[283,284,338–342] Since this share of the literature will not be treated in this review (please, refer to last paragraph of section 2.7), we direct the interested reader towards the cited literature.

Polymers-surfactants formulations are classically found in many different products such as cosmetics, paints, detergents, food, polymer melts but also in drugs or pesticides, and their properties were quite extensively studied in terms of colloidal stability, emulsification, flocculation, structuring and suspending properties, rheology control or synergistic effect.[55] For instance, the wettability of alumina, silica or coal can be greatly improved by mixing non-ionic polymers, like PEO or PVP, and anionic surfactants, like SDS, where the latter adsorbs on the hydrophobic surface and the former improves wettability by adsorbing on the hydrophilic headgroups.[465] Complex coacervation, a specific case of polyelectrolyte surfactant complexes (PESCs), is another long-studied field, where polymer-surfactant interactions occur in bulk solutions rather than on surfaces. Examples in this field were shown to synthesize chitosan-based nanocapsules serving as potential food ingredients[466] or surfactant and lipid-based DNA transfection agents for gene therapy, and where DNA is regarded to as a negative, bio-based, polyelectrolyte integrating a biological function.[1] Many examples in this field involving petrochemical surfactants can be found in a number of dedicated studies.[1,25,467–469] However, replacement of chemical surfactants by bio-based alternatives is gaining a certain interest also in the field of polymer-surfactant interactions, often involving biopolymers for fully bio-based complex system.

Biosurfactants are characterized by a rich phase behavior (Table 5, Table 6) and possess several potential applications in many domains, however, inclusion in formulations will necessarily involve further investigation of their binding behavior to multiple colloidal species or macromolecules. Indeed, there already exist some examples highlighting antimicrobial properties,[470,471] interest in formulation of care-products[472] or stabilization of emulsions.[473] Interaction with polymers and biopolymers represents a new trend in this field with a growing interest, for many consumer products do contain polymer/surfactant formulations.

This section will thus highlight the main features concerning the behavior of biosurfactants with macromolecules: polymers, biopolymers, as well as proteins or enzymes. We will briefly introduce this section by mentioning the specific interactions occurring between polymers and surfactants. Then we will present two sets of systems: in the first one, we will present biosurfactant-polymer systems relying on mutual interactions, such as hydrophobic effect or electrostatic, whereas in the second one we will present those systems where biosurfactant and biopolymers coexist without any specific interaction.

Surfactant(s)-polymer(s) systems coexist in many commercial products with a combination of different effects, going from simple synergy to specific, e.g., electrostatic, interactions, governed by many parameters such as molecular weight, ratio, nature of the molecule. The main following trends are described by Holmberg *et al.*[1] and will be shortly summarized hereafter.

*Interactions*. Interactions between surfactants and polymers are similar in nature to surfactant-surfactant interactions. Hydrophobic interaction can occur between the surfactant's tail and hydrophobic regions of block copolymers or homopolymers. These are often the main contribution to the surfactant-polymer free energy of association. Electrostatic interactions (repulsive or attractive) are also important if both the surfactant and the polymer are charged. Entropic effects in terms of counterion release (favorable) but also reduced chain mobility (unfavorable) give an important contribution to the free energy of association. The polymer stiffness is another crucial factor to take into account. A flexible polymer with a short persistence length is more likely to adapt to a surfactant micelle than a polymer with long persistence length.

*Surface tension and CMC*: in case of attractive polymer-surfactant interactions, a polymer in solution constitutes a competitive interface with respect to water-air onto which the surfactant can adsorb on. Upon increasing the surfactant concentration in solution, surfactant associates to the polymer until a critical association concentration (CAC). As the polymer is saturated with surfactant, the surfactant unimer concentration progressively increases at the air-water interface and surface tension eventually becomes smaller until the unimer concentration reaches the CMC, after which surface tension is constant and formation of micelles is initiated. As a result, the presence of a polymer in solution has a tendency to delay the lowering of surface tension and to shift the CMC towards higher surfactant concentrations.

*Mechanisms of interaction*. Surfactants have two main modes of interactions with polymers: 1) strongly cooperative association, or binding, of the surfactant unimer to the polymer backbone; this is generally observed for polymers with hydrophobic groups. 2) Micellization on, or in the vicinity of, the polymer chain, often reported for hydrophilic (nonionic or ionic) homopolymers, and leading to the so-called "pearl-necklace model", with aggregation numbers in the same order as in the absence of a polymer.

This section, which is less abundant than previous ones due to the reduced amount of published work, presents biosurfactants-polymers systems controlled by strong interactions, mainly electrostatic and/or hydrophobic, but also hydrogen bonds or Van der Waals forces. Hydrophobic interactions directly depend on the affinity between hydrophobic regions while electrostatic interactions can be tuned by parameters such as pH or ionic strength in the presence of molecules bearing functional groups.

### 4.1 Interactions with polyelectrolytes

*Sophorolipids, glucolipids*

Polyelectrolyte-surfactant complexes (PESCs) are a wide class of colloidal systems where surfactant's self-assembly is combined to the complexation properties of polyelectrolytes, with applications in food science, tissue engineering, drug and gene delivery, cosmetics or water treatment, just to cite some.[474] The driving force in PESCs formation is mainly electrostatic attraction between oppositely-charged surfactants and polyelectrolytes, but other parameters, like the packing parameter of the surfactant, rigidity of the polyelectrolyte, charge density, ionic strength or pH, play an important role, as well. Polyelectrolyte-coated dense aggregates of spheroidal micelles can be found as a solid-liquid, or liquid-liquid, phase separation and in the latter case one refers to as complex coacervates.[475] Other morphologies can however occur, like pearl-necklace or multilamellar wall vesicles.[474,476,477]

Complex coacervation, that is coacervation between two macromolecules or between a macromolecule and a colloid, is a specific case of PESCs and it remains among the more mysterious systems in colloid chemistry and some argue that it could be at the origin of life on Earth.[478] It takes place in water and under relatively mild conditions of pH and temperature (eco-friendly process) and does not require neither a special device nor extensive production steps. An increasing interest is since devoted to their preparation for applications in food,[479] tissue engineering,[480,481] drug delivery,[482] underwater adhesives,[483,484] porous materials[485] or even water treatment[486,487] among many others. Recently, Ben Messaoud and co-workers[421] investigated the challenging question of complex coacervation involving bio-based amphiphiles and biomacromolecules for the development of green complex coacervates. Their work is focused on the complex coacervation of nonacetylated acidic SL-C18:1-*cis* (**1**) micelles with an oppositely charged cationic polyelectrolytes (*i.e.* chitosan oligosaccharide lactate, poly-L-lysine or polyallylamine). The pH sensitivity of both sophorolipids and polyelectrolytes makes pH a convenient way to tune the interaction and the domain of coacervate stability is directly determined by the pKa's of both the biosurfactant and the polyelectrolyte.

By combining turbidity titration, light and scanning electron microscopy (SEM), dynamic light scattering (DLS), cryogenic transmission electron microscopy (cryo-TEM) and Small Angle X-ray Scattering (SAXS), they conclude that coacervation occurs over a broad range of pH, from 5 to 10, as a function of the cationic polyelectrolyte type and concentration. Chitosan displays the shortest coacervation plateau, between 5 and about 7, because of its pKa of 6.5 (Figure 29a). Poly-L-lysine and polyallylamine, on the contrary, show a much broader coacervation plateau, between 5 and 10, due to the higher pKa's of these macromolecules. Figure 29b shows the typical cryo-TEM images of sophorolipids-based coacervates and the corresponding turbidity profiles divided into four regions of biosurfactant-polyelectrolyte co-existence behavior, where R3 identifies the coacervation plateau (Figure 29a). Cryo-TEM allows to visualize, without artifacts due to drying, the complex coacervates as well as their core micellar structure, a description in agreement with what has been reported for more classical ionic surfactant-polyelectrolyte systems.[475] The team eventually described the ability of bio-based complex coacervates to stabilize oil-in-water emulsions through a pickering effect.[422] This is an interesting property for applications in food or cosmetics industry, among others.

A recent work explores the possible interactions of a mixture of nonacetylated acidic(**1**)/lactonic(**3**) C18:1-*cis* mixture with medium molecular weight chitosan at pH between 6 and 7. If authors do not explore neither prove the mutual interaction nor the structure of the complex and they do not specify the nature of the SL phase (most likely micellar), they do show that the presence of SL improves the viscosity and elastic properties of chitosan in water by about a factor two in the range of shear rate and frequency explored, for a chitosan concentration of 0.5 wt% and SL content of 0% and 12%.[302]

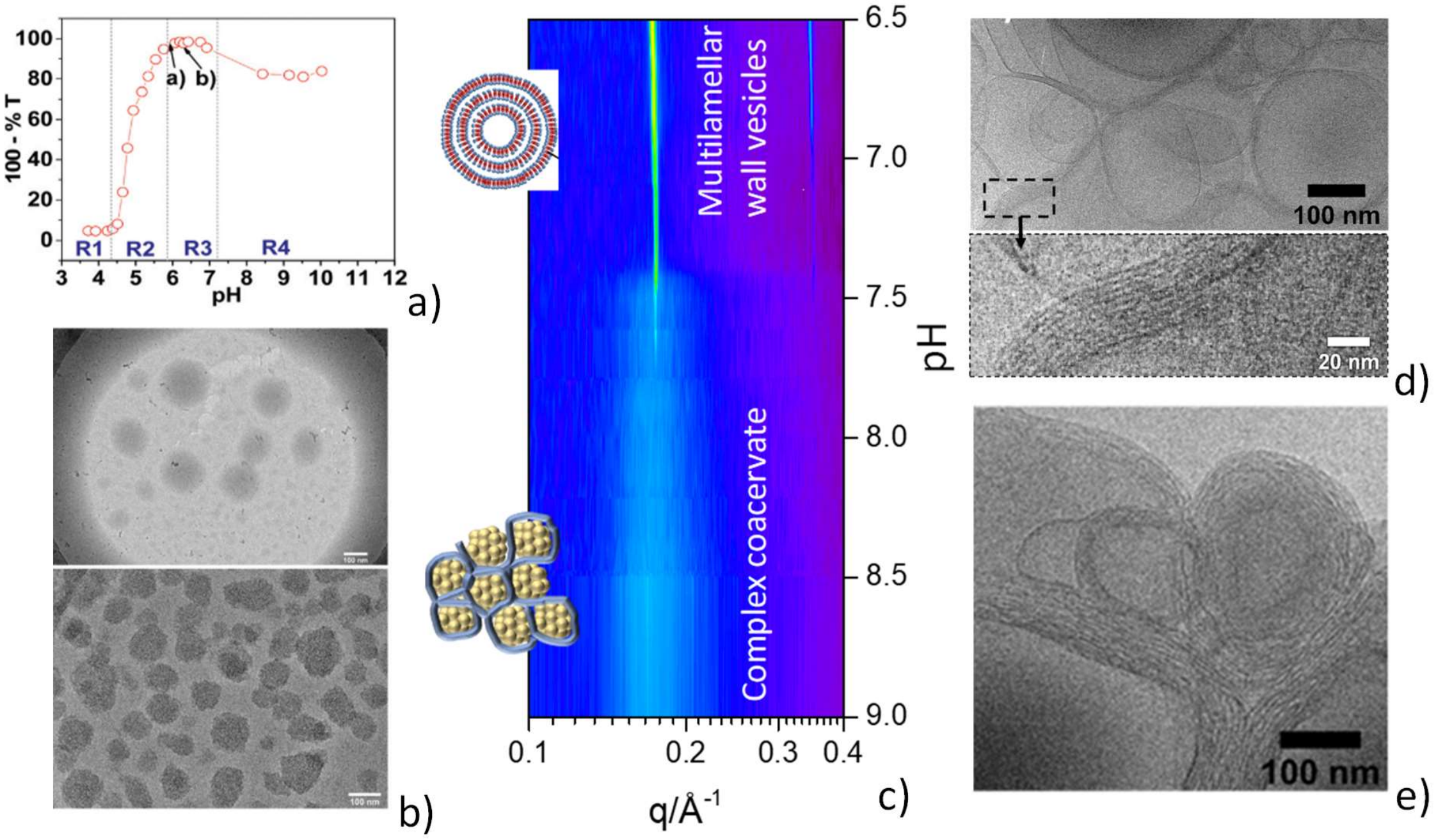

Figure 29 – a) pH-dependent turibidimetric profile and b) Cryo-TEM images (top: pH 5.94, bottom: pH 6.33) of acidic C18:1-*cis* sophorolipid(**1**)-chitosan complex coacervates (Reproduced from ref. [421] with permission of The Royal Society of Chemistry). c) pH-resolved in situ SAXS experiment showing the phase transition from complex coacervates and multilamellar wall vesicles (MLWV) for acidic C18:1-cis glucolipid (**11**) complexed with poly-L-lysine (reprinted from [488], copyright 2020, with permission from Elsevier) d-e) Cryo-TEM images of MLWV containing acidic C18:1-cis glucolipid (**11**) and d) PLL (pH 4.70) and e) PEI (pH 5.33) (adapted with permission from[489] , copyright 2020 American Chemical Society).

Complex coacervation was observed in the micellar region of the phase diagram (pH > ~7, Table 5, Table 6 and Table 10) of C18:0 sophorolipids (**13**) but also microbial C18:1 glucolipid (**11**) in combination with a number of cationic polyelectrolytes : chitosan oligosaccharide lactate, poly-L-lysine hydrobromide and polyethyleneimine hydrochloride. C18:0 sophorolipids (**13**) and C18:1-*cis* glucolipids (**11**) have a more complex phase diagram than C18:1-*cis* (**1**) sophorolipids. Both compounds undergo a pH-driven phase transition, micelle-to-fiber for sophorolipids (**13**) and micelle-to-vesicle for glucolipids (**11**) (Table 10). Since surfactant-polymer complex coacervation is generally described for surfactants having a pH-independent micellar phase diagram, the aim of this work was to demonstrate the domain of existence of a stable complex coacervate phase before the biosurfactants phase transition.

pH-resolved *in-situ* SAXS experiments employing synchrotron radiation (Figure 29c) highlighted the formation of complex coacervates only in the micellar region of the phase diagram (pH > ~7), where the charge of the micelles is negative and the charge of the polyelectrolyte positive. *In-situ* SAXS shows a broad peak at $q$= 0.174 Å$^{-1}$, corresponding to a correlation distance of 36.1 Å, a value observed in the corresponding cryo-TEM images and with a condensed network of C18:1-*cis* glucolipid (**11**) micelles. *In-situ* SAXS finally shows that the complex coacervates break apart when pH drives their self-assembly out of the micellar region of the phase diagram (Figure 29c). This was shown to occur for both biosurfactants.[489]

Following the experiment above, authors have found an unprecedented isostructural and isodimensional phase transition between complex coacervates and multilamellar wall vesicles, when the biosurfactant undergoes a micelle-to-vesicle transition below pH ~7. The strong proximity between the surfactant and polyelectrolyte in the coacevate phase was shown to lead towards a nearly quantitative formation of multilamellar wall vesicles (MLWV), as shown by optical microscopy under polarized light. The walls were shown to be constituted of alternating sandwiched layers composed of tightly packed polyelectrolyte chains and interdigitated layers of the glucolipid, as shown in Figure 29d,e.[488] On the contrary, when the C18:0 SL (**13**) undergoes a micelle-to-fiber transition, complex coacervates disassemble in favour of a mixed free polymer phase coexisting with the self-assembled fiber phase.[489] These results show how a different phase behavior of the amphiphile alone have a strong impact on the structure of colloids derived from complex coacervates.

*Rhamnolipids (mono-* (**19**)*, di-* (**18**)*)*

In a recent work, Marangon *et al*. studied the interactions between RL and chitosan, although under different experimental conditions. Instead of searching for intimate interactions between rhamnolipids and chitosan, possibly leading to complex coacervates, they designed antimicrobial rhamnolipid-rich chitosan nanoparticles (C/RL-NPs, Figure 30a). Formation of chitosan nanoparticles is driven by electrostatic interactions between chitosan and a polyphosphate (TPP), while rhamnolipids are associated to the system as an adjuvant (Figure 30b).[450] It must said that the nature of rhamnolipids (mono-, di- or mixture) nor their phase behaviour in relationship with chitosan association are clarified. Authors simply report the formation, above their CMC, of “RL nanoparticles” of undetermined structure. Considering the known RL phase behavior (*MLV* at pH below 6.8, RL paragraph in section 2.3.3, Table 8 and Table 10), it would not be outrageous to associate “RL nanoparticles” to a *MLV* phase.

Nonetheless, authors show an improved stability of chitosan nanoparticles upon inclusion of rhamnolipids; they find that the average diameter, polydispersity and zeta potential do not evolve over a period of 33 days and no precipitation occurs during 6 months of storage at 4 ºC. The particles exhibit a more positive surface charge (+60.9 ± 1.6 mV > +49.7 ± 0.3), probably due to the large presence of chitosan free amino groups on the surface of nanoparticles: chitosan bears more $NH_3^+$ groups than rhamnolipid possesses $COO^-$ groups, which results in free $NH_3^+$ groups available. Chitosan-RL NPs have shown an enhanced penetration (Figure 30c) and antimicrobial activity against *Staphylococcus* strains compared to single rhamnolipid and chitosan alone. Two parameters were taken into account in the evaluation of C/RL-NPs efficiency: their minimum inhibitory concentration (MIC) and minimum bactericidal concentration (MBC), calculated on the basis of the concentration of each individual molecule in NPs. MIC values of 14/19 µg.mL$^{-1}$ and MBC of 29/37 µg.mL$^{-1}$ were estimated for *S. aureus* DSM 1104, 29/37 µg.mL$^{-1}$ and 58/75 µg.mL$^{-1}$ against *S. aureus* ATCC 29213, and 7/9 µg.mL$^{-1}$ and 14/19 µg.mL$^{-1}$ for *S. epidermidis*. C/RL-NPs were particularly efficient in bacteria elimination according to 2D Confocal laser scanning microscopy (CLSM) images : unlike chitosan and chitosan nanoparticles, C/RL-NPs were able to eliminate more than the bacteria present in the upper parts of biofilms, to the extent that remaining viable cells could not be detected. This performance of C/RL-NPs was related to the increased concentration of chitosan and rhamnolipid at the cell surface which ensure a more efficient and targeted delivery of both compounds at the molecular scale. The design of novel nanoparticles with low cytotoxicity and huge potential for applications in pharmaceutical and food industries, among others, can thus be achieved combining chitosan and rhamnolipids.

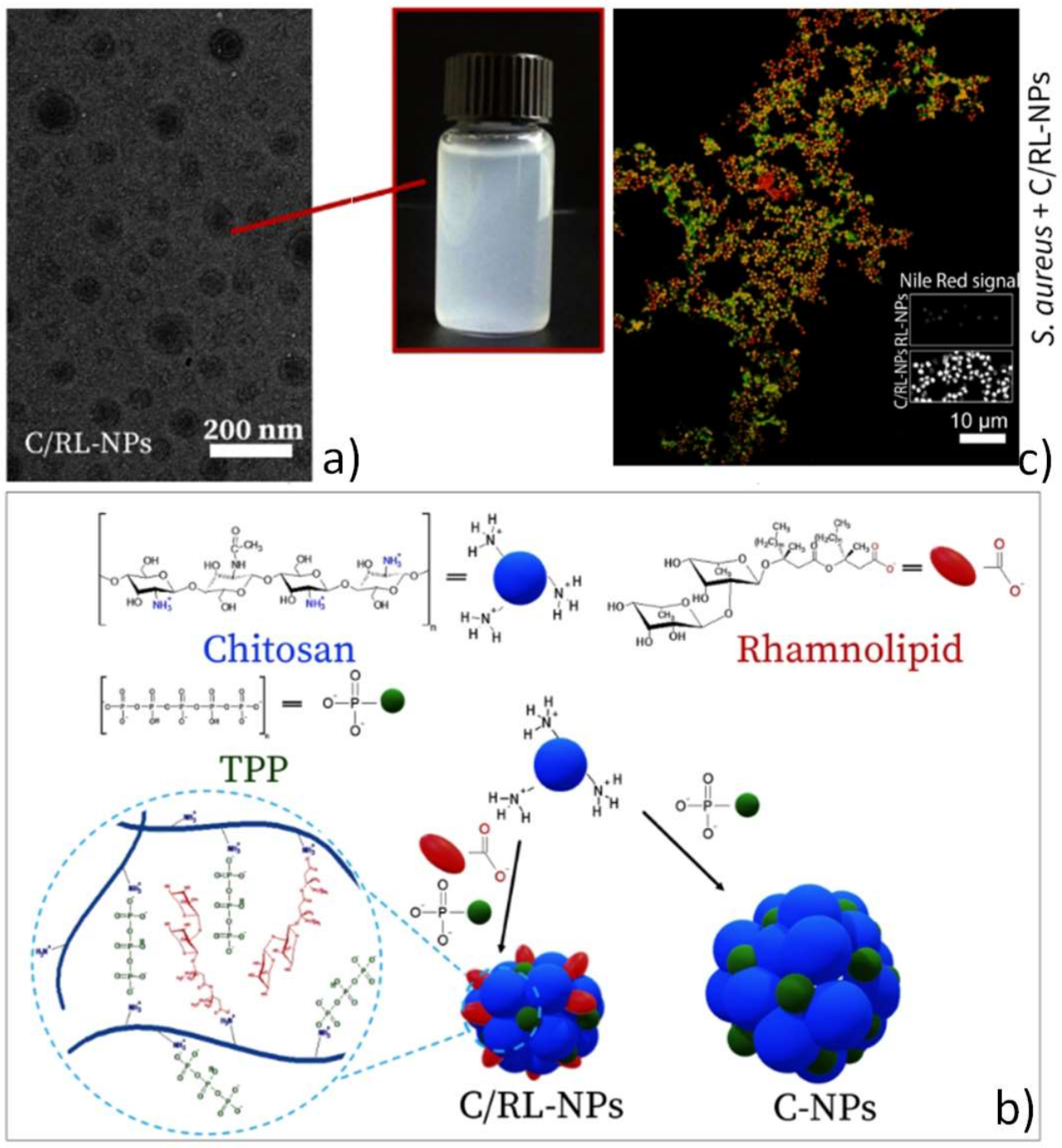

Figure 30 – a) Colloidally-stable chitosan-rhamnolipids (**18**, **19**) nanoparticles, b) their supposed structure and c) penetration into *S. aureus* (adapted with permission from [450]. Copyright 2020 American Chemical Society).

In a similar research line, and with the goal of replacing petrochemical surfactants in hair-washing formulations, Fernández-Peña *et al.*[490] recently focused on the capability of rhamnolipids (mono-RL($C_{10}$), mono-RL($C_{14}$), di-RL($C_{10}$), di-RL($C_{14}$)) to substitute sodium laureth sulfate (SLES) in binary mixtures with a cationic polyelectrolyte (PDADMAC, poly(diallyl-dimethylammonium chloride). QCM-D data showed that binary mixtures containing RL have higher *thickness* than those with SLES, and being di-RL($C_{14}$)>mono-RL($C_{10}$)>di-RL($C_{10}$)>APG. The comparison between the thickness measure by QCM-D and ellipsometry shows that the water content varies broadly from 10% to 30-80% depending on the type of RL. They concluded that for conditioning application, a balance between water content and the quantity deposited is needed, being PDADMAC-mono-RL($C_{10}$) the most appropriate with twice adsorbed molecules compared with SLES, and also increased water content.

### 4.2 Interactions with enzymes and proteins

The study of interactions between surfactants and proteins, or enzymes, is a fascinating and important topic with implications in various fields, from medical to environmental science. We address the reader to specific reviews for more information.[491,492] A number of parameters are equally important to depict a general trend, making this field highly complex. The intrinsic value of surfactant's CMC as well as its shift in the presence of proteins, the isolectric point (Ip) of proteins, and its shift in the presence of surfactants, the charged on uncharged nature of surfactants and nature of charge, they chemical nature of the charged group, the size of the protein, the solution pH and ionic strength, as well as type of counterion and so on… A somewhat general trend, valid for classical surfactants like SDS, indicates the existence of strong surfactant-protein interactions below the CMC, with possible stimulation of β-sheet formation and fibrillation. This effect seems to be triggered by both non-specific (hydrophobic effect) and specific (electrostatic) interactions. Above the CMC, on the contrary, surfactant micelles denaturate the protein, resulting in protein-decorated micelles, or bead-necklace, structures.[491,492]

When it comes to the study of biosurfactant-protein interaction, the amount of work is considerably reduced, although some have been underlined not long ago.[340] In the following, we discuss a broad set of data investigating the interactions between biosurfactants and proteins, or enzymes, and we will correlate them to the broader literature of surfactant-protein interactions. The outcome of this literature survey is outlined on Table 14.

*Rhamnolipids (mono-* (**19**)*, di-* (**18**)*)*

First studies on the effect of biosurfactants, namely RLs, on proteins stability were initiated by the group of Ortiz *et al.* as early as 2008.[493,494] Concerning RLs, they reported the thermodynamic and structural changes associated with the interaction of di-RL with bovine serum albumin (BSA),[493] a protein of which the binding affinity to small or medium sized organic hydrophobic ligands (such as fatty acids, amino acids, steroids or surfactants among the most studied) is well-known. Its low cost, stability and binding capacities in biochemical reactions make BSA a precious agent in biological assays (e.g., ELISA or western blots tests) or nutrient for cell culture. Di-RL are larger, more hydrophilic and bear a much larger polar headgroup than neutral or anionic chemical surfactants.[495] It is generally admitted that the headgroup has a strong influence on the interaction with proteins, whereas CMC and aggregation number are more sensitive to acyl chains effects. This is in agreement with isothermal titration calorimetry (ITC) and surface tension (ST) measurements, which showed that no more than two molecules of monomeric di-RL bind to a single molecule of BSA with an affinity of 2 x $10^5$ $M^{-1}$, a value being in the range of affinity constants reported for other surfactants (anionic SDS, cationic CTAB and nonionic TX-100).[496]

These data show the strong interaction between BSA and either monomer or micellar di-RL. The $\Delta H$ of the interaction in the case of di-RL (-4 Kcal $mol^{-1}$) is well below the one measured for classical anionic surfactants (70 Kcal $mol^{-1}$), and for this reason a different interaction is expected between di-RL and BSA. Despite di-RL are expected to be negatively charged at pH 7.4 used for experiments regarding its pKa (5.6), its affinity to BSA was rather similar to the one found in BSA complexes with nonionic surfactants.[496] However, the interaction was sufficiently strong, according to the affinity constant, to significantly stabilize the protein against thermal denaturation, as shown by an increase of the temperature of the thermal unfolding of BSA with dirhamnolipid (between 70 and 75°C compared to 60°C for native BSA) employing differential scanning calorimetry (DSC). FTIR further highlighted a low impact of the biosurfactant on the structure of the native protein, since most of the secondary structure remained unaffected upon interaction with the biosurfactant according to the evaluation of the amide I′ band above 1600 $cm^{-1}$,[497–500] and thermal protection of the protein over more than 10°C, followed by changes in the protein. The infrared data assess thus that the major structural changes in BSA are induced by dirhamnolipid and temperature, and further information concerning the relationships between these various temperature-induced conformational changes were obtained by 2D-correlation analysis of the FTIR spectra.

A major contribution to the field of RLs-protein interactions eventually came from the group of Otzen *et al.* from 2014 onward.[501,502] In the first quantitative study, authors studied the interactions between RL (1:0.35 mono-RL(**19**): di-RL(**18**)) and two model proteins, α-lactalbumin (αLA) and myoglobin (Mb) in buffer at pH 7.[501] RLs denature αLA below the CMC and Mb above the CMC by increasing α-helicity. Denaturation mechanism is similar to the one of anionic (e.g., SDS) and non-ionic (e.g., alkyglycosides) surfactants:RL can denature below the CMC, like SDS, but with slower kinetics, as found for APGs. Authors explain such difference in terms of the lower CMC of RL with respect to SDS, presence of carboxylate group, having less affinity towards proteins than sulfate group, and branching of RL backbone chain, having less affinity towards the protein than against itself. The number of RL per protein is about 29 against αLA and 41 against Mb. These systems were further studied by SAXS,[231] which has confirmed the formation of protein-RL complexes, whereas the protein occupies the hydrophilic shell of RL micelles (prolate ellipsoids of revolution). Authors found that a single RL micelle of about 12 molecules is associated to αLA, while each Mb is associated to a RL micellar dimer of about 10 RL molecules each. These data revealed much smaller RL-protein structures compared to SDS-protein systems, composed of not less than 40 SDS molecules per molecule. SAXS data eventually confirm the milder interactions between RL and the proteins under study.

Interestingly, the effect of RL (**18**,**19**) is unexpected on human lysozyme (HLZ),[503] a defensive protein produced by the human immune system. RL is found to bind lysozyme at a much lower extent than αLA and Mb, with only ~8 RL molecules per HLZ and authors find no direct denaturation effect. However, binding of RLs probably exposes parts of HLZ, which undergo faster degradation through proteolysis attack by bacterial and human peptidases.

Of particular interest, the effect of biosurfactants on fibrillating proteins. α-synuclein (αSN) is a protein associated to Parkinson's disease. Natively unfolded, its structure, and in particular fibrillation, can be accelerated by interactions with anionic lipid vesicles or SDS monomers. It was found that αSN can also be accelerated by interactions with RLs (**18**,**19**).[504] Interestingly, the main difference between SDS and RLs in terms of promoting faster αSN fibrillation should be found in the amount of surfactant with respect to its CMC: SDS promotes fibrillation and denaturation of this and other proteins respectively below and above its CMC.[492] On the contrary, RLs only promote fibrillation above their CMC. RLs-induced fibrils have a wormlike structure at the moment of their formation but they evolve towards linear fibrils, with similar structure as to those found with SDS. The difference between SDS and RL are coherent with the weaker binding of RLs to proteins and enzymes compared to SDS and commented above. Similar results were actually obtained on an amyloidogenic pharmaceutical peptide analogue of glucagon, where several surfactants were all tested above their respective CMC: both RL and SL induced faster fibrillation, while SDS suppress aggregation by inducing an α-helix structure.[505]

An additional set of data produced by the same group has confirmed faster fibrillation of the protein FapC induced by super-CMC concentrations of RLs.[506] Again, fibrillation was induced only below CMC when SDS is employed. Similarly, ITC could not detect specific interactions between RLs and FapC at an appreciable amount. However, SAXS could show that at RLs concentrations below 10 mM, fibrils could be described as cylinders with an ellipsoidal cross-section. Above 10 mM, this model did not held anymore and fibrils could only be described by a core-shell structures with a core radius of 6 Å and a shell of 24.1 Å. This suggests, as also described above and for SDS-protein systems,[491] presence of FapC in the hydrophilic shell of RLs. Linear combinations of the RLs and FapC signals alone could not fit the SAXS data above 10 mM. For comparison, one should keep in mind the estimated shell thickness of RL in their micelle (3 to 12 Å) and vesicle (6 Å) morphologies, as summarized in Table 7 and Table 8. It is then reasonable to consider that the excess thickness could be attributed to a protein layer.

Interestingly, other groups reported a similar trend within the framework of RLs (not clearly identified) and soy protein interactions.[424] Authors found a binding ratio of RL/protein of about 28, a value comparable to αLA[501] rather than to BSA (see Table 14)[493] and, above all, a slight increase in the content of β-sheets upon increasing the RL concentration, whereas β-sheets are associated to crystalline domains.

The group of Otzen has also studied the interactions between RLs and enzymes. The impact of RL (1:0.35 mono-RL: di-RL) against three industrially-relevant enzymes at (cellulase

Carezyme® (CZ), the phospholipase Lecitase Ultra® (LT) and the α-amylase Stainzyme® (SZ)) was studied at pH 8 and compared to SDS.[507] Authors find that RL display little, or no, binding against all enzymes, except CZ, for which about 11 RL per enzyme could be measured. RL did not affect the structural integrity nor the activity of any enzyme, differently than SDS, known, and found, to interact with proteins and enzyme in a much stronger manner. RL (1:0.35 mono-RL: di-RL) and SL (acidic(**1**):lactone(**3**) C18:1-*cis*) interactions with *T. lanuginosus* lipase (TIL) enzyme were also explored by a number of techniques.[508] Similarly as to CZ, LT and SZ enzymes, both biosurfactants show little interactions with lipase, especially at pH 8, with no impact on lipase's structure and activity. TlL's melting temperature (71.6°C) is only slightly lowered by less than 2°C in RL and SL solutions below 10 mM and activity is increased to about 150% only above the CMC of RL and SL, compared to buffer (100%). Below the CMC, the activity towards 4-methylumbelliferone butyrate (4-MUB) is reduced below 50% for both biosurfactants. However, lipase stability and activity was improved at pH 6. Thermodynamic and structural studies by ITC have shown that between 5 and 10 biosurfactant (RL or SL) molecules bind to lipase, suggesting the formation of lipase dimers (Figure 31). SEC and SAXS demonstrate the presence of elongated complexes (Figure 31), confirming the binding of low amounts of biosurfactant adsorbing onto two enzymatic units and stabilizing a dimeric complex, in agreement with previous literature studies involving SDS.[509] TIL's thermal stability and activity are multiplied up to factors of about 7 for biosurfactant concentration in the order of 2-4 mM. In terms of mechanism, authors have shown an interfacial activation of TlL by the biosurfactants, revealed by an enhanced blue shift of the positively charged triptophane residues at pH 6 upon crossing the CMC and suggesting an increasingly hydrophobic environment around the triptophane residues.

Finally, keratin–surfactant interactions may be indicative for skin irritation by surfactants. For this reason, Özdemir *et al.*[510] were interested in investigating the keratin-rhamnolipid interactions at pH 6.2 and 5.0. They supposed a direct, electrostatic-mediated interaction, between keratin and RL on the basis of the evolution of surface tension of the complex solution with respect to the individual components. Nonetheless, their study does not provide specific structural information on the nature of the keratin-RL complex.

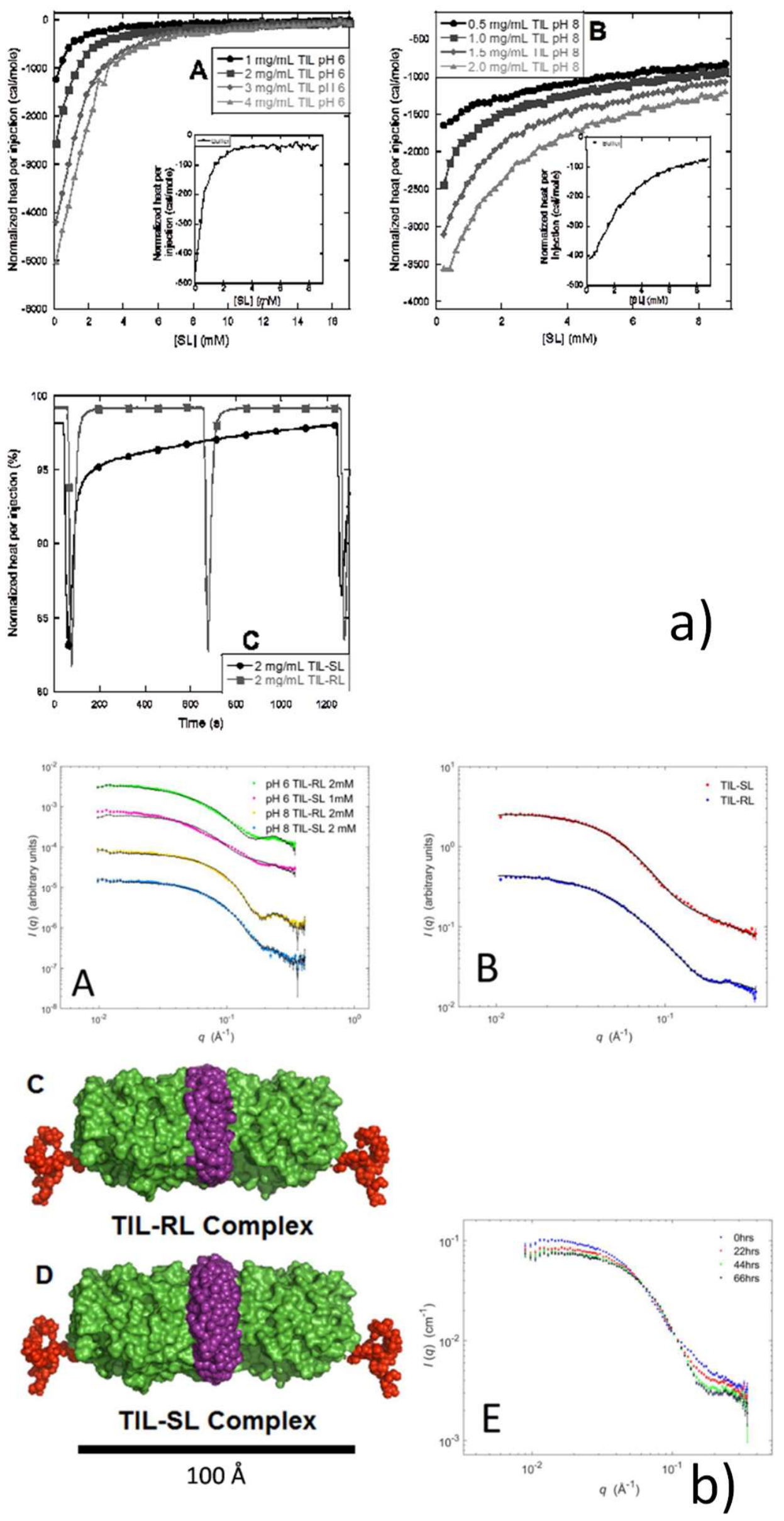

Figure 31 – pH-dependent a) ITC and b) SAXS experiments respectively demonstrating the interaction between and shape of lipase from *Thermomyces lanuginosus* (TIL) in association with sophorolipids (**1**) or rhamnolipids (**18**, **19**) (adapted with permission from [508]. Copyright 2017 American Chemical Society)

*Sophorolipids (**1-3**)*

Otzen *et al*. have studied, in parallel to RLs, the interactions between SLs and proteins and enzymes. In general, the observations are very similar for both biosurfactants, with minor differences between them. When nonacetylated acidic C18:1-cis SL (**1**) are used instead of RLs in interaction with apo-aLA,[511] authors find a similar stoichiometry (at saturation) of 29 SL (**1**) molecules per apo-aLA. SLs have an even gentler interaction with apo-aLA than RLs. Globally, acidic SL start denaturating apo-aLA with slow kinetics compared to, e.g., SDS, and only in the proximity of their CMC, while RLs show some interactions below the CMC. SLs seem to show some interactions below their CMC, but only above a certain apo-aLA threshold. As far as interactions with enzymes are concerned, SL have been studied in the presence of TIL.[508] The activity, interactions and structural data are very similar to those found for RLs, reported in detail above and in Figure 31.

Silk fibroin (SF) extracted from *Bombix mori* is a protein with well-known hydrogelation properties.[512–516] However, the hydrogelation process is quite long (from 10–16 hours below the isoelectric pH to a few days or even weeks slightly above the isoelectric pH), thus limiting its clinical applications.[517] To overcome this problem, cationic, anionic and non-ionic surfactants were found to speed up the gelation process.[518,519] However, chemical and non-biodegradable additives are not appropriate to form biologically relevant hydrogels. In this regard, Dubey and co-workers explored the possibility to use nonacetylated C18:1-*cis* sophorolipids (nonacetylated acidic, ASL (**1**), diacetylated lactonic, LSL (**3**), or 1:3 mixture of them, MSL) as SDS analogues and used them as initiators for the hydrogelation of SF.[300,301,303] Throughout their work, they show an improvement of gelation times from the order of weeks to hours, in agreement with the behaviour of other surfactants, like SDS,[518,519] which can reduce the gelation time to less than the hour. While pure silk fibroin (SF) does not gel at physiological pH during the whole experiment time (more than 20 hours), a gel is obtained in only 0.8 h upon addition of LSL (**3**) micelles. An intermediate time of 2.3 h is found using mixed LSL and ASL micelles and a longer induction time of 10 h for the case of ASL (**1**) micelles (Figure 32b). To understand these important differences in gelation times using different forms of SL, authors combined SANS with NMR experiments.

From a mechanistic point of view, the hydrophobic domains of SF have a preferential interaction with the more hydrophobic LSL (**3**), even in the mixed ASL-LSL (**1,3**) system, thus driving 3D conformational changes in SF and inducing faster gelation. An increasing amount of crystalline β-sheets, behaving as physical cross-linking regions, with respect to random coil conformation, is observed. This result, and the role of the hydrophobic effect between surfactants and proteins, is in agreement with what is known about the interaction between the SDS and SF,[518] but also between surfactants and proteins in general, including fibrillating proteins.[491,492] However, some discrepancies should be noted between the SANS data published in [303] and literature: authors propose a micellar environment for LSL (**3**) alone at 3 w/v% while the SANS data recorded by Penfold *et al*.[200] for the same compound at comparable concentration (30 mM, about 2 wt%) show a strong interface scattering and a broad peak above 0.1 $Å^{-1}$, but not micellar environment, as summarized in Table 7 (no micelles for LSL (**3**)) and Table 8 (vesicle found for LSL (**3**)). In the presence of SF, they model the LSL-SF SANS data with a pearl-necklace form factor, certainly applied for polymer-surfactant systems on the basis of SAXS arguments,[491,492] but probably slightly too ambitious for SANS data presented in this work and lacking any oscillation of the form factor.

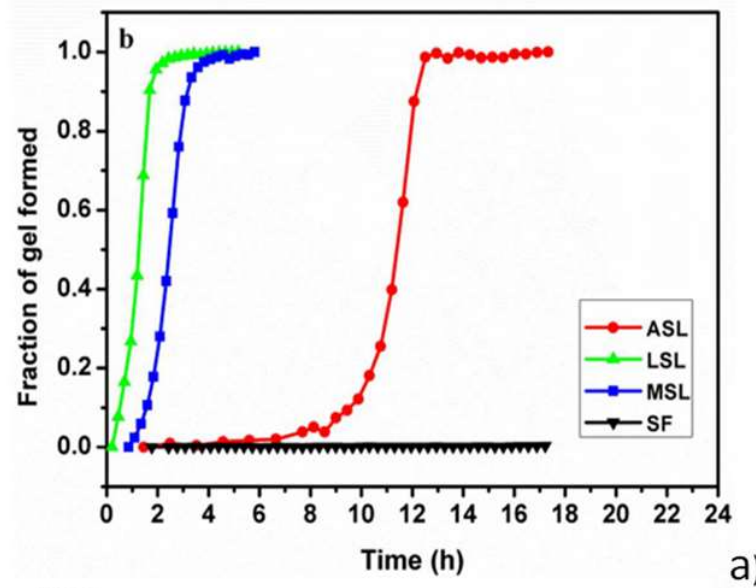

a)

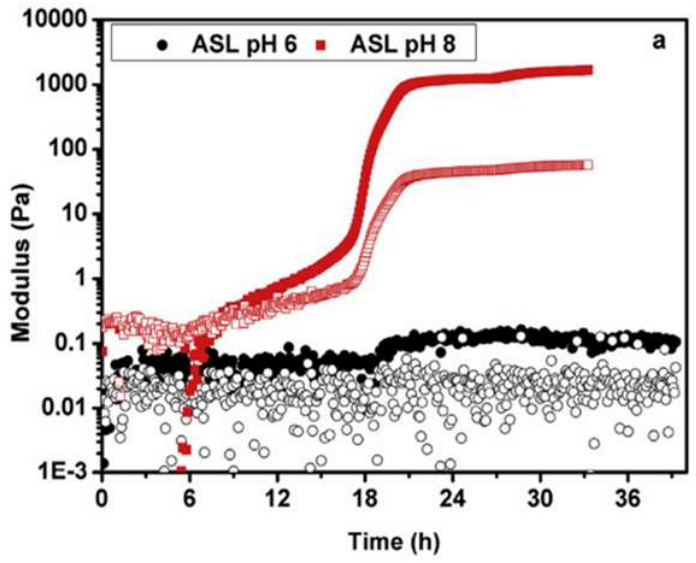

b)

Figure 32 – Formation of silk fibroin gel with time in the presence of : a) acidic SL (ASL) (**1**), lactonic SL (LSL) (**2**) and mixture of ASL and LSL (**1**, **3**) (adapted from [301], copyright 2018, with permission from Elsevier) and b) ASL (**1**) at pH 6 and pH 8. Lactonic SL (**3**) and ASL (**1**) at basic pH accelerate gelation of fibroin (adapted with permission from [303], copyright 2016 American Chemical Society). All SL are C18:1-*cis* and ASL are in the nonacetylated acidic form while LSL are in their acetylated form.

Investigation of SL-SF systems by multiple experimental techniques also reveals a pH-dependent gelation effect, where gelation is faster at higher pH (Figure 32b).[301] At pH 6, no gelation was observed for SF + ASL(**1**) system, and gelation was induced after 10 h for SF + MSL (**1**,**3**), suppositively due to non-specific interactions with LSL (**3**). On the contrary, at pH 8, a slight increase in storage modulus is observed after 10 h for SF + ASL(**1**) system, before a more pronounced increase after 18 h. The SF molecule has an isoelectric point at pH=3.9,[520] thus bearing a global negative charge at high pH values, where acidic sophorolipids are also negatively charged due to the carboxylate residue. In the sol state, SF + ASL(**1**) SANS show a micellar environment, while in the gel state, SANS data show a strong scattering with an approximate -1 q-dependence of the intensity. Authors have employed a pearl-necklace model to fit and consequently interpret these data. However, it is unclear, whether or not, the SF + ASL(**1**) signal in the gel state can be simply attributed to SF gelation, possibly without any contribution of ASL(**1**) micelles.

Concerning the faster gelation when ASL (**1**) are used in their anionic form at high pH, one should not be surprised, as this result is in agreement with the literature of both SF-surfactant and protein-surfactant systems. The main hypothesis concerns the hydrophobic effect, generally observed with ITC, between surfactants and proteins and found to be responsible for faster β-sheet formation[492] but also gelling.[518,519,521] Analogue data are reported on several RLs systems discussed above.[424,493,501,504,506] The hydrophobic effect, which generally occurs below CMC for SDS,[491,492] but above CMC for biosurfactants,[340,424,493,501,504,506] together with the important role of electrostatic repulsion between the surfactants head groups drive the formation of clusters, from which $\beta$-sheets eventually nucleate.

These works shows some interesting effect of sophorolipids on the gelation time of SF, showing that it could potentially replace other gelation triggers. In this regards, the same authors have shown that freeze-dried SF-SL scaffold can be safely employed as scaffold for fibroblast cells proliferation.[300] However, if the effect of SL on the gelation time of SF is out of doubt, it would be interesting to know the impact that SL have on the properties (mechanical, adhesion) of SF gels and dried scaffolds. Since authors do not provide data on SL-free SF gels, it is impossible to state whether SL improves or lowers both the elastic and cell adhesion properties of the SF materials.

A recent structural SANS study has been performed on a wide q-range ($10^{-3}$ < q / $Å^{-1}$ < 0.4) by Lassenberger *et al*.[522] on silk fibroin hydrogels mixed with either an acidic C18:1-*cis* (**1**) micellar phase or with a C18:0 (**13**) fibrillar hydrogel[257] (Table 12) above the respective CMC and critical aggregation concentrations. First of all, the faster gelation process of SF in the presence of SL is confirmed. However, they show that the SANS signal of the mixed SF-SL, whichever the SL phase, is always a linear combination of the SF signal (in the gel state) and SL signal (micelle or fibrillar gel) in the high-q and mid-q region. The only appreciable differences occur in the low-q region, which suggests a more compact folding of SF and a denser network structure in presence of both SL. It is curious to note that both SF and SL structures are preserved, forming an intercalating network. These findings seem to invalidate the pearl-necklace structure of SF-SL mixed hydrogels proposed earlier.[301] These data also seem to be in contrast with previously-cited work,[340,424,493,501,504,506] suggesting specific interactions between BS and proteins. The poor supposed interactions between either micells of (**1**), or fibers of (**13**), and SF found in Ref. [522] could probably be explained by a mere concentration-dependent effect, as shown in the case of RL-induced fibrillation of FapC,[506] discussed above in the RLs paragraph.

Another interesting phenomenon should be noted in [522]. In the SF gel mixed with a SL fibrous hydrogel composed of (**13**), it was possible to show that triggering a fiber-to-micelle transition from acidic to basic pH, the entire hydrogel loses its mechanical properties, while this transition does not occur in the pure SF gel. This phenomenon, which certainly needs to be repeated and studied further, suggests that: 1) pH-responsive SF hydrogels can be prepared and 2) SL micelles may not play such an important role as initially though in the fibroin gelation process.

In the field of BS-protein interactions, Tian *et al*.[523] reported a study involving the interaction between lactonic sophoro (LSL) (**3**) and di-rhamnolipids (**18**) with casein at the air/liquid interface using interfacial dilatational elasticity. The results show that the dilational elasticity for SL/casein initially increases and then decreases with sophorolipid concentration, while the opposite occurs with RL. At LSL < 100 µmol/L, the formation of LSL/casein complexes increased the elasticity, but when increasing this concentration, LSL (**3**)/casein complexes and free LSL (**3**) adsorbed competitively at the interface. The case of the two biosurfactants did show weak interactions with casein compared with the other cationic and anionic gemini surfactant surfactants studied in the work.

*Surfactin* (**20**)

Interactions between surfactin and BSA were studied by Zou *et al*. in buffer at pH 7.2.[524] Authors show a number of data which corroborate previous works studying the interactions between biosurfactants and proteins. In particular, the CMC of surfactin increases from 0.15 µM to 0.33 µM, the number of surfactin molecules per BSA is evaluated between 7.5 and 17.6, according to the BSA concentration in solution and, finally, the structure of BSA is only affected by surfactin above its CMC and in a gentle way, as shown by the moderate evolution of circular dichroism data.

Amyloid β-peptide (Aβ(1-40)) has a well-known aggregation behavior into fibrils, identified as a key pathological process at the heart of Alzheimer disease. The micellization process of biosurfactant surfactin and its influence on the aggregation behavior of Aβ(1-40) were investigated and found efficient to stimulate fibrillation in the vicinity of surfactin's CMC and prevent it above CMC.[525] AFM experiments, in agreement with ITC and circular dichroism experiments, revealed a “bell”-type effect of the concentration of surfactin: Aβ(1-40) (40 µM) arranges into globular particles of 1 to 1.5 nm diameter, as suggested by the CD spectra single minimum signal at ∼195 nm, characteristic of random coils.[526] Upon addition of a low amount of surfactin (0.01 mM), a few long fibrils that are 2.5 µm in length and 2 nm in diameter form, as shown by the shift of the positive peak at ∼201 nm and a minimum peak at ∼231 nm in the CD spectra compared to the typical β-sheet spectrum. In the presence of additional surfactin (0.1 mM), the β-sheet conformation of Aβ(1-40) is more pronounced and fibrillation is enhanced. Finally, at 0.5 mM surfactin, fibrillation is again inhibited and large amount of globular aggregates with diameters of 2.5-5 nm appear. The following mechanism is proposed: unfolding of Aβ(1-40) first results from electrostatic binding of surfactin onto Aβ(1-40) below its CMC, then hydrophobic contributions between bound surfactin monomers favors the fibrillation process, which is even reinforced by further hydrogen bonds between adjacent biosurfactant micelles and peptide chains. However, β-sheets and fibrils formation is finally inhibited by an excess of surfactin micelles and their aggregation with the separated complexes of Aβ(1-40). Interestingly, these data are in contrast with the trend depicted found for SL and RL, where fibrillation is favoured above their cmc, as discussed in the previous paragraph.[340,424,493,501,504,506]

Taira *et al*.[527] studied the effect of pH on the conformation of cyclic surfactin and its consequences on the interaction with subtilisin enzyme, an alkaline serine protease produced by many *Bacillus* species and being the most common detergent enzyme found in all major brands of European, American, and Japanese detergents. Circular dichroism experiments suggest that surfactin shows a major β-sheet conformation at low pH (4.2) although coexisting with other conformations. Upon increase in pH (6.2 < pH < 12.9), a strong negative band appears at 224 nm, indicating major conformational changes upon deprotonation of its carboxylic acids, possibly corresponding to a surfactin phase change (Table 10). Secondly, the possible interaction between surfactin and subtilisin was indirectly evaluated through the hydrolysis activity of subtilisin towards a model substrate (succinyl-Ala-Ala-Pro-Phe-p-nitroanilide (suc-AAPF-pNA)). Authors have shown that the hydrolysis kinetics of suc-AAPF-pNA is only affected by the presence of surfactin two orders of magnitude beyond the surfactin’s CMC. In comparison, the kinetics is affected at concentrations several orders of magnitude below the CMC of two model surfactants (SDS and polyoxyethylene alkyl ether).

*Mannosylerythritol lipids* (**21**)

In the study of interactions between BS and enzymes and proteins, Fan *et al*. reported several studies probing the influence of MEL-A (**21A**) on the structure and properties of β-glucosidase and β-lactoglobulin (β-lg).[528,529] They show that MEL-A (**21A**) modifies the secondary structure and physicochemical properties of β-glucosidase,[528] of which the activity is enhanced at MEL-A (**21A**) concentration in the vicinity or lower than the CMC (~20.0 µM). In the meanwhile, its activity is inhibited at higher concentrations. In addition, the midpoint temperature of β-glucosidase goes from ~74 to ~77°C according to DSC, CD and ITC data, thus indicating that MEL-A increases its thermal stability. In terms of structural data, authors find a larger proportion of β-sheet conformation compared to α -helix around the CMC of MEL-A (**21A**). When studied in interaction with β-lactoglobulin,[529] MEL-A (**21A**) contributed to a better thermal stability of the protein. They suppose that β-lactoglobulin forms an aggregated particle stabilized by an external MEL-A (**21A**) shell. The complex structure is shown to have improved interfacial and emulsifying properties if compared to MEL-A (**21A**) and β-lactoglobulin alone. Interactions were quantified by ITC, which shows a negative enthalpy and positive entropy contributions, demonstrating binding of MEL-A (**21A**) to the protein in a ratio of approximately 3-to-1 and suggesting a combination of a specific (supposed hydrogen bonding) and non-specific (hydrophobic effect) interactions.

Some years ago, Kitamoto and co-workers reported the spontaneous formation of giant MEL-based vesicles at pH 7.0 and 25 °C.[189] The vesicle surface is reasonably rich in the mannopyranoside residues, the sugar constituting the carbohydrate part of MELs, and to which mannose-binding proteins like concanavalin A (ConA) could be expected to bind easily. FITC-labelled succinyl ConA was dissolved in a phosphate buffer pH 7.0 and allowed to interact with MEL vesicles. Fluorescence microscopy shows an external fluorescent ring-like signal, taken as a proof of the ConA – mannose interaction, on those vesicles composed only by MEL-B and MEL-C, two congeners of MELs. Interestingly, both MEL-B (**21B**) and MEL-C (**21C**) bear a free OH group at C-4' or C-6' position of mannose and able to efficiently interact with Con A, according to its well-known preferential binding sites.[530] This result was in agreement with the hypothesis of a specific binding behavior on the vesicular surface reminding multidentate interactions. Interestingly, neither aggregation nor precipitation occurred in MEL-B (**21B**) and MEL-C (**21C**) – ConA assemblies. According to the authors, this can be explained by the large surface area of the vesicles compared to the receptor, ensuring the binding of the lectin to only one vesicle at the same time, although this hypothesis was not proven directly. These findings on MELs giant vesicles enhance these microbial glycolipids interest in studying surface recognition processes. Interestingly, MEL-B (**21B**) and MEL-C (**21C**) did not display any affinity towards ConA when they are in a SAM configuration, differently than MEL-A (**21A**).[378]

If the discrepancy in terms of binding affinity towards ConA between MEL-A (**21A**) vesicles (poor) and MEL-A SAMs (good) will certainly require further work to be understood, one can still formulate some hypotheses based on the literature data. Ito and co-workers[376] stress on the fact that number and position of the acetyl groups on the sugar moiety is the only structural difference between MELs. They thus attribute a key role to the acetyl on the sugar moiety in the binding between the biosurfactant monolayer and HIgG and predict a drastic decrease of the binding when only one acetyl is missing on C-4' or C-6' position (MELs B (**21B**) and C (**21C**)). In the meantime, Imura *et al.*[375] focused on the importance of the unsaturated fatty acid content and temperature, both determinant to control packing density and thus affinity, but it cannot explain differences in affinity observed for MELs with the same unsaturated fatty acid content and which were studied at the same temperature. However, a trail according to which packing density and orientation of lipids differ in monolayers and vesicles deserves attention.

According to Kitamoto *et al.*,[19] it is not surprising that structural differences lead to distinct self-assembly behaviors. Indeed, MEL-B (**21B**) and C (**21C**), both bearing two alkyl chains at C-2 and C-3 position, spontaneously form giant unilamellar vesicles of diameter larger than 10 μm[188], while MEL-A (**21A**) self-assembles into "sponge ($L_3$ or coacervate) phase" (Table 5 and Table 6).[188,279] The importance of the erythritol moiety configuration was reinforced by an interesting study of Fukuoka *et al*.:[225] they concluded that differences in MEL carbohydrate configurations significantly affect interfacial properties, self-assembly, hydrate ability and phase boundaries, after preparation and analysis of MELs diastereoisomers. In support of these allegations, one can find other cases in the literature where molecular conformation affects the physical properties of the assembled structure. SNAP-tag fusion proteins binds to β-cyclodextrin monolayers or vesicles with different affinity constants, respectively of $1.6 \times 10^{6}$ $M^{-1}$ and $1.5 \times 10^{7}$ $M^{-1}$.[531] Wettability, antiadhesive and antimicrobial properties sensitively vary for C18:1-*cis* and C18:0 nonacetylated sophorolipid monolayers.[383,385,386]

After this overview of the potential use of MELs in molecular biology, we highlight a more recent applicative example involving enzymes but in the field of environmental science, where enzymatic reactions are responsible for the degradation of biodegradable plastics (BPs) in a natural environment. Extensively used in farms, BP mulch films degrade in an uncontrolled way. To this regard, Fukuoka and co-workers have shown that preventive adsorption of MELs on poly(l-lactic acid), PLLA, modulates its degradation by PaE, a cutinase-like enzyme.[532] Indeed, adsorption of PaE to the non-treated PLLA surface was easily appreciable (341 Response Unit (RU) at 420 s after injection), whereas PaE could not adsorb onto the MEL-covered PLLA surface (37 RU at 420 s) thus inhibiting PLLA degradation. Similarly, MEL treatment on polybutylene succinate (PBS) surfaces allows the inhibition of PaE-catalyzed degradation by 1.7-fold. Authors explain that adsorbed MELs prevent the contact between PaE and BP films, thus inhibiting, and retarding, the degradation reaction by mean of an eco-friendly biosurfactant.

*Trehalolipids*

Interactions between bacterial trehalose lipid (TL) biosurfactant produced by *Rhodococcus sp.* with BSA and Cyt-*c* are reported by Ortiz *et al.*[494] They evidenced a similar behavior of TL towards BSA compared to RL,[493] but a completely different scenario was reported toward Cyt-*c*, which undergoes denaturation. This was attributed to the structures of both proteins and the interaction of TL with BSA, a α-helix rich protein, was supposed to prevent thermal unfolding, whereas interaction with Cyt-*c*, which contains more β-sheets, favors thermal unfolding.

*General comments*

Table 14 summarizes the main surfactant/protein(enzyme) systems studied in the literature and the corresponding biosurfactant molecule per protein(enzyme) ratio, generally measured in buffer by ITC titration. Some considerations are listed below:

- The non-specific hydrophobic effect is considered as the driving force for biosurfactant-protein interactions. This is in agreement with the body of work on general protein-surfactant interactions.[340,491,492] However, other specific interactions, which depend on the specific nature of the protein, biosurfactant but also pH and ionic strength, contribute to the protein stability, activity, binding affinity and molecules-per-protein.
- Biosurfactants have milder interactions with proteins, probably due to the less "aggressive" carboxylate group, which has less affinity that classical sulfate groups for oppositively-charged binding sites on the protein. However, the particular structure of biosurfactants (bolaform or branched) as well as the lower charge density distribution also contribute to reduce the strength of the interactions.
- The affinity of biosurfactants towards proteins is generally observed around and above the corresponding CMC, rather than in their monomeric state below the CMC, as classically found for SDS. ITC data recorded on several systems converge in showing negligible interactions below the biosurfactant's CMC.
- Following the above, the proteins' structure is generally not affected by low amounts of biosurfactants. However, at higher biosurfactant concentrations, proteins can integrate the biosurfactant's hydrophilic shell, provoking important structural changes, namely the increase of β-sheets.
- In this regard, fibrillation of specific proteins (e.g., α-synuclein, FapC or silk fibroin) can be induced and accelerated by the presence of biosurfactants above their CMC. This is in contrast with what it is known for SDS-protein interactions, where fibrillation is generally induced below the CMC of SDS, which eventually denaturate the protein above its CMC.
- It could be tempting to depict a trend from Table 14, for exemple as a function of the number of aminoacids-per-protein, AmA, in fact corresponding to the protein size. Such an exercise could suggest a trend in terms of BS-to-protein, according to which less biosurfactant molecules bind to large proteins, with the exception of soy protein. However, one should be cautious, because the nature of the proteins is different (both globular and fibrous proteins have been repeated in Table 14) as well as their isoelectric point. We then prefer to leave the reader interpret the data at one own's need.

Table 14 – Interaction studies between biosurfactants (N° correspond to Figure 7) and proteins or enzymes. Symbols: *= biosurfactant-to-protein(enzyme) ratio. This value is extracted from the given reference and measured by ITC at saturation (generally corresponding to the last inflection point). #: The number of aminoacids-per-protein, AmA, are taken or estimated from the literature. We used the following values of AmA: BSA= 583, αLa= apo-aLA= 123, Mb= 154, CZ= 256, HLZ= 148, TIL= 269, Soy protein= 4090 (estimated as an average value from Mw= 300-600 kDa), β-glu= 1227 (estimated from Mw= 135 kDa, Aldrich value), β-lg= 162.

| BS | Type | N° | Protein or enzyme | pH (buffer) | BS/protein(enzyme)* | Aminoacid-to-biosurfactant ratio# | Ref. |
|---|---|---|---|---|---|---|---|
| RL | Di- | 18 | Bovine serum albumin (BSA) | 7.4 | 1.3 ± 0.3 | 448 | 493 |
| RL | Di-:Mono- (0.35:1) | 18,19 | α-lactalbumin (αLA) | 7 | 29.0 ± 1.0 | 4 | 501 |
| RL | Di-:Mono- (0.35:1) | 18,19 | myoglobin (Mb) | 7 | 40.9 ± 1.6 | 4 | 501 |
| RL | Di-:Mono- (0.35:1) | 18,19 | cellulase Carezyme® (CZ) | 8 | 11.1 ± 0.6 | 23 | 507 |
| RL | Di-:Mono- (0.35:1) | 18,19 | phospholipase Lecitase Ultra® (LT) | 8 | Not detected | - | 507 |
| RL | Di-:Mono- (0.35:1) | 18,19 | α-amylase Stainzyme® (SZ) | 8 | Not detected | - | 507 |
| RL | Di-:Mono- (0.35:1) | 18,19 | human lysozyme (HLZ) | 7 | ~8 | 19 | 503 |
| RL | Di-:Mono- (1:1) | 18,19 | *T. lanuginosus* lipase (TIL) | 6, 8 | ~5-10 | 36 | 508 |
| RL | Di-:Mono- | 18,19 | amyloid-forming protein FapC | 7.5 | Not detected | - | 506 |
| RL | Not specified | 18,19 | Soy protein | 7.0 | 28 | 146 | 424 |
| SL | C18:1-*cis* | 1 | apo-α-lactalbumin (apo-aLA) | 7 | 29.5 | 4 | 511 |
| SL | Acidic/lactone C18:1-*cis* | 1,3 | *T. lanuginosus* lipase (TIL) | 6, 8 | ~5-10 | 36 | 508 |
| TL | - | - | Bovine serum albumin (BSA) | 7.4 | 0.72 ± 0.41 | 810 | 494 |
| TL | - | - | cytochrome c (Cyt-c) | 7.4 | Not detected | - | 494 |
| Surf | cyclic | 20 | Bovine serum albumin (BSA) | 7.2 | 17.6 ([BSA]= 1 µM) 7.6 ([BSA]= 3.8 µM) | 32 | 524 |
| MEL | MEL-A | 21A | β-glucosidase | 3.2 | 0.61 ± 0.05 | 77 | 528 |
| MEL | MEL-A | 21A | β-glucosidase | 7.2 | 0.70 ± 0.12 | 2012 | 528 |
| MEL | MEL-A | 21A | β-lactoglubulin (β-lg) | 7.2 | 3.16 ± 0.04 | 1753 | 529 |

### 4.3 Interactions with polymers

*Mannosylerythritol lipids* (**21**)

Park *et al.*[533] prepared polymeric nanoparticles employing poly(ethylene oxide)-b-poly(ε-caprolactone) copolymers (PEO-b-PCL) and cell-penetrating TAT peptides. They have shown that use of MELs (unclear which one) covalently modified with a maleimide peptide linker results in a more flexible polymer core, thus promoting the formation of smaller particles of a factor two and formation of a soft gel. After linking the maleimide to the YGRKKRRQRRR-cysteamine peptide, they observed improved cellular uptake by human skin fibroblasts.

*Lipopetides*

Optimisation of current flooding strategies for enhanced oil recovery is a real challenge as no more than one third of crude is efficiently recovered from reservoirs up to now. Regarding improved emulsifying abilities and oil-sweeping efficiencies of microbially-derived products such as biosurfactants and biopolymers, their production and testing worthed trying in this context. Dhanarajan *et al.*[534] used a sand-packed column and demonstrated the potential of lipopeptide biosurfactant, produced using marine *Bacillus megaterium*, and biopolymer, produced by thermophilic *Bacillus licheniformis*, were used for the enhanced oil recovery. Lipopeptide is a mixture of iturin (23%), fengycin (44%) and surfactin (33%). The injection fluid contained lipopeptide biosurfactant at varying concentration supplemented with $Ca^{2+}$ ions and whose pH was adjusted according to the 20 combinations of the three imputs of the CCD design. Emulsification index $E_{24}$ increased with increasing concentration of lipopeptide. A maximum emulsification value of 81.66% was obtained for a concentration of 2.5 $g.L^{-1}$, with no further improvement in emulsification activity. A maximum of 45% of oil was recovered using the optimal flooding solution with enhanced viscosity upon addition of 3$g.L^{-1}$ of bacterially-derived- biopolymer. Based on these preliminary results, future studies are advised to investigate the combined properties of both biosurfactant and biopolymer in the injection solution and the resulting effect in the enhanced oil recovery.

## 5. Molecular modelling and simulation of biosurfactant assemblies in solution and at interfaces

During the last decades, the applications of computational methods studies to investigate properties of surfactant systems become fairly common (see e.g. the following reviews[535–540]). Most of these studies were devoted to synthetic low molecular weight and polymeric surfactants where their molecular features to physical properties are described. More recently, researchers have also made an increasing use of modelling and simulation methods to describe properties of alkyl glycosides and biosurfactants, mainly sophorolipides, rhamnolipids and surfactin, to complement and better interpret experimental data. In the following sections, we concentrate on these recent works in relationship to the self-assembly of biosurfactants in bulk water and after absorption at liquid-air and liquid-liquid interfaces. It should be mentioned that numerical modelling studies of biosurfactants within biological membranes also exist,[339,541–543] but these will not be discussed in detail in this review.

### 5.1 Simulation methods

Simulation methods used to study biomolecular (including proteins, nucleic acids, membrane lipids and surfactants) systems are quite diverse and are often categorized into either particle-based or field-based approaches. The particle-based methods incorporate particles to represent the building blocks of polymers such as atoms, molecules, monomers, or even an entire polymer chain [535]. The combination of these particles in the form of bonds, angles, dihedrals often interact with each other through certain forces form a potential energy function (also called a force field (FF) [544]). By the application of a statistical mechanical sampling method and the use of the laws of classical mechanic (i.e. Newton second's law) or stochastic (Monte Carlo) techniques, the particles are allowed to move within a certain thermodynamic ensemble and hence simulate a desired process [545]. Perhaps, the most well-known particle-based techniques to study surfactant systems are molecular dynamics (MD) and its coarser mesoscopic version Dissipative Particle Dynamics (DPD) approach [546,547]. In the second category, i.e., the field-based approaches, the system is typically described in terms of effective potentials, collective dynamic variables, and density fields which determine the degrees of freedom of the model.[548,549] Therefore, a reduced representation of the system is developed based on some phenomenological approximation [550] (e.g. Flory approximation of the free energy of a polymer [551]). Others field-based methods also exist such as polymer reference interaction site model (PRISM) which attempts to realize the polymer structure in terms of density correlation functions [552] or quantum density functional theory (DFT) [545,553] or self-consistent field theory (SCFT) [554,555]. The reader interested by detailed information on the above-mentioned simulation techniques may refer to the above-cited references or to excellent books [556,557]. In the following sections, we will only focus on works using MD and DPD simulations since these approaches were the most used to study biosurfactant assemblies properties.

### 5.2 Choice of the Simulation Methods for Simulation of Surfactant Solutions

The choice of the computational approach used to study a molecular system depends on the system size and representation (atomistic vc. coarse grained), the simulation approach (i.e., using classical or quantum physics) and the timescale of phenomenon of interest. Figure 33 provides a rough estimate of the various time and length scales of the different phenomena observed in a solution of sodium dodecyl sulfate in water [540]. For this surfactant, the time and length scales vary from fs to s and 1 Å – 0.1 mm. The information about the properties can be extracted vary from the level of approximation and the time and length scales of observed phenomenon and required different simulations techniques. For instance, explicit electronic description of atoms using quantum mechanics (QM) calculations is lost when using methods such as classical MD or DPD simulations. The upper limits are dictated by the processing power of current computers and the manner in which each method or algoritms is implemented in different simulation programs. In the course of time, these methods became overgrown with variants and were transformed into large groups. Moreover, the fields of applicability of different methods began to substantially overlap, and hybrid approaches were developed, which used combinations of different algorithms, such as quantum chemistry with classical methods (e.g[558,559]) or MD with DPD simulations (e.g. [560]). At present, complex approaches using simultaneously experimental and simulation methods are applied more often (e.g. [561]).

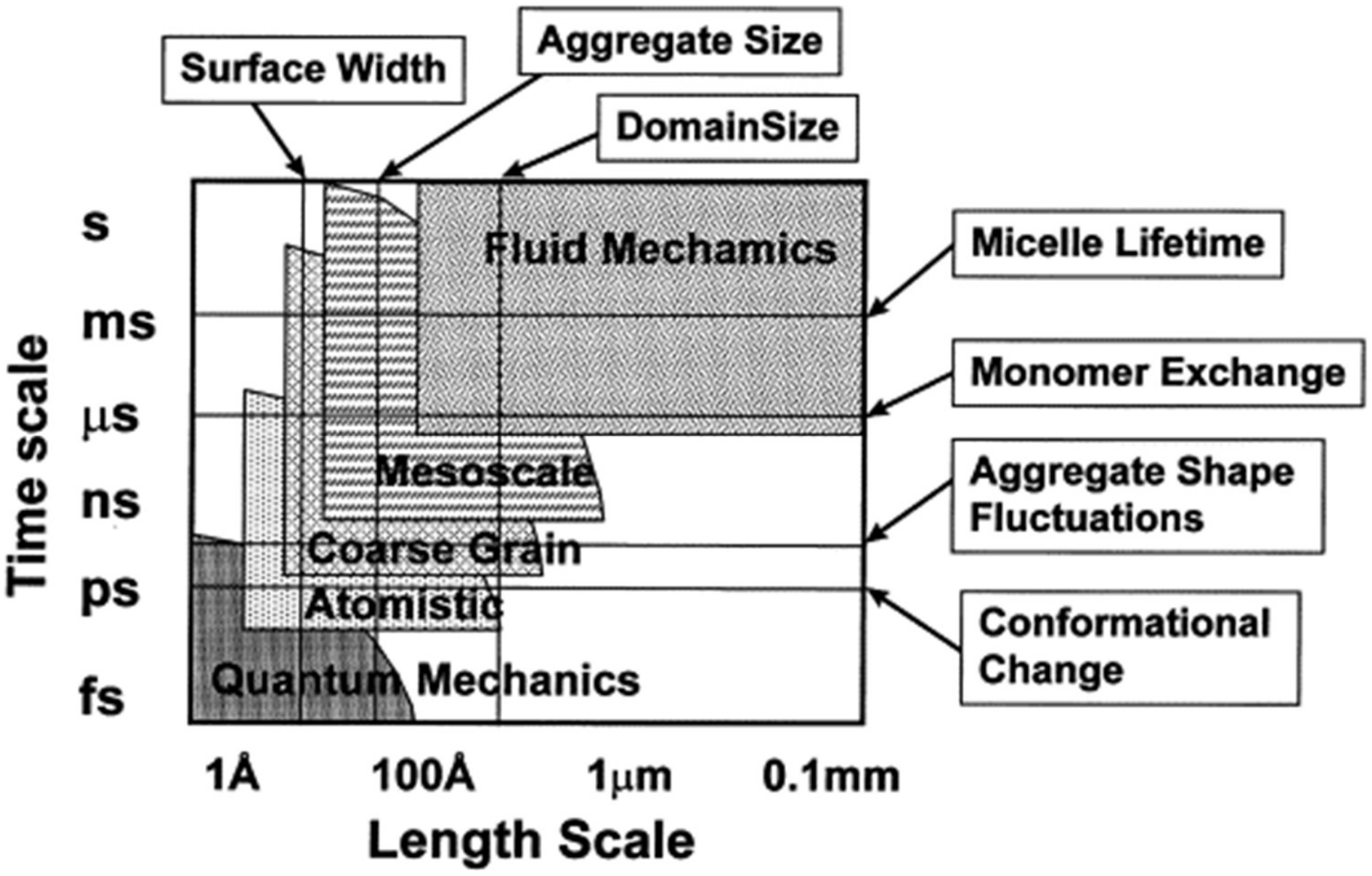

Figure 33 - Time and length scales in surfactant solutions. Reprinted from Reprinted from 540. Copyright © 2000 Elselvier.

Recent years have seen enormous progress in the development of new methods for modeling and studying molecular systems (see the following recent reviews [562,563]). The introduction of massively parallelized computer architectures with CPUs[564,565] and/or graphic processing units (GPUs)[566,567], together with the availability of more efficient algorithms (e.g. enhanced sampling methods[568–571]) or mix of quantum and molecular mechanic (QM/MM) approaches[558,559]) has allowed for a spectacular increase in terms of the size (up to millions of atoms [572,573]) and complexity of the molecular system (e.g. entire membrane of organelles with proteins [572–574] or entire viruses [575–577]) of the systems studied as well as the length of the simulations that can be performed (μs [578,579] and, more recently, even MD simulations of ms [580] in length).

## 5.3 Empirical force fields

One of the key factors that dictate the size, complexity and the accuracy of any simulations of biomolecular systems is the force field (FF). In general, the FF parameters are based generally upon experimental measurements (e.g. free energy of solvation [581]) of small molecules or fragments (i.e. amino acids, polymer unit) and quantum chemistry calculations (e.g. [582,583] and references cited therein). There are large variety of FF that can be used to model surfactant molecules with different degree of resolution, each with their specific strengths and weaknesses (see discussions in the following reviews [584–587]). In the united atoms FFs, for instance, based on GROMOS [588], the interactions between atoms of molecules are modeled explicitly or with particles that incorporate a group of atoms (e.g. all hydrogens on an aliphatic carbon are united with carbons). Identical or groups of atoms can be also represented as a bead in the case of the coarse grained (CG) FF such as using the well-known MARTINI FF[589, 590]. More detailed and expensive model (in term of computation can be used using all-atom FFs such as CHARMM [591], AMBER [592] or OPLS/AA [593] or by adding terms of polarization to improve the treatment of electrostatics and take fully account to the response of the charge distribution to environment [594,595]. Moreover, each FF was typically developed with the use of specific water model for instance, GROMOS with simple (fixed) charge model (SPC) [596], CHARMM and AMBER with TIP3P [597] and modifications of these models [598,599]. Although a plethora of water models or polarizable for molecular simulation exist [599], only the simple ones have gained popularity for biomolecular simulation even if for a large of them they are not capable to reproduce accurately many experimental data such as dynamic and dielectric properties [597,600].

### 5.4 Structure of biosurfactant assemblies in the solution and at the interfaces

Surfactant micelles can be characterized using two parameters; the critical micelle concentration (CMC) and the aggregation number. The overwhelming majority of the works have focused on understanding these parameters and on linking them to the micellar structure and relating this back to experimental observation and functional/physicochemical properties. In Table 15, we provide a list of the recent simulations of sophorolipid, rhamnolipids and surfactin systems, that will be discussed in the following sections of this review.

Table 15 - Recent simulations of biosurfactant in solution or at interfaces. For each simulation, the author group, publication year, basis force field, force-field type, simulation environment and literature reference are indicated. [b] AA, all atom; UA, united atom and CG, coarse grained reprepesentations. [c] MD, Molecular dynamics; MC, Monte Carlo, DPD, dissipative particle dynamics simulations. [e] continuous medium which can be subjected to elastic deformations. [d] Based on CHARMM36 parameters with a mapping scheme based on Flory Huggins solubility theory.

| Author | Year | Basis force field | Repres[b] | Methods[c] | Applications | Ref. |
|---|---|---|---|---|---|---|
| **Sophorolipids** | | | | | | |
| Dhasaiyan | 2014 | GROMOS53A6 [588] | UA | MD | Micelle/Vesicle | 214 |
| Manet | 2015 | | UA | MD | Micelle | 237 |
| Pandey | 2016 | | UA | MD | Micelle | 603 |
| Sarkar | 2018 | CG beads[601] | CG[d] | DPD | Micelle/Vesicle/membrane | 604 |
| **Rhamnolipids** | | | | | | |
| Zu | 2011 | - | CG | DPD | water/octane interface | 605 |
| Abbasi | 2013 | GROMOS96 43a1 [588] | UA | MD | air/water interface | 606 |
| Munusamy | 2017 | CHARMM27[607] | AA | MD | In water | 608 |
| Munusamy | 2018 | CGENFF[582] | AA | MD | air/water and water/decane interfaces | 609 |
| Eismin | 2017 | CGENFF[582] | AA | MD | Micelle | 610 |
| Monnier | 2019 | AMBER/GLYCAM [611,612] | AA | MD | Membrane | 339 |
| Oliva | 2020 | CHARMM36[601] | AA | MD/Exp | membrane | 541 |
| Luft | 2018 | CGENFF[582] | AA | MD | water with decane | 613 |
| Luft | 2020 | CHARMM[591] CGENFF[582] | AA | MD | air/water interface in water | 614 |
| **Surfactin** | | | | | | |
| Bonmatin | 1994 | GROMOS [588] | UA | MD | In vacuum | 181 |
| Ishigami | 1995 | CHARMM22[591] | AA | - | In vaccum | 157 |
| Gallet | 1999 | Same as Ref. [181] | UA | MD/Exp | Hydrophobic/Hydrophilic interface (model) | 615 |
| Nicholas | 2003 | CHARMM27 [607] | AA | MD | water/hexane interface | 616 |
| Gang | 2010 | MMFF94 [617–620] | AA | MD | water/decane interface | 621 |
| Gang | 2011 | OPLS [622] | UA | MD | Air/water interface | 623 |
| She | 2010/2012 | GROMOS96 45a3 [624] | UA | MD | Micelle | 625,626 |
| Iglesias-Fernández | 2015 | CHARMM27 [607] | AA | MD | air/water Interface Micelle | 627 |
| Gang | 2015/2020 | OPLS-AA [628] MARTINI [590] | AA/CG | MD | air/water and water/octane interfaces | 629,630 |

### 5.4.1 Self-assembly in the bulk

*Sophorolipids*

There have been several MD works during the recent years on sophorolipids (SL) that have studied their micellization processes as well as the structure of their assemblies, or phase diagrams, depending on the variation of the chemical structure of the molecule.

Dhasaiyan *et al.* [214] and Pandey *et al.* [603] combined experimental (scanning electron microscope (SEM) or transmission electron microscope (TEM)) with atomistic MD approaches to compare the assembly structure of nonacetylated acidic SL assemblies containing oleic (C18:1-*cis*) (**1**), linoleic (C18:2-*cis*), and linolenic (C18:3-*cis*) (**15**) acyl chains (referred to as OASL, LOSL, and LNSL, respectively). The SL molecules were modeled with the united atom force field based on the GROMOS53A6 parameters [588]. The authors performed long MD simulations (< 700 ns) where the same lipids, or their mixture, were placed randomly simulation box filled with SPC water. The simulations showed that the number of double bonds in alkyl chain has a great influence on the self-assembled structures. OASL (**1**) self-assembled into an aggregate after 700 ns of simulation without well-defined organization whereas LNSL (**15**) self-assembled into vesicle-like structures with water molecules entrapped in the vesicle lumen, even with presence of OASL in the box (LNSL(**15**):OASL(**1**) 1:3) [214,603]. H-bonding interactions among head or tail groups were investigated and results suggested that H-bonding mainly occurs among the head group sophorose moieties. The order of H-bonding among different groups is head-head > head-tail > tail-tail for all the systems. The simulations also showed that in the vesicle, the LNSL (**15**) molecules prefer a bilayer-like arrangement, with a head–head organization, rather than any other arrangements (i.e., tail-tail and head-tail[214,603]).

The formation of different microstructures of SL depend on the surfactant concentration. This is essentially a macroscopic phenomenon in terms of length and timescale and therefore it requires costly computational resources, in particular when the molecular system is investigated using an atomistic representation. This limitation can be overcome as mentioned earlier by applying a coarse-grained representation of the surfactant molecules and the solvent combined with MD, or mesoscopic, simulation using DPD approaches. Sarkar et al. [604] adopted the latter approach to study the structure of the assemblies of acid LNSL (**15**) lipids at different concentrations. The interaction parameters between DPD beads were calculated from atomistic MD simulations using CHARMM36 force field [601] and experimental data using a mapping scheme based on Flory Huggins solubility theory [602]. They simulated a LNSL(**15**)/water concentrations varying from 5% (v/v) to 50% (v/v) with 5% (v/v) interval. Their DPD simulations provided complete phase diagram of this binary system and showed different self-assembled morphologies. Consistent with the above cited atomistic MD, the SL self-assemble into spherical morphologies at low LNSL(**15**)/water concentrations (up to 15% (v/v), cylindrical aggregate at 20% (v/v) and distorted cylindrical at 30% (v/v) or curved and double bilayer 40% and 50% (v/v). The flipping frequencies of the LNSL (**15**) chains were also monitored and compared between the different aggregates and were to be to be less in spherical morphology in comparison to others. In conclusion to these works, it must be however said that correlation between the simulated and experimental phase diagram of LNSL (**15**) is not yet established and requires further studies. LNSL (**15**) were show to form vesicles at acidic pH on the basis of SEM, TEM and fluorescence microscopy studies, but the experimental volume fraction was 0.3 wt%, [214] while bilayer vesicles are only described above 50 wt% by numerical simulations. [604]

To go further and correlate more precisely simulation and experiments, Manet et al.[237] combined SAXS, anomalous SAXS, contrast matched SANS and MD simulation approaches to better understand the structure of C18:1-*cis* SL (**1**) micelles within comparable volume fractions. Their experimental results showed that the micelles can be modelled using a core–shell model better described by a nonconventional "coffee bean"-like structure (Figure 34a). The MD are performed with a large range ($N_{agg}$ = 28 – 112) of protonated SL placed randomly in the large water box and modeled with the same GROMOS FF as in [214,603]. The MD results showed that the shape of SL assemblies changed with their SL contents, from (quasi)sphere to an elongated ellipsoidal shape with the increase of the $N_{agg}$ value (Figure 34b). Small micelles tend to be spherical, corroborating experimental data.[213] The SL (**1**) acyl chains organize in bent conformations, where the sugar and carboxyl groups are located at the micellar surface and the alkyl chain in the micelle core, a structure closer to the one observed in classical head-tail surfactant micelles (Figure 34b). With the larger number of SL (**1**) (112) in the box, the lipids formed a cylindrical rod and the SL (**1**) acyl chains adopted a wider range of conformations in the micelle hydrophobic core. Moreover, all the aggregates present a rough surface with the sophorose and COOH headgroups in direct contact with the solvent leading to a larger hydration of these groups compared to the alkyl chain moiety (Figure 34c). The hydration behavior of the glycolipid sugar headgroup present similar features with other micelles formed with glycolipids with similar structure headgroup such as dodecyl maltoside (DDM) detergent [631]: water molecules solvate primarily the hydroxyl oxygen of the sugar and the glucose bonded to the alky chain is, on average, less hydrated than the outermost glucose.

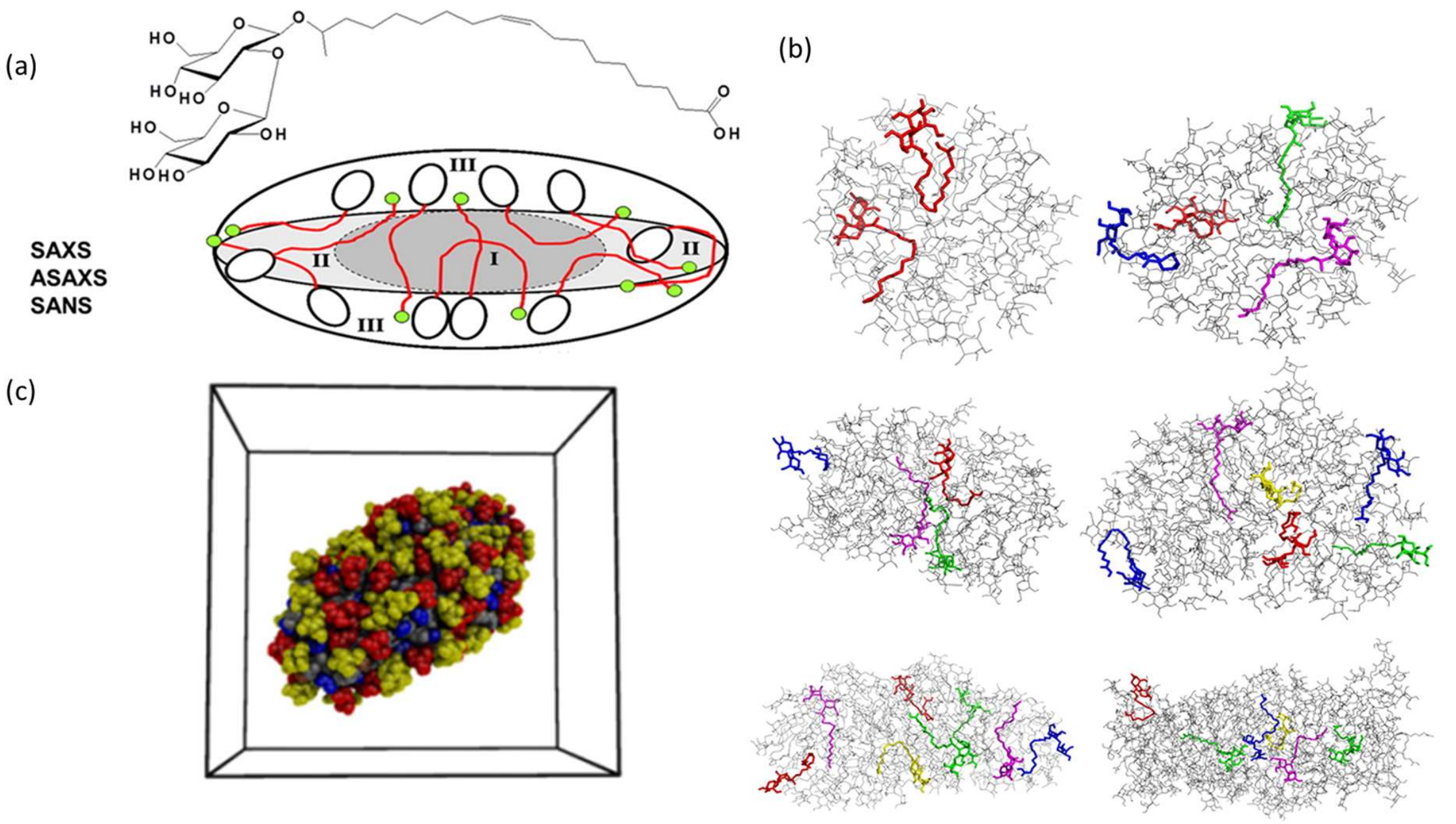

Figure 34 - (a) Structure of C18:1-cis (LNSL) sophorolipid (**1**) with the proposed "coffee-bean" core shell model structure of the micelle pH < 5. Region I = full aliphatic; Region II = mixed sophorose/water/COOH/aliphatic; Region III = sophorose/water/COOH. (b) Representative conformation of the LNSL in the aggregates with the increase of the number of surfactant in the aggregates. Red, blue, green, and yellow colors corresponds to the surfactant conformations where the sophorose and the COOH headgroups are located at the micelle surface with the rest of the surfactant in the hydrophobic core (red), the LNSL reside entirely at the micelle surface (blue), alkyl chain is extended and cross the hydrophobic core and with the sophorose and headgroups at the surface (magenta), where COOH in the hydrophobic core (green), where the overall LNSL are deeply buried in the hydrophobic core (yellow). (c) micelle surface with Nagg = 80 in the simulation cubic box delimited by black line. The outermost and innermost glucose are colored in yellow and red, respectively, whereas the alkyl chain, double bonds and COOH groups are in grey, cyan and blue colors, respectively. Reprinted from Ref. [237]. Copyright (2015), American Chemical Society.

### *Rhamnolipids (mono-* (**19**)*, di-* (**18**)*)*

Munusamy *et al*. [608] and Eismin *et al*. [610] combined experimental approaches (surface tension, dynamic light scattering and fluorescence measurements) with atomistic MD simulations with the CHARMM[601] and CGENFF[582] parameters to obtain some clues about the effects on carboxylic acid charge on the structure and stability of the nonionic and anionic mono-RL assemblies in water. They showed that mono-RL aggregates strongly differ between the two forms, neutral and charged, of RL, which are related to the difference in inter and intra H-bonding interactions and the effective alkyl chain length in the aggregate hydrophobic core. In particular, nonionic RL do not form micellar aggregates larger than ~40 molecules per aggregate with a large variety of shape ranging from spherical to ellipsoidal, torus-like, and unilamellar vesicle (ULV) (Figure 35ab).[608] This is contrast to the anionic counterpart, which can form micellar aggregates larger than 40 molecules per aggregate, with long tubular shape in addition to spherical and ellipsoidal one [608,610]. The headgroup conformations in the aggregate are mostly open with a small percentage in the folded state in which the carboxylic acid group can forms intramolecular H-bonds with the hydroxyl group of the rhamnose ring. A lack of inter mono-RL H-bonding and the effective alkyl chain length being shorter restrict the population of nonionic micellar structures larger than the aggregation number, $N_{agg}$ ~40. These data are in somewhat agreement with the experimental solution self-assembly studies of RL, showing a prevalent micellar behavior at basic pH and vesicle formation at acidic pH (section 2.3, Table 5, Table 6).

In addition to the examination of the micellar assemblies, the authors also investigated the structure of torus-like aggregate and small unilamellar vesicle (SUV) structures since the latter structure have a great interest in nano- and biotechnology for encapsulation and delivery of drug molecules (e.g. [335]). To this aim, they performed additional MD simulations of preformed torus-like and SUV aggregate with 285 and 810 nonionic mono-RL monomers. The simulations show torus-like aggregate and SUVs are highly stable due to a bilayer structure tightly packed. SUV has also additional stability from the constrained water molecules present inside them that have a structure close to the bulk.

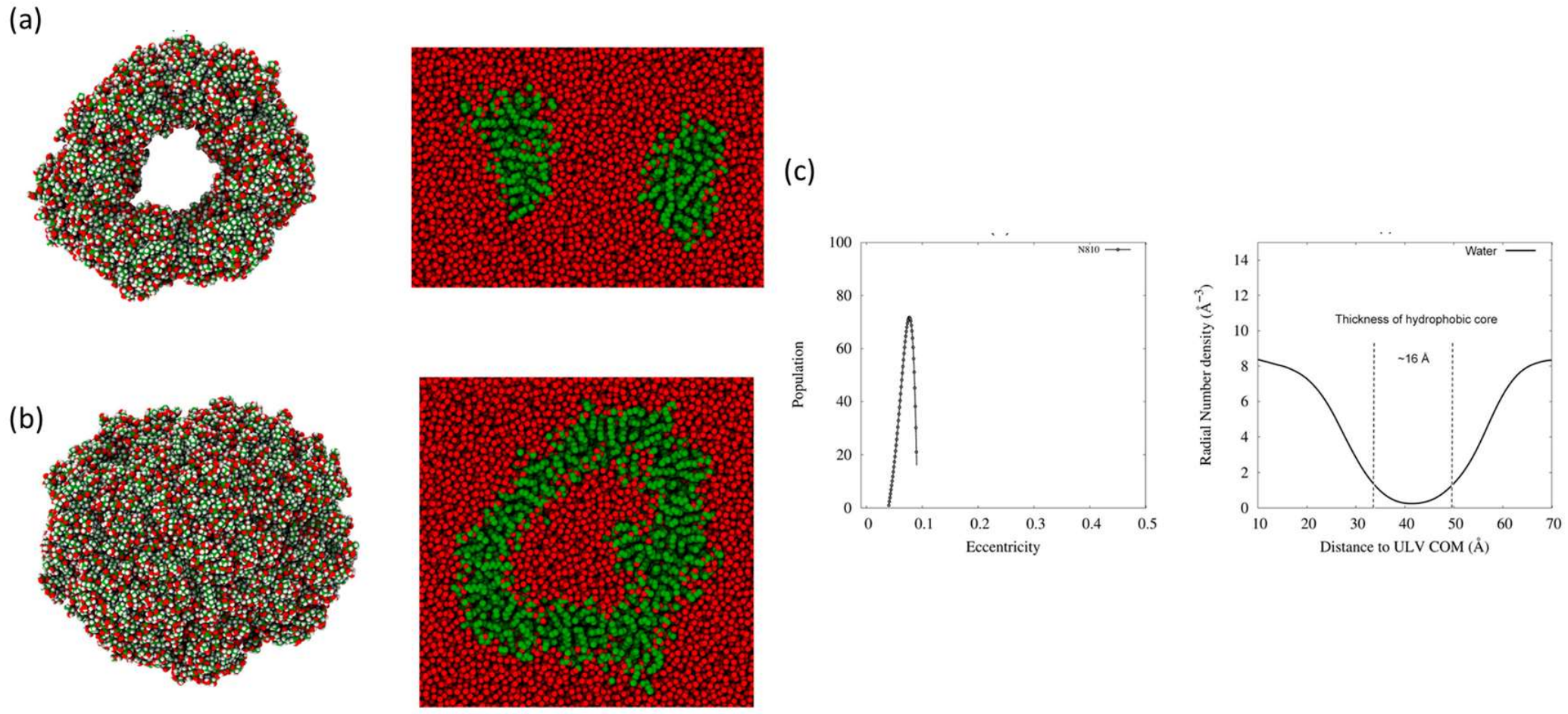

Figure 35 - Snapshots of the (a) torus aggregate and (b) unilamellar vesicle (ULV) obtained from MD simulations formed by mono-RL (**19**). Cross-sectional view of the (b) torus aggregate and (c) SUV in water. RL and water are green and red colors, respectively. (c) Population distribution of eccentricities of the SUV and density of the water molecule. Reprinted from Ref. [608]. Copyright (2017), American Chemical Society.

In another work, Luft et al. [614] investigated the effect of the addition of the rhamnose group and the length of the alkyl chain by simulating dirhamnolipid (di-RL) in water and at the air water/interface. Their all-atom and MD simulations with the CHARMM[591] and CGENFF[582] FFs show that aggregation of dirhamnolipid in bulk is affected by the presence of a second rhamnose group and by the short alkyl chain and form micellar structure with $N_{agg}$ ~22, lower than for monorhamnolipid ($N_{agg}$ ~ 40) [608]. In the same study, authors demonstrate that di-RL with longer chain (C14 to C18) can form micellar aggregates larger than $N_{agg}$ ~22 (up to ~30). These overall results show that presence of one rhamnose group and C10 alkyl chain is adequate for mono-RL to form spherical and cylindrical micellar aggregate in water and that the second rhamnose group in dirhamnolipid creates an imbalance between hydrophilic and hydrophobic components, resulting in the limitation to form micellar aggregates in water. These authors finally noticed that the structural differences in mono- and dirhamnolipid have no significant effect on the aggregation at the air–water interface [614].

Finally, because of their strong surfactant properties, rhamnolipids (as many other surfactants [632]) have also been a focus for environmental recovery operations, most notably heavy-metal (such as cadmium, lead, and zinc) chelation in contaminated soil and oil spill remediation [633]. This particular use of rhamnolipids in this latter context was subject to a MD simulation study by Luft et al. [613]. In this work, the authors simulated the capture of decane molecules by nonionic mono-RL aggregates in water and showed this process is highly favorable. The mono-RL/oil formed stable assemblies with an ellipsoidal or cylindrical shapes. Computations of the binding free energy of decane with mono-RL aggregates using free-energy perturbation method showed that the trapped alkane molecules are strongly bound to the Rha-C10-C10 chain and thus provide an additional stability factor to stabilize larger mono-RL aggregates (with $N_{agg} > 50$) known be unstable in bulk water [614].

*Surfactin* (**20**)

She et al. [625,626] examined, using MD using the united atom parameters based on GROMOS[624], the structure of surfactin micelles in water depending on the length of the surfactin alkyl chain or the temperature. They found that the micelle size, shape and surfactant hydration of the micelles formed with surfactin with n-C15 and iso-C16 alkyl chain are similar. The micelle has a size of around 22 - 25 Å and have core-shell structure, where the peptide rings occupy most of the micelle surface. With the increase of the temperature (up to 343 K), the shape of the micelle changes to an ellipsoidal aggregate with reduction of the solvent accessible surface area and the water-surfactant hydrogen bonds.[616] The high temperature also has a significant effect on the mobility of the lipid alkyl chain by increasing their flip-flops. Using computations of the formation of hydrogen bonds (intramolecular, and short-lived intermolecular between surfactin monomers, and with the solvent) the authors also noted a small proportion of *γ*- and *β*-turns structures in the peptide rings, which would not appear to be a sufficient major driving force for micelle formation or stabilization,[525] in particular at higher temperature.

In a more recent works, Gang et al. [630] using a CG simulation and model based on MARTINI force field [590] that allow microseconds long simulations and larger system sizes have examined the self-assembly of C15 surfactin molecules. From randomly distributed configurations, they were able to characterized the kinetics of micellization in different stages and found that aggregation numbers of the dominant aggregates are in range of 11 - 30 after 300 ns which is in accordance with the SANS experimental data (20 ± 5). [182]

As mentioned earlier, surfactin contains two acidic amino acid residues with carboxyl groups in the peptide moiety, Glu1 and Asp5 and the protonate state of the two residues are known dramatically affected surfactin properties, such as spatial arrangement of surfactin in solution or at the interfaces which were relative and to change the critical micelle concentration (CMC) and to lowering surface tension. Iglesias-Fernández [627] examined the effect of the pH and the deprotonation states of these two residues using explicit MD and compared with neutron reflectivity data [212]. They showed that the micelle properties (micelle aggregation number, size, fatty acyl chain tail and amino acid distributions) respond to pH variability as described in experiments [212] (Figure 36). Surfactin forms spherical-like aggregates in all ionization states studied. However, at high values of pH, where both acidic residues are deprotonated, the size of the aggregates decreases due to the increase in electrostatic repulsion between the head groups. In contrast to at lower pH values, a unique structure is formed. The authors also showed that the orientation of the lipid chain with respect to the cyclic polypeptide is pH independent and can adopt different conformations, from fully extended to folded, where the hydrophobic tail interacts with the side chains of leucine and valine residues.

Concerning the match between the experimental and simulation data, surfactin was shown to self-assemble into micelles having a equatorial radius of 25 Å (Table 7),[182] in agreement with the data published by She et al.[625,626] The ambiguous role played by leucines on the peptide orientation and hydrophilicity was also pointed out before,[182] as also illustrated by the long-date debate on the peptide conformation, previously discussed in section 3.1 (Figure 21). However, some disagreement seems to occurs at lower pH, where simulation seems to favour micellar aggregates, while experiments suggest the formation of bilayer membranes.[182,212] That may suggest that force field and/or simulation parameters used in this study are not enough acurate to model this type of system.

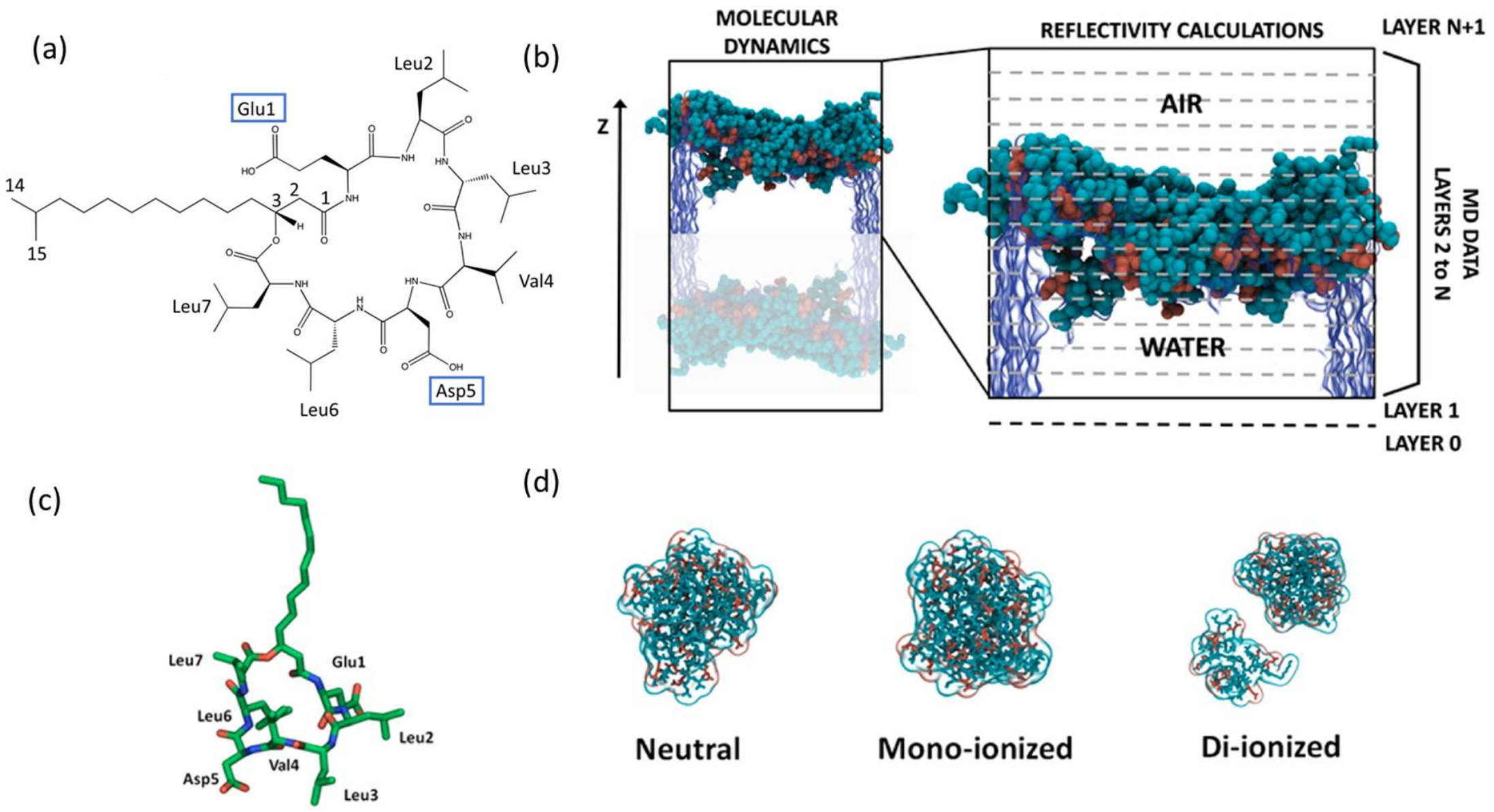

Figure 36 - (a) Chemical composition of surfactin (**20**) with the two possibly charged residues highlighted with a blue box. (b) Three-dimensional structure of C15-surfactin. C atoms are depicted in green, N atoms in blue, and O atoms in red. Hydrogen atoms are not displayed for clarity. (c) Representative snapshot of the di-ionized surfactin system at the water/air interface, obtained from the MD simulation (left side). Acidic and nonpolar residues are colored in red and cyan, respectively. Water is shown in a continuous transparent representation. Only half of the simulation box (one interface) is used for the calculation of the reflectivity and volume profiles. A scheme of the layers representation used for the calculation of the reflectivity profiles is also shown. Layers 0, 1, and N + 1 are used to complement the MD data in order to mimic the experimental setup (useful in the case of supporting surface experiments). In the present case, layers 0/1 and layer N + 1 are used respectively to extend the water and air regions of the MD data to consider their larger sizes in the experimental setup. Layers 2 to N correspond the MD simulation. (d) Snapshots of the molecular aggregates of the three systems studied taken at the end of the simulation. Acidic residues are displayed in red and nonpolar in cyan. Modified from Ref. [627]. Copyright (2015), American Chemical Society.

### 5.4.2 Self-assembly at interfaces

*Rhamnolipids (mono-* (**19**)*, di-* (**18**)*)*

The interactions properties of rhamnolipids at various biological and materials' surfaces are quite limited. For instance, there are some MD works that have studied the interactions properties of RLs with oil and solid material surfaces such as quartz[634] or calcite[635] and within phospholipidic membrane,[339,541,542] the latter of particular interest for the better understanding of the antibacterial properties of RL.[636,637] In the following, we will mainly focus on MD simulations of rhamnolipids at the air/water and oil/water interfaces.

Abbasi *et al.*,[606] using atomistic MD, investigated the effect of the deprotonation of the carboxylic group of mono-RL above its pKa (pH = 5.9). They showed that mono-RL in the protonated state have a higher diffusion coefficient than the anionic form, indicating that the negative charge greatly reduced its mobility in the monolayer surface. The presence of the negative charge in the monolayer results in a more diffuse and disrupted lipid/water interface, as well as decreased lipid–lipid interactions. This was illustrated, for instance, by a different distribution of negatively charged carboxylic acid groups within the interfacial layer, where at pH 5.9 they have forced to move into the water phase from the air phase. The calculated surface area per molecule obtained experimentally using surface tension measurements in the same study is found slightly higher (0.63 $nm^2$) at pH 4 than at pH 7 (0.57 $nm^2$) due to the interfacial rearrangement of the lipid chain containing the free carboxyl. The presence of the $Na^+$ ions are localized within the interface, interacting with the carboxylate groups of mono-RL would facilitate micelle formation, which could result in a lower value for the cmc as observed experimentally.[606]

In other study, Zhu *et al*.[605] have investigated the absorption properties of mono-RL and di-RL at the octane/water interface with conventional anionic surfactants such as sodium dodecyl sulfate (SDS), sodium dodecyl benzene sulfonate (SDBS). Their results confirmed the experimental results[144,354] that the lower adsorption of di-RL to the octane/water interface is due to the larger size of the hydrophilic head group of di-RL. In addition to the larger head group of di-RL leading to a lower ability to compete with anionic surfactants, for di- RL there was also a very noticeable phase separation into di- RL rich and anionic surfactant rich regions at the interface which was absent in the mono- RL plus anionic surfactant simulations.

*Surfactin* (**20**)

In addition to its solution properties, surfactin is also known to adsorb to and self-assemble at air/water and oil/water interfaces. One of the first molecular models of surfactin at an air/water interface based on primary structure from mass spectroscopy and amphiphilic properties was done by Ishigami *et al.*[157] in the middle of the 90's. Following their opinions, surfactin molecules form dimers by hydrophobic interactions between their aliphatic chains. Without compression, molecules orientate themselves with the peptide ring and aliphatic chains lying flat on the air/water surface. When compressed, they take up a close packing configuration with the aliphatic chains normal to the peptide ring lying on the air/water surface. This is in contradiction with the assumption of Bonmatin *et al.*[181], which presents both the peptide ring and the aliphatic chains standing vertically on the water surface when molecules are under compression conditions. Later, by means of molecular modeling and by fitting NMR data, Gallet *et al.* [615] examined the absorption a complete model of surfactin at a hydrophobic/hydrophilic interface modeled using two dielectric constants. Their modelling work indicated that surfactin adopts a conformation in which the peptide ring stays parallel to the interface, with the two acidic residues extending to the aqueous solution, and the β-hydroxy fatty acid folded back permitting the interaction with a leucine side chain consistent with the neutron reflectometry techniques.[182,212] Such discrepancies were discussed before in Section 3.1 (Figure 21).

Iglesias-Fernández et *al.* [627] also showed that at air/water interface, surfactin molecules form films independently of the environmental pH, in agreement with experimental reports[212] and also confirms that the surfactin peptidic domain tends to be found parallel to the water/air interface, and fatty acyl chain tails show high flexibility.

In another study with systems with up to 32 molecules of surfactin modeled with the atomistic CHARMM27 [607] FF, Nicolas *et al.* [616] demonstrated that at the water/hexane interface, the surfactin peptidic backbone can exhibits a large flexibility and that conformational motions of the surfactin molecules depend strongly on the number of surfactin absorbed in the interface concentration. For instance in a crowded environment, surfactin molecules are associated such that the interactions between hydrophobic residues and the hydrophilic medium are minimized. This conformations are stabilized by van der Waals interactions and sporadicaly by intermolecular hydrogen bonds involving carboxylic lateral chains. In addition when hydrophilic residues are shielded from the environment, a complete tumbling over of the peptidic part can occur. This can be related to the ability of a surfactin molecule to go across a hydrophobic medium as a lipid membrane. Existence of γ- and β-turns structures in the peptide rings were also found as in ref. [626].

Finally, the effect of the nature on counterions on surfactin/ions bindings at air/water interface were also examined by Gang *et al.*[629]. Their results showed that surfactin exhibits higher binding affinity to divalent counterions, such as $Ca^{2+}$, and $Ba^{2+}$, and smaller monovalent counterion, $Li^{+}$, than $Na^{+}$ and $K^{+}$. Both carboxyl groups in surfactin are accessible for counterions, but the carboxyl group in Glu1 is easier to access by counterions than Asp5. The bound counterions induce the dehydration of carboxyl groups and disturb the hydrogen bonds built between carboxyl group and hydration water. These results are in good agreement with neutron scattering data recorded on surfactin in solution in the presence of mono and divalent ions.[212]

## 6 Concluding remarks and perspectives

Microbial biosurfactants are a class of amphiphiles with a rich phase behaviour (Table 5, Table 6), which has only been partially studied so far. Besides classical temperature (Table 11) and concentration-dependent self-assembly, microbial biosurfactants have a strong pH dependency (Table 10), generally not being the case in classical synthetic head-tail amphiphiles. Their uncommon chemical structure has a huge impact on their solution (section 2), as well as their adsorption properties at intefaces (section 3) and interactions with macromolecules (seciton 4). These aspects have all been reviewed in this work. Below, we provide some general considerations.

*Biosurfactants or bioamphiphiles?* One of the most important aspects of these molecules is their double role, surfactant- or lipid-like, according to pH (Table 10). In their deionized form at basic pH, most microbial amphiphiles do behave as surfactants, for which CMC and evolution of surface tension can be measured and where the micellar phase (Table 7) is commonly found. The use of the word *biosurfactants* is then fully justified. However, in their neutral form at basic pH, they often show a lipid-like behavior, with the formation of curved, flat or condensed bilayers (Table 8), or more exotic phases like fibrillar and columnar. Due to such double role, the word *bioamphiphile* is probably more approapriate to address this class of compounds. Nonetheless, a sudden change in habits is not the scope of this work and for this reason we kept addressing to these compounds as biosurfactants throughout this document.

*Structure-property relationship*. Despite some effort in terms of numerical modelling of these molecules (section 5), a clear-cut correlation between the microbial amphiphile's molecular structure and properties is still far from being fully understood. In this regard, the classical theory of packing parameter cannot be safely and systematically used to predict, or even understand, the type of self-assembled structures formed by the most common microbial amphiphiles, as largely discussed in section 2.5 (Table 9).

*Anionic-non ionic mixture*. Despite few rare exceptions, practically all microbial biosurfactants have a free-standing COOH group. Since most studies and possible applications are generally tested between pH 5 and 8, one must always consider a biosurfactant as an actual mixture of its ionic and non-ionic form at a proportion which is not always obvious to determine. The actual properties, both in solution and at interfaces, but also with macromolecules, should then be considered in light of an actual mixture of an anionic and non-anionic form of the same molecule, whereas mixture of ionic and non-ionic surfactants are known to behave differently that each form taken individually.[1,2]

*High-end applications*. Antimicrobial, detergency, emulsifying or bioremediation applications have been considered as the applications since three decades, often without enough targeted information on the structure of the given biosurfactant in solution. The impressively complex and broad phase and interfacial behavior of biosurfactants will certainly help improving the chance to develop these classical applications, but, and above all, suggest new high-end applications, such as biocompatible surface stabilizers for colloids, hydrogels, solid foams, mild modulators of proteins' activity, 2D lubricants, encapsulating and delivery hosts, rheomodifiers, and so on. These, and other fields, will be certainly developed more and more in the future after a better knowledge of their behavior at simple and complex interfaces and after their interactions with macromolecules, but also lipids and surfactants, will be better understood.

*Perspectives*. Compared to the field of petrochemical surfactants, research on biosurfactants properties is only at the beginning and should constitute an important field of research for the future years. Despite the work summarized in this document, so much is still lacking that making an exhaustive list is pointless. Some directions are listed below as bullet points:

- Have a precise idea of the self-assembly and phase behavior for all biosurfactants and derivatives. This should include multiparametric phase diagrams, including effects of concentration, pH, temperature, cosolvent, counterions, performed in a much broader and more systematic approach as done so far.
- Biosurfactants behave as lipids in their neutral form. Then, similarly to phospholipids, one expects to know both simple and advanced physicochemical parameters, like melting temperature ($T_m$), pKa but also bending rigidity of membranes ($\kappa/K_bT$), head-to-head membrane thickness (δ), area-per-headgroup and type of lipid packing within membranes, just to cite some.
- In soft matter and material science, it could be interesting to know which molecules form hydrogels. Good understanding of the supramolecular structure of the hydrogel and control the viscoelastic properties through thermodynamic and kinetic studies is also very important.
- Knowledge about interactions between biosurfactants and other amphiphiles, including other biosurfactants, is still in its young infancy. Complex multiparametric studies of these systems will also be welcome.
- Similarly, studies between biosurfactants and macromolecules are only starting. These systems are expected to behave differently than classical surfactant-polymer systems, for the presence of functional groups on biosurfactants, allowing variable specific (electrostatic, H-bonding) interactions with macromolecules. Such differences have already been outlined in recent studies on biosurfactant-proteins[504] and biosurfactant polyelectrolytes.[489]
- Biosurfactants-based colloids, such as coated nanoparticles, are very interesting biocompatible nanosystems, especially for the possibility to perform one-step systnthesis. However, future works should invest more in the rigour of in their characterization and reproducibility.
- The effect of typical processing methods and conditions, such as dip/spin coating, wet spraying and spray drying, freeze-drying and freeze-casting, microfluidics, shearing and much much more, should be studied.
- As with other types of surfactant systems, the use of molecular modeling and simulation approaches is of great interest to have a good understanding of the physico-chemical parameters governing the structural properties of biosurfactant assemblies. However, due to the high complexity of the observed assemblies depending on the type of molecules, the experimental conditions, and the current limitations of the molecular models (in particular for the treatment of the electrostatic interactions) used in the simulations, these studies must be done in close collaboration with the experimenalists to develop accurate models and obtain good results. This will be achieved through a better tracking of the electostatic interactions between the surfactants or the surrounding ions present in the solution. [594,595]
- Finally, the structure of biosurfactants at interfaces has deserved some attention so far, but most studies concentrated on surfactin and rhamnolipids. This effort should be extended to other biosurfactants.

The authors of this work believe that success in the diffusion and eventual commercialization of microbial biosurfactants strongly depend on the continuity of the development-characterization-application chain. The constant search of new molecules, may them be obtained from wild type or modified organisms or by chemical modification of existing molecules, is at the forefront of this field. However, rigorous characterization of their phase behviour in solution followed by study of the interaction with their chemical environment (ions, pH, low molecular weight and macro- molecules) is a must to expect the unexpected and design successful applictions. A tight collaboration between microbiologists, genetic engineers, chemists, physicochemists, colloids' and materials' scientists is strongly suggested as being the key to cross the phase between research and successful innovation (also known as the "valley of death").

## 7 Acknowledgements

We kindly acknowledge Prof. D. Otzen (Aarhus Unversity, Denmark) for helpful comments.

## 8 Funding

This work could be possible due to funding from: French Agence Nationale de la Recherche (ANR), Project SELFAMPHI - 19-CE43-0012-01; ANR within the Investissements d'Avenir program under Reference ANR-11-IDEX-0004-02 and more specifically within the framework of the Cluster of Excellence MATISSE; Ecole Doctorale ED397, Sorbonne Université, Paris, France.